\documentclass[a4paper,11pt]{article}
\pdfoutput=1

\usepackage{jheppub} 
\usepackage[utf8]{inputenc}
\usepackage{float}
\usepackage{mathtools}
\usepackage{dsfont}
\usepackage{cancel}
\usepackage{amsmath}
\usepackage{amsfonts}
\usepackage{amssymb}
\usepackage{ytableau}
\usepackage{cellspace}
\usepackage{array}
\usepackage{makecell}
\usepackage{pgfplots}
\usepackage{tikz}
\usepackage{tikz-cd}
\usetikzlibrary{shapes.misc,shadows,fit,decorations.pathmorphing,positioning,trees,decorations.markings,decorations.pathreplacing,calc,shapes,patterns,arrows}
\usepackage{xcolor}
\usepackage{hyperref}
\usepackage{hhline}
\usepackage[capitalize,sort]{cleveref}
\usepackage{comment}
\usepackage{graphicx}
\usepackage{subcaption}
\usepackage{environ}
\usepackage{enumitem}
\usepackage[section]{placeins}
\usepackage{tabularx}
\usepackage{longtable}

\addtocounter{MaxMatrixCols}{1}
\renewcommand{\arraystretch}{1}

\hypersetup{
	pdftitle={Spin(7) Orientifolds and 2d $\mathcal{N}= (0,1)$ Triality},    
	pdfauthor={\textcopyright\ Sebastian Franco, Alessandro Mininno, Angel Uranga, Xingyang Yu},     
	pdfsubject={HEP},   
	pdfcreator={pdfLaTex},   
	pdfproducer={LaTex}, 
	pdfkeywords={},
	colorlinks=true,
	linktocpage=true
}

\newcolumntype{M}[1]{>{\centering\arraybackslash}m{#1}}

\makeatletter
\newsavebox{\measure@tikzpicture}
\NewEnviron{scaletikzpicturetowidth}[1]{%
	\def\tikz@width{#1}%
	\def\tikzscale{1}\begin{lrbox}{\measure@tikzpicture}%
		\BODY
	\end{lrbox}%
	\pgfmathparse{#1/\wd\measure@tikzpicture}%
	\edef\tikzscale{\pgfmathresult}%
	\BODY
}
\makeatother

\makeatletter
\def\squarecorner#1{
	\pgf@x=\the\wd\pgfnodeparttextbox%
	\pgfmathsetlength\pgf@xc{\pgfkeysvalueof{/pgf/inner xsep}}%
	\advance\pgf@x by 2\pgf@xc%
	\pgfmathsetlength\pgf@xb{\pgfkeysvalueof{/pgf/minimum width}}%
	\ifdim\pgf@x<\pgf@xb%
	\pgf@x=\pgf@xb%
	\fi%
	\pgf@y=\ht\pgfnodeparttextbox%
	\advance\pgf@y by\dp\pgfnodeparttextbox%
	\pgfmathsetlength\pgf@yc{\pgfkeysvalueof{/pgf/inner ysep}}%
	\advance\pgf@y by 2\pgf@yc%
	\pgfmathsetlength\pgf@yb{\pgfkeysvalueof{/pgf/minimum height}}%
	\ifdim\pgf@y<\pgf@yb%
	\pgf@y=\pgf@yb%
	\fi%
	\ifdim\pgf@x<\pgf@y%
	\pgf@x=\pgf@y%
	\else
	\pgf@y=\pgf@x%
	\fi
	\pgf@x=#1.5\pgf@x%
	\advance\pgf@x by.5\wd\pgfnodeparttextbox%
	\pgfmathsetlength\pgf@xa{\pgfkeysvalueof{/pgf/outer xsep}}%
	\advance\pgf@x by#1\pgf@xa%
	\pgf@y=#1.5\pgf@y%
	\advance\pgf@y by-.5\dp\pgfnodeparttextbox%
	\advance\pgf@y by.5\ht\pgfnodeparttextbox%
	\pgfmathsetlength\pgf@ya{\pgfkeysvalueof{/pgf/outer ysep}}%
	\advance\pgf@y by#1\pgf@ya%
}
\makeatother

\pgfdeclareshape{square}{
	\savedanchor\northeast{\squarecorner{}}
	\savedanchor\southwest{\squarecorner{-}}
	
	\foreach \x in {east,west} \foreach \y in {north,mid,base,south} {
		\inheritanchor[from=rectangle]{\y\space\x}
	}
	\foreach \x in {east,west,north,mid,base,south,center,text} {
		\inheritanchor[from=rectangle]{\x}
	}
	\inheritanchorborder[from=rectangle]
	\inheritbackgroundpath[from=rectangle]
}

\DeclareMathOperator{\Res}{Res}
\DeclareMathOperator{\U}{U}
\DeclareMathOperator{\SU}{SU}
\DeclareMathOperator{\SO}{SO}
\DeclareMathOperator{\SL}{SL}
\DeclareMathOperator{\USp}{USp}
\DeclareMathOperator{\Spin}{Spin}
\DeclareMathOperator{\GL}{GL}
\DeclareMathOperator{\Sp}{Sp}
\DeclareMathOperator{\EE}{E}
\DeclareMathOperator{\str}{STr}
\DeclareMathOperator{\PE}{PE}
\DeclareMathOperator{\PL}{PL}
\DeclareMathOperator{\HS}{HS}
\DeclareMathOperator{\spa}{Span}
\DeclareMathOperator{\adj}{Ad}
\DeclareMathOperator{\spec}{Spec}
\DeclareMathOperator{\sign}{sign}
\DeclareMathOperator{\perm}{perm}
\newcommand{\de}{\partial}
\newcommand{\CC}{\mathbb{C}}
\newcommand{\PP}{\mathbb{P}}
\newcommand{\RR}{\mathbb{R}}
\newcommand{\ZZ}{\mathbb{Z}}
\newcommand{\ID}{\mathds{1}}
\newcommand{\IM}{\mathbf{M}}
\newcommand{\IX}{\mathbf{X}}
\newcommand{\IZ}{\mathbf{Z}}
\newcommand{\coma}{\, , \,}
\newcommand{\fstop}{\, .}
\newcommand{\hc}{\text{ h.c.}}
\newcommand{\with}{\,\text{ with }\,}
\newcommand{\aand}{\,\text{ and }\,}
\def\quadro{\rotatebox[origin=c]{45}{$\blacksquare$}}
\def\simrep{\mbox{\ydiagram{2}}}
\def\asimrep{\mbox{\ydiagram{1,1}}}
\newcommand{\AM}[1]{{zzzz \color{blue}{\bf{AM}}: #1}}
\newcommand{\AU}[1]{{zzzz \color{red}{\bf{AU}}: #1}}
\newcommand{\SF}[1]{{zzzz \color{red}{\bf{SF}}: #1}}
\newcommand{\XY}[1]{{zzzz \color{blue}{\bf{XY}}: #1}}

\definecolor{redX}{RGB}{218,59,38}
\definecolor{blueX}{RGB}{71,159,248}
\definecolor{yellowX}{RGB}{239,189,64}
\definecolor{greenX}{HTML}{54ae32}
\definecolor{cyanX}{HTML}{90e0ef}

\newcommand{\tzstar}{   \begin{tikzpicture}[overlay]
    \node[star,star points=5, star point ratio=2.25, inner sep=1pt, fill=black, draw=black] at (-0.3ex,0.6ex) {};
\end{tikzpicture}
}

\newcommand*\centermathcell[1]{\omit\hfil$\displaystyle#1$\hfil\ignorespaces}

\allowdisplaybreaks[1] 

\ytableausetup{smalltableaux,centertableaux}

\preprint{ZMP-HH/21-21, IFT-UAM/CSIC-21-142}

\title{Spin(7) Orientifolds and 2d $\mathcal{N}= (0,1)$ Triality}

\author[a,b,c]{Sebasti\'an Franco,}
\author[d]{Alessandro Mininno,}
\author[e]{\'{A}ngel M. Uranga,}
\author[f]{Xingyang Yu}

\affiliation[a]{Physics Department, The City College of the CUNY\\
	160 Convent Avenue, New York, NY 10031, USA}
\affiliation[b]{Physics Program and \textsuperscript{$c$}Initiative for the Theoretical Sciences\\
	The Graduate School and University Center, The City University of New York\\
	365 Fifth Avenue, New York NY 10016, USA}
\affiliation[d]{II. Institut f\"ur Theoretische Physik, Universit\"at Hamburg,\\
Luruper Chaussee 149, 22607 Hamburg, Germany}
\affiliation[e]{Instituto de F\'{\i}sica Te\'orica IFT-UAM/CSIC,\\
	C/ Nicol\'as Cabrera 13-15, 
	Campus de Cantoblanco, 28049 Madrid, Spain}
\affiliation[f]{Center for Cosmology and Particle Physics,\\
	Department of Physics, New York University,\\
	726 Broadway, New York, NY 10003, USA}

\emailAdd{sfranco@ccny.cuny.edu}
\emailAdd{alessandro.mininno@desy.de}
\emailAdd{angel.uranga@csic.es}
\emailAdd{xy1038@nyu.edu}

\abstract{We present a new, geometric perspective on the recently proposed triality of 2d $\mathcal{N}=(0,1)$ gauge theories, based on its engineering in terms of D1-branes probing Spin(7) orientifolds. In this context, triality translates into the fact that multiple gauge theories correspond to the same underlying orientifold. We show how Spin(7) orientifolds based on a particular involution, which we call the universal involution, give rise to precisely the original version of $\mathcal{N}=(0,1)$ triality. Interestingly, our work also shows that the space of possibilities is significantly richer. Indeed, general Spin(7) orientifolds extend triality to theories that can be regarded as consisting of coupled $\mathcal{N}=(0,2)$ and $(0,1)$ sectors. The geometric construction of 2d gauge theories in terms of D1-branes at singularities therefore leads to extensions of triality that interpolate between the pure $\mathcal{N}=(0,2)$ and $(0,1)$ cases.
}

\begin{document}
	
	\maketitle

	\flushbottom
	
\section{Introduction}
\label{sec:Intro}
	
2d $\mathcal{N}=(0,1)$ quantum field theories are extremely interesting, since they are barely supersymmetric and live at the borderline between non-SUSY theories and others with higher amounts of SUSY, for which powerful tools such as holomorphy become applicable. Due to the reduced SUSY, they enjoy a broad range of interesting dynamics. While there has been recent progress in their understanding, they remain relatively unexplored.

In \cite{Gadde:2013lxa}, it was discovered that $2$d $\mathcal{N}= (0, 2)$ theories exhibit IR dualities reminiscent of Seiberg duality in 4d $\mathcal{N}=1$ gauge theories \cite{Seiberg:1994pq}. This low-energy equivalence was dubbed {\it  triality} since, in its simplest incarnation, three SQCD-like theories become equivalent at low energies. Recently, an IR {\it triality} between 2d $\mathcal{N} = (0,1)$ theories with $\SO$ and $\USp$ gauge groups was proposed in \cite{Gukov:2019lzi}. Evidence supporting the proposal includes matching of anomalies and elliptic genera. This new triality can be regarded as a relative of its $\mathcal{N}= (0, 2)$ counterpart. 

The geometric engineering of $2$d $\mathcal{N} = (0, 1)$ gauge theories on D1-branes probing singularities was initiated in \cite{Franco:2021ixh}, where a new class of backgrounds denoted {\it Spin(7) orientifolds} was introduced. These orientifolds are quotients of Calabi-Yau (CY) 4-folds by a combination of an anti-holomorphic involution leading to a Spin(7) cone and worldsheet parity. They provide a beautiful correspondence between the perspective of $\mathcal{N} = (0, 1)$ theories as real slices of $\mathcal{N} = (0, 2)$  theories and Joyce’s geometric construction of Spin(7) manifolds starting from CY 4-folds. This geometric perspective provides a new approach for studying $2$d $\mathcal{N}=(0,1)$ theories.

For branes at singularities, a single geometry often corresponds to multiple gauge theories. Such non-uniqueness is the manifestation of gauge theory dualities in this context. Examples of this phenomenon abound in different dimensions. The various 4d $\mathcal{N}=1$ gauge theories on D3-branes over the same CY 3-fold are related by Seiberg duality \cite{Seiberg:1994pq,Beasley:2001zp,Feng:2001bn}. The triality of 2d $\mathcal{N}=(0,2)$ gauge theories on D1-branes over CY 4-folds and the quadrality of 0d $\mathcal{N}=1$ gauge theories on D$(-1)$-branes over CY 5-folds can be similarly understood \cite{Franco:2016nwv,Franco:2016tcm}. These ideas were further extended to the $(m +1)$-dualities of the $m$-graded quivers that describe the open string sector of the topological B-model on CY $(m + 2)$-folds for arbitrary $m\geq 0$ \cite{Franco:2017lpa,Closset:2018axq,Franco:2020ijt}. In this paper, we will show that the engineering of $2$d $\mathcal{N}=(0,1)$ gauge theories in terms of D1-branes probing Spin(7) orientifolds leads to a similar perspective on $\mathcal{N}=(0,1)$ triality.

The paper is organized as follows. In Section \ref{sec:N02N01trial} we review $\mathcal{N}=(0,2)$ and $\mathcal{N}=(0,1)$ trialities in their original formulations and comment on their generalizations to quivers. We discuss Spin(7) orientifolds and the corresponding $2$d $\mathcal{N}=(0,1)$ field theories arising on D$1$-branes probing them in Section \ref{sec:Spin7Orient}. In Section \ref{sec:01trial-UI} we explain how the basic $\mathcal{N}=(0,1)$ triality arises from the \textit{universal involution}. In Section \ref{sec:BeyonUI} we investigate how (generalizations of) $\mathcal{N}=(0,1)$ triality arise in the case of Spin(7) orientifolds based on more general involutions. We present our conclusions in Section \ref{sec:conclusions}. There are, also, three appendices that may help the reader to follow the discussion in the main text. In Appendix \ref{app:N01multiplets} we review the $\mathcal{N}=(0,1)$ formalism for $2$d gauge theories, and in Appendix \ref{app:AnomalyContr} we list the possible contributions to $2$d gauge anomalies for the groups and representations that we will encounter in the main text. Finally, in Appendix \ref{app:Q111Z2-details} we give all the necessary details for the phases of $Q^{1,1,1}/\ZZ_2$ involved in the triality web introduced in Section \ref{sec:Q111Z2}.


\section{$\mathcal{N}=(0,2)$ and $\mathcal{N}=(0,1)$ Triality}
\label{sec:N02N01trial}

In this section, we review the trialities of 2d $\mathcal{N}=(0,2)$ \cite{Gadde:2013lxa} and $\mathcal{N}=(0,1)$ \cite{Gukov:2019lzi} gauge theories. Discussing $\mathcal{N}=(0,2)$ triality first is not only useful for setting the stage since both trialities share various features, but it is also convenient since Spin(7) orientifolds connect them.

\subsection{$\mathcal{N}=(0,2)$ Triality}
\label{sec:N02triality}

Here we present a quick review of 2d $\mathcal{N}=(0,2)$ triality. A detailed discussion can be found in \cite{Gadde:2013lxa}. Additional developments, including connections to 4d, its realization in terms of D1-branes at CY$_4$ singularities, brane brick models and mirror symmetry, appear in \cite{Gadde:2015wta,Franco:2016nwv,Franco:2016qxh,Franco:2016fxm,Franco:2018qsc}.

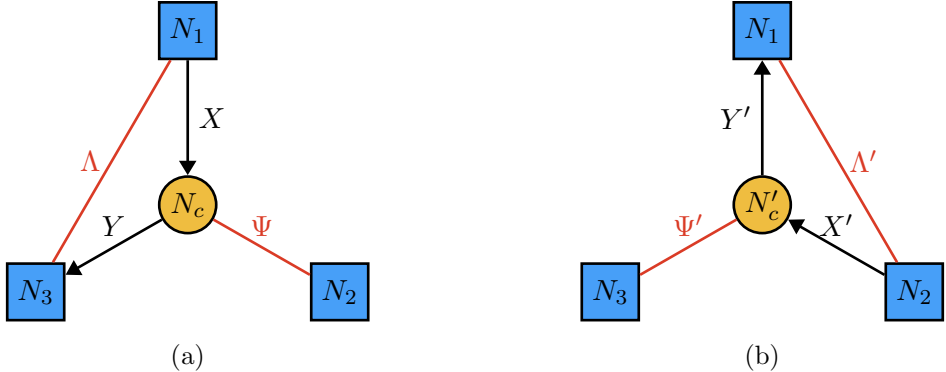
\begin{figure}[H]
	\centering
	\begin{subfigure}{0.5\textwidth}
		\centering
		\begin{tikzpicture}[scale=2]
			\def\L{2};
			\node[draw=black,line width=1pt,circle,fill=yellowX,minimum width=0.75cm,inner sep=1pt] (Nc) at (0,-0.289) {$N_c$};
			\node[draw=black,line width=1pt,fill=blueX,square,minimum width=0.75cm,inner sep=1pt] (N1) at (0,\L*0.433) {$N_1$};
			\node[draw=black,line width=1pt,fill=blueX,square,minimum width=0.75cm,inner sep=1pt] (N2) at (\L/2,-\L*0.433) {$N_2$};
			\node[draw=black,line width=1pt,fill=blueX,square,minimum width=0.75cm,inner sep=1pt] (N3) at (-\L/2,-\L*0.433) {$N_3$};
			\draw[line width=1pt,Triangle-] (Nc) -- node[right,midway] {$X$}  (N1);
			\draw[line width=1pt,redX] (Nc) -- node[above,midway] {$\Psi$} (N2);
			\draw[line width=1pt,redX] (N1) -- node[left,midway] {$\Lambda$} (N3);
			\draw[line width=1pt,-Triangle] (Nc) -- node[above,midway] {$Y$} (N3);
		\end{tikzpicture}
		\caption{}
		\label{fig:SQCDN02}
	\end{subfigure}\hfill
	\begin{subfigure}{0.5\textwidth}
		\centering
		\begin{tikzpicture}[scale=2]
			\def\L{2};
			\node[draw=black,line width=1pt,circle,fill=yellowX,minimum width=0.75cm,inner sep=1pt] (Nc) at (0,-0.289) {$N_c^\prime$};
			\node[draw=black,line width=1pt,fill=blueX,square,minimum width=0.75cm,inner sep=1pt] (N1) at (0,\L*0.433) {$N_1$};
			\node[draw=black,line width=1pt,fill=blueX,square,minimum width=0.75cm,inner sep=1pt] (N2) at (\L/2,-\L*0.433) {$N_2$};
			\node[draw=black,line width=1pt,fill=blueX,square,minimum width=0.75cm,inner sep=1pt] (N3) at (-\L/2,-\L*0.433) {$N_3$};
			\draw[line width=1pt,-Triangle] (Nc) -- node[left,midway] {$Y^\prime$}  (N1);
			\draw[line width=1pt,Triangle-] (Nc) -- node[above,midway] {$X^\prime$} (N2);
			\draw[line width=1pt,redX] (N1) -- node[right,midway] {$\Lambda^\prime$} (N2);
			\draw[line width=1pt,redX] (Nc) -- node[above,midway] {$\Psi^\prime$} (N3);
		\end{tikzpicture}
		\caption{}
		\label{fig:SQCDN02prime}
	\end{subfigure}
	\caption{$2$d $\mathcal{N}=(0,2)$ SQCD and its triality dual. The central nodes have ranks given in \eqref{eq:N02NcNcprank}.}
	\label{fig:SQCDN02transfor}
\end{figure}

Without loss of generality, we can focus on the quiver shown in Figure \ref{fig:SQCDN02}, which can be regarded as 2d $\mathcal{N}=(0,2)$ SQCD. The yellow node represents the $\SU(N_c)$ gauge group that undergoes triality, while the blue nodes are flavor $\SU(N_i)$ groups, $i=1,\ldots,3$.\footnote{More generally, as in theories arising on D1-branes probing CY$_4$ singularities, such groups can have additional matter charged under them and be gauged.} We have absorbed the multiplicities of flavor fields in the ranks of the flavor nodes. In $\mathcal{N}=(0,2)$ quivers, we adopt the convention that the head and tail of the arrow associated to a chiral field correspond to fundamental and antifundamental representations, respectively. A Fermi field connecting the flavor nodes 1 and 3 has been included to make the original and dual theories more similar.

The triality dual is shown in Figure \ref{fig:SQCDN02prime}. The rank of the central node in both theories is determined by anomaly cancellation to be
\begin{equation}
	N_c = {N_1 +N_3 - N_2 \over 2} \coma
	N_c' =  {N_2 +N_1 - N_3 \over 2} \fstop
	\label{eq:N02NcNcprank}
\end{equation}
The transformation of the rank can also be written as
\begin{equation}
	N_c'= N_1 -N_c \fstop
	\label{triality_transformation_rank}
\end{equation}
Both theories in Figure \ref{fig:SQCDN02transfor} have $J$-/$E$- terms associated to the triangular loops in the quivers.

\begin{figure}[H]
	\centering
	\scalebox{0.69}{
		\begin{tikzpicture}[scale=9]
			\def\R{1};
			\def\d{1};
			\node (v1) at (-0.02,0.4) {
				\scalebox{\d}{
					\begin{tikzpicture}[scale=2]
						\def\L{2};
						\node[draw=black,line width=1pt,circle,fill=yellowX,minimum width=0.75cm,inner sep=1pt] (Nc) at (0,-0.289) {$N_c$};
						\node[draw=black,line width=1pt,fill=blueX,square,minimum width=0.75cm,inner sep=1pt] (N1) at (0,\L*0.433) {$N_1$};
						\node[draw=black,line width=1pt,fill=blueX,square,minimum width=0.75cm,inner sep=1pt] (N2) at (\L/2,-\L*0.433) {$N_2$};
						\node[draw=black,line width=1pt,fill=blueX,square,minimum width=0.75cm,inner sep=1pt] (N3) at (-\L/2,-\L*0.433) {$N_3$};
						\draw[line width=1pt,Triangle-] (Nc) -- node[right,midway] {$X$}  (N1);
						\draw[line width=1pt,redX] (Nc) -- node[above,midway] {$\Psi$} (N2);
						\draw[line width=1pt,redX] (N1) -- node[left,midway] {$\Lambda$} (N3);
						\draw[line width=1pt,-Triangle] (Nc) -- node[above,midway] {$Y$} (N3);
						\node at (1.4,\L*0.11) {$\displaystyle{N_c=\frac{N_1+N_3-N_2}{2}}$};
					\end{tikzpicture}
				}    
			};
			\node (v2) at (0.25,-0.42) {
				\scalebox{\d}{
					\begin{tikzpicture}[scale=2]
						\def\L{2};
						\node[draw=black,line width=1pt,circle,fill=yellowX,minimum width=0.75cm,inner sep=1pt] (Nc) at (0,-0.289) {$N_c^\prime$};
						\node[draw=black,line width=1pt,fill=blueX,square,minimum width=0.75cm,inner sep=1pt] (N1) at (0,\L*0.433) {$N_1$};
						\node[draw=black,line width=1pt,fill=blueX,square,minimum width=0.75cm,inner sep=1pt] (N2) at (\L/2,-\L*0.433) {$N_2$};
						\node[draw=black,line width=1pt,fill=blueX,square,minimum width=0.75cm,inner sep=1pt] (N3) at (-\L/2,-\L*0.433) {$N_3$};
						\draw[line width=1pt,-Triangle] (Nc) -- node[left,midway] {$Y^\prime$}  (N1);
						\draw[line width=1pt,Triangle-] (Nc) -- node[above,midway] {$X^\prime$} (N2);
						\draw[line width=1pt,redX] (N1) -- node[right,midway] {$\Lambda^\prime$} (N2);
						\draw[line width=1pt,redX] (Nc) -- node[above,midway] {$\Psi^\prime$} (N3);
						\node at (-1.2,\L*0.15) {$\displaystyle{N_c^\prime=\frac{N_2+N_1-N_3}{2}}$};
					\end{tikzpicture}       
				}
			};
			\node (v3) at (-0.62,-0.42) {
				\scalebox{\d}{
					\begin{tikzpicture}[scale=2]
						\def\L{2};
						\node[draw=black,line width=1pt,circle,fill=yellowX,minimum width=0.75cm,inner sep=1pt] (Nc) at (0,-0.289) {$N_c^{\prime\prime}$};
						\node[draw=black,line width=1pt,fill=blueX,square,minimum width=0.75cm,inner sep=1pt] (N1) at (0,\L*0.433) {$N_1$};
						\node[draw=black,line width=1pt,fill=blueX,square,minimum width=0.75cm,inner sep=1pt] (N2) at (\L/2,-\L*0.433) {$N_2$};
						\node[draw=black,line width=1pt,fill=blueX,square,minimum width=0.75cm,inner sep=1pt] (N3) at (-\L/2,-\L*0.433) {$N_3$};
						\draw[line width=1pt,redX] (Nc) -- node[right,midway] {$\Psi^{\prime\prime}$}  (N1);
						\draw[line width=1pt,-Triangle] (Nc) -- node[above,midway] {$Y^{\prime\prime}$} (N2);
						\draw[line width=1pt,redX] (N3) -- node[below,midway] {$\Lambda^{\prime\prime}$} (N2);
						\draw[line width=1pt,Triangle-] (Nc) -- node[above,midway] {$X^{\prime\prime}$} (N3);
						\node at (1.4,\L*0.11) {$\displaystyle{N_c^{\prime\prime}=\frac{N_3+N_2-N_1}{2}}$};
					\end{tikzpicture}
				}
			};
			\draw[line width=1mm, greenX,-Triangle] (-0.51,-0.22) -- (-0.3,0.1);
			\draw[line width=1mm, greenX,Triangle-] (0.21,-0.22) -- (0,0.1);
			\draw[line width=1mm, greenX,-Triangle] (0.08,-0.48) -- (-0.38,-0.48);
		\end{tikzpicture}  
	}
	\caption{Triality loop for $2$d $\mathcal{N}=(0,2)$ SQCD.}
	\label{fig:SCQDN02trialoop}
\end{figure}
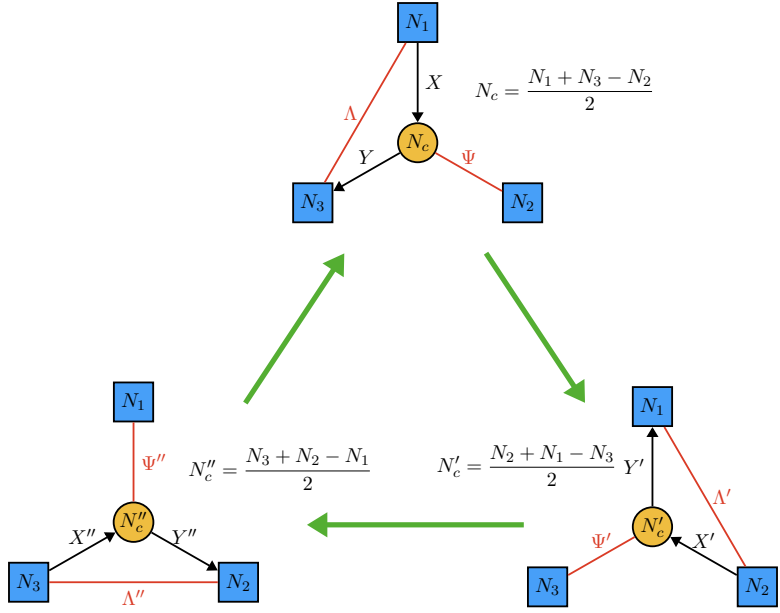

Taking the dual theory as the new starting point and acting on it with triality, we obtain the theory shown on the bottom left of Figure \ref{fig:SCQDN02trialoop}. Applying triality a third time takes us back to the original theory. We can therefore think about this second dual as connected to the original theory by inverse triality.\footnote{The distinction between triality and inverse triality is just a convention.} The triality among these three theories can be viewed as a cyclic permutation of $N_1$, $N_2$ and $N_3$.

We will later use $\mathcal{N}=(0,2)$ gauge theories engineered on D1-branes probing CY 4-folds as starting points of orientifold constructions. Such theories have $\U(N)$ gauge groups. A $\U(N_c)$ version of $\mathcal{N}=(0,2)$ triality was also introduced in \cite{Gadde:2013lxa}. It only differs from the $\SU(N_c)$ triality depicted in Figure \ref{fig:SCQDN02trialoop} by the presence of additional Fermi fields in the determinant representation of the gauge group, which are necessary for the cancellation of the Abelian anomaly. It is expected that Abelian anomalies of gauge theories on D1-branes are cancelled via a generalized Green-Schwarz mechanism (see \cite{Ibanez:1998qp,Mohri:1997ef} for 4d ${\cal N}=1$ and $2$d ${\cal N}=(0,2)$ theories realized on D-branes probing orbifolds/orientifolds singularities). For this reason, the determinant Fermi fields are not present in such theories and triality reduces to the one considered in this section.

$\mathcal{N}=(0,2)$ triality can be extended to general quivers (see e.g. \cite{Gadde:2013lxa,Franco:2016nwv,Franco:2016qxh,Franco:2016fxm,Franco:2018qsc,Franco:2021elb}). It acts as a local operation on the dualized node, with the part of the quiver that is not connected to it acting as a spectator. The transformation of such a theory under triality on a gauge node $k$ can be summarized as follows. The rank of node $k$ changes according to
\begin{align}
	N'_k = \sum_{j\neq k} n_{jk}^\chi N_j - N_k \,, 
	\label{rank-rule}
\end{align}
where $n^\chi_{jk}$ is the number of chiral fields from node $j$ to node $k$. All other ranks remain the same. The field content around node $k$ changes according to the following rules:

\begin{enumerate}[label={(R.\arabic*)}]
	\item\label{rule1} {\bf Dual Flavors}. Replace each of $(\rightarrow k)$, $(\leftarrow k)$, $(\textcolor{red}{\text{ --- }}  k)$ by  $(\leftarrow k)$, $( \textcolor{red}{\text{ --- }}  k)$, $(\rightarrow k)$, respectively.
	\item\label{rule2} {\bf Chiral-Chiral Mesons}. For each subquiver $i\rightarrow k \rightarrow j$, add a new chiral field $i\rightarrow j$.
	\item\label{rule3} {\bf Chiral-Fermi Mesons}. For each subquiver $i\rightarrow k \textcolor{red}{\text{ --- }} j$, add a new Fermi field $i \textcolor{red}{\text{ --- }}  j$.
	\item\label{rule4} Remove all chiral-Fermi massive pairs generated in the previous steps.
\end{enumerate}

For a detailed discussion of the transformation of $J$- and $E$-terms, see e.g., \cite{Franco:2017lpa}.

\subsection{$\mathcal{N}=(0,1)$ Triality}
\label{sec:N01triality}

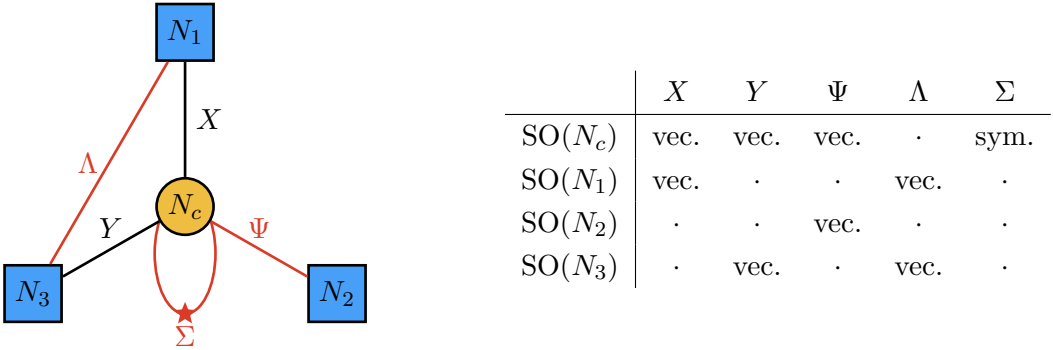
\begin{figure}[H]
	\centering
	\begin{tikzpicture}[scale=2]
		\node (Q) at (-3.2,0) {
			\begin{tikzpicture}[scale=2]
				\def\L{2};
				\draw[line width=1pt,redX] (0,-0.6) ellipse (0.2 and 0.4) node[yshift=-1.1cm] {$\Sigma$} node[xshift=0cm,yshift=-0.8cm,star,star points=5, star point ratio=2.25, inner sep=1pt, fill=redX, draw=redX] {};
				\node[draw=black,line width=1pt,circle,fill=yellowX,minimum width=0.75cm,inner sep=1pt] (Nc) at (0,-0.289) {$N_c$};
				\node[draw=black,line width=1pt,fill=blueX,square,minimum width=0.75cm,inner sep=1pt] (N1) at (0,\L*0.433) {$N_1$};
				\node[draw=black,line width=1pt,fill=blueX,square,minimum width=0.75cm,inner sep=1pt] (N2) at (\L/2,-\L*0.433) {$N_2$};
				\node[draw=black,line width=1pt,fill=blueX,square,minimum width=0.75cm,inner sep=1pt] (N3) at (-\L/2,-\L*0.433) {$N_3$};
				\draw[line width=1pt] (Nc) -- node[right,midway] {$X$}  (N1);
				\draw[line width=1pt,redX] (Nc) -- node[above,midway] {$\Psi$} (N2);
				\draw[line width=1pt,redX] (N1) -- node[left,midway] {$\Lambda$} (N3);
				\draw[line width=1pt] (Nc) -- node[above,midway] {$Y$} (N3);
			\end{tikzpicture}
		};        
		\node (T) at (0.7,0) {
			\begin{tabular}{c|ccccc}
				              & $X$     &  $Y$     & $\Psi$ & $\Lambda$ & $\Sigma$ \\
				\hline                              
				$\SO(N_c)$    & vec.    &  vec.    & vec. & $\cdot$ & sym. \\
				$\SO(N_1)$    & vec.    &  $\cdot$ & $\cdot$ & vec. & $\cdot$ \\
				$\SO(N_2)$    & $\cdot$ &  $\cdot$ & vec. & $\cdot$ & $\cdot$ \\
				$\SO(N_3)$    & $\cdot$ &  vec.    & $\cdot$ & vec. & $\cdot$ \\
			\end{tabular}
		};
	\end{tikzpicture}
	\caption{$2$d $\mathcal{N}=(0,1)$ SQCD. $N_c$ is given in Eq.~\eqref{eq:Nc01}.}
	\label{fig:SQCDN01}
\end{figure}

A similar triality for 2d $\mathcal{N}=(0,1)$ gauge theories was introduced in \cite{Gukov:2019lzi}. The primary example in which the proposal was investigated is $2$d $\mathcal{N}=(0,1)$ SQCD with $\SO(N_c)$ gauge group, whose quiver diagram is shown in Figure \ref{fig:SQCDN01}.\footnote{When drawing $\mathcal{N}=(0,1)$ quivers, black and red lines correspond to real $\mathcal{N}=(0,1)$ scalar and Fermi fields, respectively. In addition, we indicate symmetric and antisymmetric representations with star and diamond symbols, respectively.} The theory has $N_1+N_3$ scalar multiplets in the vector representation of $\SO(N_c)$. These scalar fields are further divided into two sets, $X$ and $Y$, transforming under $\SO(N_1)$ and $\SO(N_3)$ flavor groups, respectively. A bifundamental Fermi multiplet $\Lambda$ connects $\SO(N_1)$ and $\SO(N_3)$.\footnote{We will use the term bifundamental in the case of matter fields that connect pairs of nodes, even when one or both of them is either $\SO$ or $\USp$.} There are also $N_2$ Fermi multiplets $\Psi$ in the vector representation of $\SO(N_c)$ and a Fermi multiplet $\Sigma$ in the symmetric representation of $\SO(N_c)$.

Anomaly cancellation for the $\SO(N_c)$ gauge group requires that\footnote{The anomaly contributions of $\mathcal{N}=(0,1)$ multiplets in various representations are listed in Appendix \ref{app:AnomalyContr}.}
\begin{equation}
	N_c=\frac{N_1+N_3-N_2}{2}\fstop
	\label{eq:Nc01}
\end{equation} 	

The theory also has the following superpotential consistent with its symmetries
\begin{equation}
	\label{eq:SDCD superpotential}
	W^{(0,1)}=\sum_{\alpha,\beta=1}^{N_c}\Sigma_{\alpha\beta}\left(\sum_{a=1}^{N_1}X^a_{\alpha}X^a_{\beta}+\sum_{b=1}^{N_3}Y^b_{\alpha}Y^b_{\beta}-\delta_{\alpha\beta}\right)+\sum_{a=1}^{N_1}\sum_{b=1}^{N_3}\sum_{\alpha=1}^{N_c}\Lambda_{ab}X^a_\alpha Y^b_\alpha\fstop
\end{equation}

\begin{figure}[H]
	\centering
	\begin{tikzpicture}[scale=2]
		\node (Q) at (-3.2,0) {
			\begin{tikzpicture}[scale=2]
				\def\L{2};
				\draw[line width=1pt,redX] (0,-0.6) ellipse (0.2 and 0.4) node[yshift=-1.1cm] {$\Sigma^\prime$} node[xshift=0cm,yshift=-0.8cm,star,star points=5, star point ratio=2.25, inner sep=1pt, fill=redX, draw=redX] {};
				\node[draw=black,line width=1pt,circle,fill=yellowX,minimum width=0.75cm,inner sep=1pt] (Nc) at (0,-0.289) {$N_c^\prime$};
				\node[draw=black,line width=1pt,fill=blueX,square,minimum width=0.75cm,inner sep=1pt] (N1) at (0,\L*0.433) {$N_1$};
				\node[draw=black,line width=1pt,fill=blueX,square,minimum width=0.75cm,inner sep=1pt] (N2) at (\L/2,-\L*0.433) {$N_2$};
				\node[draw=black,line width=1pt,fill=blueX,square,minimum width=0.75cm,inner sep=1pt] (N3) at (-\L/2,-\L*0.433) {$N_3$};
				\draw[line width=1pt] (Nc) -- node[left,midway] {$Y^\prime$}  (N1);
				\draw[line width=1pt] (Nc) -- node[above,midway] {$X^\prime$} (N2);
				\draw[line width=1pt,redX] (N1) -- node[right,midway] {$\Lambda^\prime$} (N2);
				\draw[line width=1pt,redX] (Nc) -- node[above,midway] {$\Psi^\prime$} (N3);
			\end{tikzpicture}
		};        
		\node (T) at (0.7,0) {
			\begin{tabular}{c|ccccc}
				              & $X^\prime$ &  $Y^\prime$ & $\Psi^\prime$ & $\Lambda^\prime$ & $\Sigma^\prime$ \\
				\hline                                     
				$\SO(N_c)$    & vec.       &  vec.       & vec. & $\cdot$ & sym. \\
				$\SO(N_1)$    & $\cdot$    &  vec.       & $\cdot$ & vec. & $\cdot$ \\
				$\SO(N_2)$    & vec.       &  $\cdot$    & $\cdot$ & vec. & $\cdot$ \\
				$\SO(N_3)$    & $\cdot$    &  $\cdot$    & vec. & $\cdot$ & $\cdot$ \\
			\end{tabular}
		};
	\end{tikzpicture}
	\caption{2d $\mathcal{N}=(0,1)$ triality dual of the theory in Figure~\ref{fig:SQCDN01}. $N^\prime_c$ is given in Eq.~\eqref{eq:Ncp01}.}
	\label{fig:SQCDN01-dual}
\end{figure}
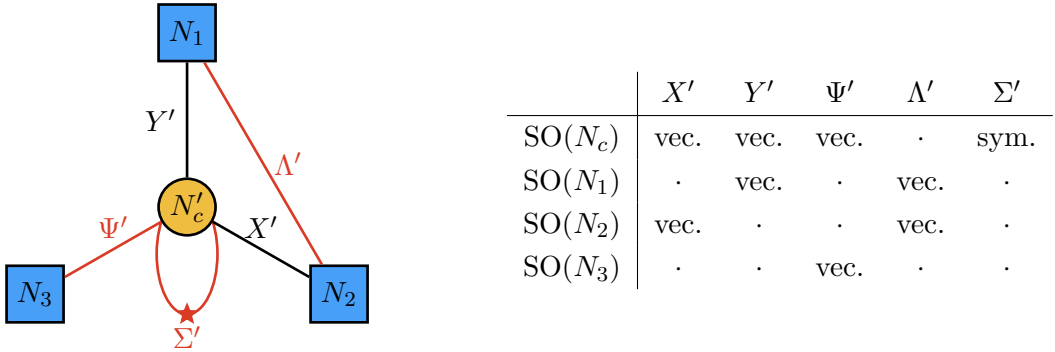

Figure \ref{fig:SQCDN01-dual} shows the dual under triality. The transformation is rather similar to the $\mathcal{N}=(0,2)$ triality discussed in the previous section. Once again, in this simple example, the structure of the dual theory is identical to the original one up to a cyclic permutation of $N_1$, $N_2$ and $N_3$. For the flavors, scalar multiplets $X$, $Y$ and Fermi multiplets $\Psi$ are replaced by scalar multiplets $Y^\prime$, Fermi multiplets $\Psi^\prime$, and scalar multiplets $X^\prime$, respectively. The new theory also contains a Fermi field $\Sigma'$ in the symmetric representation of the gauge group.

The gauge group is $\SO(N_c^\prime)$, with the rank determined by anomaly cancellation
\begin{equation}
	N_c^\prime=\frac{N_2+N_1-N_3}{2} \coma
	\label{eq:Ncp01}
\end{equation}
which can be expressed as
\begin{equation}
	N_c^\prime=N_1-N_c \fstop
\end{equation}

Very much like Rule \ref{rule3} of the previous section,
the new Fermi $\Lambda^\prime$ in the bifundamental representation of $\SO(N_1)\times \SO(N_2)$ can be regarded as a scalar-Fermi meson in terms of the fields in the initial theory, i.e. $\Lambda^\prime=X\Psi$. Similarly, we can interpret the disappearance of the original Fermi $\Lambda$ between Figures \ref{fig:SQCDN01} and \ref{fig:SQCDN01-dual} as the result of it becoming massive via its superpotential coupling to the scalar-scalar meson $X Y$, which is analogous to the chiral-chiral mesons of Rule \ref{rule2}. An interesting difference with respect to $\mathcal{N}=(0,2)$ SQCD follows from the fact that SO representations are real. Equivalently, the quivers under consideration are not oriented. It is therefore natural to ask why, in addition to $\Lambda^\prime=X\Psi$, Figure \ref{fig:SQCDN01-dual} does not {\it simultaneously} have another scalar-Fermi meson $Y \Psi$ in the bifundamental representation of $\SO(N_2)\times \SO(N_3)$. Its absence can be interpreted as descending from $\mathcal{N}=(0,2)$ triality, in which the orientation of chiral fields prevent the formation of such a gauge invariant. Additional thoughts on the connection between $\mathcal{N}=(0,2)$ and $\mathcal{N}=(0,1)$ trialities will be presented in Section \ref{sec:N01triality_quivers}. Also related to this issue, in the coming section, we will discuss scalar-Fermi mesons in more general quivers.

The superpotential is identical to \eqref{eq:SDCD superpotential} upon replacing all fields by the primed counterparts and permuting $N_1$, $N_2$ and $N_3$.

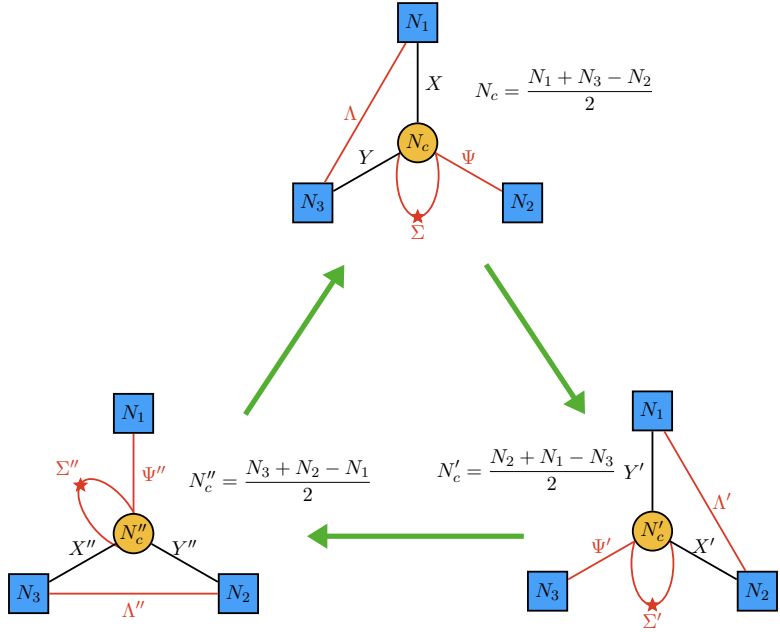
\begin{figure}[H]
	\centering
	\scalebox{0.69}{
		\begin{tikzpicture}[scale=9]
			\def\R{1};
			\def\d{1};
			\node (v1) at (-0.02,0.4) {
				\scalebox{\d}{     
					\begin{tikzpicture}[scale=2]
						\def\L{2};
						\draw[line width=1pt,redX] (0,-0.6) ellipse (0.2 and 0.4) node[yshift=-1.1cm] {$\Sigma$} node[xshift=0cm,yshift=-0.8cm,star,star points=5, star point ratio=2.25, inner sep=1pt, fill=redX, draw=redX] {};
						\node[draw=black,line width=1pt,circle,fill=yellowX,minimum width=0.75cm,inner sep=1pt] (Nc) at (0,-0.289) {$N_c$};
						\node[draw=black,line width=1pt,fill=blueX,square,minimum width=0.75cm,inner sep=1pt] (N1) at (0,\L*0.433) {$N_1$};
						\node[draw=black,line width=1pt,fill=blueX,square,minimum width=0.75cm,inner sep=1pt] (N2) at (\L/2,-\L*0.433) {$N_2$};
						\node[draw=black,line width=1pt,fill=blueX,square,minimum width=0.75cm,inner sep=1pt] (N3) at (-\L/2,-\L*0.433) {$N_3$};
						\draw[line width=1pt] (Nc) -- node[right,midway] {$X$}  (N1);
						\draw[line width=1pt,redX] (Nc) -- node[above,midway] {$\Psi$} (N2);
						\draw[line width=1pt,redX] (N1) -- node[left,midway] {$\Lambda$} (N3);
						\draw[line width=1pt] (Nc) -- node[above,midway] {$Y$} (N3);
						\node at (1.4,\L*0.11) {$\displaystyle{N_c=\frac{N_1+N_3-N_2}{2}}$};
					\end{tikzpicture}
				} 
			};
			\node (v2) at (0.25,-0.43) {
				\scalebox{\d}{
					\begin{tikzpicture}[scale=2]
						\def\L{2};
						\draw[line width=1pt,redX] (0,-0.6) ellipse (0.2 and 0.4) node[yshift=-1.1cm] {$\Sigma^\prime$} node[xshift=0cm,yshift=-0.8cm,star,star points=5, star point ratio=2.25, inner sep=1pt, fill=redX, draw=redX] {};
						\node[draw=black,line width=1pt,circle,fill=yellowX,minimum width=0.75cm,inner sep=1pt] (Nc) at (0,-0.289) {$N_c^\prime$};
						\node[draw=black,line width=1pt,fill=blueX,square,minimum width=0.75cm,inner sep=1pt] (N1) at (0,\L*0.433) {$N_1$};
						\node[draw=black,line width=1pt,fill=blueX,square,minimum width=0.75cm,inner sep=1pt] (N2) at (\L/2,-\L*0.433) {$N_2$};
						\node[draw=black,line width=1pt,fill=blueX,square,minimum width=0.75cm,inner sep=1pt] (N3) at (-\L/2,-\L*0.433) {$N_3$};
						\draw[line width=1pt] (Nc) -- node[left,midway] {$Y^\prime$}  (N1);
						\draw[line width=1pt] (Nc) -- node[above,midway] {$X^\prime$} (N2);
						\draw[line width=1pt,redX] (N1) -- node[right,midway] {$\Lambda^\prime$} (N2);
						\draw[line width=1pt,redX] (Nc) -- node[above,midway] {$\Psi^\prime$} (N3);
						\node at (-1.2,\L*0.15) {$\displaystyle{N_c^\prime=\frac{N_2+N_1-N_3}{2}}$};
					\end{tikzpicture}       
				}
			};
			\node (v3) at (-0.62,-0.42) {
				\scalebox{\d}{
					\begin{tikzpicture}[scale=2]
						\def\L{2};
						\draw[line width=1pt,redX,rotate around={39:(-0.23,-0.1)}] (-0.23,-0.1) ellipse (0.2 and 0.4) node[xshift=-0.8cm,yshift=0.9cm] {$\Sigma^{\prime\prime}$} node[xshift=-0.55cm,yshift=0.55cm,star,star points=5, star point ratio=2.25, inner sep=1pt, fill=redX, draw=redX] {};
						\node[draw=black,line width=1pt,circle,fill=yellowX,minimum width=0.75cm,inner sep=1pt] (Nc) at (0,-0.289) {$N_c^{\prime\prime}$};
						\node[draw=black,line width=1pt,fill=blueX,square,minimum width=0.75cm,inner sep=1pt] (N1) at (0,\L*0.433) {$N_1$};
						\node[draw=black,line width=1pt,fill=blueX,square,minimum width=0.75cm,inner sep=1pt] (N2) at (\L/2,-\L*0.433) {$N_2$};
						\node[draw=black,line width=1pt,fill=blueX,square,minimum width=0.75cm,inner sep=1pt] (N3) at (-\L/2,-\L*0.433) {$N_3$};
						\draw[line width=1pt,redX] (Nc) -- node[right,midway] {$\Psi^{\prime\prime}$}  (N1);
						\draw[line width=1pt] (Nc) -- node[above,midway] {$Y^{\prime\prime}$} (N2);
						\draw[line width=1pt,redX] (N3) -- node[below,midway] {$\Lambda^{\prime\prime}$} (N2);
						\draw[line width=1pt] (Nc) -- node[above,midway] {$X^{\prime\prime}$} (N3);
						\node at (1.4,\L*0.11) {$\displaystyle{N_c^{\prime\prime}=\frac{N_3+N_2-N_1}{2}}$};
					\end{tikzpicture}
				}
			};
			\draw[line width=1mm, greenX,-Triangle] (-0.51,-0.22) -- (-0.3,0.1);
			\draw[line width=1mm, greenX,Triangle-] (0.21,-0.22) -- (0,0.1);
			\draw[line width=1mm, greenX,-Triangle] (0.08,-0.48) -- (-0.38,-0.48);
		\end{tikzpicture}  
	}
	\caption{Triality loop for $2$d $\mathcal{N}=(0,1)$ SQCD.} 
	\label{fig:SCQDN01trialoop}
\end{figure}

Acting with triality again gives rise to the theory shown on the bottom left of Figure \ref{fig:SCQDN01trialoop}. A third triality takes us back to the original theory. 

There is also a symplectic version of $\mathcal{N}=(0,1)$ triality \cite{Gukov:2019lzi}. The corresponding SQCD has $\USp(N_c)$ gauge group and $\USp(N_1)\times \USp(N_2)\times \USp(N_3)$ global symmetry.\footnote{Differently from~\cite{Gukov:2019lzi}, we adopt the convention $\USp(2)\simeq \SU(2)$ in order to be consistent with the notation of the orientifold theories we construct later.} The matter content is almost the same as in the $\SO(N_c)$ SQCD quiver shown in Figure \ref{fig:SQCDN01}, with the exception that the Fermi field $\Sigma$ instead transforms in the antisymmetric representation of $\USp(N_c)$. The rank of the gauge group is $N_c=\frac{N_1+N_3-N_2}{2}$ to cancel gauge anomalies. In this case, the triality loop is identical to the one shown in Figure \ref{fig:SCQDN01trialoop}.

Evidence for the $\mathcal{N}=(0,1)$ triality proposal includes matching of anomalies and elliptic genera \cite{Gukov:2019lzi}. In the coming sections, we will provide further support for this idea, by realizing 2d $\mathcal{N}=(0,1)$ theories via $\Spin(7)$ orientifolds.

\subsection{$\mathcal{N}=(0,1)$ Triality for Quiver Gauge Theories}
\label{sec:N01triality_quivers}

Let us consider the extension of $\mathcal{N}=(0,1)$ triality to general quivers. To do so, it is useful to first draw some lessons from Seiberg duality and $\mathcal{N}=(0,2)$ triality. In both cases, incoming chiral fields at the dualized gauge group play a special role.\footnote{This is a general phenomenon that applies e.g. to the order $(m+1)$ dualities of $m$-graded quivers \cite{Franco:2017lpa}. Seiberg duality and $\mathcal{N}=(0,2)$ triality correspond to the $m=1$ and $2$ cases, respectively.} They control the rank of the dual gauge group and, for triality, determine which mesons are formed. Since $\mathcal{N}=(0,1)$ quivers are unoriented, how to split the scalar fields terminating on a dualized node into two sets analogous to ``incoming" and ``outgoing" flavors is not clear. This issue was hinted in our discussion in the previous section.

In \cite{Gukov:2019lzi}, a generalization of triality to a simple class of quiver theories with $\SO(N_{c_1})\times \SO(N_{c_2})\times \ldots$ gauge group (or the symplectic counterpart) was briefly discussed. Theories in this family are obtained by combining various $\mathcal{N}=(0,1)$ SQCD building blocks, which are glued by identifying any of the three global nodes of a given theory with the gauge node of another one. Locally, the resulting theories have the same structure of basic SQCD. Namely, every gauge node is connected to three other nodes, to two of them via scalar fields and to the remaining one via Fermi fields. Due to this simple structure, the dualization of any of the gauge groups is unambiguous and proceeds as in basic triality. For every node, the two possible choices of scalar fields acting as ``incoming" or ``outgoing" corresponds to acting with triality or inverse triality.   

For general quivers, in which a given node can be connected to multiple others, how to separate the flavor scalar fields at every gauge group into two sets is an open question. All the theories that we will construct later using $\Spin(7)$ orientifolds are indeed beyond the above special class. However, this ambiguity is resolved in them by inheriting the separation of flavors from the parent $\mathcal{N}=(0,2)$ theories.

\section{Spin(7) Orientifolds}
\label{sec:Spin7Orient}

In this section we review the construction of Spin(7) orientifolds introduced in \cite{Franco:2021ixh} and the 2d $\mathcal{N}=(0,1)$ field theories arising on D1-branes probing them. We focus the overview on a few key points relevant for subsequent sections, and refer the reader to this reference for additional details. 

Our starting point is a toric CY $4$-fold singularity $\IM_8$. When probed by a stack of D1-branes at the singular point, the worldvolume theory corresponds to an $\mathcal{N}=(0,2)$ quiver gauge field theory. When $\IM_8$ is toric, the structure of gauge groups, matter content and interactions of these theories is nicely encoded by brane brick models \cite{Franco:2015tya,Franco:2016qxh,Franco:2017cjj} (see \cite{GarciaCompean:1998kh} for an early related construction). Nevertheless, for our purposes it suffices to use the quiver description, supplemented by the explicit expression of the interaction terms ($J$- and $E$-terms).

We then perform an orientifold quotient by the action $\Omega \sigma$, where $\Omega$ is worldsheet parity and $\sigma$ is an anti-holomorphic involution of $\IM_8$ leaving a specific 4-form, that we call $\Omega^{(4)}$, invariant. Such $4$-form is constructed from the CY holomorphic $4$-form $\Omega^{(4,0)}$ and the K\"ahler form $J^{(1,1)}$ as 
\begin{equation}
    \Omega^{(4)}=\text{Re}\left(\Omega^{(4,0)}\right)+\frac{1}{2}J^{(1,1)}\wedge J^{(1,1)}\fstop
\end{equation}
If the quotient did not involve worldsheet parity, this quotient corresponds to Joyce's construction of Spin(7) geometries, with $\Omega^{(4)}$ defining the invariant Cayley 4-form of such varieties. To keep this connection in mind, the above orientifold quotients were dubbed \textit{Spin(7) orientifolds} in \cite{Franco:2021ixh}.

This orientifold quotient has a natural counterpart on the D1-brane systems, and naturally realizes a ``real projection" of the $2$d $\mathcal{N}=(0,2)$ theories in~\cite{Gukov:2019lzi}, resulting in a 2d $\mathcal{N}=(0,1)$ gauge field theory. Its structure is determined by a set of rules analogous to  those of orientifold field theories in higher dimensions (see e.g. \cite{Franco:2007ii} in the 4d context), and which were explicitly determined in \cite{Franco:2021ixh}. Morally, it corresponds to identifying the gauge factors and matter fields of the parent $\mathcal{N}=(0,2)$ theory under an involution symmetry ${\tilde \sigma}$ of the quiver, compatible with the set of interactions.

To describe the orientifold action on the field theory in more detail, we label the different nodes by an index $i$, and their orientifold images by $i'$ (with $i'=i$ corresponding to nodes mapped to themselves under the orientifold action), and denote $X_{ij}$ and $\Lambda_{ij}$ the bifundamental $\mathcal{N}=(0,2)$ chiral or Fermi multiplets charged under the gauge factors $i$ and $j$ (with $j=i$ corresponding to adjoints). The results of \cite{Franco:2021ixh} are:

\begin{enumerate}[label=1\alph*.,ref=1\alph*]
\item\label{rule:1a}  Two gauge factors $\U(N)_i$, $\U(N)_{i'}$ mapped to each other under the orientifold action (namely $i\neq i'$) are identified and give rise to a single $\U(N)$ factor in the orientifold theory. 
\item\label{rule:1b} On the other hand, a gauge factor $\U(N)_i$ mapped to itself (namely, $i'=i$) is projected down to $\SO(N)$ or $\USp(N)$. 
\end{enumerate}

\begin{enumerate}[label=2\alph*.,ref=2\alph*]
\item\label{rule:2a}   Two {\em different} chiral or Fermi bifundamental fields $X_{ij}$  and $X_{i'j'}$, mapped to each other under the orientifold action, become identified\footnote{\label{the-eta-issue} In the presence of multiple sets of fields in these representations, the mapping may include a non-trivial action on the flavor index, encoded in a matrix $\eta$. As explained in \cite{Franco:2021ixh}, the choice can impact on the orientifold projection of the relevant gauge factors. We will encounter a non-trivial use of this freedom in the example in Section \ref{sec:Q111}.} and lead to a single (chiral or Fermi) bifundamental field. This holds even in special cases for the gauge factors, such as $i'=i$, or simultaneously $i'=i$ and $j'=j$, and for the special case of fields in the adjoint, $j=i$, $j'=i'$. 

\item\label{rule:2b} Two {\em different} chiral or Fermi bifundamental fields $X_{ii'}$ and $Y_{i'i}$, related each to the (conjugate of the) other under the orientifold action, give rise to one field in the two-index symmetric and one field in the two-index antisymmetric representation of the corresponding $\SO/\USp$ $i^{th}$ gauge factor in the orientifold quotient. The rule holds also in the case of adjoint fields, namely $i'=i$.  
\end{enumerate}

\begin{enumerate}[label=3\alph*.,ref=3\alph*]
\item\label{rule:3a}  A bifundamental field $X_{ij}$ that is mapped to itself by the orientifold action gives rise to a real $\mathcal{N}=(0,1)$ field  transforming under the bifundamental of $G_i\times G_j$, where $G_i$ and $G_j$ are the same type of $\SO$ or $\USp$ gauge group.

\item\label{rule:3b} A bifundamental Fermi field $\Lambda_{ii'}$ can only be mapped to itself (resp. minus itself) in the case of a holomorphic transformation, and gives rise to a complex Fermi superfield in the symmetric (resp. antisymmetric) representation of the resulting $\U(n)_i$ group. 

\item\label{rule:3c} Closely related to Rule~\ref{rule:3b}, an adjoint complex Fermi field $\Lambda_{ii}$ that is mapped to itself (resp. minus itself) via a holomorphic transformation, gives rise to a complex Fermi field in the symmetric/antisymmetric (resp. antisymmetric/symmetric) representation of $\SO/\USp$. 

\item\label{rule:3d} An adjoint complex scalar or Fermi field that is mapped to itself gives rise to two real scalar or Fermi fields, one symmetric and one antisymmetric.
\end{enumerate}

\begin{enumerate}[label=4\alph*.,ref=4\alph*]
\item\label{rule:4a}  A real Fermi $\Lambda^R_{ii}$ which transforms into $\Lambda^R_{i'i'}$, with $i'\neq i$, are projected down to a single real Fermi $\Lambda^R_{ii}$.
\item\label{rule:4b} A real Fermi $\Lambda^R_{ii}$ mapped to itself (resp. minus itself), with $i'\neq i$, gives rise to a symmetric (resp. antisymmetric) real Fermi for an $\SO$ (resp. $\USp$) projection of the node $i$.   
\end{enumerate}

These rules suffice to construct large classes of examples of 2d $\mathcal{N}=(0,1)$ field theories, in particular the explicit examples in coming sections.

Note that the $\mathcal{N}=(0,1)$ theory obtained upon orientifolding the parent $\mathcal{N}=(0,2)$ may have non-abelian gauge anomalies. In such cases, the models require the introduction of extra flavor branes (namely, D5- or D9-branes extending in the non-compact dimensions of the CY 4-fold) for consistency. As already remarked in \cite{Franco:2021ixh}, very often the orientifolded theories happen to be non-anomalous, and hence do not require flavor branes. This will be the case in our examples later on.

\subsection*{The Universal Involution}

We would like to conclude this overview by recalling from \cite{Franco:2021ixh} that any $\mathcal{N}=(0,2)$ quiver gauge theory from D1-branes at toric CY 4-fold singularities admits a universal anti-holomorphic involution. It corresponds to mapping each gauge factor to itself (maintaining all with the same $\SO$ or $\USp$ projection), and mapping every $\mathcal{N}=(0,2)$ chiral or Fermi field  to itself anti-holomorphically.

To be more explicit, let us introduce a set of matrices $\gamma_{\Omega_i}$ implementing the action of the orientifold on the gauge degrees of freedom of the $i^{th}$ node.\footnote{Actually, the matrices $\gamma_{\Omega_i}$ are a useful ingredient in implementing the orientifold projection, even in examples beyond the universal involution, as we will exploit in explicit examples in later sections.\label{foot:gamma-omega}} Then, the orientifold projections for the universal involution read
\begin{equation}
    X_{ij}\rightarrow \gamma_{\Omega_i}\bar{X}_{ij}\gamma_{\Omega_j}^{-1} \coma \Lambda_{ij}\rightarrow \gamma_{\Omega_i}\bar{\Lambda}_{ij}\gamma_{\Omega_j}^{-1}\coma
    \label{eq:genUImap}
\end{equation}
where, by $X_{ij}$ or $\Lambda_{ij}$ we mean any chiral or Fermi field present in the gauge theory. In addition, the $\mathcal{N}=(0,1)$ adjoint Fermi fields coming from $\mathcal{N}=(0,2)$ vector multiplets transform as 
\begin{equation}
\Lambda_i^R \to \gamma_{\Omega_{i}} \Lambda_{i'}^{R \, \, T} \gamma_{\Omega_{i}}^{-1} \, .
\label{Lambda^R_projection}
\end{equation}
There is relative sign between this projection and the one for gauge fields, which implies that an $\SO$ or $\USp$ projection of the gauge group is correlated with a projection of $\Lambda_i^R$ into a symmetric or antisymmetric representation, respectively. These projections are consistent with the invariance of the $\mathcal{N}=(0,1)$ superpotential. Modding out by this orientifold action, the resulting $\mathcal{N}=(0,1)$ field theory is determined by applying the above rules.

From the geometric perspective, this universal involution corresponds to the conjugation of all generators of the toric CY 4-fold. The action on the holomorphic 4-form is $\Omega^{(4,0)}\rightarrow \bar{\Omega}^{(0,4)}$, suitable for the realization of an Spin(7) orientifold. The following section focuses on models obtained via the universal involution.

\section{$\mathcal{N}=(0,1)$ Triality and the Universal Involution}
\label{sec:01trial-UI}

Let us consider what happens when the universal involution is applied to two gauge theories associated to the same CY 4-fold, which are therefore related by $\mathcal{N}=(0,2)$ triality. Remarkably, we obtain two theories connected by precisely $\mathcal{N}=(0,1)$ triality. By construction, the two theories correspond to the same underlying Spin(7) orientifold, realizing the general idea of $\mathcal{N}=(0,1)$ triality arising from the non-uniqueness of the map between Spin(7) orientifolds and gauge theories.

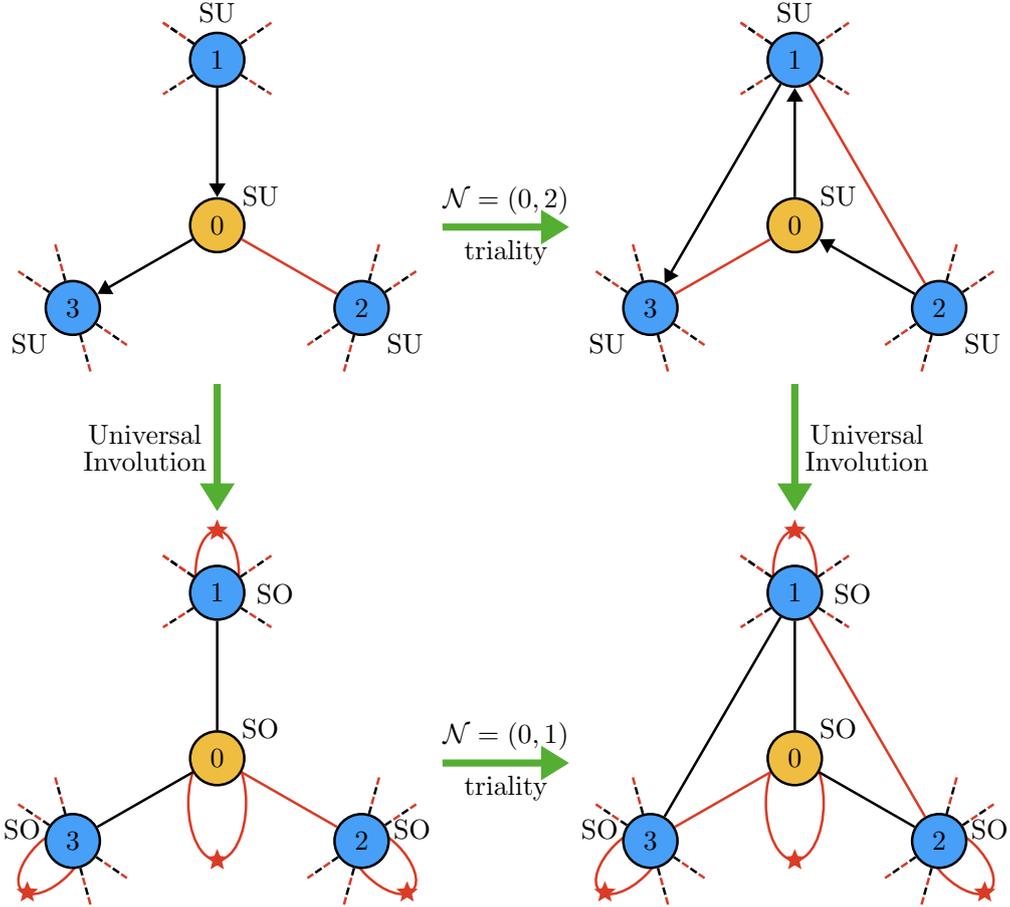
\begin{figure}[H]
    \centering
    \scalebox{0.95}{
    \begin{tikzpicture}[scale=4]
    \node (NW) at (0,1.85) {
    \begin{tikzpicture}[scale=2]
	    \def\L{2};
	     \node[draw=black,line width=1pt,circle,fill=yellowX,minimum width=0.75cm,inner sep=1pt,label={[xshift=0.6cm,yshift=-0.25cm]:$\SU$}] (Nc) at (0,-0.289) {$0$};
	     \node[draw=black,line width=1pt,fill=blueX,circle,minimum width=0.75cm,inner sep=1pt,label={[xshift=0.0cm,yshift=0cm]:$\SU$}] (N1) at (0,\L*0.433) {$1$};
	     \node[draw=black,line width=1pt,fill=blueX,circle,minimum width=0.75cm,inner sep=1pt,label={[xshift=0.6cm,yshift=-1.15cm]:$\SU$}] (N2) at (\L/2,-\L*0.433) {$2$};
	     \node[draw=black,line width=1pt,fill=blueX,circle,minimum width=0.75cm,inner sep=1pt,label={[xshift=-0.6cm,yshift=-1.15cm]:$\SU$}] (N3) at (-\L/2,-\L*0.433) {$3$};
	     \draw[line width=1pt,-Triangle] (N1) --  (Nc);
	     \draw[line width=1pt,redX] (Nc) -- (N2);
	     \draw[line width=1pt,-Triangle] (Nc) -- (N3);
	     \draw[line width=1pt,postaction={draw,redX,dash pattern= on 3pt off 5pt,dash phase=4pt}] [line width=1pt,black,dash pattern= on 3pt off 5pt] (N1) -- (-0.375,1.125);
	     \draw[line width=1pt,postaction={draw,redX,dash pattern= on 3pt off 5pt,dash phase=4pt}] [line width=1pt,black,dash pattern= on 3pt off 5pt] (N1) -- (-0.375,1.125);
	     \draw[line width=1pt,postaction={draw,redX,dash pattern= on 3pt off 5pt,dash phase=4pt}] [line width=1pt,black,dash pattern= on 3pt off 5pt] (N1) -- (-0.375,0.625);
	     \draw[line width=1pt,postaction={draw,redX,dash pattern= on 3pt off 5pt,dash phase=4pt}] [line width=1pt,black,dash pattern= on 3pt off 5pt] (N1) -- (0.375,1.125);
	     \draw[line width=1pt,postaction={draw,redX,dash pattern= on 3pt off 5pt,dash phase=4pt}] [line width=1pt,black,dash pattern= on 3pt off 5pt] (N1) -- (0.375,0.625);
	     \draw[line width=1pt,postaction={draw,redX,dash pattern= on 3pt off 5pt,dash phase=4pt}] [line width=1pt,black,dash pattern= on 3pt off 5pt] (N2) -- (1.125,-0.41);
	     \draw[line width=1pt,postaction={draw,redX,dash pattern= on 3pt off 5pt,dash phase=4pt}] [line width=1pt,black,dash pattern= on 3pt off 5pt] (N2) -- (1.378,-0.61);
	     \draw[line width=1pt,postaction={draw,redX,dash pattern= on 3pt off 5pt,dash phase=4pt}] [line width=1pt,black,dash pattern= on 3pt off 5pt] (N2) -- (0.875,-1.32);
	     \draw[line width=1pt,postaction={draw,redX,dash pattern= on 3pt off 5pt,dash phase=4pt}] [line width=1pt,black,dash pattern= on 3pt off 5pt] (N2) -- (0.625,-1.125);
	     \draw[line width=1pt,postaction={draw,redX,dash pattern= on 3pt off 5pt,dash phase=4pt}] [line width=1pt,black,dash pattern= on 3pt off 5pt] (N3) -- (-1.125,-0.41);
	     \draw[line width=1pt,postaction={draw,redX,dash pattern= on 3pt off 5pt,dash phase=4pt}] [line width=1pt,black,dash pattern= on 3pt off 5pt] (N3) -- (-1.378,-0.61);
	     \draw[line width=1pt,postaction={draw,redX,dash pattern= on 3pt off 5pt,dash phase=4pt}] [line width=1pt,black,dash pattern= on 3pt off 5pt] (N3) -- (-0.875,-1.32);
	     \draw[line width=1pt,postaction={draw,redX,dash pattern= on 3pt off 5pt,dash phase=4pt}] [line width=1pt,black,dash pattern= on 3pt off 5pt] (N3) -- (-0.625,-1.125);
	    \end{tikzpicture}
    };
    \node (NE) at (2,1.85) {
    \begin{tikzpicture}[scale=2]
	    \def\L{2};
	     \node[draw=black,line width=1pt,circle,fill=yellowX,minimum width=0.75cm,inner sep=1pt,label={[xshift=0.6cm,yshift=-0.25cm]:$\SU$}] (Nc) at (0,-0.289) {$0$};
	     \node[draw=black,line width=1pt,fill=blueX,circle,minimum width=0.75cm,inner sep=1pt,label={[xshift=0.0cm,yshift=0cm]:$\SU$}] (N1) at (0,\L*0.433) {$1$};
	     \node[draw=black,line width=1pt,fill=blueX,circle,minimum width=0.75cm,inner sep=1pt,label={[xshift=0.6cm,yshift=-1.15cm]:$\SU$}] (N2) at (\L/2,-\L*0.433) {$2$};
	     \node[draw=black,line width=1pt,fill=blueX,circle,minimum width=0.75cm,inner sep=1pt,label={[xshift=-0.6cm,yshift=-1.15cm]:$\SU$}] (N3) at (-\L/2,-\L*0.433) {$3$};
	     \draw[line width=1pt,-Triangle] (Nc) --  (N1);
	     \draw[line width=1pt,-Triangle] (N2) -- (Nc);
	     \draw[line width=1pt,-Triangle] (N1) -- (N3);
	     \draw[line width=1pt,redX] (Nc) -- (N3);
	     \draw[line width=1pt,redX] (N2) -- (N1);
	     \draw[line width=1pt,postaction={draw,redX,dash pattern= on 3pt off 5pt,dash phase=4pt}] [line width=1pt,black,dash pattern= on 3pt off 5pt] (N1) -- (-0.375,1.125);
	     \draw[line width=1pt,postaction={draw,redX,dash pattern= on 3pt off 5pt,dash phase=4pt}] [line width=1pt,black,dash pattern= on 3pt off 5pt] (N1) -- (-0.375,1.125);
	     \draw[line width=1pt,postaction={draw,redX,dash pattern= on 3pt off 5pt,dash phase=4pt}] [line width=1pt,black,dash pattern= on 3pt off 5pt] (N1) -- (-0.375,0.625);
	     \draw[line width=1pt,postaction={draw,redX,dash pattern= on 3pt off 5pt,dash phase=4pt}] [line width=1pt,black,dash pattern= on 3pt off 5pt] (N1) -- (0.375,1.125);
	     \draw[line width=1pt,postaction={draw,redX,dash pattern= on 3pt off 5pt,dash phase=4pt}] [line width=1pt,black,dash pattern= on 3pt off 5pt] (N1) -- (0.375,0.625);
	     \draw[line width=1pt,postaction={draw,redX,dash pattern= on 3pt off 5pt,dash phase=4pt}] [line width=1pt,black,dash pattern= on 3pt off 5pt] (N2) -- (1.125,-0.41);
	     \draw[line width=1pt,postaction={draw,redX,dash pattern= on 3pt off 5pt,dash phase=4pt}] [line width=1pt,black,dash pattern= on 3pt off 5pt] (N2) -- (1.378,-0.61);
	     \draw[line width=1pt,postaction={draw,redX,dash pattern= on 3pt off 5pt,dash phase=4pt}] [line width=1pt,black,dash pattern= on 3pt off 5pt] (N2) -- (0.875,-1.32);
	     \draw[line width=1pt,postaction={draw,redX,dash pattern= on 3pt off 5pt,dash phase=4pt}] [line width=1pt,black,dash pattern= on 3pt off 5pt] (N2) -- (0.625,-1.125);
	     \draw[line width=1pt,postaction={draw,redX,dash pattern= on 3pt off 5pt,dash phase=4pt}] [line width=1pt,black,dash pattern= on 3pt off 5pt] (N3) -- (-1.125,-0.41);
	     \draw[line width=1pt,postaction={draw,redX,dash pattern= on 3pt off 5pt,dash phase=4pt}] [line width=1pt,black,dash pattern= on 3pt off 5pt] (N3) -- (-1.378,-0.61);
	     \draw[line width=1pt,postaction={draw,redX,dash pattern= on 3pt off 5pt,dash phase=4pt}] [line width=1pt,black,dash pattern= on 3pt off 5pt] (N3) -- (-0.875,-1.32);
	     \draw[line width=1pt,postaction={draw,redX,dash pattern= on 3pt off 5pt,dash phase=4pt}] [line width=1pt,black,dash pattern= on 3pt off 5pt] (N3) -- (-0.625,-1.125);
	    \end{tikzpicture}
    
    };
    \node (SW) at (0,0) {
      \begin{tikzpicture}[scale=2]
	    \def\L{2};
	    \draw[line width=1pt,redX] (0,-0.6) ellipse (0.2 and 0.4)  node[xshift=0cm,yshift=-0.8cm,star,star points=5, star point ratio=2.25, inner sep=1pt, fill=redX, draw=redX] {};
	    \draw[line width=1pt,redX,rotate around ={45:(\L/2*1.14,-\L*0.5)}] (\L/2*1.14,-\L*0.5) ellipse (0.15 and 0.3)  node[xshift=0.35cm,yshift=-0.45cm,star,star points=5, star point ratio=2.25, inner sep=1pt, fill=redX, draw=redX] {};
	    \draw[line width=1pt,redX,rotate around ={-45:(-\L/2*1.14,-\L*0.5)}] (-\L/2*1.14,-\L*0.5) ellipse (0.15 and 0.3)  node[xshift=-0.35cm,yshift=-0.45cm,star,star points=5, star point ratio=2.25, inner sep=1pt, fill=redX, draw=redX] {};
	    \draw[line width=1pt,redX] (0,\L*0.5) ellipse (0.15 and 0.3)  node[xshift=0cm,yshift=0.6cm,star,star points=5, star point ratio=2.25, inner sep=1pt, fill=redX, draw=redX] {};
	     \node[draw=black,line width=1pt,circle,fill=yellowX,minimum width=0.75cm,inner sep=1pt,label={[xshift=0.6cm,yshift=-0.25cm]:$\SO$}] (Nc) at (0,-0.289) {$0$};
	     \node[draw=black,line width=1pt,fill=blueX,circle,minimum width=0.75cm,inner sep=1pt,label={[xshift=0.8cm,yshift=-0.68cm]:$\SO$}] (N1) at (0,\L*0.433) {$1$};
	     \node[draw=black,line width=1pt,fill=blueX,circle,minimum width=0.75cm,inner sep=1pt,label={[xshift=0.7cm,yshift=-0.5cm]:$\SO$}] (N2) at (\L/2,-\L*0.433) {$2$};
	     \node[draw=black,line width=1pt,fill=blueX,circle,minimum width=0.75cm,inner sep=1pt,label={[xshift=-0.7cm,yshift=-0.5cm]:$\SO$}] (N3) at (-\L/2,-\L*0.433) {$3$};
	     \draw[line width=1pt,] (N1) --  (Nc);
	     \draw[line width=1pt,redX] (Nc) -- (N2);
	     \draw[line width=1pt,] (Nc) -- (N3);
	     \draw[line width=1pt,postaction={draw,redX,dash pattern= on 3pt off 5pt,dash phase=4pt}] [line width=1pt,black,dash pattern= on 3pt off 5pt] (N1) -- (-0.375,1.125);
	     \draw[line width=1pt,postaction={draw,redX,dash pattern= on 3pt off 5pt,dash phase=4pt}] [line width=1pt,black,dash pattern= on 3pt off 5pt] (N1) -- (-0.375,1.125);
	     \draw[line width=1pt,postaction={draw,redX,dash pattern= on 3pt off 5pt,dash phase=4pt}] [line width=1pt,black,dash pattern= on 3pt off 5pt] (N1) -- (-0.375,0.625);
	     \draw[line width=1pt,postaction={draw,redX,dash pattern= on 3pt off 5pt,dash phase=4pt}] [line width=1pt,black,dash pattern= on 3pt off 5pt] (N1) -- (0.375,1.125);
	     \draw[line width=1pt,postaction={draw,redX,dash pattern= on 3pt off 5pt,dash phase=4pt}] [line width=1pt,black,dash pattern= on 3pt off 5pt] (N1) -- (0.375,0.625);
	     \draw[line width=1pt,postaction={draw,redX,dash pattern= on 3pt off 5pt,dash phase=4pt}] [line width=1pt,black,dash pattern= on 3pt off 5pt] (N2) -- (1.125,-0.41);
	     \draw[line width=1pt,postaction={draw,redX,dash pattern= on 3pt off 5pt,dash phase=4pt}] [line width=1pt,black,dash pattern= on 3pt off 5pt] (N2) -- (1.378,-0.61);
	     \draw[line width=1pt,postaction={draw,redX,dash pattern= on 3pt off 5pt,dash phase=4pt}] [line width=1pt,black,dash pattern= on 3pt off 5pt] (N2) -- (0.875,-1.32);
	     \draw[line width=1pt,postaction={draw,redX,dash pattern= on 3pt off 5pt,dash phase=4pt}] [line width=1pt,black,dash pattern= on 3pt off 5pt] (N2) -- (0.625,-1.125);
	     \draw[line width=1pt,postaction={draw,redX,dash pattern= on 3pt off 5pt,dash phase=4pt}] [line width=1pt,black,dash pattern= on 3pt off 5pt] (N3) -- (-1.125,-0.41);
	     \draw[line width=1pt,postaction={draw,redX,dash pattern= on 3pt off 5pt,dash phase=4pt}] [line width=1pt,black,dash pattern= on 3pt off 5pt] (N3) -- (-1.378,-0.61);
	     \draw[line width=1pt,postaction={draw,redX,dash pattern= on 3pt off 5pt,dash phase=4pt}] [line width=1pt,black,dash pattern= on 3pt off 5pt] (N3) -- (-0.875,-1.32);
	     \draw[line width=1pt,postaction={draw,redX,dash pattern= on 3pt off 5pt,dash phase=4pt}] [line width=1pt,black,dash pattern= on 3pt off 5pt] (N3) -- (-0.625,-1.125);
	    \end{tikzpicture}
    
    };
    \node (SE) at (2,0) {
    \begin{tikzpicture}[scale=2]
	    \def\L{2};
	      \draw[line width=1pt,redX] (0,-0.6) ellipse (0.2 and 0.4)  node[xshift=0cm,yshift=-0.8cm,star,star points=5, star point ratio=2.25, inner sep=1pt, fill=redX, draw=redX] {};
	    \draw[line width=1pt,redX,rotate around ={45:(\L/2*1.14,-\L*0.5)}] (\L/2*1.14,-\L*0.5) ellipse (0.15 and 0.3)  node[xshift=0.35cm,yshift=-0.45cm,star,star points=5, star point ratio=2.25, inner sep=1pt, fill=redX, draw=redX] {};
	    \draw[line width=1pt,redX,rotate around ={-45:(-\L/2*1.14,-\L*0.5)}] (-\L/2*1.14,-\L*0.5) ellipse (0.15 and 0.3)  node[xshift=-0.35cm,yshift=-0.45cm,star,star points=5, star point ratio=2.25, inner sep=1pt, fill=redX, draw=redX] {};
	    \draw[line width=1pt,redX] (0,\L*0.5) ellipse (0.15 and 0.3)  node[xshift=0cm,yshift=0.6cm,star,star points=5, star point ratio=2.25, inner sep=1pt, fill=redX, draw=redX] {};
	     \node[draw=black,line width=1pt,circle,fill=yellowX,minimum width=0.75cm,inner sep=1pt,label={[xshift=0.6cm,yshift=-0.25cm]:$\SO$}] (Nc) at (0,-0.289) {$0$};
	     \node[draw=black,line width=1pt,fill=blueX,circle,minimum width=0.75cm,inner sep=1pt,label={[xshift=0.8cm,yshift=-0.68cm]:$\SO$}] (N1) at (0,\L*0.433) {$1$};
	     \node[draw=black,line width=1pt,fill=blueX,circle,minimum width=0.75cm,inner sep=1pt,label={[xshift=0.7cm,yshift=-0.5cm]:$\SO$}] (N2) at (\L/2,-\L*0.433) {$2$};
	     \node[draw=black,line width=1pt,fill=blueX,circle,minimum width=0.75cm,inner sep=1pt,label={[xshift=-0.7cm,yshift=-0.5cm]:$\SO$}] (N3) at (-\L/2,-\L*0.433) {$3$};
	     \draw[line width=1pt] (Nc) --  (N1);
	     \draw[line width=1pt] (N2) -- (Nc);
	     \draw[line width=1pt] (N1) -- (N3);
	     \draw[line width=1pt,redX] (Nc) -- (N3);
	     \draw[line width=1pt,redX] (N2) -- (N1);
	     \draw[line width=1pt,postaction={draw,redX,dash pattern= on 3pt off 5pt,dash phase=4pt}] [line width=1pt,black,dash pattern= on 3pt off 5pt] (N1) -- (-0.375,1.125);
	     \draw[line width=1pt,postaction={draw,redX,dash pattern= on 3pt off 5pt,dash phase=4pt}] [line width=1pt,black,dash pattern= on 3pt off 5pt] (N1) -- (-0.375,1.125);
	     \draw[line width=1pt,postaction={draw,redX,dash pattern= on 3pt off 5pt,dash phase=4pt}] [line width=1pt,black,dash pattern= on 3pt off 5pt] (N1) -- (-0.375,0.625);
	     \draw[line width=1pt,postaction={draw,redX,dash pattern= on 3pt off 5pt,dash phase=4pt}] [line width=1pt,black,dash pattern= on 3pt off 5pt] (N1) -- (0.375,1.125);
	     \draw[line width=1pt,postaction={draw,redX,dash pattern= on 3pt off 5pt,dash phase=4pt}] [line width=1pt,black,dash pattern= on 3pt off 5pt] (N1) -- (0.375,0.625);
	     \draw[line width=1pt,postaction={draw,redX,dash pattern= on 3pt off 5pt,dash phase=4pt}] [line width=1pt,black,dash pattern= on 3pt off 5pt] (N2) -- (1.125,-0.41);
	     \draw[line width=1pt,postaction={draw,redX,dash pattern= on 3pt off 5pt,dash phase=4pt}] [line width=1pt,black,dash pattern= on 3pt off 5pt] (N2) -- (1.378,-0.61);
	     \draw[line width=1pt,postaction={draw,redX,dash pattern= on 3pt off 5pt,dash phase=4pt}] [line width=1pt,black,dash pattern= on 3pt off 5pt] (N2) -- (0.875,-1.32);
	     \draw[line width=1pt,postaction={draw,redX,dash pattern= on 3pt off 5pt,dash phase=4pt}] [line width=1pt,black,dash pattern= on 3pt off 5pt] (N2) -- (0.625,-1.125);
	     \draw[line width=1pt,postaction={draw,redX,dash pattern= on 3pt off 5pt,dash phase=4pt}] [line width=1pt,black,dash pattern= on 3pt off 5pt] (N3) -- (-1.125,-0.41);
	     \draw[line width=1pt,postaction={draw,redX,dash pattern= on 3pt off 5pt,dash phase=4pt}] [line width=1pt,black,dash pattern= on 3pt off 5pt] (N3) -- (-1.378,-0.61);
	     \draw[line width=1pt,postaction={draw,redX,dash pattern= on 3pt off 5pt,dash phase=4pt}] [line width=1pt,black,dash pattern= on 3pt off 5pt] (N3) -- (-0.875,-1.32);
	     \draw[line width=1pt,postaction={draw,redX,dash pattern= on 3pt off 5pt,dash phase=4pt}] [line width=1pt,black,dash pattern= on 3pt off 5pt] (N3) -- (-0.625,-1.125);
	    \end{tikzpicture}
    };
    \draw[-Triangle,greenX,line width=1mm] (0.78,1.70) -- node[above,midway,black] {$\mathcal{N}=(0,2)$} node[below,midway,black] {triality} (1.22,1.70);
    \draw[-Triangle,greenX,line width=1mm] (0.78,-0.17) -- node[above,midway,black] {$\mathcal{N}=(0,1)$} node[below,midway,black] {triality} (1.22,-0.17);
    \draw[-Triangle,greenX,line width=1mm] (NW) -- node[xshift=-1cm,midway,black] {\shortstack{Universal\\Involution}} (SW);
    \draw[-Triangle,greenX,line width=1mm] (NE) -- node[xshift=1cm,midway,black] {\shortstack{Universal\\Involution}} (SE);
    \end{tikzpicture}
    }
 \caption{The universal involution on $\mathcal{N}=(0,2)$ triality results in $\mathcal{N}=(0,1)$ triality.}
\label{universal_involution_triality}
\end{figure}

We illustrate this projection in Figure \ref{universal_involution_triality}, which shows the neighborhood of the quiver around a dualized node $0$.\footnote{The universal involution with $\USp$ projection is analogous.} As in the previous section, nodes 1, 2 and 3 represent possibly multiple nodes which, in turn, might be connected to node $0$ by different multiplicities of fields. The red and black dashed lines represent the rest of the quiver, which might include fields stretching between nodes 1, 2 and 3. If triality generates massive fields, they can be integrated out.

An explicit example of a triality pairs associated to the universal involution will be presented in Section \ref{section_H4}. However, in Section \ref{sec:BeyonUI}, we will show how more general orientifold actions lead to interesting generalizations of the basic $\mathcal{N}=(0,1)$ triality. The general strategy will be to focus on parent CY$_4$ geometries with more than one $\mathcal{N}=(0,2)$ triality dual toric phases\footnote{We refer to a toric phase as one associated to a brane brick model \cite{Franco:2015tya}, for which the connection to the underlying CY$_4$ is considerably simplified.} (see e.g. \cite{Franco:2016nwv,Franco:2018qsc}) and to consider anti-holomorphic involutions leading to the same $\Spin(7)$ orientifold.

\subsection{The Universal Involution of $H_4$}
\label{section_H4}

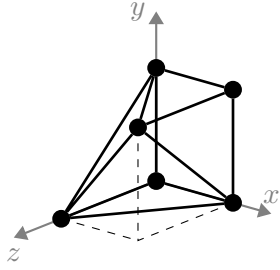
\begin{figure}[H]
			\centering
			\begin{tikzpicture}[scale=1.5, rotate around y=-30]
			\draw[thick,gray,-Triangle] (0,0,0) -- node[above,pos=1] {$x$} (1.5,0,0);
			\draw[thick,gray,-Triangle] (0,0,0) -- node[left,pos=1] {$y$} (0,1.5,0);
			\draw[thick,gray,-Triangle] (0,0,0) -- node[below,pos=1] {$z$} (0,0,1.5);
			\node[draw=black,line width=1pt,circle,fill=black,minimum width=0.2cm,inner sep=1pt] (O) at (0,0,0) {};
			\node[draw=black,line width=1pt,circle,fill=black,minimum width=0.2cm,inner sep=1pt] (A) at (1,0,0) {};
			\node[draw=black,line width=1pt,circle,fill=black,minimum width=0.2cm,inner sep=1pt] (B) at (0,1,0) {};
			\node[draw=black,line width=1pt,circle,fill=black,minimum width=0.2cm,inner sep=1pt] (C) at (0,0,1) {};
			\node[draw=black,line width=1pt,circle,fill=black,minimum width=0.2cm,inner sep=1pt] (D) at (1,1,1) {};
			\node[draw=black,line width=1pt,circle,fill=black,minimum width=0.2cm,inner sep=1pt] (E) at (1,1,0) {};
			\draw[line width=1pt] (O)--(A)--(E)--(B)--(O);
			\draw[line width=1pt] (O)--(C)--(B)--(D)--(A)--(C)--(D)--(E);
			\draw[thin,dashed] (1,0,0) -- (1,0,1) -- (1,1,1);
			\draw[thin,dashed] (0,0,1) -- (1,0,1);
			\end{tikzpicture}
			\caption{Toric diagram for $H_4$.}
			\label{fig:H4toricdiagram}
	\end{figure}

As explained above, the universal involution works for every CY$_4$. Therefore, it is sufficient to present one example to illustrate the main features of the construction. Let us consider the CY$_4$ with toric diagram shown in Figure \ref{fig:H4toricdiagram}, which is often referred to as $H_4$. Below we consider two toric phases for D1-branes probing $H_4$ and construct the $\mathcal{N}=(0,1)$ theories that correspond to them via the universal involution.

\subsubsection{Phase A}
	\label{sec:H4PhaseA}

\begin{figure}[H]
		\centering
			\begin{tikzpicture}[scale=2]
			\draw[line width=1pt,decoration={markings, mark=at position 0.75 with{\arrow{Triangle}}}, postaction={decorate}] (2.22,-0.22) circle (0.25);
			\draw[line width=1pt,redX] (-0.22,2.22) circle (0.25) node[fill=white,text opacity=1,fill opacity=1,draw=black,rectangle,xshift=-0.35cm,yshift=0.35cm,thin] {\color{redX}{$3$}};
			\node[draw=black,line width=1pt,circle,fill=yellowX,minimum width=0.75cm,inner sep=1pt] (A) at (0,0) {$4$};
			\node[draw=black,line width=1pt,circle,fill=yellowX,minimum width=0.75cm,inner sep=1pt] (B) at (2,0) {$3$};
			\node[draw=black,line width=1pt,circle,fill=yellowX,minimum width=0.75cm,inner sep=1pt] (C) at (2,2) {$2$};
			\node[draw=black,line width=1pt,circle,fill=yellowX,minimum width=0.75cm,inner sep=1pt] (D) at (0,2) {$1$};
			\path[Triangle-] (A) edge[line width=1pt] (B);
			\path[-Triangle] (B) edge[line width=1pt] (C);
			\path[Triangle-] (C) edge[line width=1pt] (D);
			\path[-Triangle] (D) edge[line width=1pt] (A);
			\draw[line width=1pt,redX] (A) to[bend right=20] (B);
			\draw[line width=1pt,redX] (B) to[bend right=20] (C);
			\draw[line width=1pt,-Triangle] (C) to[bend right=20] node[fill=white,text opacity=1,fill opacity=1,draw=black,rectangle,thin,pos=0.5] {$3$} (D);
			\draw[line width=1pt,Triangle-] (D) to[bend right=20] node[fill=white,text opacity=1,fill opacity=1,draw=black,rectangle,thin,pos=0.5] {$3$} (A);
			\draw[line width=1pt,redX] (A) to node[fill=white,text opacity=1,fill opacity=1,draw=black,rectangle,thin,pos=0.75] {\color{redX}{$2$}} (C);
			\draw[line width=1pt,redX] (D) to[bend right=10] node[fill=white,text opacity=1,fill opacity=1,draw=black,rectangle,thin,pos=0.25] {\color{redX}{$2$}} (B);
			\draw[line width=1pt,-Triangle] (D) to[bend left=10] node[fill=white,text opacity=1,fill opacity=1,draw=black,rectangle,thin,pos=0.75] {$2$} (B);
			\node at (0,-0.6) {};
			\end{tikzpicture}
		\caption{Quiver diagram for phase A of $H_4$.}
		\label{fig:H4AquivN02}
	\end{figure}
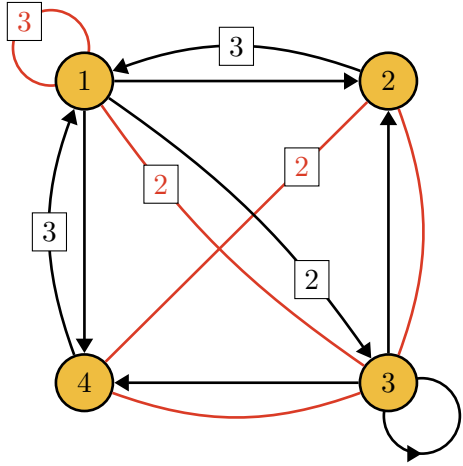

Figure \ref{fig:H4AquivN02} shows the quiver diagram for one of the toric phases of $H_4$, which we denote phase A. This theory was first introduced in \cite{Franco:2017cjj}.
	
The corresponding $J$- and $E$-terms are 
\begin{alignat}{4}
	\renewcommand{\arraystretch}{1.1}
    & \centermathcell{J}                           &\text{\hspace{.5cm}}& \centermathcell{E                               }\nonumber \\
	\Lambda_{11}^1 \,:\, & \centermathcell{X_{14}X_{41} - X_{13}X_{32}Z_{21}  }&  & \centermathcell{Y_{13}X_{34}Z_{41}-X_{12}Y_{21}        }\nonumber \\
	\Lambda_{11}^2 \,:\, & \centermathcell{X_{14}Y_{41} - Y_{13}X_{32}Z_{21}  }&  & \centermathcell{X_{12}X_{21}  - X_{13}X_{34}Z_{41}     }\nonumber \\
	\Lambda_{11}^3 \,:\, & \centermathcell{X_{14}Z_{41} - X_{12}Z_{21}        }&  & \centermathcell{X_{13}X_{32}Y_{21} - Y_{13}X_{34}X_{41}}\nonumber \\
	\Lambda_{13}^1 \,:\, & \centermathcell{X_{32}X_{21}  - X_{34}X_{41}       }&  & \centermathcell{Y_{13}X_{33} - X_{14}Z_{41}Y_{13}      } \label{eq:H4PhaseAJEterms}\\
	\Lambda_{13}^2 \,:\, & \centermathcell{X_{32}Y_{21}  - X_{34}Y_{41}       }&  & \centermathcell{X_{12}Z_{21}X_{13} - X_{13}X_{33}      }\nonumber \\
	\Lambda_{42}^1 \,:\, & \centermathcell{X_{21}X_{14}  - Z_{21}X_{13}X_{34} }&  & \centermathcell{Z_{41}Y_{13}X_{32} -Y_{41}X_{12}       }\nonumber \\
	\Lambda_{42}^2 \,:\, & \centermathcell{Y_{21}X_{14}  - Z_{21}Y_{13}X_{34} }&  & \centermathcell{X_{41}X_{12} - Z_{41}X_{13}X_{32}      }\nonumber \\
	\Lambda_{23} \,:\,   & \centermathcell{X_{33}X_{32}  - X_{32}Z_{21}X_{12} }&  & \centermathcell{Y_{21}X_{13} -X_{21}Y_{13}             }\nonumber \\
	\Lambda_{43} \,:\,   & \centermathcell{X_{33}X_{34}  - X_{34}Z_{41}X_{14} }&  & \centermathcell{X_{41}Y_{13} - Y_{41}X_{13}            }\nonumber 
	\end{alignat}

The $\mathcal{N}=(0,1)$ superpotential is then
	\begin{equation}
	\begin{split}
	W^{(0,1)}=&\,W^{(0,2)}+ \Lambda_{11}^{4R}( X^\dagger_{12}X_{12}+ X^\dagger_{14}X_{14}+ X^\dagger_{21}X_{21}+ Y^\dagger_{21}Y_{21}+ Z^\dagger_{21}Z_{21}+\\
	&+ X^\dagger_{41}X_{41}+ Y^\dagger_{41}Y_{41}+ Z^\dagger_{41}Z_{41}+ X^\dagger_{13}X_{13}+ Y^\dagger_{13}Y_{13})+\\
	&+\Lambda_{22}^{R}( X^\dagger_{12}X_{12}+ X^\dagger_{32}X_{32}+ X^\dagger_{21}X_{21}+ Y^\dagger_{21}Y_{21}+ Z^\dagger_{21}Z_{21})+\\
	&+\Lambda_{33}^{R}( X^\dagger_{33}X_{33}+ X^\dagger_{32}X_{32}+ X^\dagger_{34}X_{34}+ X^\dagger_{13}X_{13}+ Y^\dagger_{13}Y_{13})+\\
	&+\Lambda_{44}^{R}( X^\dagger_{14}X_{14}+ X^\dagger_{34}X_{34}+ X^\dagger_{41}X_{41}+ Y^\dagger_{41}Y_{41}+ Z^\dagger_{41}Z_{41})\fstop
	\end{split}
	\end{equation}

The generators of $H_4$, which arises as the moduli space of the gauge theory, can be determined for instance using the Hilbert Series (HS) \cite{Benvenuti:2006qr,Feng:2007ur,Franco:2015tya} (see also \cite{Franco:2021ixh}). We list them in Table \ref{tab:GenerH4PhaseA}, together with their expressions as mesons in terms of chiral fields in phase A.
	\begin{table}[H]
		\centering
		\renewcommand{\arraystretch}{1.1}
		\begin{tabular}{c|c}
			Meson & Chiral superfields  \\
			\hline
			$M_1$ & $X_{33}=X_{14}Z_{41}=Z_{21}X_{12}$ \\
			$M_2$ & $Y_{21}X_{12}=Z_{41}Y_{13}X_{34}$ \\
			$M_3$ & $X_{14}Y_{41}=Z_{21}Y_{13}X_{32}$ \\
			$M_4$ & $X_{32}Y_{21}Y_{13}=X_{34}Y_{41}Y_{13}$ \\
			$M_5$ & $X_{21}X_{12}=Z_{41}X_{13}X_{34}$ \\
			$M_6$ & $X_{14}X_{41}=Z_{21}X_{13}X_{32}$ \\
			$M_7$ & $X_{32}Y_{21}X_{13}=X_{32}X_{21}Y_{13}=X_{34}Y_{41}X_{13}=X_{34}X_{41}Y_{13}$ \\
			$M_8$ & $X_{32}X_{21}X_{13}=X_{34}X_{41}X_{13}$ 
		\end{tabular}
		\caption{Generators of $H_4$ in terms of fields in phase A.}
		\label{tab:GenerH4PhaseA}
	\end{table}

The generators satisfy the following relations
	\begin{equation}
	\begin{split}
	\mathcal{I} = &\left\langle M_1M_4=M_2M_3\coma M_1M_7=M_2M_6\coma M_1M_7=M_3M_5\coma M_2M_7=M_4M_5\coma\right.\\
	&\left.M_3M_7=M_4M_6\coma M_1M_8=M_5M_6\coma M_2M_8=M_5M_7\coma M_3M_8=M_6M_7\coma\right.\\
	&\left.M_4M_8=M_7^2\right\rangle\fstop
	\label{eq:H4HSrel}
	\end{split}
	\end{equation}
	
The universal involution acts on the fields of any theory as \eqref{eq:genUImap}. This results in the expected map of the generators	
	\begin{equation}
     \begin{array}{cccccccccccc}
	M_{1}  & \rightarrow & \bar{M}_{1} \coma &
	M_{2}  & \rightarrow & \bar{M}_{2} \coma &
	M_{3}  & \rightarrow & \bar{M}_{3} \coma &
	M_{4}  & \rightarrow & \bar{M}_{4} \coma \\
	M_{5}  & \rightarrow & \bar{M}_{5} \coma &
	M_{6}  & \rightarrow & \bar{M}_{6} \coma &
	M_{7}  & \rightarrow & \bar{M}_{7} \coma &
	M_{8}  & \rightarrow & \bar{M}_{8} \fstop
	\end{array} 
	\label{eq:H4PhaseA-invol}
	\end{equation}
	
	The quiver for the $2$d $\mathcal{N}=(0,1)$ orientifold theory is shown in Figure \ref{(0,1) theory of Spin(7) orientifold from h4 phase A}. It is rather straightforward to write the projected superpotential but, for brevity, we will omit it here and in the examples that follow.
	
		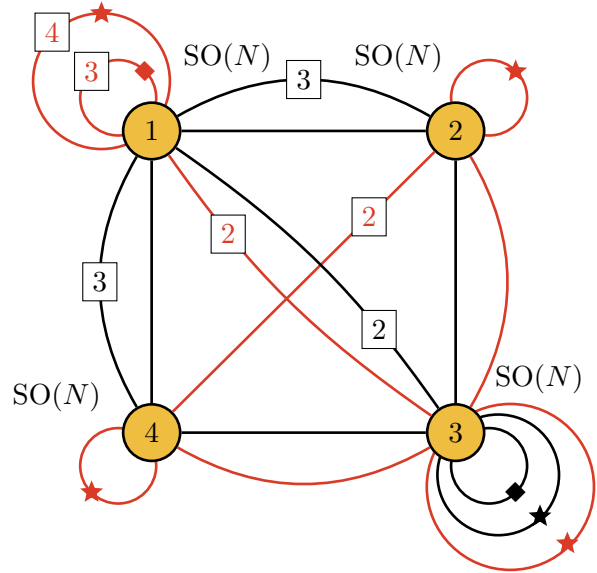
\begin{figure}[H]
	    \centering
	   	\begin{tikzpicture}[scale=2]
			\draw[line width=1pt] (2.22,-0.22) circle (0.25) node[xshift=0.35cm,yshift=-0.35cm] {\scriptsize{$\quadro$}};
			\draw[line width=1pt] (2.28,-0.28) circle (0.4)  node[xshift=0.55cm,yshift=-0.55cm,star,star points=5, star point ratio=2.25, inner sep=1pt, fill=black, draw=black] {};
			\draw[line width=1pt,redX] (2.36,-0.36) circle (0.55)  node[xshift=0.75cm,yshift=-0.75cm,star,star points=5, star point ratio=2.25, inner sep=1pt, fill=redX, draw=redX] {};
			\draw[line width=1pt,redX] (-0.22,-0.22) circle (0.25) node[xshift=-0.35cm,yshift=-0.35cm,star,star points=5, star point ratio=2.25, inner sep=1pt, fill=redX, draw=redX] {};
			\draw[line width=1pt,redX] (2.22,2.22) circle (0.25) node[xshift=0.35cm,yshift=0.35cm,star,star points=5, star point ratio=2.25, inner sep=1pt, fill=redX, draw=redX] {};
			\draw[line width=1pt,redX] (-0.22,2.22) circle (0.25)  node[xshift=0.35cm,yshift=0.35cm] {\color{redX}{\scriptsize{$\quadro$}}} node[fill=white,text opacity=1,fill opacity=1,draw=black,rectangle,xshift=-0.35cm,yshift=0.35cm,thin] {\color{redX}{$3$}};
			\draw[line width=1pt,redX] (-0.33,2.33) circle (0.45)  node[yshift=0.9cm,star,star points=5, star point ratio=2.25, inner sep=1pt, fill=redX, draw=redX] {} node[fill=white,text opacity=1,fill opacity=1,draw=black,rectangle,xshift=-0.65cm,yshift=0.65cm,thin] {\color{redX}{$4$}};
			\node[draw=black,line width=1pt,circle,fill=yellowX,minimum width=0.75cm,inner sep=1pt,label={[xshift=-1.25cm,yshift=-0.25cm]:$\SO(N)$}] (A) at (0,0) {$4$};
			\node[draw=black,line width=1pt,circle,fill=yellowX,minimum width=0.75cm,inner sep=1pt,label={[xshift=1.1cm,yshift=0cm]:$\SO(N)$}] (B) at (2,0) {$3$};
			\node[draw=black,line width=1pt,circle,fill=yellowX,minimum width=0.75cm,inner sep=1pt,label={[xshift=-0.75cm,yshift=0.25cm]:$\SO(N)$}] (C) at (2,2) {$2$};
			\node[draw=black,line width=1pt,circle,fill=yellowX,minimum width=0.75cm,inner sep=1pt,label={[xshift=1cm,yshift=0.25cm]:$\SO(N)$}] (D) at (0,2) {$1$};
			\path (A) edge[line width=1pt] (B);
			\path (B) edge[line width=1pt]  (C);
			\path (C) edge[line width=1pt] (D);
			\path (D) edge[line width=1pt] (A);
			\draw[line width=1pt,redX] (A) to[bend right=30] (B);
			\draw[line width=1pt,redX] (B) to[bend right=30] (C);
			\draw[line width=1pt] (C) to[bend right=30] node[fill=white,text opacity=1,fill opacity=1,draw=black,rectangle,thin,pos=0.5] {$3$} (D);
			\draw[line width=1pt] (D) to[bend right=30] node[fill=white,text opacity=1,fill opacity=1,draw=black,rectangle,thin,pos=0.5] {$3$} (A);
			\draw[line width=1pt,redX] (A) to node[fill=white,text opacity=1,fill opacity=1,draw=black,rectangle,thin,pos=0.75] {\color{redX}{$2$}} (C);
			\draw[line width=1pt,redX] (D) to[bend right=10] node[fill=white,text opacity=1,fill opacity=1,draw=black,rectangle,thin,pos=0.25] {\color{redX}{$2$}} (B);
			\draw[line width=1pt] (D) to[bend left=10] node[fill=white,text opacity=1,fill opacity=1,draw=black,rectangle,thin,pos=0.75] {$2$} (B);
			\end{tikzpicture}
	    \caption{Quiver diagram for the Spin(7) orientifold of phase A of $H_4$ using the universal involution.}
		\label{(0,1) theory of Spin(7) orientifold from h4 phase A}
	\end{figure}

\subsubsection{Phase B}
	\label{sec:H4PhaseB}

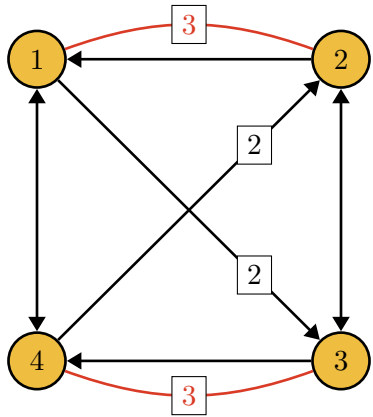
\begin{figure}[H]
		\centering
			\begin{tikzpicture}[scale=2]
			\node[draw=black,line width=1pt,circle,fill=yellowX,minimum width=0.75cm,inner sep=1pt] (A) at (0,0) {$4$};
			\node[draw=black,line width=1pt,circle,fill=yellowX,minimum width=0.75cm,inner sep=1pt] (B) at (2,0) {$3$};
			\node[draw=black,line width=1pt,circle,fill=yellowX,minimum width=0.75cm,inner sep=1pt] (C) at (2,2) {$2$};
			\node[draw=black,line width=1pt,circle,fill=yellowX,minimum width=0.75cm,inner sep=1pt] (D) at (0,2) {$1$};
			\path[-Triangle] (A) edge[line width=1pt] node[fill=white,text opacity=1,fill opacity=1,draw=black,rectangle,thin,pos=0.75] {$2$} (C);
			\path[-Triangle] (B) edge[line width=1pt] (A);
			\path[-Triangle] (D) edge[line width=1pt] node[fill=white,text opacity=1,fill opacity=1,draw=black,rectangle,thin,pos=0.75] {$2$} (B);
			\path[-Triangle] (C) edge[line width=1pt] (D);
			\draw[line width=1pt,redX] (A) to[bend right=20] node[fill=white,text opacity=1,fill opacity=1,draw=black,rectangle,thin,pos=0.5] {\color{redX}{$3$}} (B);
			\draw[line width=1pt,redX] (C) to[bend right=20] node[fill=white,text opacity=1,fill opacity=1,draw=black,rectangle,thin,pos=0.5] {\color{redX}{$3$}} (D);
			\draw[line width=1pt,Triangle-Triangle] (A) to  (D);
			\draw[line width=1pt,Triangle-Triangle] (B) to  (C);
			\node at (0,-0.42) {};
			\end{tikzpicture}
		\caption{Quiver diagram for phase B of $H_4$.}
			\label{fig:H4BN02}
	\end{figure}

Let us now consider the so-called phase $B$ of $H_4$ \cite{Franco:2017cjj}. Its quiver diagram is shown in Figure \ref{fig:H4BN02}.

\vspace{0.6cm}

The $J$- and $E$-terms are 
\begin{alignat}{4}
	\renewcommand{\arraystretch}{1.1}
    & \centermathcell{J}                           &\text{\hspace{.5cm}}& \centermathcell{E                               }\nonumber \\
\Lambda_{21} \,:\,     & \centermathcell{X_{1 3}X_{34}Y_{42} - Y_{1 3}X_{34}X_{42}       }       &    & \centermathcell{ X_{21}X_{1 4}X_{41} - X_{23}X_{32}X_{21}}\nonumber \\
\Lambda_{1 2}^{1} \,:\, & \centermathcell{X_{23}X_{34}Y_{42}X_{21} - X_{21}Y_{1 3}X_{34}X_{41} } &    & \centermathcell{ X_{1 3}X_{32} - X_{1 4}X_{42}            } \nonumber\\
\Lambda_{1 2}^{2} \,:\, & \centermathcell{X_{21}X_{1 3}X_{34}X_{41} - X_{23}X_{34}X_{42}X_{21}  }&    & \centermathcell{ Y_{1 3}X_{32} - X_{1 4}Y_{42}             } \label{eq:H4PhaseB-JEterms}\\
\Lambda_{34} \,:\,     & \centermathcell{Y_{42}X_{21}X_{1 3} - X_{42}X_{21}Y_{1 3}             } &    & \centermathcell{ X_{34}X_{41}X_{1 4} - X_{32}X_{23}X_{34} }\nonumber\\
\Lambda_{43}^{1} \,:\, & \centermathcell{X_{34}Y_{42}X_{21}X_{1 4} - X_{32}X_{21}Y_{1 3}X_{34} } &    & \centermathcell{ X_{42}X_{23} - X_{41}X_{1 3}            } \nonumber\\         
\Lambda_{43}^{2} \,:\, & \centermathcell{X_{32}X_{21}X_{1 3}X_{34} - X_{34}X_{42}X_{21}X_{1 4} } &    & \centermathcell{ Y_{42}X_{23} - X_{41}Y_{1 3}    } \nonumber
	\end{alignat}
	
The corresponding $W^{(0,1)}$ is
	\begin{equation}
	\begin{split}
	W^{(0,1)}=&\,W^{(0,2)}+ \Lambda^R_{1 1} (X_{21}  X^\dagger_{21}+X_{41}  X^\dagger_{41}+X_{1 4}  X^\dagger_{1 4}+X_{1 3}  X^\dagger_{1 3}+Y_{1 3}  Y^\dagger_{1 3})+\\
&+\Lambda^R_{22} (X_{23}  X^\dagger_{23}+X_{21}  X^\dagger_{21}+X_{42}  X^\dagger_{42}+X_{32}  X^\dagger_{32}+Y_{42}  Y^\dagger_{42})+\\
&+\Lambda^R_{33} (X_{23}  X^\dagger_{23}+X_{32}  X^\dagger_{32}+X_{34}  X^\dagger_{34}+X_{1 3}  X^\dagger_{1 3}+Y_{1 3}  Y^\dagger_{1 3})+\\	
&+\Lambda^R_{44} (X_{42}  X^\dagger_{42}+X_{41}  X^\dagger_{41}+X_{34}  X^\dagger_{34}+X_{1 4}  X^\dagger_{1 4}+Y_{42}  Y^\dagger_{42})\fstop
\end{split}
	\end{equation}

Table \ref{tab:GenerH4PhaseB} lists the generators of $H_4$, this time expressed in terms of chiral fields in phase B. They satisfy the same relations we presented in \eqref{eq:H4HSrel} when discussing Phase A.

\begin{table}[H]
		\centering
		\renewcommand{\arraystretch}{1.1}
		\begin{tabular}{c|c}
			Meson    & Chiral superfields  \\
			\hline
$M_1$ & $X_{23}X_{32}=X_{41}X_{1 4}$ \\
$M_2$ & $X_{34}Y_{42}X_{23}=X_{34}X_{41}Y_{1 3}$ \\
$M_3$ & $X_{21}X_{1 4}Y_{42}=X_{21}Y_{1 3}X_{32}$ \\
$M_4$ & $X_{34}Y_{42}X_{21}Y_{1 3}$ \\
$M_5$ & $X_{34}X_{42}X_{23}=X_{34}X_{41}X_{1 3}$ \\
$M_6$ & $X_{21}X_{1 4}X_{42}=X_{21}X_{1 3}X_{32}$ \\
$M_7$ & $X_{42}X_{21}Y_{1 3}X_{34}=Y_{42}X_{21}X_{1 3}X_{34}$ \\
$M_8$ & $X_{42}X_{21}X_{1 3}X_{34}$ \\
		\end{tabular}
		\caption{Generators of $H_4$ in terms of fields in phase B.}
		\label{tab:GenerH4PhaseB}
	\end{table}

Once again, we consider the universal involution, which acts on the fields of phase B as in \eqref{eq:genUImap}. This, in turn, maps the generators as in \eqref{eq:H4PhaseA-invol}. 

Figure \ref{(0,1) theory of Spin(7) orientifold from h4 phase B}, shows the resulting quiver for the orientifold theory. By construction, this gauge theory corresponds to the same Spin(7) orientifold as the one constructed from phase A in the previous section. In Section \ref{sec:H4ABtrial}, we will elaborate on the connection between both theories.

 		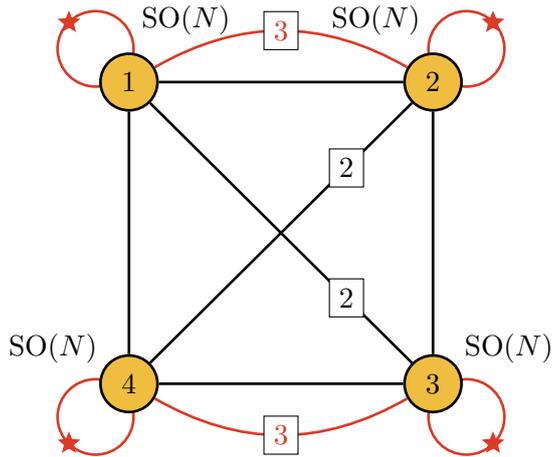
\begin{figure}[H]
	    \centering
	        	\begin{tikzpicture}[scale=2]
			\draw[line width=1pt,redX] (-0.22,-0.22) circle (0.25) node[xshift=-0.35cm,yshift=-0.35cm,star,star points=5, star point ratio=2.25, inner sep=1pt, fill=redX, draw=redX] {};
			\draw[line width=1pt,redX] (-0.22,2.22) circle (0.25) node[xshift=-0.35cm,yshift=0.35cm,star,star points=5, star point ratio=2.25, inner sep=1pt, fill=redX, draw=redX] {};
			\draw[line width=1pt,redX] (2.22,-0.22) circle (0.25) node[xshift=0.35cm,yshift=-0.35cm,star,star points=5, star point ratio=2.25, inner sep=1pt, fill=redX, draw=redX] {};
			\draw[line width=1pt,redX] (2.22,2.22) circle (0.25) node[xshift=0.35cm,yshift=0.35cm,star,star points=5, star point ratio=2.25, inner sep=1pt, fill=redX, draw=redX] {};
			\node[draw=black,line width=1pt,circle,fill=yellowX,minimum width=0.75cm,inner sep=1pt,label={[xshift=-1cm,yshift=-0.25cm]:$\SO(N)$}] (A) at (0,0) {$4$};
			\node[draw=black,line width=1pt,circle,fill=yellowX,minimum width=0.75cm,inner sep=1pt,label={[xshift=1cm,yshift=-0.25cm]:$\SO(N)$}] (B) at (2,0) {$3$};
			\node[draw=black,line width=1pt,circle,fill=yellowX,minimum width=0.75cm,inner sep=1pt,label={[xshift=-0.75cm,yshift=0.1cm]:$\SO(N)$}] (C) at (2,2) {$2$};
			\node[draw=black,line width=1pt,circle,fill=yellowX,minimum width=0.75cm,inner sep=1pt,label={[xshift=0.75cm,yshift=0.1cm]:$\SO(N)$}] (D) at (0,2) {$1$};
			\path (A) edge[line width=1pt] node[fill=white,text opacity=1,fill opacity=1,draw=black,rectangle,thin,pos=0.75] {$2$} (C);
			\path (B) edge[line width=1pt] (A);
			\path (D) edge[line width=1pt] node[fill=white,text opacity=1,fill opacity=1,draw=black,rectangle,thin,pos=0.75] {$2$} (B);
			\path (C) edge[line width=1pt] (D);
			\draw[line width=1pt,redX] (A) to[bend right=30] node[fill=white,text opacity=1,fill opacity=1,draw=black,rectangle,thin,pos=0.5] {\color{redX}{$3$}} (B);
			\draw[line width=1pt,redX] (C) to[bend right=30] node[fill=white,text opacity=1,fill opacity=1,draw=black,rectangle,thin,pos=0.5] {\color{redX}{$3$}} (D);
			\draw[line width=1pt] (A) to (D);
			\draw[line width=1pt] (B) to (C);
			\end{tikzpicture}
	    \caption{Quiver diagram for the Spin(7) orientifold of phase B of $H_4$ using the universal involution.}
		\label{(0,1) theory of Spin(7) orientifold from h4 phase B}
	\end{figure}

\subsubsection{Triality Between the Orientifolded Theories}
\label{sec:H4ABtrial}

Let us now elaborate on the connection between the two theories that we have constructed via the universal involution. Both of them correspond to the same Spin(7) orientifold of $H_4$. The parent theories, phases A and B of $H_4$, are related by $\mathcal{N}=(0,2)$ triality on either node 2 or 4 of phase A (equivalently, by inverse triality on the same nodes of phase B). This leads to a similar connection between the two orientifolded theories, this time via $\mathcal{N}=(0,1)$ triality on node 2 or 4. Figure \ref{(0,1) triality between orientifold theories from h4} summarizes the interplay between triality and orientifolding. This was expected, given our general discussion of the universal involution in Section \ref{sec:01trial-UI}.

\begin{figure}[ht!]
    \centering
    \scalebox{0.95}{
    \begin{tikzpicture}[scale=4]
    \node (H4A) at (0,2) {
    
    	\begin{tikzpicture}[scale=2]
			\draw[line width=1pt,decoration={markings, mark=at position 0.75 with{\arrow{Triangle}}}, postaction={decorate}] (2.22,-0.22) circle (0.25);
			\draw[line width=1pt,redX] (-0.22,2.22) circle (0.25) node[fill=white,text opacity=1,fill opacity=1,draw=black,rectangle,xshift=-0.35cm,yshift=0.35cm,thin] {\color{redX}{$3$}};
			\node[draw=black,line width=1pt,circle,fill=yellowX,minimum width=0.75cm,inner sep=1pt] (A) at (0,0) {$4$};
			\node[draw=black,line width=1pt,circle,fill=yellowX,minimum width=0.75cm,inner sep=1pt] (B) at (2,0) {$3$};
			\node[draw=black,line width=1pt,circle,fill=greenX,minimum width=0.75cm,inner sep=1pt] (C) at (2,2) {$2$};
			\node[draw=black,line width=1pt,circle,fill=yellowX,minimum width=0.75cm,inner sep=1pt] (D) at (0,2) {$1$};
			\path[Triangle-] (A) edge[line width=1pt] (B);
			\path[-Triangle] (B) edge[line width=1pt] (C);
			\path[Triangle-] (C) edge[line width=1pt] (D);
			\path[-Triangle] (D) edge[line width=1pt] (A);
			\draw[line width=1pt,redX] (A) to[bend right=20] (B);
			\draw[line width=1pt,redX] (B) to[bend right=20] (C);
			\draw[line width=1pt,-Triangle] (C) to[bend right=20] node[fill=white,text opacity=1,fill opacity=1,draw=black,rectangle,thin,pos=0.5] {$3$} (D);
			\draw[line width=1pt,Triangle-] (D) to[bend right=20] node[fill=white,text opacity=1,fill opacity=1,draw=black,rectangle,thin,pos=0.5] {$3$} (A);
			\draw[line width=1pt,redX] (A) to node[fill=white,text opacity=1,fill opacity=1,draw=black,rectangle,thin,pos=0.75] {\color{redX}{$2$}} (C);
			\draw[line width=1pt,redX] (D) to[bend right=10] node[fill=white,text opacity=1,fill opacity=1,draw=black,rectangle,thin,pos=0.25] {\color{redX}{$2$}} (B);
			\draw[line width=1pt,-Triangle] (D) to[bend left=10] node[fill=white,text opacity=1,fill opacity=1,draw=black,rectangle,thin,pos=0.75] {$2$} (B);
			\end{tikzpicture}
    
    };
    \node (H4B) at (2,2) {
    	\begin{tikzpicture}[scale=2]
			\node[draw=black,line width=1pt,circle,fill=yellowX,minimum width=0.75cm,inner sep=1pt] (A) at (0,0) {$4$};
			\node[draw=black,line width=1pt,circle,fill=yellowX,minimum width=0.75cm,inner sep=1pt] (B) at (2,0) {$3$};
			\node[draw=black,line width=1pt,circle,fill=greenX,minimum width=0.75cm,inner sep=1pt] (C) at (2,2) {$2$};
			\node[draw=black,line width=1pt,circle,fill=yellowX,minimum width=0.75cm,inner sep=1pt] (D) at (0,2) {$1$};
			\path[-Triangle] (A) edge[line width=1pt] node[fill=white,text opacity=1,fill opacity=1,draw=black,rectangle,thin,pos=0.75] {$2$} (C);
			\path[-Triangle] (B) edge[line width=1pt] (A);
			\path[-Triangle] (D) edge[line width=1pt] node[fill=white,text opacity=1,fill opacity=1,draw=black,rectangle,thin,pos=0.75] {$2$} (B);
			\path[-Triangle] (C) edge[line width=1pt] (D);
			\draw[line width=1pt,redX] (A) to[bend right=20] node[fill=white,text opacity=1,fill opacity=1,draw=black,rectangle,thin,pos=0.5] {\color{redX}{$3$}} (B);
			\draw[line width=1pt,redX] (C) to[bend right=20] node[fill=white,text opacity=1,fill opacity=1,draw=black,rectangle,thin,pos=0.5] {\color{redX}{$3$}} (D);
			\draw[line width=1pt,Triangle-Triangle] (A) to  (D);
			\draw[line width=1pt,Triangle-Triangle] (B) to  (C);
			\end{tikzpicture}
    
    };
    \node (H4OA) at (-0.01,0.24) {
    
    	\begin{tikzpicture}[scale=2]
			\draw[line width=1pt] (2.22,-0.22) circle (0.25) node[xshift=0.35cm,yshift=-0.35cm] {\scriptsize{$\quadro$}};
			\draw[line width=1pt] (2.28,-0.28) circle (0.4)  node[xshift=0.55cm,yshift=-0.55cm,star,star points=5, star point ratio=2.25, inner sep=1pt, fill=black, draw=black] {};
			\draw[line width=1pt,redX] (2.36,-0.36) circle (0.55)  node[xshift=0.75cm,yshift=-0.75cm,star,star points=5, star point ratio=2.25, inner sep=1pt, fill=redX, draw=redX] {};
			\draw[line width=1pt,redX] (-0.22,-0.22) circle (0.25) node[xshift=-0.35cm,yshift=-0.35cm,star,star points=5, star point ratio=2.25, inner sep=1pt, fill=redX, draw=redX] {};
			\draw[line width=1pt,redX] (2.22,2.22) circle (0.25) node[xshift=0.35cm,yshift=0.35cm,star,star points=5, star point ratio=2.25, inner sep=1pt, fill=redX, draw=redX] {};
			\draw[line width=1pt,redX] (-0.22,2.22) circle (0.25)  node[xshift=0.35cm,yshift=0.35cm] {\color{redX}{\scriptsize{$\quadro$}}} node[fill=white,text opacity=1,fill opacity=1,draw=black,rectangle,xshift=-0.35cm,yshift=0.35cm,thin] {\color{redX}{$3$}};
			\draw[line width=1pt,redX] (-0.33,2.33) circle (0.45)  node[yshift=0.9cm,star,star points=5, star point ratio=2.25, inner sep=1pt, fill=redX, draw=redX] {} node[fill=white,text opacity=1,fill opacity=1,draw=black,rectangle,xshift=-0.65cm,yshift=0.65cm,thin] {\color{redX}{$4$}};
			\node[draw=black,line width=1pt,circle,fill=yellowX,minimum width=0.75cm,inner sep=1pt,label={[xshift=-1.25cm,yshift=-0.25cm]:$\SO(N)$}] (A) at (0,0) {$4$};
			\node[draw=black,line width=1pt,circle,fill=yellowX,minimum width=0.75cm,inner sep=1pt,label={[xshift=1.1cm,yshift=0cm]:$\SO(N)$}] (B) at (2,0) {$3$};
			\node[draw=black,line width=1pt,circle,fill=greenX,minimum width=0.75cm,inner sep=1pt,label={[xshift=-0.75cm,yshift=0.25cm]:$\SO(N)$}] (C) at (2,2) {$2$};
			\node[draw=black,line width=1pt,circle,fill=yellowX,minimum width=0.75cm,inner sep=1pt,label={[xshift=1cm,yshift=0.25cm]:$\SO(N)$}] (D) at (0,2) {$1$};
			\path (A) edge[line width=1pt] (B);
			\path (B) edge[line width=1pt]  (C);
			\path (C) edge[line width=1pt] (D);
			\path (D) edge[line width=1pt] (A);
			\draw[line width=1pt,redX] (A) to[bend right=30] (B);
			\draw[line width=1pt,redX] (B) to[bend right=30] (C);
			\draw[line width=1pt] (C) to[bend right=30] node[fill=white,text opacity=1,fill opacity=1,draw=black,rectangle,thin,pos=0.5] {$3$} (D);
			\draw[line width=1pt] (D) to[bend right=30] node[fill=white,text opacity=1,fill opacity=1,draw=black,rectangle,thin,pos=0.5] {$3$} (A);
			\draw[line width=1pt,redX] (A) to node[fill=white,text opacity=1,fill opacity=1,draw=black,rectangle,thin,pos=0.75] {\color{redX}{$2$}} (C);
			\draw[line width=1pt,redX] (D) to[bend right=10] node[fill=white,text opacity=1,fill opacity=1,draw=black,rectangle,thin,pos=0.25] {\color{redX}{$2$}} (B);
			\draw[line width=1pt] (D) to[bend left=10] node[fill=white,text opacity=1,fill opacity=1,draw=black,rectangle,thin,pos=0.75] {$2$} (B);
			\end{tikzpicture}
    
    };
    \node (H4OB) at (2,0.30) {
    	\begin{tikzpicture}[scale=2]
			\draw[line width=1pt,redX] (-0.22,-0.22) circle (0.25) node[xshift=-0.35cm,yshift=-0.35cm,star,star points=5, star point ratio=2.25, inner sep=1pt, fill=redX, draw=redX] {};
			\draw[line width=1pt,redX] (-0.22,2.22) circle (0.25) node[xshift=-0.35cm,yshift=0.35cm,star,star points=5, star point ratio=2.25, inner sep=1pt, fill=redX, draw=redX] {};
			\draw[line width=1pt,redX] (2.22,-0.22) circle (0.25) node[xshift=0.35cm,yshift=-0.35cm,star,star points=5, star point ratio=2.25, inner sep=1pt, fill=redX, draw=redX] {};
			\draw[line width=1pt,redX] (2.22,2.22) circle (0.25) node[xshift=0.35cm,yshift=0.35cm,star,star points=5, star point ratio=2.25, inner sep=1pt, fill=redX, draw=redX] {};
			\node[draw=black,line width=1pt,circle,fill=yellowX,minimum width=0.75cm,inner sep=1pt,label={[xshift=-1cm,yshift=-0.25cm]:$\SO(N)$}] (A) at (0,0) {$4$};
			\node[draw=black,line width=1pt,circle,fill=yellowX,minimum width=0.75cm,inner sep=1pt,label={[xshift=1cm,yshift=-0.25cm]:$\SO(N)$}] (B) at (2,0) {$3$};
			\node[draw=black,line width=1pt,circle,fill=greenX,minimum width=0.75cm,inner sep=1pt,label={[xshift=-0.75cm,yshift=0.25cm]:$\SO(N)$}] (C) at (2,2) {$2$};
			\node[draw=black,line width=1pt,circle,fill=yellowX,minimum width=0.75cm,inner sep=1pt,label={[xshift=0.75cm,yshift=0.25cm]:$\SO(N)$}] (D) at (0,2) {$1$};
			\path (A) edge[line width=1pt] node[fill=white,text opacity=1,fill opacity=1,draw=black,rectangle,thin,pos=0.75] {$2$} (C);
			\path (B) edge[line width=1pt] (A);
			\path (D) edge[line width=1pt] node[fill=white,text opacity=1,fill opacity=1,draw=black,rectangle,thin,pos=0.75] {$2$} (B);
			\path (C) edge[line width=1pt] (D);
			\draw[line width=1pt,redX] (A) to[bend right=30] node[fill=white,text opacity=1,fill opacity=1,draw=black,rectangle,thin,pos=0.5] {\color{redX}{$3$}} (B);
			\draw[line width=1pt,redX] (C) to[bend right=30] node[fill=white,text opacity=1,fill opacity=1,draw=black,rectangle,thin,pos=0.5] {\color{redX}{$3$}} (D);
			\draw[line width=1pt] (A) to (D);
			\draw[line width=1pt] (B) to (C);
			\end{tikzpicture}
    
    };
    \draw[Triangle-Triangle,greenX,line width=1mm] (0.8,2) -- node[above,midway,black] {$\mathcal{N}=(0,2)$} node[below,midway,black] {triality} (1.35,2);
    \draw[Triangle-Triangle,greenX,line width=1mm] (0.8,0.25) -- node[above,midway,black] {$\mathcal{N}=(0,1)$} node[below,midway,black] {triality} (1.35,0.25);
    \end{tikzpicture}
    }
 \caption{Phases A and B of $H_4$ are connected by $\mathcal{N}=(0,2)$ triality on node 2 (shown in green). Upon orientifolding with the universal involution, the resulting theories are similarly connected by $\mathcal{N}=(0,1)$ triality.}
		\label{(0,1) triality between orientifold theories from h4}
\end{figure}
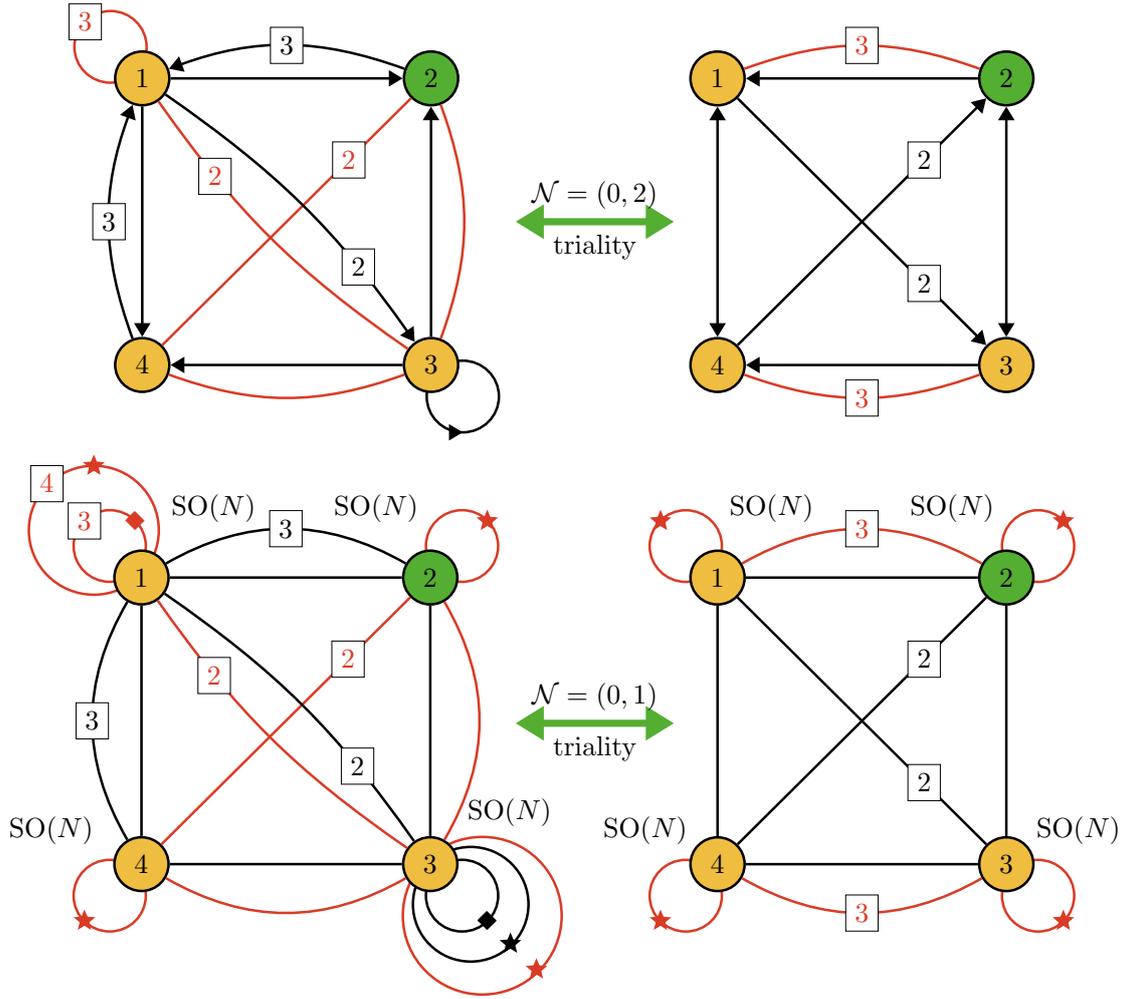

It is important to emphasize that it is possible for two $\Spin(7)$ orientifolds to correspond to the same geometric involution while differing in the choice of vector structure. In practical terms, the appearance of the choices of vector structure in orientifolds arises when, for a given geometry, there are different $\mathbb{Z}_2$  symmetries on the underlying quiver gauge theory, which differ in the action on the quiver nodes. Such a discrete choice generalizes beyond orbifold singularities, and it was studied in detail in \cite{Franco:2021ixh}, in anticipation of the application of $\Spin(7)$ orientifolds to triality that we carry out in this paper. In order for equivalent orientifold geometric involutions to actually produce dual theories, it is necessary that they also agree on the choice of vector structure they implicitly define. This is the case for all the examples considered in this paper.

Finally, it is interesting to note that, as we discussed in Section \ref{sec:N01triality_quivers}, in orientifold theories the number of ``incoming flavors” at the dualized node is inherited from the parent.

\section{Beyond the Universal Involution}
\label{sec:BeyonUI}

In this section, we present theories that are obtained from $\mathcal{N}=(0,2)$ triality dual parents by Spin(7) orientifolds that do not correspond to the universal involution. We will see that they lead to interesting generalizations of the basic $\mathcal{N}=(0,1)$ triality.\footnote{We will rightfully continue referring to the resulting equivalences between theories as trialities, due to their connections to the basic trialities of SQCD-type theories. It is reasonable to expect that we can indeed perform these transformations three times on the same quiver node. However, the three transformations, can sometimes fall outside our analysis, provided they actually exist. This is due to our restriction to the class of theories obtained as Spin(7) orientifolds of toric phases.}

\subsection{$Q^{1,1,1}$}
\label{sec:Q111}

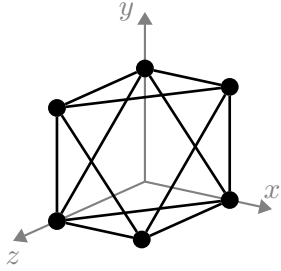
\begin{figure}[H]
		\centering
		\begin{tikzpicture}[scale=1.5, rotate around y=-25]
		\draw[thick,gray,-Triangle] (0,0,0) -- node[above,pos=1] {$x$} (1.5,0,0);
		\draw[thick,gray,-Triangle] (0,0,0) -- node[left,pos=1] {$y$} (0,1.5,0);
		\draw[thick,gray,-Triangle] (0,0,0) -- node[below,pos=1] {$z$} (0,0,1.5);
		\node[draw=black,line width=1pt,circle,fill=black,minimum width=0.2cm,inner sep=1pt] (A) at (1,0,0) {};
		\node[draw=black,line width=1pt,circle,fill=black,minimum width=0.2cm,inner sep=1pt] (B) at (0,1,0) {};
		\node[draw=black,line width=1pt,circle,fill=black,minimum width=0.2cm,inner sep=1pt] (C) at (0,0,1) {};
		\node[draw=black,line width=1pt,circle,fill=black,minimum width=0.2cm,inner sep=1pt] (D) at (1,1,0) {};
		\node[draw=black,line width=1pt,circle,fill=black,minimum width=0.2cm,inner sep=1pt] (E) at (0,1,1) {};
		\node[draw=black,line width=1pt,circle,fill=black,minimum width=0.2cm,inner sep=1pt] (F) at (1,0,1) {};
		\draw[line width=1pt] (A)--(B)--(C)--(A);
		\draw[line width=1pt] (D)--(E)--(F)--(D);
		\draw[line width=1pt] (A)--(D)--(B);
		\draw[line width=1pt] (A)--(F)--(C);
		\draw[line width=1pt] (B)--(E)--(C);
		\end{tikzpicture}
		\caption{Toric diagram for $Q^{1,1,1}$.}
		\label{fig:Q111toricdiagram}
\end{figure}

Let us now consider the cone over $Q^{1,1,1}$, or $Q^{1,1,1}$ for short, whose toric diagram is shown in Figure \ref{fig:Q111toricdiagram}. The $\mathcal{N}=(0,2)$ gauge theories, brane brick models and the triality web relating the toric phases for this geometry have been studied at length \cite{Franco:2015tna,Franco:2015tya,Franco:2016nwv}. However, none of its Spin(7) orientifolds has been presented in the literature. Below, we construct an orientifold based on a non-universal involution.

\subsubsection{Phase A}
	\label{sec:Q111phaseA}

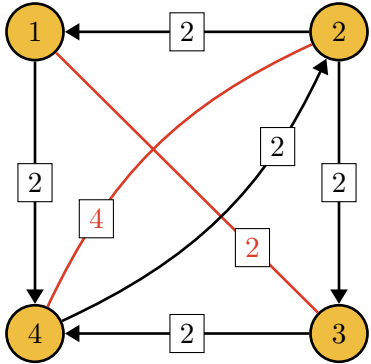
\begin{figure}[H]
	\centering
	\begin{tikzpicture}[scale=2]
	\node[draw=black,line width=1pt,circle,fill=yellowX,minimum width=0.75cm,inner sep=1pt] (A) at (0,0) {$4$};
	\node[draw=black,line width=1pt,circle,fill=yellowX,minimum width=0.75cm,inner sep=1pt] (B) at (2,0) {$3$};
	\node[draw=black,line width=1pt,circle,fill=yellowX,minimum width=0.75cm,inner sep=1pt] (C) at (2,2) {$2$};
	\node[draw=black,line width=1pt,circle,fill=yellowX,minimum width=0.75cm,inner sep=1pt] (D) at (0,2) {$1$};
	\draw[line width=1pt,-Triangle] (C) to node[fill=white,text opacity=1,fill opacity=1,draw=black,rectangle,thin,pos=0.5] {$2$} (D);
	\draw[line width=1pt,-Triangle] (C) to node[fill=white,text opacity=1,fill opacity=1,draw=black,rectangle,thin,pos=0.5] {$2$} (B);
	\draw[line width=1pt,-Triangle] (B) to node[fill=white,text opacity=1,fill opacity=1,draw=black,rectangle,thin,pos=0.5] {$2$} (A);
	\draw[line width=1pt,-Triangle] (D) to node[fill=white,text opacity=1,fill opacity=1,draw=black,rectangle,thin,pos=0.5] {$2$} (A);
	\draw[line width=1pt,redX] (D) to node[fill=white,text opacity=1,fill opacity=1,draw=black,rectangle,thin,pos=0.75] {\color{redX}{$2$}} (B);
	\draw[line width=1pt,-Triangle] (A) to[bend right=20] node[fill=white,text opacity=1,fill opacity=1,draw=black,rectangle,thin,pos=0.75] {$2$} (C);
	\draw[line width=1pt,redX] (A) to[bend left=20] node[fill=white,text opacity=1,fill opacity=1,draw=black,rectangle,thin,pos=0.25] {\color{redX}{$4$}} (C);
	\node at (0,-0.42) {};
	\end{tikzpicture}
	\caption{Quiver diagram for phase A of $Q^{1,1,1}$.}
		\label{fig:Q111AN02}
\end{figure}

The toric phases for $Q^{1,1,1}$ were studied in \cite{Franco:2016nwv}. Figure \ref{fig:Q111AN02} shows the quiver for the so-called phase A.

	The $J$- and $E$-terms are
	\begin{alignat}{4}
	\renewcommand{\arraystretch}{1.1}
    & \centermathcell{J}                           &\text{\hspace{.5cm}}& \centermathcell{E                               }\nonumber \\
	\Lambda_{24}^1 \,:\, & \centermathcell{Y_{42} X_{23} Y_{34}X_{42} -Y_{42}X_{21} X_{14} X_{42} } &                     &\centermathcell{  Y_{23}X_{34}-Y_{21}Y_{14}                }\nonumber\\
	\Lambda_{24}^2 \,:\, & \centermathcell{Y_{42} Y_{21} Y_{14} X_{42}-Y_{42} Y_{23} X_{34} X_{42}} &                     &\centermathcell{ X_{23} Y_{34}-X_{21}X_{14}                }\nonumber\\
	\Lambda_{24}^3 \,:\, & \centermathcell{Y_{42} X_{14} Y_{21} X_{42}-Y_{42} X_{23} X_{34} X_{42}} &                     &\centermathcell{ X_{21} Y_{14}-Y_{23}Y_{34}                }\\
	\Lambda_{24}^4 \,:\, & \centermathcell{Y_{42} X_{21} Y_{14}X_{42} -Y_{42}Y_{23} Y_{34} X_{42} } &                     &\centermathcell{ X_{23}X_{34} -Y_{21}X_{14}                }\nonumber\\
	\Lambda_{31}^1 \,:\, & \centermathcell{Y_{14}X_{42} X_{23} -X_{14} X_{42} Y_{23}              } &                     &\centermathcell{ X_{34}Y_{42}  X_{21}-Y_{34}Y_{42}  Y_{21} }\nonumber\\
	\Lambda_{31}^2 \,:\, & \centermathcell{X_{14}Y_{42}  Y_{23}- Y_{14}Y_{42} X_{23}              } &                     &\centermathcell{  X_{34} X_{42} X_{21}-Y_{34}X_{42}  Y_{21}}\nonumber
	\end{alignat}
Finding the corresponding $W^{(0,1)}$ is a simple exercise, but we omit it here for brevity. Table~\ref{tab:GenerQ111phaseA} lists the generators for $Q^{1,1,1}$ written in terms of the gauge theory.

	\begin{table}[H]
		\centering
		\renewcommand{\arraystretch}{1.1}
		\begin{tabular}{c|c}
			Field    & Chiral superfields  \\
			\hline
			$M_1$    & $Y_{42}Y_{23}X_{34}=Y_{42}Y_{21}Y_{14}$ \\
			$M_2$    & $X_{42}Y_{23}X_{34}=X_{42}Y_{21}Y_{14}$ \\
			$M_3$    & $Y_{42}X_{23}X_{34}=Y_{42}Y_{21}X_{14}$ \\
			$M_4$    & $X_{42}X_{23}X_{34}=X_{42}Y_{21}X_{14}$ \\
			$M_5$    & $Y_{42}Y_{23}Y_{34}=Y_{42}X_{21}Y_{14}$ \\
			$M_6$    & $X_{42}Y_{23}Y_{34}=X_{42}X_{21}Y_{14}$ \\
			$M_7$    & $Y_{42}X_{23}Y_{34}=Y_{42}X_{21}X_{14}$ \\
			$M_8$    & $X_{42}X_{23}Y_{34}=X_{42}X_{21}X_{14}$ 
		\end{tabular}
		\caption{Generators of $Q^{1,1,1}$ in terms of fields in phase A.}
		\label{tab:GenerQ111phaseA}
	\end{table}

The generators satisfy the following relations
	\begin{equation}
	\begin{split}
	\mathcal{I} = &\left\langle M_1M_7=M_3M_5\coma M_3M_8=M_4M_7\coma M_1M_4=M_2M_3\coma M_5M_8=M_7M_6\coma\right.\\
	&\left.M_1M_8=M_2M_7\coma M_3M_6=M_4M_5\coma M_1M_8=M_4M_5\coma M_1M_6=M_5M_2\coma\right.\\
	&\left.M_2M_8=M_4M_6\right\rangle\fstop
	\label{eq:Q111HSrel}
	\end{split}
	\end{equation}

Let us now consider the involution that maps all the four gauge groups to themselves and has the following action on chiral fields
\begin{equation}
	\begin{array}{cccccccccccc}
	Y_{42} &\rightarrow&  -\gamma_{\Omega_4}\bar{X}_{42}\gamma_{\Omega_2}^{-1}\coma & 
	X_{42} &\rightarrow& \gamma_{\Omega_4}\bar{Y}_{42}\gamma_{\Omega_2}^{-1} \coma & 
	X_{34}&\rightarrow& \gamma_{\Omega_3}\bar{Y}_{34}\gamma_{\Omega_4}^{-1} \coma &
	Y_{34}&\rightarrow& -\gamma_{\Omega_3}\bar{X}_{34}\gamma_{\Omega_4}^{-1}\coma \\
	X_{21}&\rightarrow& -\gamma_{\Omega_2}\bar{Y}_{21}\gamma_{\Omega_1}^{-1}\coma & 
	Y_{21}&\rightarrow& \gamma_{\Omega_2}\bar{X}_{21}\gamma_{\Omega_1}^{-1}\coma & 
	Y_{23}&\rightarrow& \gamma_{\Omega_2}\bar{Y}_{23}\gamma_{\Omega_3}^{-1}\coma &  
	X_{23}&\rightarrow& \gamma_{\Omega_2}\bar{X}_{23}\gamma_{\Omega_3}^{-1}\coma \\
	& & & Y_{14}&\rightarrow& \gamma_{\Omega_1}\bar{Y}_{14}\gamma_{\Omega_4}^{-1}\coma & X_{14}&\rightarrow& \gamma_{\Omega_1}\bar{X}_{14}\gamma_{\Omega_4}^{-1}\coma
	\end{array}
	\label{eq:Q111A-chiral-invol}
	\end{equation}
	where we have used the $\gamma_{\Omega_i}$ matrices mentioned in Footnote \ref{foot:gamma-omega}.
	
	Invariance of $W^{(0,1)}$ further implies that the involution acts on Fermi fields as follows  
	\begin{equation}
	\begin{array}{ccccccccc}
	\Lambda_{24}^1&\rightarrow& -\gamma_{\Omega_2}\bar{\Lambda}^3_{24}\gamma_{\Omega_4}^{-1} \coma &
	\Lambda_{24}^2&\rightarrow& -\gamma_{\Omega_2}\bar{\Lambda}^4_{24}\gamma_{\Omega_4}^{-1} \coma &
	\Lambda_{24}^3&\rightarrow& \gamma_{\Omega_2}\bar{\Lambda}^1_{24}\gamma_{\Omega_4}^{-1} \coma \\ 
	\Lambda_{24}^4&\rightarrow& \gamma_{\Omega_2}\bar{\Lambda}^2_{24}\gamma_{\Omega_4}^{-1}\coma &
	\Lambda_{31}^1&\rightarrow& -\gamma_{\Omega_3}\bar{\Lambda}_{31}^2\gamma_{\Omega_1}^{-1}\coma & 
	\Lambda_{31}^2&\rightarrow& \gamma_{\Omega_3}\Lambda_{31}^1\gamma_{\Omega_1}^{-1}\coma
	\end{array}
	\label{eq:Q111A-Fermi-invol}
	\end{equation}
	and
	\begin{equation}
    \Lambda_{11}^R\rightarrow \gamma_{\Omega_1}\Lambda_{11}^{R\,\,T}\gamma_{\Omega_1}^{-1} \coma \Lambda_{22}^R\rightarrow \gamma_{\Omega_2}\Lambda_{22}^{R\,\,T}\gamma_{\Omega_2}^{-1} \coma \Lambda_{33}^R\rightarrow \gamma_{\Omega_3}\Lambda_{33}^{R\,\,T}\gamma_{\Omega_3}^{-1} \coma \Lambda_{44}^R\rightarrow \gamma_{\Omega_4}\Lambda_{44}^{R\,\,T}\gamma_{\Omega_4}^{-1} \fstop
    \label{eq:Q111A-RFermi-invol}
	\end{equation}
	
Interestingly, the involution in \eqref{eq:Q111A-chiral-invol} and \eqref{eq:Q111A-Fermi-invol} involves a non-trivial action on flavor indices (see e.g. the action on pairs of fields such as $(X_{21},Y_{21})$). As briefly mentioned in Section \ref{sec:Spin7Orient}, this leads to a constraint on the matrices $\gamma_{\Omega_i}$ that encode the action of the orientifold group on the gauge groups, which reads 
\begin{equation}
    \gamma_{\Omega_1}= \gamma_{\Omega_4}\neq  \gamma_{\Omega_2}= \gamma_{\Omega_3}\fstop
    \label{eq:Q111A-Omega-cond}
\end{equation}
This constraint follows for requiring that the involution squares to the identity. For a detailed discussion of this constraint and additional explicit examples, we refer the interested reader to our previous work \cite{Franco:2021ixh}. 

For concreteness, we will focus on the following solution to the constraint
\begin{equation}
\begin{array}{ccccc}
\gamma_{\Omega_1} & = & \gamma_{\Omega_4} & = & J \coma \\
\gamma_{\Omega_2} & = & \gamma_{\Omega_3} & = & \ID_N \coma
\end{array}
\label{gamma_matrices_phase_A}
\end{equation}
where $J=i\epsilon_{N/2}$ is the symplectic matrix, and $\ID_N$ is the identity matrix.
	
Using Table~\ref{tab:GenerQ111phaseA}, the involution in \eqref{eq:Q111A-chiral-invol} translates into the following action at the level of the geometry
\begin{equation}
     \begin{array}{cccccccccccc}
	M_{1}  & \rightarrow & -\bar{M}_{6} \coma &
	M_{2}  & \rightarrow & \bar{M}_{5} \coma &
	M_{3}  & \rightarrow & -\bar{M}_{8} \coma &
	M_{4}  & \rightarrow & \bar{M}_{7} \coma \\
	M_{5}  & \rightarrow & \bar{M}_{2} \coma &
	M_{6}  & \rightarrow & -\bar{M}_{1} \coma &
	M_{7}  & \rightarrow & \bar{M}_{4} \coma &
	M_{8}  & \rightarrow & -\bar{M}_{3} \coma
	\end{array} 
	\label{eq:Q111phaseA-nonuni-geom}
	\end{equation}
which is clearly not the universal involution. 

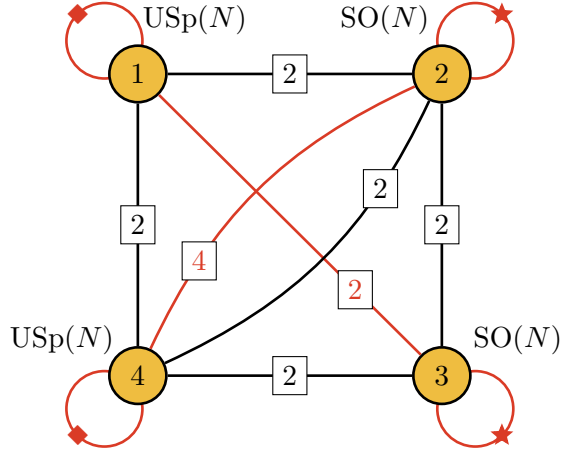
\begin{figure}[H]
	    \centering
	  	\begin{tikzpicture}[scale=2]
		\draw[line width=1pt,redX] (-0.22,-0.22) circle (0.25) node[xshift=-0.35cm,yshift=-0.35cm] {\scriptsize{$\quadro$}};
		\draw[line width=1pt,redX] (-0.22,2.22) circle (0.25) node[xshift=-0.35cm,yshift=0.35cm] {\scriptsize{$\quadro$}};
		\draw[line width=1pt,redX] (2.22,-0.22) circle (0.25) node[xshift=0.35cm,yshift=-0.35cm,star,star points=5, star point ratio=2.25, inner sep=1pt, fill=redX, draw=redX] {};
		\draw[line width=1pt,redX] (2.22,2.22) circle (0.25) node[xshift=0.35cm,yshift=0.35cm,star,star points=5, star point ratio=2.25, inner sep=1pt, fill=redX, draw=redX] {};
			\node[draw=black,line width=1pt,circle,fill=yellowX,minimum width=0.75cm,inner sep=1pt,label={[xshift=-1cm,yshift=-0.25cm]:$\USp(N)$}] (A) at (0,0) {$4$};
		\node[draw=black,line width=1pt,circle,fill=yellowX,minimum width=0.75cm,inner sep=1pt,label={[xshift=1cm,yshift=-0.25cm]:$\SO(N)$}] (B) at (2,0) {$3$};
		\node[draw=black,line width=1pt,circle,fill=yellowX,minimum width=0.75cm,inner sep=1pt,label={[xshift=-0.75cm,yshift=0cm]:$\SO(N)$}] (C) at (2,2) {$2$};
		\node[draw=black,line width=1pt,circle,fill=yellowX,minimum width=0.75cm,inner sep=1pt,label={[xshift=0.75cm,yshift=0cm]:$\USp(N)$}] (D) at (0,2) {$1$};
		\draw[line width=1pt] (C) to node[fill=white,text opacity=1,fill opacity=1,draw=black,rectangle,thin,pos=0.5] {$2$} (D);
		\draw[line width=1pt] (C) to node[fill=white,text opacity=1,fill opacity=1,draw=black,rectangle,thin,pos=0.5] {$2$} (B);
		\draw[line width=1pt] (B) to node[fill=white,text opacity=1,fill opacity=1,draw=black,rectangle,thin,pos=0.5] {$2$} (A);
		\draw[line width=1pt] (D) to node[fill=white,text opacity=1,fill opacity=1,draw=black,rectangle,thin,pos=0.5] {$2$} (A);
		\draw[line width=1pt,redX] (D) to node[fill=white,text opacity=1,fill opacity=1,draw=black,rectangle,thin,pos=0.75] {\color{redX}{$2$}} (B);
		\draw[line width=1pt] (A) to[bend right=20] node[fill=white,text opacity=1,fill opacity=1,draw=black,rectangle,thin,pos=0.75] {$2$} (C);
		\draw[line width=1pt,redX] (A) to[bend left=20] node[fill=white,text opacity=1,fill opacity=1,draw=black,rectangle,thin,pos=0.25] {\color{redX}{$4$}} (C);
		\end{tikzpicture}
	    \caption{Quiver diagram for the Spin(7) orientifold of phase A of $Q^{1,1,1}$ using the involution in \cref{eq:Q111A-chiral-invol,eq:Q111A-Fermi-invol,eq:Q111A-RFermi-invol}, together with our choice of $\gamma_{\Omega_i}$ matrices.}
		\label{o theory phase A q111}
	\end{figure}

Figure \ref{o theory phase A q111} shows the quiver for the orientifold theory, which is free of gauge anomalies. %

\newpage

\subsubsection{Phase S}
	\label{sec:Q111phaseS} 
	
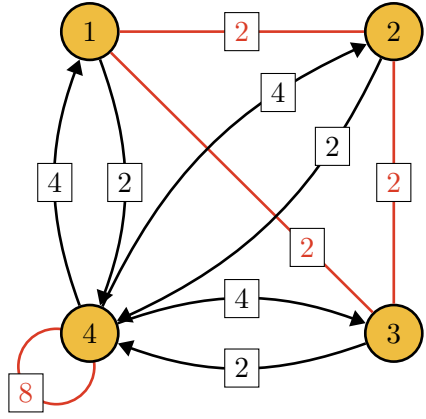
\begin{figure}[H]
	\centering
		\begin{tikzpicture}[scale=2]
		\draw[line width=1pt,redX] (-0.22,-0.22) circle (0.25) node[fill=white,text opacity=1,fill opacity=1,draw=black,rectangle,xshift=-0.4cm,yshift=-0.35cm,thin] {\color{redX}{$8$}};
		\node[draw=black,line width=1pt,circle,fill=yellowX,minimum width=0.75cm,inner sep=1pt] (A) at (0,0) {$4$};
		\node[draw=black,line width=1pt,circle,fill=yellowX,minimum width=0.75cm,inner sep=1pt] (B) at (2,0) {$3$};
		\node[draw=black,line width=1pt,circle,fill=yellowX,minimum width=0.75cm,inner sep=1pt] (C) at (2,2) {$2$};
		\node[draw=black,line width=1pt,circle,fill=yellowX,minimum width=0.75cm,inner sep=1pt] (D) at (0,2) {$1$};
		\draw[line width=1pt,redX] (C) to node[fill=white,text opacity=1,fill opacity=1,draw=black,rectangle,thin,pos=0.5] {\color{redX}{$2$}} (D);
		\draw[line width=1pt,redX] (B) to node[fill=white,text opacity=1,fill opacity=1,draw=black,rectangle,thin,pos=0.5] {\color{redX}{$2$}} (C);
		\draw[line width=1pt,redX] (D) to node[fill=white,text opacity=1,fill opacity=1,draw=black,rectangle,thin,pos=0.75] {\color{redX}{$2$}} (B);
		\draw[line width=1pt,Triangle-] (A) to[bend right=20] node[fill=white,text opacity=1,fill opacity=1,draw=black,rectangle,thin,pos=0.5] {$2$} (B);
		\draw[line width=1pt,-Triangle] (A) to[bend left=20] node[fill=white,text opacity=1,fill opacity=1,draw=black,rectangle,thin,pos=0.5] {$4$} (B);
		\draw[line width=1pt,Triangle-] (A) to[bend right=20] node[fill=white,text opacity=1,fill opacity=1,draw=black,rectangle,thin,pos=0.5] {$2$} (D);
		\draw[line width=1pt,-Triangle] (A) to[bend left=20] node[fill=white,text opacity=1,fill opacity=1,draw=black,rectangle,thin,pos=0.5] {$4$} (D);
		\draw[line width=1pt,Triangle-] (A) to[bend right=20] node[fill=white,text opacity=1,fill opacity=1,draw=black,rectangle,thin,pos=0.75] {$2$} (C);
		\draw[line width=1pt,-Triangle] (A) to[bend left=20] node[fill=white,text opacity=1,fill opacity=1,draw=black,rectangle,thin,pos=0.75] {$4$} (C);
		\node at (0,-0.65) {};
		\end{tikzpicture}
	\caption{Quiver diagram for phase S of $Q^{1,1,1}$.}
	\label{fig:Q111SN02}
\end{figure}

Figure \ref{fig:Q111SN02} shows the quiver for phase S of $Q^{1,1,1}$ \cite{Franco:2016nwv}.

The $J$- and $E$-terms are
		\begin{alignat}{4}
	\renewcommand{\arraystretch}{1.1}
    & \centermathcell{J}                           &\text{\hspace{.5cm}}& \centermathcell{E                               }\nonumber \\
	\Lambda_{23}^1 \,:\,& \centermathcell{X_{34}Y_{42} -Y_{34}W_{42}  }& & \centermathcell{Y_{24}X_{43} -X_{24} Z_{43}}\nonumber \\
	\Lambda_{23}^2 \,:\,& \centermathcell{X_{34}X_{42}-Y_{34} Z_{42}  }& & \centermathcell{X_{24}W_{43}-Y_{24}Y_{43}  }\nonumber \\
	\Lambda_{31}^1 \,:\,& \centermathcell{X_{14} Z_{43}-Y_{14}W_{43}  }& & \centermathcell{X_{34}X_{41} -Y_{34}Z_{41} }\nonumber \\
	\Lambda_{31}^2 \,:\,& \centermathcell{X_{14}X_{43} -Y_{14} Y_{43} }& & \centermathcell{Y_{34}W_{41} -X_{34}Y_{41} }\nonumber \\
	\Lambda_{12}^1 \,:\,& \centermathcell{Y_{24}Z_{41}-X_{24}W_{41}   }& & \centermathcell{X_{14}X_{42}-Y_{14} Y_{42} }\nonumber \\
	\Lambda_{12}^2 \,:\,& \centermathcell{Y_{24}X_{41} -X_{24}Y_{41}  }& & \centermathcell{Y_{14}W_{42}-X_{14}Z_{42}  }\nonumber \\
	\Lambda_{44}^1 \,:\,& \centermathcell{Y_{43}Y_{34} -X_{41} X_{14} }& & \centermathcell{W_{41} Y_{14}-Z_{42}Y_{24} }\label{Q111PhaseSJEterms} \\
	\Lambda_{44}^2 \,:\,& \centermathcell{W_{41}Y_{14}-Z_{43}X_{34}   }& & \centermathcell{X_{41}X_{14}-Y_{42}X_{24}  }\nonumber \\
	\Lambda_{44}^3 \,:\,& \centermathcell{Y_{41}Y_{14} -X_{42}Y_{24}  }& & \centermathcell{W_{42} X_{24}-Y_{43}X_{34} }\nonumber \\
	\Lambda_{44}^4 \,:\,& \centermathcell{W_{42} X_{24}-Z_{41}X_{14}  }& & \centermathcell{X_{42} Y_{24}-Z_{43}Y_{34} }\nonumber \\
	\Lambda_{44}^5 \,:\,& \centermathcell{Z_{42}X_{24}-X_{43} X_{34}  }& & \centermathcell{Y_{24}Y_{42}-Y_{41} X_{14} }\nonumber \\
	\Lambda_{44}^6 \,:\,& \centermathcell{W_{43} Y_{34}-Y_{42}Y_{24}  }& & \centermathcell{X_{24} Z_{42}-Z_{41}Y_{14} }\nonumber \\
	\Lambda_{44}^7 \,:\,& \centermathcell{W_{43} X_{34}-W_{41} X_{14} }& & \centermathcell{X_{41}Y_{14}-X_{42} X_{24} }\nonumber \\
	\Lambda_{44}^8 \,:\,& \centermathcell{X_{41}Y_{14}-X_{43} Y_{34}  }& & \centermathcell{W_{41} X_{14}-W_{42}Y_{24} }\nonumber
	\end{alignat}
	
Table~\ref{tab:GenerQ111phaseS} shows the generators of $Q^{1,1,1}$ in terms of the gauge theory. They satisfy the same relations given in \eqref{eq:Q111HSrel}.

\begin{table}[H]
		\centering
		\renewcommand{\arraystretch}{1.1}
		\begin{tabular}{c|c}
			Field    & Chiral superfields  \\
			\hline
			$M_1$    & $Z_{42}Y_{24}=Z_{43}X_{34}=W_{41}Y_{14}$ \\
			$M_2$    & $Z_{42}X_{24}=X_{43}X_{34}=Z_{41}Y_{14}$ \\
			$M_3$    & $W_{42}Y_{24}=W_{43}X_{34}=W_{41}X_{14}$ \\
			$M_4$    & $W_{42}X_{24}=Y_{43}X_{34}=Z_{41}X_{14}$ \\
			$M_5$    & $X_{42}Y_{24}=Z_{43}Y_{34}=Y_{41}Y_{14}$ \\
			$M_6$    & $X_{42}X_{24}=X_{43}Y_{34}=X_{41}Y_{14}$ \\
			$M_7$    & $Y_{42}Y_{24}=W_{43}Y_{34}=Y_{41}X_{14}$ \\
			$M_8$    & $Y_{42}X_{24}=Y_{43}Y_{34}=X_{41}X_{14}$
		\end{tabular}
		\caption{Generators of $Q^{1,1,1}$ in terms of fields in phase S.}
		\label{tab:GenerQ111phaseS}
	\end{table}	

Let us consider the involution that maps all gauge groups to themselves and acts on chiral fields as follows
\begin{equation}
	\begin{array}{cccccccccccc}
	Z_{42} & \rightarrow & \gamma_{\Omega_4}\bar{X}_{42}\gamma_{\Omega_2}^{-1}\coma &
	X_{42} & \rightarrow & -\gamma_{\Omega_4}\bar{Z}_{42}\gamma_{\Omega_2}^{-1}\coma &
	Y_{24} & \rightarrow & -\gamma_{\Omega_2}\bar{X}_{24}\gamma_{\Omega_4}^{-1}\coma &
	X_{24} & \rightarrow & \gamma_{\Omega_2}\bar{Y}_{24}\gamma_{\Omega_4}^{-1}\coma \\ 
	Z_{43} & \rightarrow & -\gamma_{\Omega_4}\bar{X}_{43}\gamma_{\Omega_3}^{-1}\coma &
	X_{43} & \rightarrow & \gamma_{\Omega_4}\bar{Z}_{43}\gamma_{\Omega_3}^{-1}\coma &
	X_{34} & \rightarrow & \gamma_{\Omega_3}\bar{Y}_{34}\gamma_{\Omega_4}^{-1}\coma &
	Y_{34} & \rightarrow & -\gamma_{\Omega_3}\bar{X}_{34}\gamma_{\Omega_4}^{-1}\coma \\
	W_{41} & \rightarrow & -\gamma_{\Omega_4}\bar{X}_{41}\gamma_{\Omega_1}^{-1}\coma &
	X_{41} & \rightarrow & -\gamma_{\Omega_4}\bar{W}_{41}\gamma_{\Omega_1}^{-1}\coma &
	Z_{41} & \rightarrow & \gamma_{\Omega_4}\bar{Y}_{41}\gamma_{\Omega_1}^{-1}\coma &
	Y_{41} & \rightarrow & \gamma_{\Omega_4}\bar{Z}_{41}\gamma_{\Omega_1}^{-1}\coma \\
	W_{42} & \rightarrow & \gamma_{\Omega_4}\bar{Y}_{42}\gamma_{\Omega_2}^{-1}\coma &
	Y_{42} & \rightarrow & -\gamma_{\Omega_4}\bar{W}_{42}\gamma_{\Omega_2}^{-1}\coma &
	W_{43} & \rightarrow & -\gamma_{\Omega_4}\bar{Y}_{43}\gamma_{\Omega_3}^{-1}\coma &
	Y_{43} & \rightarrow & \gamma_{\Omega_4}\bar{W}_{43}\gamma_{\Omega_3}^{-1}\coma \\
	& & & 
	Y_{14} & \rightarrow & \gamma_{\Omega_1}\bar{Y}_{14}\gamma_{\Omega_4}^{-1}\coma &
	X_{14} & \rightarrow & \gamma_{\Omega_1}\bar{X}_{14}\gamma_{\Omega_4}^{-1}\fstop 
	\end{array}
	\label{eq:Q111S-chiral-invol}
	\end{equation}
As we will explain shortly, we have chosen this involution in order to connect to the orientifold of phase A that we constructed in the previous section.
	
Invariance of $W^{(0,1)}$ implies the following action on Fermi fields
	\begin{equation}
	\begin{array}{cccccccccccc}
	\Lambda_{23}^1 & \rightarrow & \gamma_{\Omega_2}\bar{\Lambda}_{23}^{1}\gamma_{\Omega_3}^{-1}\coma & 
	\Lambda_{23}^2 & \rightarrow & \gamma_{\Omega_2}\bar{\Lambda}_{23}^{2}\gamma_{\Omega_3}^{-1}\coma & 
	\Lambda_{31}^1 & \rightarrow &  -\gamma_{\Omega_3}\bar{\Lambda}_{31}^{2}\gamma_{\Omega_1}^{-1}\coma & 
	\Lambda_{31}^2 & \rightarrow &  \gamma_{\Omega_3}\bar{\Lambda}_{31}^{1}\gamma_{\Omega_1}^{-1}\coma \\ 
	\Lambda_{12}^1 & \rightarrow & -\gamma_{\Omega_1}\bar{\Lambda}_{12}^{2}\gamma_{\Omega_2}^{-1}\coma & 
	\Lambda_{12}^2 & \rightarrow & \gamma_{\Omega_1}\bar{\Lambda}_{12}^{1}\gamma_{\Omega_2}^{-1}\coma & 
	\Lambda_{44}^1 & \rightarrow &  -\gamma_{\Omega_4}\bar{\Lambda}_{44}^{7}\gamma_{\Omega_4}^{-1}\coma & 
	\Lambda_{44}^7 & \rightarrow &  -\gamma_{\Omega_4}\bar{\Lambda}_{44}^{1}\gamma_{\Omega_4}^{-1}\coma \\ 
	\Lambda_{44}^2 & \rightarrow &  -\gamma_{\Omega_4}\bar{\Lambda}_{44}^{8}\gamma_{\Omega_4}^{-1}\coma & 
	\Lambda_{44}^8 & \rightarrow &  -\gamma_{\Omega_4}\bar{\Lambda}_{44}^{2}\gamma_{\Omega_4}^{-1}\coma & 
	\Lambda_{44}^3 & \rightarrow & -\gamma_{\Omega_4}\Lambda_{44}^{6\,\,T} \gamma_{\Omega_4}^{-1}\coma & 
	\Lambda_{44}^6 & \rightarrow & -\gamma_{\Omega_4}\Lambda_{44}^{3\,\,T} \gamma_{\Omega_4}^{-1}\coma \\
	& & & 
	\Lambda_{44}^4 & \rightarrow &  \gamma_{\Omega_4}\Lambda_{44}^{5\,\,T} \gamma_{\Omega_4}^{-1}\coma & 
	\Lambda_{44}^5 & \rightarrow &  \gamma_{\Omega_4}\Lambda_{44}^{4\,\,T} \gamma_{\Omega_4}^{-1}\coma 
	\end{array}
	\label{eq:Q111S-Fermi-invol}
	\end{equation}
	and
	\begin{equation}
    \Lambda_{11}^R\rightarrow \gamma_{\Omega_1}\Lambda_{11}^{R\,\,T}\gamma_{\Omega_1}^{-1} \coma \Lambda_{22}^R\rightarrow \gamma_{\Omega_2}\Lambda_{22}^{R\,\,T}\gamma_{\Omega_2}^{-1} \coma \Lambda_{33}^R\rightarrow \gamma_{\Omega_3}\Lambda_{33}^{R\,\,T}\gamma_{\Omega_3}^{-1} \coma \Lambda_{44}^R\rightarrow \gamma_{\Omega_4}\Lambda_{44}^{R\,\,T}\gamma_{\Omega_4}^{-1} \fstop
    \label{eq:Q111S-RFermi-invol}
	\end{equation}

The involution on bifundamental fields leads to the same constraints on the $\gamma_{\Omega_i}$ matrices as in~\eqref{eq:Q111A-Omega-cond}. As for phase A, we pick
\begin{equation}
\begin{array}{ccccc}
\gamma_{\Omega_1} & = & \gamma_{\Omega_4} & = & J \coma \\
\gamma_{\Omega_2} & = & \gamma_{\Omega_3} & = & \ID_N \fstop
\end{array}
\end{equation}	

\newpage

\begin{figure}[H]
	    \centering
	    	\begin{tikzpicture}[scale=2]
		\draw[line width=1pt,redX] (-0.22,-0.22) circle (0.25) node[xshift=-0.35cm,yshift=0.35cm,star,star points=5, star point ratio=2.25, inner sep=1pt, fill=redX, draw=redX] {} node[fill=white,text opacity=1,fill opacity=1,draw=black,rectangle,xshift=-0.4cm,yshift=-0.35cm,thin] {\color{redX}{$8$}};
		\draw[line width=1pt,redX] (-0.33,-0.33) circle (0.45) node[xshift=-0.65cm,yshift=0.6cm,redX] {\scriptsize{$\quadro$}} node[fill=white,text opacity=1,fill opacity=1,draw=black,rectangle,xshift=-0.65cm,yshift=-0.65cm,thin] {\color{redX}{$9$}} ;
		\draw[line width=1pt,redX] (-0.22,2.22) circle (0.25) node[xshift=-0.35cm,yshift=0.35cm] {\scriptsize{$\quadro$}};
		\draw[line width=1pt,redX] (2.22,-0.22) circle (0.25) node[xshift=0.35cm,yshift=-0.35cm,star,star points=5, star point ratio=2.25, inner sep=1pt, fill=redX, draw=redX] {};
		\draw[line width=1pt,redX] (2.22,2.22) circle (0.25) node[xshift=0.35cm,yshift=0.35cm,star,star points=5, star point ratio=2.25, inner sep=1pt, fill=redX, draw=redX] {};
		\node[draw=black,line width=1pt,circle,fill=yellowX,minimum width=0.75cm,inner sep=1pt,label={[xshift=-1cm,yshift=0cm]:$\USp(N)$}] (A) at (0,0) {$4$};
		\node[draw=black,line width=1pt,circle,fill=yellowX,minimum width=0.75cm,inner sep=1pt,label={[xshift=1cm,yshift=-0.25cm]:$\SO(N)$}] (B) at (2,0) {$3$};
		\node[draw=black,line width=1pt,circle,fill=yellowX,minimum width=0.75cm,inner sep=1pt,label={[xshift=-0.75cm,yshift=0cm]:$\SO(N)$}] (C) at (2,2) {$2$};
		\node[draw=black,line width=1pt,circle,fill=yellowX,minimum width=0.75cm,inner sep=1pt,label={[xshift=0.75cm,yshift=0cm]:$\USp(N)$}] (D) at (0,2) {$1$};
		\draw[line width=1pt,redX] (C) to node[fill=white,text opacity=1,fill opacity=1,draw=black,rectangle,thin,pos=0.5] {\color{redX}{$2$}} (D);
		\draw[line width=1pt,redX] (B) to node[fill=white,text opacity=1,fill opacity=1,draw=black,rectangle,thin,pos=0.5] {\color{redX}{$2$}} (C);
		\draw[line width=1pt,redX] (D) to node[fill=white,text opacity=1,fill opacity=1,draw=black,rectangle,thin,pos=0.75] {\color{redX}{$2$}} (B);
		\draw[line width=1pt] (A) to[bend right=20] node[fill=white,text opacity=1,fill opacity=1,draw=black,rectangle,thin,pos=0.5] {$2$} (B);
		\draw[line width=1pt] (A) to[bend left=20] node[fill=white,text opacity=1,fill opacity=1,draw=black,rectangle,thin,pos=0.5] {$4$} (B);
		\draw[line width=1pt] (A) to[bend right=20] node[fill=white,text opacity=1,fill opacity=1,draw=black,rectangle,thin,pos=0.5] {$2$} (D);
		\draw[line width=1pt] (A) to[bend left=20] node[fill=white,text opacity=1,fill opacity=1,draw=black,rectangle,thin,pos=0.5] {$4$} (D);
		\draw[line width=1pt] (A) to[bend right=20] node[fill=white,text opacity=1,fill opacity=1,draw=black,rectangle,thin,pos=0.75] {$2$} (C);
		\draw[line width=1pt] (A) to[bend left=20] node[fill=white,text opacity=1,fill opacity=1,draw=black,rectangle,thin,pos=0.75] {$4$} (C);
		\end{tikzpicture}
		
		\vspace{0.9cm} 
		
	    \caption{Quiver  diagram  for  the  Spin(7)  orientifold  of  phase  S of $Q^{1,1,1}$ using the involution in \cref{eq:Q111S-chiral-invol,eq:Q111S-Fermi-invol,eq:Q111S-RFermi-invol}, together with our choice of $\gamma_{\Omega_i}$ matrices.}
		\label{fig:o_theory_q111_S}
	\end{figure}
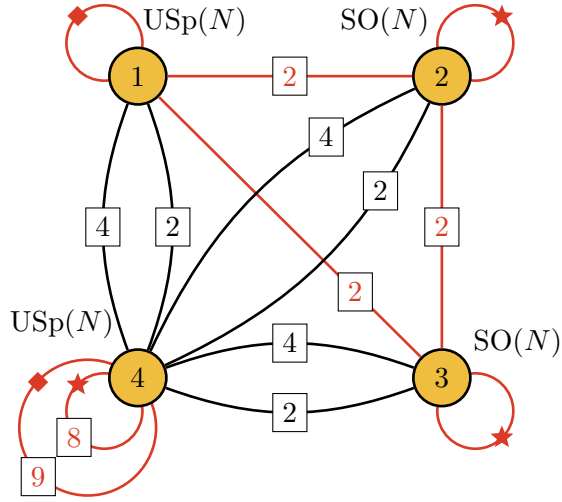

Using Table~\ref{tab:GenerQ111phaseS}, \eqref{eq:Q111S-chiral-invol} translates into the following action on the generators
\begin{equation}
 \begin{array}{cccccccccccc}
	M_{1}  & \rightarrow & -\bar{M}_{6} \coma &
	M_{2}  & \rightarrow & \bar{M}_{5} \coma &
	M_{3}  & \rightarrow & -\bar{M}_{8} \coma &
	M_{4}  & \rightarrow & \bar{M}_{7} \coma \\
	M_{5}  & \rightarrow & \bar{M}_{2} \coma &
	M_{6}  & \rightarrow & -\bar{M}_{1} \coma &
	M_{7}  & \rightarrow & \bar{M}_{4} \coma &
	M_{8}  & \rightarrow & -\bar{M}_{3} \coma
	\end{array} 
\label{eq:Q111phaseS-nonuni-geom}
\end{equation} 
which is the same geometric involution that we found for phase A in \eqref{eq:Q111phaseA-nonuni-geom}. Therefore, the involutions considered on these two phases correspond to the same Spin(7) orientifold of $Q^{1,1,1}$. 
Figure \ref{fig:o_theory_q111_S} shows the quiver for the orientifold theory, which is free of gauge anomalies.

\subsubsection{Triality Between the Orientifolded Theories}
\label{sec:Q111triality}

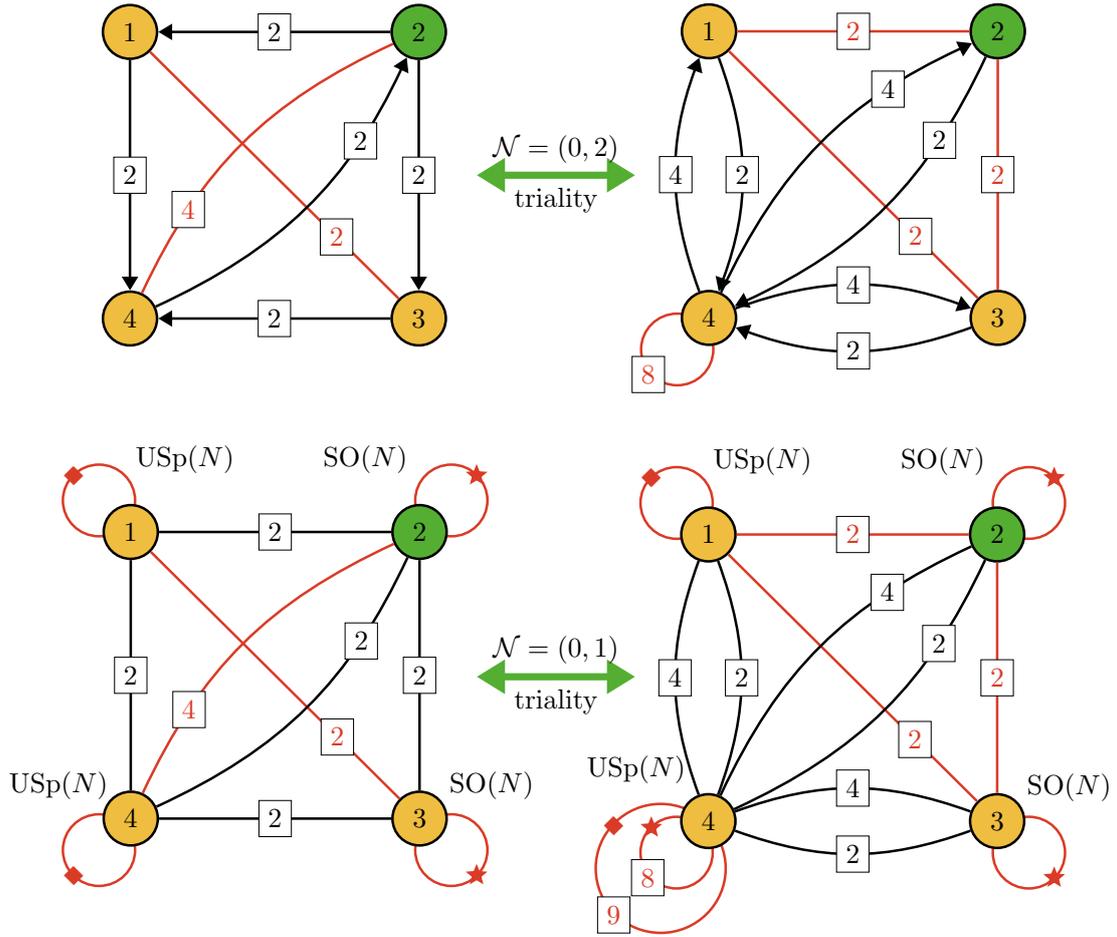
\begin{figure}[H]
    \centering
    \scalebox{0.95}{
    \begin{tikzpicture}[scale=4]
    \node (Q111A) at (0,2) {
    
    		\begin{tikzpicture}[scale=2]
	\node[draw=black,line width=1pt,circle,fill=yellowX,minimum width=0.75cm,inner sep=1pt] (A) at (0,0) {$4$};
	\node[draw=black,line width=1pt,circle,fill=yellowX,minimum width=0.75cm,inner sep=1pt] (B) at (2,0) {$3$};
	\node[draw=black,line width=1pt,circle,fill=greenX,minimum width=0.75cm,inner sep=1pt] (C) at (2,2) {$2$};
	\node[draw=black,line width=1pt,circle,fill=yellowX,minimum width=0.75cm,inner sep=1pt] (D) at (0,2) {$1$};
	\draw[line width=1pt,-Triangle] (C) to node[fill=white,text opacity=1,fill opacity=1,draw=black,rectangle,thin,pos=0.5] {$2$} (D);
	\draw[line width=1pt,-Triangle] (C) to node[fill=white,text opacity=1,fill opacity=1,draw=black,rectangle,thin,pos=0.5] {$2$} (B);
	\draw[line width=1pt,-Triangle] (B) to node[fill=white,text opacity=1,fill opacity=1,draw=black,rectangle,thin,pos=0.5] {$2$} (A);
	\draw[line width=1pt,-Triangle] (D) to node[fill=white,text opacity=1,fill opacity=1,draw=black,rectangle,thin,pos=0.5] {$2$} (A);
	\draw[line width=1pt,redX] (D) to node[fill=white,text opacity=1,fill opacity=1,draw=black,rectangle,thin,pos=0.75] {\color{redX}{$2$}} (B);
	\draw[line width=1pt,-Triangle] (A) to[bend right=20] node[fill=white,text opacity=1,fill opacity=1,draw=black,rectangle,thin,pos=0.75] {$2$} (C);
	\draw[line width=1pt,redX] (A) to[bend left=20] node[fill=white,text opacity=1,fill opacity=1,draw=black,rectangle,thin,pos=0.25] {\color{redX}{$4$}} (C);
	\end{tikzpicture}
    
    };
    \node (Q111S) at (1.92,1.92) {
    
    	\begin{tikzpicture}[scale=2]
		\draw[line width=1pt,redX] (-0.22,-0.22) circle (0.25) node[fill=white,text opacity=1,fill opacity=1,draw=black,rectangle,xshift=-0.4cm,yshift=-0.35cm,thin] {\color{redX}{$8$}};
		\node[draw=black,line width=1pt,circle,fill=yellowX,minimum width=0.75cm,inner sep=1pt] (A) at (0,0) {$4$};
		\node[draw=black,line width=1pt,circle,fill=yellowX,minimum width=0.75cm,inner sep=1pt] (B) at (2,0) {$3$};
		\node[draw=black,line width=1pt,circle,fill=greenX,minimum width=0.75cm,inner sep=1pt] (C) at (2,2) {$2$};
		\node[draw=black,line width=1pt,circle,fill=yellowX,minimum width=0.75cm,inner sep=1pt] (D) at (0,2) {$1$};
		\draw[line width=1pt,redX] (C) to node[fill=white,text opacity=1,fill opacity=1,draw=black,rectangle,thin,pos=0.5] {\color{redX}{$2$}} (D);
		\draw[line width=1pt,redX] (B) to node[fill=white,text opacity=1,fill opacity=1,draw=black,rectangle,thin,pos=0.5] {\color{redX}{$2$}} (C);
		\draw[line width=1pt,redX] (D) to node[fill=white,text opacity=1,fill opacity=1,draw=black,rectangle,thin,pos=0.75] {\color{redX}{$2$}} (B);
		\draw[line width=1pt,Triangle-] (A) to[bend right=20] node[fill=white,text opacity=1,fill opacity=1,draw=black,rectangle,thin,pos=0.5] {$2$} (B);
		\draw[line width=1pt,-Triangle] (A) to[bend left=20] node[fill=white,text opacity=1,fill opacity=1,draw=black,rectangle,thin,pos=0.5] {$4$} (B);
		\draw[line width=1pt,Triangle-] (A) to[bend right=20] node[fill=white,text opacity=1,fill opacity=1,draw=black,rectangle,thin,pos=0.5] {$2$} (D);
		\draw[line width=1pt,-Triangle] (A) to[bend left=20] node[fill=white,text opacity=1,fill opacity=1,draw=black,rectangle,thin,pos=0.5] {$4$} (D);
		\draw[line width=1pt,Triangle-] (A) to[bend right=20] node[fill=white,text opacity=1,fill opacity=1,draw=black,rectangle,thin,pos=0.75] {$2$} (C);
		\draw[line width=1pt,-Triangle] (A) to[bend left=20] node[fill=white,text opacity=1,fill opacity=1,draw=black,rectangle,thin,pos=0.75] {$4$} (C);
		\end{tikzpicture}
    
    };
    \node (Q111OA) at (-0.01,0.29) {
    
    	\begin{tikzpicture}[scale=2]
		\draw[line width=1pt,redX] (-0.22,-0.22) circle (0.25) node[xshift=-0.35cm,yshift=-0.35cm] {\scriptsize{$\quadro$}};
		\draw[line width=1pt,redX] (-0.22,2.22) circle (0.25) node[xshift=-0.35cm,yshift=0.35cm] {\scriptsize{$\quadro$}};
		\draw[line width=1pt,redX] (2.22,-0.22) circle (0.25) node[xshift=0.35cm,yshift=-0.35cm,star,star points=5, star point ratio=2.25, inner sep=1pt, fill=redX, draw=redX] {};
		\draw[line width=1pt,redX] (2.22,2.22) circle (0.25) node[xshift=0.35cm,yshift=0.35cm,star,star points=5, star point ratio=2.25, inner sep=1pt, fill=redX, draw=redX] {};
			\node[draw=black,line width=1pt,circle,fill=yellowX,minimum width=0.75cm,inner sep=1pt,label={[xshift=-1cm,yshift=-0.25cm]:$\USp(N)$}] (A) at (0,0) {$4$};
		\node[draw=black,line width=1pt,circle,fill=yellowX,minimum width=0.75cm,inner sep=1pt,label={[xshift=1cm,yshift=-0.25cm]:$\SO(N)$}] (B) at (2,0) {$3$};
		\node[draw=black,line width=1pt,circle,fill=greenX,minimum width=0.75cm,inner sep=1pt,label={[xshift=-0.75cm,yshift=0.3cm]:$\SO(N)$}] (C) at (2,2) {$2$};
		\node[draw=black,line width=1pt,circle,fill=yellowX,minimum width=0.75cm,inner sep=1pt,label={[xshift=0.75cm,yshift=0.3cm]:$\USp(N)$}] (D) at (0,2) {$1$};
		\draw[line width=1pt] (C) to node[fill=white,text opacity=1,fill opacity=1,draw=black,rectangle,thin,pos=0.5] {$2$} (D);
		\draw[line width=1pt] (C) to node[fill=white,text opacity=1,fill opacity=1,draw=black,rectangle,thin,pos=0.5] {$2$} (B);
		\draw[line width=1pt] (B) to node[fill=white,text opacity=1,fill opacity=1,draw=black,rectangle,thin,pos=0.5] {$2$} (A);
		\draw[line width=1pt] (D) to node[fill=white,text opacity=1,fill opacity=1,draw=black,rectangle,thin,pos=0.5] {$2$} (A);
		\draw[line width=1pt,redX] (D) to node[fill=white,text opacity=1,fill opacity=1,draw=black,rectangle,thin,pos=0.75] {\color{redX}{$2$}} (B);
		\draw[line width=1pt] (A) to[bend right=20] node[fill=white,text opacity=1,fill opacity=1,draw=black,rectangle,thin,pos=0.75] {$2$} (C);
		\draw[line width=1pt,redX] (A) to[bend left=20] node[fill=white,text opacity=1,fill opacity=1,draw=black,rectangle,thin,pos=0.25] {\color{redX}{$4$}} (C);
		\end{tikzpicture}
    
    };
    \node (Q111OS) at (1.99,0.168) {
    	  	\begin{tikzpicture}[scale=2]
	\draw[line width=1pt,redX] (-0.22,-0.22) circle (0.25) node[xshift=-0.35cm,yshift=0.35cm,star,star points=5, star point ratio=2.25, inner sep=1pt, fill=redX, draw=redX] {} node[fill=white,text opacity=1,fill opacity=1,draw=black,rectangle,xshift=-0.4cm,yshift=-0.35cm,thin] {\color{redX}{$8$}};
		\draw[line width=1pt,redX] (-0.33,-0.33) circle (0.45) node[xshift=-0.65cm,yshift=0.6cm,redX] {\scriptsize{$\quadro$}} node[fill=white,text opacity=1,fill opacity=1,draw=black,rectangle,xshift=-0.65cm,yshift=-0.65cm,thin] {\color{redX}{$9$}} ;
		\draw[line width=1pt,redX] (-0.22,2.22) circle (0.25) node[xshift=-0.35cm,yshift=0.35cm] {\scriptsize{$\quadro$}};
		\draw[line width=1pt,redX] (2.22,-0.22) circle (0.25) node[xshift=0.35cm,yshift=-0.35cm,star,star points=5, star point ratio=2.25, inner sep=1pt, fill=redX, draw=redX] {};
		\draw[line width=1pt,redX] (2.22,2.22) circle (0.25) node[xshift=0.35cm,yshift=0.35cm,star,star points=5, star point ratio=2.25, inner sep=1pt, fill=redX, draw=redX] {};
		\node[draw=black,line width=1pt,circle,fill=yellowX,minimum width=0.75cm,inner sep=1pt,label={[xshift=-1cm,yshift=0cm]:$\USp(N)$}] (A) at (0,0) {$4$};
		\node[draw=black,line width=1pt,circle,fill=yellowX,minimum width=0.75cm,inner sep=1pt,label={[xshift=1cm,yshift=-0.25cm]:$\SO(N)$}] (B) at (2,0) {$3$};
		\node[draw=black,line width=1pt,circle,fill=greenX,minimum width=0.75cm,inner sep=1pt,label={[xshift=-0.75cm,yshift=0.3cm]:$\SO(N)$}] (C) at (2,2) {$2$};
		\node[draw=black,line width=1pt,circle,fill=yellowX,minimum width=0.75cm,inner sep=1pt,label={[xshift=0.75cm,yshift=0.3cm]:$\USp(N)$}] (D) at (0,2) {$1$};
		\draw[line width=1pt,redX] (C) to node[fill=white,text opacity=1,fill opacity=1,draw=black,rectangle,thin,pos=0.5] {\color{redX}{$2$}} (D);
		\draw[line width=1pt,redX] (B) to node[fill=white,text opacity=1,fill opacity=1,draw=black,rectangle,thin,pos=0.5] {\color{redX}{$2$}} (C);
		\draw[line width=1pt,redX] (D) to node[fill=white,text opacity=1,fill opacity=1,draw=black,rectangle,thin,pos=0.75] {\color{redX}{$2$}} (B);
		\draw[line width=1pt] (A) to[bend right=20] node[fill=white,text opacity=1,fill opacity=1,draw=black,rectangle,thin,pos=0.5] {$2$} (B);
		\draw[line width=1pt] (A) to[bend left=20] node[fill=white,text opacity=1,fill opacity=1,draw=black,rectangle,thin,pos=0.5] {$4$} (B);
		\draw[line width=1pt] (A) to[bend right=20] node[fill=white,text opacity=1,fill opacity=1,draw=black,rectangle,thin,pos=0.5] {$2$} (D);
		\draw[line width=1pt] (A) to[bend left=20] node[fill=white,text opacity=1,fill opacity=1,draw=black,rectangle,thin,pos=0.5] {$4$} (D);
		\draw[line width=1pt] (A) to[bend right=20] node[fill=white,text opacity=1,fill opacity=1,draw=black,rectangle,thin,pos=0.75] {$2$} (C);
		\draw[line width=1pt] (A) to[bend left=20] node[fill=white,text opacity=1,fill opacity=1,draw=black,rectangle,thin,pos=0.75] {$4$} (C);
		\node at (0,-0.92) {};
		\end{tikzpicture}
    
    };
    \draw[Triangle-Triangle,greenX,line width=1mm] (0.7,2) -- node[above,midway,black] {$\mathcal{N}=(0,2)$} node[below,midway,black] {triality} (1.25,2);
    \draw[Triangle-Triangle,greenX,line width=1mm] (0.7,0.25) -- node[above,midway,black] {$\mathcal{N}=(0,1)$} node[below,midway,black] {triality} (1.25,0.25);
    \end{tikzpicture}
    }
 \caption{Phases A and S of $Q^{1,1,1}$ are connected by $\mathcal{N}= (0,2)$ triality on node 2 (shown in green). The orientifolded theories are similarly connected by $\mathcal{N}= (0,1)$ triality.}
		\label{01_02_triality_q111}
\end{figure}

Figure \ref{01_02_triality_q111} summarizes the connections between the theories considered in this section. Again, we observe that the two theories we constructed for the same Spin(7) orientifold are related by $\mathcal{N}=(0,1)$ triality. More precisely, they are related by a simple generalization of the basic triality reviewed in Section \ref{sec:N01triality}. First, in this case, triality is applied to quivers with multiple gauge nodes. More importantly, some of the nodes that act as flavor groups are of a different type (in this example, USp) than the dualized node. As in previous examples, the orientifold construction leads to a clear prescription on how to treat scalar flavors, which is inherited from the parent theories.

\subsection{Theories with Unitary Gauge Groups: $Q^{1,1,1}/\mathbb{Z}_2$}
\label{sec:Q111Z2}

All $\mathcal{N}=(0,1)$ triality examples we constructed so far contain only $\SO(N)$ and $\USp(N)$ gauge groups. Namely, the anti-holomorphic involutions of the parent $\mathcal{N}=(0,2)$ theories, universal or not,  map all gauge groups to themselves. In this section we will construct Spin(7) orientifolds giving rise to gauge theories that include $\U(N)$ gauge groups. To do so, we focus on $Q^{1,1,1}/\mathbb{Z}_2$, whose toric diagram is shown in Figure~\ref{fig:Q111Z2toricdiagram}.\footnote{More precisely, this is the $\mathbb{Z}_2$ orbifold of the real cone over $Q^{1,1,1}$.} This CY$_4$ has a rich family of 14 toric phases. They were classified in \cite{Franco:2018qsc}, whose nomenclature we will follow. We will restrict to a subset consisting of 5 of these toric phases. In order to streamline our discussion, several details about these theories are collected in Appendix \ref{app:Q111Z2-details}.

	\begin{figure}[H]
		\centering
		\begin{tikzpicture}[scale=1.5]
		\draw[thick,gray,-Triangle] (-1.5,0,0) -- node[above,pos=1] {$x$} (1.5,0,0);
		\draw[thick,gray,-Triangle] (0,-1.5,0) -- node[left,pos=1] {$y$} (0,1.5,0);
		\draw[thick,gray,-Triangle] (0,0,-1.5) -- node[below,pos=1] {$z$} (0,0,1.5);
		\node[draw=black,line width=1pt,circle,fill=black,minimum width=0.2cm,inner sep=1pt] (p1) at (1,0,0) {};
		\node[draw=black,line width=1pt,circle,fill=black,minimum width=0.2cm,inner sep=1pt] (p2) at (-1,0,0) {};
		\node[draw=black,line width=1pt,circle,fill=black,minimum width=0.2cm,inner sep=1pt] (p3) at (0,1,0) {};
		\node[draw=black,line width=1pt,circle,fill=black,minimum width=0.2cm,inner sep=1pt] (p4) at (0,-1,0) {};
		\node[draw=black,line width=1pt,circle,fill=black,minimum width=0.2cm,inner sep=1pt] (p5) at (0,0,1) {};
		\node[draw=black,line width=1pt,circle,fill=black,minimum width=0.2cm,inner sep=1pt] (p6) at (0,0,-1) {};
		\node[draw=black,line width=1pt,circle,fill=black,minimum width=0.2cm,inner sep=1pt] (si) at (0,0,0) {};
		\draw[line width=1pt] (p1)--(p3)--(p2)--(p4)--(p1);
		\draw[line width=1pt] (p1)--(p6)--(p2);
		\draw[line width=1pt] (p4)--(p6)--(p3);
		\draw[line width=1pt] (p1)--(p5)--(p2);
		\draw[line width=1pt] (p4)--(p5)--(p3);
		\end{tikzpicture}
		\caption{Toric diagram for $Q^{1,1,1}/\ZZ_2$.}
		\label{fig:Q111Z2toricdiagram}
	\end{figure}	

Let us first consider phase D, whose quiver diagram is shown in Figure \ref{fig:Q111Z2quiverD}. We provide a 3d representation of the quiver in order to make the action of the anti-holomorphic involution that we will use to construct a Spin(7) orientifold more manifest. 

			\begin{figure}[H]
		\centering
		\begin{tikzpicture}[scale=2, rotate around y = 0]
	\def\R{1.75};
	        \node[draw=black,line width=1pt,circle,fill=yellowX,minimum width=0.75cm,inner sep=1pt] (v1) at (-\R,0,-\R) {$1$};
	        \node[draw=black,line width=1pt,circle,fill=yellowX,minimum width=0.75cm,inner sep=1pt] (v2) at (\R,0,-\R) {$7$};
	        \node[draw=black,line width=1pt,circle,fill=yellowX,minimum width=0.75cm,inner sep=1pt] (v3) at (0,0,-\R) {$3$};
	        \node[draw=black,line width=1pt,circle,fill=yellowX,minimum width=0.75cm,inner sep=1pt] (v4) at (0,\R,0) {$4$};
	        \node[draw=black,line width=1pt,circle,fill=yellowX,minimum width=0.75cm,inner sep=1pt] (v5) at (0,-\R,0) {$5$};
	        \node[draw=black,line width=1pt,circle,fill=yellowX,minimum width=0.75cm,inner sep=1pt] (v6) at (0,0,\R) {$6$};
	        \node[draw=black,line width=1pt,circle,fill=yellowX,minimum width=0.75cm,inner sep=1pt] (v7) at (\R,0,\R) {$2$};
	        \node[draw=black,line width=1pt,circle,fill=yellowX,minimum width=0.75cm,inner sep=1pt] (v8) at (-\R,0,\R) {$8$};
	        \draw[line width=1pt,redX] (v1) to node[fill=white,text opacity=1,fill opacity=1,draw=black,rectangle,thin,pos=0.25] {\color{redX}{$2$}} 
	        (v3);
	        \draw[line width=1pt,redX] (v3) to node[fill=white,text opacity=1,fill opacity=1,draw=black,rectangle,thin,pos=0.5] {\color{redX}{$2$}} 
	        (v2);
	        \draw[line width=1pt,redX] (v4) to node[fill=white,text opacity=1,fill opacity=1,draw=black,rectangle,thin,pos=0.5] {\color{redX}{$8$}} 
	        (v5);
	        \draw[line width=1pt,redX] (v8) to node[fill=white,text opacity=1,fill opacity=1,draw=black,rectangle,thin,pos=0.5] {\color{redX}{$2$}} 
	        (v6);
	        \draw[line width=1pt,redX] (v6) to node[fill=white,text opacity=1,fill opacity=1,draw=black,rectangle,thin,pos=0.75] {\color{redX}{$2$}} 
	        (v7);
	        \draw[line width=1pt,-Triangle] (v1) to node[fill=white,text opacity=1,fill opacity=1,draw=black,rectangle,thin,pos=0.5] {$2$} 
	        (v8);
	        \draw[line width=1pt,-Triangle] (v3) to node[fill=white,text opacity=1,fill opacity=1,draw=black,rectangle,thin,pos=0.5] {$4$} 
	        (v4);
	        \draw[line width=1pt,-Triangle] (v4) to node[fill=white,text opacity=1,fill opacity=1,draw=black,rectangle,thin,pos=0.5] {$2$} 
	        (v1);
	        \draw[line width=1pt,-Triangle] (v4) to node[fill=white,text opacity=1,fill opacity=1,draw=black,rectangle,thin,pos=0.5] {$2$} 
	        (v2);
	        \draw[line width=1pt,-Triangle] (v2) to node[fill=white,text opacity=1,fill opacity=1,draw=black,rectangle,thin,pos=0.5] {$2$} 
	        (v7);
	        \draw[line width=1pt,-Triangle] (v8) to node[fill=white,text opacity=1,fill opacity=1,draw=black,rectangle,thin,pos=0.5] {$2$} 
	        (v5);
	        \draw[line width=1pt,-Triangle] (v5) to node[fill=white,text opacity=1,fill opacity=1,draw=black,rectangle,thin,pos=0.6] {$2$} 
	        (v3);
	        \draw[line width=1pt,-Triangle] (v5) to node[fill=white,text opacity=1,fill opacity=1,draw=black,rectangle,thin,pos=0.5] {$4$} 
	        (v6);
	        \draw[line width=1pt,-Triangle] (v6) to node[fill=white,text opacity=1,fill opacity=1,draw=black,rectangle,thin,pos=0.4] {$2$} 
	        (v4);
	        \draw[line width=1pt,-Triangle] (v7) to node[fill=white,text opacity=1,fill opacity=1,draw=black,rectangle,thin,pos=0.5] {$2$} 
	        (v5);
	\end{tikzpicture}
	\caption{Quiver diagram for phase D of $Q^{1,1,1}/\ZZ_2$.}
		\label{fig:Q111Z2quiverD}
	\end{figure}

The $J$- and $E$-terms for this theory are 
		\begin{alignat}{4}
	\renewcommand{\arraystretch}{1.1}
    & \centermathcell{J}                           &\text{\hspace{.5cm}}& \centermathcell{E                               }\nonumber \\
 \Lambda^1_{13}  \,:\, &  \centermathcell{W_{34} X_{41}-Y_{41} Z_{34}} & & \centermathcell{X_{18} X_{85} Y_{53}-X_{53} X_{85} Y_{18} }\nonumber\\
 \Lambda^2_{13}  \,:\, &  \centermathcell{X_{41} Y_{34}-X_{34} Y_{41} }& & \centermathcell{X_{53} Y_{18} Y_{85}-X_{18} Y_{53} Y_{85} }\nonumber\\
 \Lambda^1_{37}  \,:\, &  \centermathcell{X_{72} Y_{53} Y_{25}-X_{53} Y_{72} Y_{25} }&  & \centermathcell{X_{47} Z_{34}-X_{34} Y_{47}} \nonumber\\
 \Lambda^2_{37}  \,:\, &  \centermathcell{X_{72} X_{25} Y_{53}-X_{53} X_{25} Y_{72} }&  & \centermathcell{Y_{34} Y_{47}-W_{34} X_{47}}\nonumber \\
 \Lambda^1_{86}  \,:\, &  \centermathcell{X_{64} Y_{18} Y_{41}-X_{18} Y_{41} Y_{64} }&  & \centermathcell{X_{56} Y_{85}-X_{85} Z_{56}}\nonumber \\
 \Lambda^2_{86}  \,:\, &  \centermathcell{X_{41} X_{64} Y_{18}-X_{18} X_{41} Y_{64} }&  & \centermathcell{W_{56} X_{85}-Y_{56} Y_{85}}\nonumber \\
 \Lambda^1_{62}  \,:\, &  \centermathcell{Y_{25} Z_{56}-W_{56} X_{25} }&  & \centermathcell{X_{47} X_{64} Y_{72}-X_{72} X_{47} Y_{64}} \nonumber\\
 \Lambda^2_{62}  \,:\, &  \centermathcell{X_{25} Y_{56}-X_{56} Y_{25} }&  & \centermathcell{X_{64} Y_{72} Y_{47}-X_{72} Y_{47} Y_{64}} \label{eq:Q111Z2JEtermsD}\\
 \Lambda^1_{45}  \,:\, &  \centermathcell{W_{56} Y_{64}-W_{34} Y_{53} }&  & \centermathcell{X_{18} X_{41} X_{85}-X_{72} X_{47} X_{25}}\nonumber \\
 \Lambda^2_{45}  \,:\, &  \centermathcell{W_{56} X_{64}-W_{34} X_{53} }&  & \centermathcell{X_{47} X_{25} Y_{72}-X_{41} X_{85} Y_{18}}\nonumber \\
 \Lambda^3_{45}  \,:\, &  \centermathcell{Y_{64} Z_{56}-Y_{53} Z_{34} }&  & \centermathcell{X_{72} X_{47} Y_{25}-X_{18} X_{85} Y_{41}}\nonumber \\
 \Lambda^4_{45}  \,:\, &  \centermathcell{X_{64} Z_{56}-X_{53} Z_{34} }&  & \centermathcell{X_{85} Y_{18} Y_{41}-X_{47} Y_{72} Y_{25}}\nonumber \\
 \Lambda^5_{45}  \,:\, &  \centermathcell{Y_{56} Y_{64}-Y_{34} Y_{53} }&  & \centermathcell{X_{72} X_{25} Y_{47}-X_{18} X_{41} Y_{85}}\nonumber \\
 \Lambda^6_{45}  \,:\, &  \centermathcell{X_{64} Y_{56}-X_{53} Y_{34} }&  & \centermathcell{X_{41} Y_{18} Y_{85}-X_{25} Y_{72} Y_{47}}\nonumber \\
 \Lambda^7_{45}  \,:\, &  \centermathcell{X_{56} Y_{64}-X_{34} Y_{53} }&  & \centermathcell{X_{18} Y_{41} Y_{85}-X_{72} Y_{47} Y_{25}}\nonumber \\
 \Lambda^8_{45}  \,:\, &  \centermathcell{X_{56} X_{64}-X_{34} X_{53} }&  & \centermathcell{Y_{72} Y_{47} Y_{25}-Y_{18} Y_{41} Y_{85}}\nonumber 
	\end{alignat}

The generators of $Q^{1,1,1}/\mathbb{Z}_2$ in terms of the chiral fields in phase D are listed in Table~\ref{tab:GenQ111Z2PhaseD}.  Note that the generators and their relations are common to all the phases, but their realizations in terms of chiral superfields in each of them are different. Let us consider an anti-holomorphic involution of phase D which acts on Figure~\ref{fig:Q111Z2quiverD} as a reflection with respect to the vertical plane that contains nodes 3, 4, 5 and 6. The nodes on the plane map to themselves, while the following pairs $1\leftrightarrow 7$ and $2\leftrightarrow 8$ get identified. This leads to the anticipated mixture of $\SO/\USp$ and $\U$ gauge groups.

The involution on chiral fields is 
   \begin{equation}
     \begin{array}{cccccccccccc}
	X_{18} & \rightarrow & \gamma_{\Omega_7}\bar{Y}_{72}\gamma_{\Omega_2}^{-1}\coma &
	Y_{18} & \rightarrow & \gamma_{\Omega_7}\bar{X}_{72}\gamma_{\Omega_2}^{-1}\coma &
	X_{72} & \rightarrow & \gamma_{\Omega_1}\bar{Y}_{18}\gamma_{\Omega_8}^{-1}\coma &
	Y_{72} & \rightarrow & \gamma_{\Omega_1}\bar{X}_{18}\gamma_{\Omega_8}^{-1}\coma \\ 
	X_{34} & \rightarrow & \gamma_{\Omega_3}\bar{X}_{34}\gamma_{\Omega_4}^{-1}\coma &
	Y_{34} & \rightarrow & \gamma_{\Omega_3}\bar{Z}_{34}\gamma_{\Omega_4}^{-1}\coma &
	Z_{34} & \rightarrow & \gamma_{\Omega_3}\bar{Y}_{34}\gamma_{\Omega_4}^{-1}\coma &
	W_{34} & \rightarrow & \gamma_{\Omega_3}\bar{W}_{34}\gamma_{\Omega_4}^{-1}\coma \\
	X_{41} & \rightarrow & \gamma_{\Omega_4}\bar{X}_{47}\gamma_{\Omega_7}^{-1}\coma &
	Y_{41} & \rightarrow & \gamma_{\Omega_4}\bar{Y}_{47}\gamma_{\Omega_7}^{-1}\coma &
	X_{47} & \rightarrow & \gamma_{\Omega_4}\bar{X}_{41}\gamma_{\Omega_1}^{-1}\coma &
	Y_{47} & \rightarrow & \gamma_{\Omega_4}\bar{Y}_{41}\gamma_{\Omega_1}^{-1}\coma \\
	X_{53} & \rightarrow & \gamma_{\Omega_5}\bar{Y}_{53}\gamma_{\Omega_3}^{-1}\coma &
	Y_{53} & \rightarrow & \gamma_{\Omega_5}\bar{X}_{53}\gamma_{\Omega_3}^{-1}\coma &
	X_{56} & \rightarrow & \gamma_{\Omega_5}\bar{X}_{56}\gamma_{\Omega_6}^{-1}\coma &
	Y_{56} & \rightarrow & \gamma_{\Omega_5}\bar{Z}_{56}\gamma_{\Omega_6}^{-1}\coma \\
	Z_{56} & \rightarrow & \gamma_{\Omega_5}\bar{Y}_{56}\gamma_{\Omega_6}^{-1}\coma &
	W_{56} & \rightarrow & \gamma_{\Omega_5}\bar{W}_{56}\gamma_{\Omega_6}^{-1}\coma &
	X_{64} & \rightarrow & \gamma_{\Omega_6}\bar{Y}_{64}\gamma_{\Omega_4}^{-1}\coma &
	Y_{64} & \rightarrow & \gamma_{\Omega_6}\bar{X}_{64}\gamma_{\Omega_4}^{-1}\coma \\
	X_{25} & \rightarrow & \gamma_{\Omega_8}\bar{X}_{85}\gamma_{\Omega_5}^{-1}\coma &
	Y_{25} & \rightarrow & \gamma_{\Omega_8}\bar{Y}_{85}\gamma_{\Omega_5}^{-1}\coma &
	X_{85} & \rightarrow & \gamma_{\Omega_2}\bar{X}_{25}\gamma_{\Omega_5}^{-1}\coma &
	Y_{85} & \rightarrow & \gamma_{\Omega_2}\bar{Y}_{25}\gamma_{\Omega_5}^{-1}\fstop
	\end{array}   
	\label{eq:Q111Z2D-chiral-invol}
    \end{equation}

Invariance of $W^{(0,1)}$ implies the following action on Fermi fields
\begin{equation}
     \begin{array}{cccccccccccc}
	\Lambda^1_{13} & \rightarrow & -\gamma_{\Omega_3}\Lambda^2_{37}\gamma_{\Omega_7}^{-1}\coma &
	\Lambda^2_{13} & \rightarrow & \gamma_{\Omega_3}\Lambda^1_{37}\gamma_{\Omega_7}^{-1}\coma &
	\Lambda^1_{37} & \rightarrow & \gamma_{\Omega_1}\Lambda^2_{13}\gamma_{\Omega_3}^{-1}\coma &
	\Lambda^2_{37} & \rightarrow & -\gamma_{\Omega_3}\Lambda^1_{37}\gamma_{\Omega_7}^{-1}\coma \\ 
	\Lambda^1_{86} & \rightarrow & -\gamma_{\Omega_6}\Lambda^2_{62}\gamma_{\Omega_8}^{-1}\coma &
	\Lambda^2_{86} & \rightarrow & -\gamma_{\Omega_6}\Lambda^1_{62}\gamma_{\Omega_8}^{-1}\coma &
	\Lambda^1_{62} & \rightarrow & -\gamma_{\Omega_8}\Lambda^2_{86}\gamma_{\Omega_6}^{-1}\coma &
	\Lambda^2_{62} & \rightarrow & -\gamma_{\Omega_8}\Lambda^1_{86}\gamma_{\Omega_6}^{-1}\coma \\
	\Lambda^1_{45} & \rightarrow & \gamma_{\Omega_4}\bar{\Lambda}^2_{45}\gamma_{\Omega_5}^{-1}\coma &
	\Lambda^2_{45} & \rightarrow & \gamma_{\Omega_4}\bar{\Lambda}^1_{45}\gamma_{\Omega_5}^{-1}\coma &
	\Lambda^3_{45} & \rightarrow & \gamma_{\Omega_4}\bar{\Lambda}^6_{45}\gamma_{\Omega_5}^{-1}\coma &
	\Lambda^4_{45} & \rightarrow & \gamma_{\Omega_4}\bar{\Lambda}^5_{45}\gamma_{\Omega_5}^{-1}\coma \\
	\Lambda^5_{45} & \rightarrow & \gamma_{\Omega_5}\bar{\Lambda}^4_{45}\gamma_{\Omega_5}^{-1}\coma &
	\Lambda^6_{45} & \rightarrow & \gamma_{\Omega_5}\bar{\Lambda}^3_{45}\gamma_{\Omega_5}^{-1}\coma &
	\Lambda^7_{45} & \rightarrow & \gamma_{\Omega_5}\bar{\Lambda}^8_{45}\gamma_{\Omega_5}^{-1}\coma &
	\Lambda^8_{45} & \rightarrow & \gamma_{\Omega_5}\bar{\Lambda}^7_{45}\gamma_{\Omega_5}^{-1}\coma
	\end{array}   
	\label{eq:Q111Z2D-Fermi-invol}
    \end{equation}
and
	\begin{equation}
	 \begin{array}{cccccccccccc}
	\Lambda_{11}^R & \rightarrow & \gamma_{\Omega_7}\Lambda_{77}^{R\,\,T}\gamma_{\Omega_7}^{-1} \coma &
	\Lambda_{22}^R & \rightarrow & \gamma_{\Omega_8}\Lambda_{88}^{R\,\,T}\gamma_{\Omega_8}^{-1} \coma &
	\Lambda_{33}^R & \rightarrow & \gamma_{\Omega_3}\Lambda_{33}^{R\,\,T}\gamma_{\Omega_3}^{-1} \coma &
	\Lambda_{44}^R & \rightarrow & \gamma_{\Omega_4}\Lambda_{44}^{R\,\,T}\gamma_{\Omega_4}^{-1} \coma \\
	\Lambda_{55}^R & \rightarrow & \gamma_{\Omega_5}\Lambda_{55}^{R\,\,T}\gamma_{\Omega_5}^{-1} \coma &
	\Lambda_{66}^R & \rightarrow & \gamma_{\Omega_6}\Lambda_{66}^{R\,\,T}\gamma_{\Omega_6}^{-1} \coma &
	\Lambda_{77}^R & \rightarrow & \gamma_{\Omega_1}\Lambda_{11}^{R\,\,T}\gamma_{\Omega_1}^{-1} \coma &
	\Lambda_{88}^R & \rightarrow & \gamma_{\Omega_2}\Lambda_{22}^{R\,\,T}\gamma_{\Omega_2}^{-1} \fstop
	\end{array}
	\label{eq:Q111Z2D-RFermi-invol}
	\end{equation}
	
Using Table \ref{tab:GenQ111Z2PhaseD}, we find the corresponding geometric involution on the generators of $Q^{1,1,1}/\mathbb{Z}_2$
\begin{equation}
     \begin{array}{ccccccccccccccc}
	M_{1}  & \rightarrow & \bar{M}_{27} \coma &
	M_{2}  & \rightarrow & \bar{M}_{24} \coma &
	M_{3}  & \rightarrow & \bar{M}_{21} \coma &
	M_{4}  & \rightarrow & \bar{M}_{18} \coma &
	M_{5}  & \rightarrow & \bar{M}_{12} \coma \\
	M_{6}  & \rightarrow & \bar{M}_{6} \coma &
	M_{7}  & \rightarrow & \bar{M}_{26} \coma &
	M_{8}  & \rightarrow & \bar{M}_{23} \coma &
	M_{9}  & \rightarrow & \bar{M}_{20} \coma &
	M_{10} & \rightarrow & \bar{M}_{17} \coma \\
	M_{11} & \rightarrow & \bar{M}_{11} \coma &
	M_{12} & \rightarrow & \bar{M}_{5} \coma &
	M_{13} & \rightarrow & \bar{M}_{25} \coma &
	M_{14} & \rightarrow & \bar{M}_{22} \coma &
	M_{15} & \rightarrow & \bar{M}_{19} \coma \\
	M_{16} & \rightarrow & \bar{M}_{16} \coma &
	M_{17} & \rightarrow & \bar{M}_{10} \coma &
	M_{18} & \rightarrow & \bar{M}_{4} \coma &
	M_{19} & \rightarrow & \bar{M}_{15} \coma &
	M_{20} & \rightarrow & \bar{M}_{9} \coma \\ 
	M_{21} & \rightarrow & \bar{M}_{3}\coma &
	M_{22} & \rightarrow & \bar{M}_{14}\coma &
	M_{23} & \rightarrow & \bar{M}_{8}\coma &
	M_{24} & \rightarrow & \bar{M}_{2}\coma &
	M_{25} & \rightarrow & \bar{M}_{13}\coma \\
	\multicolumn{15}{c}{M_{26}   \rightarrow  \bar{M}_{7}\coma 
	M_{27}  \rightarrow  \bar{M}_{1}\fstop}
	\end{array}   
	\label{eq:Q111Z2D-Meson-invol}
    \end{equation}
    
The orientifolded theory has gauge group $\U(N)_1\times \U(N)_2\times \prod_{i=3}^6G_i(N)$. The involution of fields connecting nodes 3, 4, 5 and 6 gives rise to the constraint
\begin{equation}
\gamma_{\Omega_3}=\gamma_{\Omega_4}=\gamma_{\Omega_5}=\gamma_{\Omega_6} \fstop 
\label{gamma_matrices_Q111/Z2}
\end{equation}
Let us set the four matrices equal to $\ID_N$, i.e. project the corresponding gauge groups to $\SO(N)$. Figure~\ref{fig:orien_Q111Z2quiverD} shows the quiver for the resulting theory, which is free of gauge anomalies.

	\begin{figure}[H]
		\centering
	  	\begin{tikzpicture}[scale=2]
	\def\R{1.5};
	     \draw[line width=1pt,redX] (120:\R*1.15) circle (0.25);
	\draw[line width=1pt,redX] (60:\R*1.15) circle (0.25);
	    \draw[line width=1pt,redX] (0:\R*1.15) circle (0.25) node[xshift=0.5cm,star,star points=5, star point ratio=2.25, inner sep=1pt, fill=redX, draw] {};
	    \draw[line width=1pt,redX] (300:\R*1.15) circle (0.25) node[xshift=0.34cm,yshift=-0.34cm,star,star points=5, star point ratio=2.25, inner sep=1pt, fill=redX, draw] {};
	    \draw[line width=1pt,redX] (240:\R*1.15) circle (0.25) node[xshift=-0.34cm,yshift=-0.34cm,star,star points=5, star point ratio=2.25, inner sep=1pt, fill=redX, draw] {};
	    \draw[line width=1pt,redX] (180:\R*1.15) circle (0.25) node[xshift=-0.5cm,star,star points=5, star point ratio=2.25, inner sep=1pt, fill=redX, draw] {};
	       \node[draw=black,line width=1pt,circle,fill=yellowX,minimum width=0.75cm,inner sep=1pt,label={[xshift=0.7cm,yshift=-0cm]:$\U(N)$}] (v1) at  (120:\R) {$1$};
	        \node[draw=black,line width=1pt,circle,fill=yellowX,minimum width=0.75cm,inner sep=1pt,label={[xshift=0.75cm,yshift=0cm]:$\SO(N)$}] (v3) at (0:\R) {$3$};
	        \node[draw=black,line width=1pt,circle,fill=yellowX,minimum width=0.75cm,inner sep=1pt,label={[xshift=1cm,yshift=-0.3cm]:$\SO(N)$}] (v4) at (300:\R) {$4$};
	        \node[draw=black,line width=1pt,circle,fill=yellowX,minimum width=0.75cm,inner sep=1pt,label={[xshift=-1cm,yshift=-0.3cm]:$\SO(N)$}] (v5) at (240:\R) {$5$};
	        \node[draw=black,line width=1pt,circle,fill=yellowX,minimum width=0.75cm,inner sep=1pt,label={[xshift=-0.75cm,yshift=0cm]:$\SO(N)$}] (v6) at (180:\R) {$6$};
	        \node[draw=black,line width=1pt,circle,fill=yellowX,minimum width=0.75cm,inner sep=1pt,label={[xshift=-0.7cm,yshift=-0cm]:$\U(N)$}] (v7) at  (60:\R) {$2$};
	        \draw[line width=1pt,redX] (v1) to node[fill=white,text opacity=1,fill opacity=1,draw=black,rectangle,thin,pos=0.75] {\color{redX}{$4$}} (v3);
	        \draw[line width=1pt,redX] (v4) to node[fill=white,text opacity=1,fill opacity=1,draw=black,rectangle,thin,pos=0.5] {\color{redX}{$8$}} (v5);
	        \draw[line width=1pt,redX] (v6) to node[fill=white,text opacity=1,fill opacity=1,draw=black,rectangle,thin,pos=0.25] {\color{redX}{$4$}} (v7);
	        \draw[line width=1pt,-Triangle] (v1) to node[fill=white,text opacity=1,fill opacity=1,draw=black,rectangle,thin,pos=0.5] {$4$} (v7);
	        \draw[line width=1pt] (v3) to node[fill=white,text opacity=1,fill opacity=1,draw=black,rectangle,thin,pos=0.5] {$4$} (v4);
	        \draw[line width=1pt,-Triangle] (v4) to node[fill=white,text opacity=1,fill opacity=1,draw=black,rectangle,thin,pos=0.7] {$4$} (v1);
	        \draw[line width=1pt] (v5) to node[fill=white,text opacity=1,fill opacity=1,draw=black,rectangle,thin,pos=0.7] {$2$} (v3);
	        \draw[line width=1pt] (v5) to node[fill=white,text opacity=1,fill opacity=1,draw=black,rectangle,thin,pos=0.5] {$4$} (v6);
	        \draw[line width=1pt] (v6) to node[fill=white,text opacity=1,fill opacity=1,draw=black,rectangle,thin,pos=0.3] {$2$} (v4);
	        \draw[line width=1pt,-Triangle] (v7) to node[fill=white,text opacity=1,fill opacity=1,draw=black,rectangle,thin,pos=0.3] {$4$} (v5);
	\end{tikzpicture}
	\caption{Quiver  diagram  for  the  Spin(7)  orientifold  of  phase  D  of $Q^{1,1,1}/\ZZ_2$ using  the involution in \cref{eq:Q111Z2D-chiral-invol,eq:Q111Z2D-Fermi-invol,eq:Q111Z2D-RFermi-invol}, together with our choice of $\gamma_{\Omega_i}$ matrices.}
		\label{fig:orien_Q111Z2quiverD}
	\end{figure}

Let us pause for a moment to think about a possible interpretation on this theory. We note that it has two distinct types of nodes. First, we have $\U(N)$ nodes with adjoint Fermi fields, which can be combined into $\mathcal{N}=(0,2)$ vector multiplets. Second, there are $\SO(N)$ nodes with symmetric Fermi fields which, contrary to the previous case, are inherently $\mathcal{N}=(0,1)$. This is because the adjoint of $\SO(N)$ is instead the antisymmetric representation. 
We can similarly consider whether it is possible to combine the bifundamental fields into $\mathcal{N}=(0,2)$ multiplets, which may or may not be broken by the superpotential. In this example, all bifundamental fields come in pairs so, leaving the superpotential aside, they can form $\mathcal{N}=(0,2)$ multiplets. Broadly speaking, we can therefore regard this theory as consisting of coupled $\mathcal{N}=(0,1)$ and $\mathcal{N}=(0,2)$ sectors.\footnote{A similar interpretation in terms of coupled $\mathcal{N}=(0,1)$ and $\mathcal{N}=(0,2)$ sectors was proposed in the analysis of non-compact models in \cite{Gukov:2019lzi}.} This discussion extends to the other orientifolds of $Q^{1,1,1}/\mathbb{Z}_2$ considered in this section and is a generic phenomenon. Interestingly, we will see below that Spin(7) orientifolds produce theories in which triality acts on either of these two sectors.

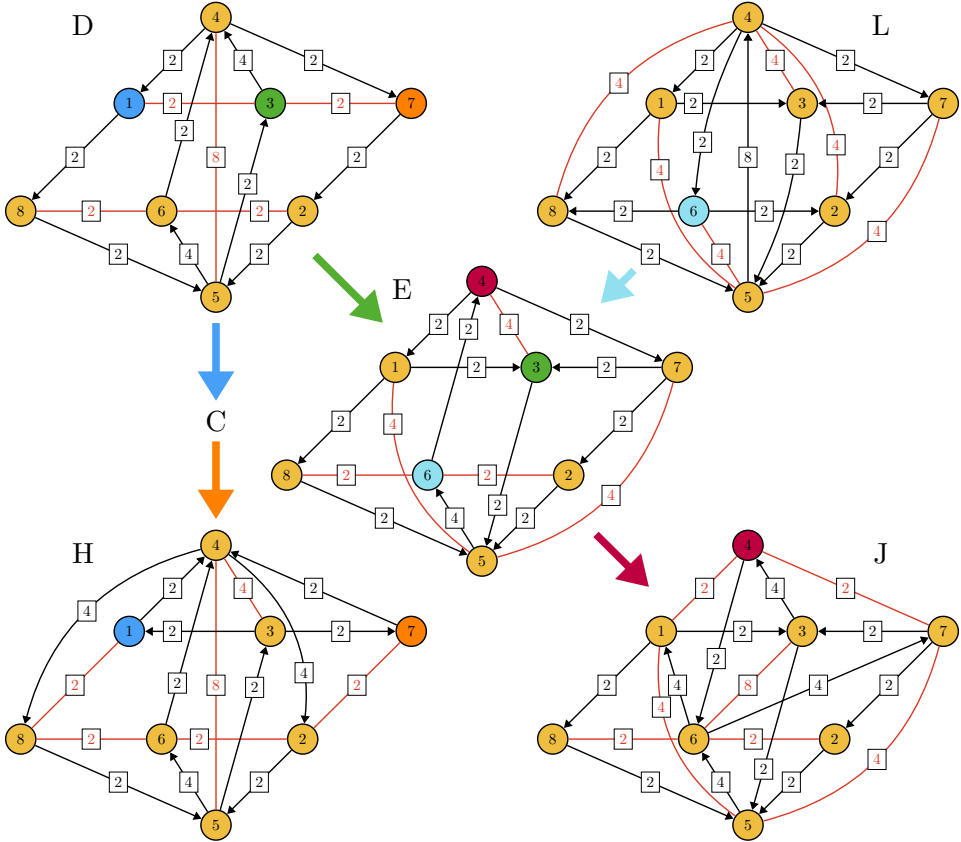
\begin{figure}[H]
    \centering
	      \begin{tikzpicture}[scale=2]
	   \def\L{1.75};
	   \def\d{0.53};
	   \node (Q111Z2D) at (-\L,\L) {
	   \scalebox{\d}{
	   \begin{tikzpicture}[scale=2]
	        \def\R{1.75};
	        \node[draw=black,line width=1pt,circle,fill=blueX,minimum width=0.75cm,inner sep=1pt] (v1) at (-\R,0,-\R) {$1$};
	        \node[draw=black,line width=1pt,circle,fill=orange,minimum width=0.75cm,inner sep=1pt] (v2) at (\R,0,-\R) {$7$};
	        \node[draw=black,line width=1pt,circle,fill=greenX,minimum width=0.75cm,inner sep=1pt] (v3) at (0,0,-\R) {$3$};
	        \node[draw=black,line width=1pt,circle,fill=yellowX,minimum width=0.75cm,inner sep=1pt] (v4) at (0,\R,0) {$4$};
	        \node[draw=black,line width=1pt,circle,fill=yellowX,minimum width=0.75cm,inner sep=1pt] (v5) at (0,-\R,0) {$5$};
	        \node[draw=black,line width=1pt,circle,fill=yellowX,minimum width=0.75cm,inner sep=1pt] (v6) at (0,0,\R) {$6$};
	        \node[draw=black,line width=1pt,circle,fill=yellowX,minimum width=0.75cm,inner sep=1pt] (v7) at (\R,0,\R) {$2$};
	        \node[draw=black,line width=1pt,circle,fill=yellowX,minimum width=0.75cm,inner sep=1pt] (v8) at (-\R,0,\R) {$8$};
	        \draw[line width=1pt,redX] (v1) to node[fill=white,text opacity=1,fill opacity=1,draw=black,rectangle,thin,pos=0.25] {\color{redX}{$2$}} 
	        (v3);
	        \draw[line width=1pt,redX] (v3) to node[fill=white,text opacity=1,fill opacity=1,draw=black,rectangle,thin,pos=0.5] {\color{redX}{$2$}} 
	        (v2);
	        \draw[line width=1pt,redX] (v4) to node[fill=white,text opacity=1,fill opacity=1,draw=black,rectangle,thin,pos=0.5] {\color{redX}{$8$}} 
	        (v5);
	        \draw[line width=1pt,redX] (v8) to node[fill=white,text opacity=1,fill opacity=1,draw=black,rectangle,thin,pos=0.5] {\color{redX}{$2$}} 
	        (v6);
	        \draw[line width=1pt,redX] (v6) to node[fill=white,text opacity=1,fill opacity=1,draw=black,rectangle,thin,pos=0.75] {\color{redX}{$2$}} 
	        (v7);
	        \draw[line width=1pt,-Triangle] (v1) to node[fill=white,text opacity=1,fill opacity=1,draw=black,rectangle,thin,pos=0.5] {$2$} 
	        (v8);
	        \draw[line width=1pt,-Triangle] (v3) to node[fill=white,text opacity=1,fill opacity=1,draw=black,rectangle,thin,pos=0.5] {$4$} 
	        (v4);
	        \draw[line width=1pt,-Triangle] (v4) to node[fill=white,text opacity=1,fill opacity=1,draw=black,rectangle,thin,pos=0.5] {$2$} 
	        (v1);
	        \draw[line width=1pt,-Triangle] (v4) to node[fill=white,text opacity=1,fill opacity=1,draw=black,rectangle,thin,pos=0.5] {$2$} 
	        (v2);
	        \draw[line width=1pt,-Triangle] (v2) to node[fill=white,text opacity=1,fill opacity=1,draw=black,rectangle,thin,pos=0.5] {$2$} 
	        (v7);
	        \draw[line width=1pt,-Triangle] (v8) to node[fill=white,text opacity=1,fill opacity=1,draw=black,rectangle,thin,pos=0.5] {$2$} 
	        (v5);
	        \draw[line width=1pt,-Triangle] (v5) to node[fill=white,text opacity=1,fill opacity=1,draw=black,rectangle,thin,pos=0.6] {$2$} 
	        (v3);
	        \draw[line width=1pt,-Triangle] (v5) to node[fill=white,text opacity=1,fill opacity=1,draw=black,rectangle,thin,pos=0.5] {$4$} 
	        (v6);
	        \draw[line width=1pt,-Triangle] (v6) to node[fill=white,text opacity=1,fill opacity=1,draw=black,rectangle,thin,pos=0.4] {$2$} 
	        (v4);
	        \draw[line width=1pt,-Triangle] (v7) to node[fill=white,text opacity=1,fill opacity=1,draw=black,rectangle,thin,pos=0.5] {$2$} 
	        (v5);
	        \end{tikzpicture}
	        }
	   };
	   \node (Q111Z2E) at (0,0) {
	   \scalebox{\d}{
	   \begin{tikzpicture}[scale=2]
	\def\R{1.75};
	\node[draw=black,line width=1pt,circle,fill=yellowX,minimum width=0.75cm,inner sep=1pt] (v1) at (-\R,0,-\R) {$1$};
	\node[draw=black,line width=1pt,circle,fill=yellowX,minimum width=0.75cm,inner sep=1pt] (v2) at (\R,0,-\R) {$7$};
    \node[draw=black,line width=1pt,circle,fill=greenX,minimum width=0.75cm,inner sep=1pt] (v3) at (0,0,-\R) {$3$};
    \node[draw=black,line width=1pt,circle,fill=purple,minimum width=0.75cm,inner sep=1pt] (v4) at (0,\R,0) {$4$};
    \node[draw=black,line width=1pt,circle,fill=yellowX,minimum width=0.75cm,inner sep=1pt] (v5) at (0,-\R,0) {$5$};
    \node[draw=black,line width=1pt,circle,fill=cyanX,minimum width=0.75cm,inner sep=1pt] (v6) at (0,0,\R) {$6$};
    \node[draw=black,line width=1pt,circle,fill=yellowX,minimum width=0.75cm,inner sep=1pt] (v7) at (\R,0,\R) {$2$};
    \node[draw=black,line width=1pt,circle,fill=yellowX,minimum width=0.75cm,inner sep=1pt] (v8) at (-\R,0,\R) {$8$};
    \draw[line width=1pt,redX] (v8) to node[fill=white,text opacity=1,fill opacity=1,draw=black,rectangle,thin,pos=0.4] {\color{redX}{$2$}} (v6);
    \draw[line width=1pt,redX] (v1) to[bend right] node[fill=white,text opacity=1,fill opacity=1,draw=black,rectangle,thin,pos=0.2] {\color{redX}{$4$}} (v5);
    \draw[line width=1pt,redX] (v3) to node[fill=white,text opacity=1,fill opacity=1,draw=black,rectangle,thin,pos=0.5] {\color{redX}{$4$}} (v4);
    \draw[line width=1pt,redX] (v2) to[bend left] node[fill=white,text opacity=1,fill opacity=1,draw=black,rectangle,thin,pos=0.5] {\color{redX}{$4$}} (v5);
    \draw[line width=1pt,redX] (v6) to node[fill=white,text opacity=1,fill opacity=1,draw=black,rectangle,thin,pos=0.4] {\color{redX}{$2$}} (v7);
    \draw[line width=1pt,Triangle-] (v1) to node[fill=white,text opacity=1,fill opacity=1,draw=black,rectangle,thin,pos=0.5] {$2$} (v4);
    \draw[line width=1pt,Triangle-] (v3) to node[fill=white,text opacity=1,fill opacity=1,draw=black,rectangle,thin,pos=0.4] {$2$} (v1);
    \draw[line width=1pt,Triangle-] (v3) to node[fill=white,text opacity=1,fill opacity=1,draw=black,rectangle,thin,pos=0.5] {$2$} (v2);
    \draw[line width=1pt,Triangle-] (v4) to node[fill=white,text opacity=1,fill opacity=1,draw=black,rectangle,thin,pos=0.2] {$2$} (v6);
    \draw[line width=1pt,Triangle-] (v2) to node[fill=white,text opacity=1,fill opacity=1,draw=black,rectangle,thin,pos=0.5] {$2$} (v4);
    \draw[line width=1pt,Triangle-] (v8) to node[fill=white,text opacity=1,fill opacity=1,draw=black,rectangle,thin,pos=0.5] {$2$} (v1);
    \draw[line width=1pt,Triangle-] (v5) to node[fill=white,text opacity=1,fill opacity=1,draw=black,rectangle,thin,pos=0.5] {$2$} (v8);
    \draw[line width=1pt,Triangle-] (v5) to node[fill=white,text opacity=1,fill opacity=1,draw=black,rectangle,thin,pos=0.25] {$2$} (v3);
    \draw[line width=1pt,Triangle-] (v5) to node[fill=white,text opacity=1,fill opacity=1,draw=black,rectangle,thin,pos=0.5] {$2$} (v7);
    \draw[line width=1pt,Triangle-] (v6) to node[fill=white,text opacity=1,fill opacity=1,draw=black,rectangle,thin,pos=0.5] {$4$} (v5);
    \draw[line width=1pt,Triangle-] (v7) to node[fill=white,text opacity=1,fill opacity=1,draw=black,rectangle,thin,pos=0.5] {$2$} (v2);
	\end{tikzpicture}
	}
	   };
	   \node (Q111Z2H) at (-\L,-\L) {
	   \scalebox{\d}{
	   	\begin{tikzpicture}[scale=2]
	\def\R{1.75};
	 \node[draw=black,line width=1pt,circle,fill=blueX,minimum width=0.75cm,inner sep=1pt] (v1) at (-\R,0,-\R) {$1$};
	        \node[draw=black,line width=1pt,circle,fill=orange,minimum width=0.75cm,inner sep=1pt] (v2) at (\R,0,-\R) {$7$};
	        \node[draw=black,line width=1pt,circle,fill=yellowX,minimum width=0.75cm,inner sep=1pt] (v3) at (0,0,-\R) {$3$};
	        \node[draw=black,line width=1pt,circle,fill=yellowX,minimum width=0.75cm,inner sep=1pt] (v4) at (0,\R,0) {$4$};
	        \node[draw=black,line width=1pt,circle,fill=yellowX,minimum width=0.75cm,inner sep=1pt] (v5) at (0,-\R,0) {$5$};
	        \node[draw=black,line width=1pt,circle,fill=yellowX,minimum width=0.75cm,inner sep=1pt] (v6) at (0,0,\R) {$6$};
	        \node[draw=black,line width=1pt,circle,fill=yellowX,minimum width=0.75cm,inner sep=1pt] (v7) at (\R,0,\R) {$2$};
	        \node[draw=black,line width=1pt,circle,fill=yellowX,minimum width=0.75cm,inner sep=1pt] (v8) at (-\R,0,\R) {$8$};
	        \draw[line width=1pt,redX] (v7) to node[fill=white,text opacity=1,fill opacity=1,draw=black,rectangle,thin,pos=0.8] {\color{redX}{$2$}} (v6);
	        \draw[line width=1pt,redX] (v7) to node[fill=white,text opacity=1,fill opacity=1,draw=black,rectangle,thin,pos=0.5] {\color{redX}{$2$}} (v2);
	        \draw[line width=1pt,redX] (v6) to node[fill=white,text opacity=1,fill opacity=1,draw=black,rectangle,thin,pos=0.5] {\color{redX}{$2$}} (v8);
	        \draw[line width=1pt,redX] (v8) to node[fill=white,text opacity=1,fill opacity=1,draw=black,rectangle,thin,pos=0.5] {\color{redX}{$2$}} (v1);
	        \draw[line width=1pt,redX] (v5) to node[fill=white,text opacity=1,fill opacity=1,draw=black,rectangle,thin,pos=0.5] {\color{redX}{$8$}} (v4);
	        \draw[line width=1pt,redX] (v4) to node[fill=white,text opacity=1,fill opacity=1,draw=black,rectangle,thin,pos=0.5] {\color{redX}{$4$}} (v3);
	        \draw[line width=1pt,-Triangle] (v7) to node[fill=white,text opacity=1,fill opacity=1,draw=black,rectangle,thin,pos=0.5] {$2$} (v5);
	        \draw[line width=1pt,-Triangle] (v8) to node[fill=white,text opacity=1,fill opacity=1,draw=black,rectangle,thin,pos=0.5] {$2$} (v5);
	        \draw[line width=1pt,-Triangle] (v5) to node[fill=white,text opacity=1,fill opacity=1,draw=black,rectangle,thin,pos=0.5] {$4$} (v6);
	        \draw[line width=1pt,-Triangle] (v5) to node[fill=white,text opacity=1,fill opacity=1,draw=black,rectangle,thin,pos=0.75] {$2$} (v3);
	        \draw[line width=1pt,-Triangle] (v6) to node[fill=white,text opacity=1,fill opacity=1,draw=black,rectangle,thin,pos=0.25] {$2$} (v4);
	        \draw[line width=1pt,-Triangle] (v3) to node[fill=white,text opacity=1,fill opacity=1,draw=black,rectangle,thin,pos=0.75] {$2$} (v1);
	        \draw[line width=1pt,-Triangle] (v3) to node[fill=white,text opacity=1,fill opacity=1,draw=black,rectangle,thin,pos=0.5] {$2$} (v2);
	        \draw[line width=1pt,-Triangle] (v4) to[bend left] node[fill=white,text opacity=1,fill opacity=1,draw=black,rectangle,thin,pos=0.75] {$4$} (v7);
	        \draw[line width=1pt,-Triangle] (v4) to[bend right] node[fill=white,text opacity=1,fill opacity=1,draw=black,rectangle,thin,pos=0.5] {$4$} (v8);
	        \draw[line width=1pt,-Triangle] (v1) to node[fill=white,text opacity=1,fill opacity=1,draw=black,rectangle,thin,pos=0.5] {$2$} (v4);
	        \draw[line width=1pt,-Triangle] (v2) to node[fill=white,text opacity=1,fill opacity=1,draw=black,rectangle,thin,pos=0.5] {$2$} (v4);
	\end{tikzpicture}
	}
	   };
	   \node (Q111Z2J) at (\L,-\L) {
	   \scalebox{\d}{
	   	\begin{tikzpicture}[scale=2]
	\def\R{1.75};
	        \node[draw=black,line width=1pt,circle,fill=yellowX,minimum width=0.75cm,inner sep=1pt] (v1) at (-\R,0,-\R) {$1$};
	\node[draw=black,line width=1pt,circle,fill=yellowX,minimum width=0.75cm,inner sep=1pt] (v2) at (\R,0,-\R) {$7$};
    \node[draw=black,line width=1pt,circle,fill=yellowX,minimum width=0.75cm,inner sep=1pt] (v3) at (0,0,-\R) {$3$};
    \node[draw=black,line width=1pt,circle,fill=purple,minimum width=0.75cm,inner sep=1pt] (v4) at (0,\R,0) {$4$};
    \node[draw=black,line width=1pt,circle,fill=yellowX,minimum width=0.75cm,inner sep=1pt] (v5) at (0,-\R,0) {$5$};
    \node[draw=black,line width=1pt,circle,fill=yellowX,minimum width=0.75cm,inner sep=1pt] (v6) at (0,0,\R) {$6$};
    \node[draw=black,line width=1pt,circle,fill=yellowX,minimum width=0.75cm,inner sep=1pt] (v7) at (\R,0,\R) {$2$};
    \node[draw=black,line width=1pt,circle,fill=yellowX,minimum width=0.75cm,inner sep=1pt] (v8) at (-\R,0,\R) {$8$};
	        \draw[line width=1pt,redX] (v1) to node[fill=white,text opacity=1,fill opacity=1,draw=black,rectangle,thin,pos=0.5] {\color{redX}{$2$}} (v4);
	        \draw[line width=1pt,redX] (v1) to[bend right] node[fill=white,text opacity=1,fill opacity=1,draw=black,rectangle,thin,pos=0.3] {\color{redX}{$4$}} (v5);
	        \draw[line width=1pt,redX] (v6) to node[fill=white,text opacity=1,fill opacity=1,draw=black,rectangle,thin,pos=0.5] {\color{redX}{$8$}} (v3);
	        \draw[line width=1pt,redX] (v6) to node[fill=white,text opacity=1,fill opacity=1,draw=black,rectangle,thin,pos=0.4] {\color{redX}{$2$}} (v7);
	        \draw[line width=1pt,redX] (v8) to node[fill=white,text opacity=1,fill opacity=1,draw=black,rectangle,thin,pos=0.5] {\color{redX}{$2$}} (v6);
	        \draw[line width=1pt,redX] (v4) to node[fill=white,text opacity=1,fill opacity=1,draw=black,rectangle,thin,pos=0.5] {\color{redX}{$2$}} (v2);
	        \draw[line width=1pt,redX] (v5) to[bend right] node[fill=white,text opacity=1,fill opacity=1,draw=black,rectangle,thin,pos=0.5] {\color{redX}{$4$}} (v2);
	        \draw[line width=1pt,-Triangle] (v1) to node[fill=white,text opacity=1,fill opacity=1,draw=black,rectangle,thin,pos=0.5] {$2$} (v8);
	        \draw[line width=1pt,-Triangle] (v1) to node[fill=white,text opacity=1,fill opacity=1,draw=black,rectangle,thin,pos=0.6] {$2$} (v3);
	        \draw[line width=1pt,-Triangle] (v6) to node[fill=white,text opacity=1,fill opacity=1,draw=black,rectangle,thin,pos=0.5] {$4$} (v1);
	        \draw[line width=1pt,-Triangle] (v6) to node[fill=white,text opacity=1,fill opacity=1,draw=black,rectangle,thin,pos=0.5] {$4$} (v2);
	        \draw[line width=1pt,-Triangle] (v8) to node[fill=white,text opacity=1,fill opacity=1,draw=black,rectangle,thin,pos=0.5] {$2$} (v5);
	        \draw[line width=1pt,-Triangle] (v4) to node[fill=white,text opacity=1,fill opacity=1,draw=black,rectangle,thin,pos=0.6] {$2$} (v6);
	        \draw[line width=1pt,-Triangle] (v5) to node[fill=white,text opacity=1,fill opacity=1,draw=black,rectangle,thin,pos=0.5] {$4$} (v6);
	        \draw[line width=1pt,-Triangle] (v2) to node[fill=white,text opacity=1,fill opacity=1,draw=black,rectangle,thin,pos=0.5] {$2$} (v7);
	        \draw[line width=1pt,-Triangle] (v2) to node[fill=white,text opacity=1,fill opacity=1,draw=black,rectangle,thin,pos=0.5] {$2$} (v3);
	        \draw[line width=1pt,-Triangle] (v7) to node[fill=white,text opacity=1,fill opacity=1,draw=black,rectangle,thin,pos=0.5] {$2$} (v5);
	        \draw[line width=1pt,-Triangle] (v3) to node[fill=white,text opacity=1,fill opacity=1,draw=black,rectangle,thin,pos=0.5] {$4$} (v4);
	        \draw[line width=1pt,-Triangle] (v3) to node[fill=white,text opacity=1,fill opacity=1,draw=black,rectangle,thin,pos=0.75] {$2$} (v5);
	\end{tikzpicture}
	}
	   };
	   \node (Q111Z2L) at (\L,\L) {
	   \scalebox{\d}{
	   	\begin{tikzpicture}[scale=2]
	\def\R{1.75};
	 \node[draw=black,line width=1pt,circle,fill=yellowX,minimum width=0.75cm,inner sep=1pt] (v1) at (-\R,0,-\R) {$1$};
	        \node[draw=black,line width=1pt,circle,fill=yellowX,minimum width=0.75cm,inner sep=1pt] (v2) at (\R,0,-\R) {$7$};
	        \node[draw=black,line width=1pt,circle,fill=yellowX,minimum width=0.75cm,inner sep=1pt] (v3) at (0,0,-\R) {$3$};
	        \node[draw=black,line width=1pt,circle,fill=yellowX,minimum width=0.75cm,inner sep=1pt] (v4) at (0,\R,0) {$4$};
	        \node[draw=black,line width=1pt,circle,fill=yellowX,minimum width=0.75cm,inner sep=1pt] (v5) at (0,-\R,0) {$5$};
	        \node[draw=black,line width=1pt,circle,fill=cyanX,minimum width=0.75cm,inner sep=1pt] (v6) at (0,0,\R) {$6$};
	        \node[draw=black,line width=1pt,circle,fill=yellowX,minimum width=0.75cm,inner sep=1pt] (v7) at (\R,0,\R) {$2$};
	        \node[draw=black,line width=1pt,circle,fill=yellowX,minimum width=0.75cm,inner sep=1pt] (v8) at (-\R,0,\R) {$8$};
	        \draw[line width=1pt,redX] (v1) to[bend right] node[fill=white,text opacity=1,fill opacity=1,draw=black,rectangle,thin,pos=0.25] {\color{redX}{$4$}} (v5);
	        \draw[line width=1pt,redX] (v2) to[bend left] node[fill=white,text opacity=1,fill opacity=1,draw=black,rectangle,thin,pos=0.5] {\color{redX}{$4$}} (v5);
	        \draw[line width=1pt,redX] (v3) to node[fill=white,text opacity=1,fill opacity=1,draw=black,rectangle,thin,pos=0.5] {\color{redX}{$4$}} (v4);
	        \draw[line width=1pt,redX] (v4) to[bend right] node[fill=white,text opacity=1,fill opacity=1,draw=black,rectangle,thin,pos=0.5] {\color{redX}{$4$}} (v8);
	        \draw[line width=1pt,redX] (v4) to[bend left] node[fill=white,text opacity=1,fill opacity=1,draw=black,rectangle,thin,pos=0.75] {\color{redX}{$4$}} (v7);
	        \draw[line width=1pt,redX] (v5) to node[fill=white,text opacity=1,fill opacity=1,draw=black,rectangle,thin,pos=0.5] {\color{redX}{$4$}} (v6);
	        \draw[line width=1pt,-Triangle] (v1) to node[fill=white,text opacity=1,fill opacity=1,draw=black,rectangle,thin,pos=0.15] {$2$} (v3);
	        \draw[line width=1pt,-Triangle] (v1) to node[fill=white,text opacity=1,fill opacity=1,draw=black,rectangle,thin,pos=0.5] {$2$} (v8);
	        \draw[line width=1pt,-Triangle] (v2) to node[fill=white,text opacity=1,fill opacity=1,draw=black,rectangle,thin,pos=0.5] {$2$} (v3);
	        \draw[line width=1pt,-Triangle] (v2) to node[fill=white,text opacity=1,fill opacity=1,draw=black,rectangle,thin,pos=0.5] {$2$} (v7);
	        \draw[line width=1pt,-Triangle] (v3) to[bend left=10] node[fill=white,text opacity=1,fill opacity=1,draw=black,rectangle,thin,pos=0.25] {$2$} (v5);
	        \draw[line width=1pt,-Triangle] (v4) to node[fill=white,text opacity=1,fill opacity=1,draw=black,rectangle,thin,pos=0.5] {$2$} (v1);
	        \draw[line width=1pt,-Triangle] (v4) to node[fill=white,text opacity=1,fill opacity=1,draw=black,rectangle,thin,pos=0.5] {$2$} (v2);
	        \draw[line width=1pt,-Triangle] (v4) to[bend right=10] node[fill=white,text opacity=1,fill opacity=1,draw=black,rectangle,thin,pos=0.7] {$2$} (v6);
	        \draw[line width=1pt,-Triangle] (v5) to node[fill=white,text opacity=1,fill opacity=1,draw=black,rectangle,thin,pos=0.5] {$8$} (v4);
	        \draw[line width=1pt,-Triangle] (v6) to node[fill=white,text opacity=1,fill opacity=1,draw=black,rectangle,thin,pos=0.5] {$2$} (v7);
	        \draw[line width=1pt,-Triangle] (v6) to node[fill=white,text opacity=1,fill opacity=1,draw=black,rectangle,thin,pos=0.5] {$2$} (v8);
	        \draw[line width=1pt,-Triangle] (v7) to node[fill=white,text opacity=1,fill opacity=1,draw=black,rectangle,thin,pos=0.5] {$2$} (v5);
	        \draw[line width=1pt,-Triangle] (v8) to node[fill=white,text opacity=1,fill opacity=1,draw=black,rectangle,thin,pos=0.5] {$2$} (v5);
	\end{tikzpicture}
	}
	   };
	   \node (Q111Z2C) at (-\L,0) {C};
	   \node at (-\L*1.5,\L*1.5) {D};
	   \node at (-\L*0.3,\L*0.5) {E};
	   \node at (-\L*1.5,-\L*0.5) {H};
	   \node at (\L*1.5,-\L*0.5) {J};
	   \node at (\L*1.5,\L*1.5) {L};
	   \draw[line width=1mm,blueX,-Triangle] (Q111Z2D) -- 
	   (Q111Z2C);
	   \draw[line width=1mm,orange,-Triangle] (Q111Z2C) -- 
	   (Q111Z2H);
	   \draw[line width=1mm,greenX,Triangle-] (Q111Z2D) -- 
	   (Q111Z2E);
	   \draw[line width=1mm,cyanX,-Triangle] (1,1) -- 
	   (0.78,0.78);
	   \draw[line width=1mm,purple,-Triangle] (0.75,-0.75) -- 
	   (1.1,-1.1);
	    \end{tikzpicture}
	    \caption{$\mathcal{N}=(0,2)$ triality web for phases D, E, H, J and L of $Q^{1,1,1}/\ZZ_2$.}
	    \label{fig:02trialityq111z2}
\end{figure}

In order to find other $\mathcal{N}=(0,1)$ theories associated with the same $\Spin(7)$ orientifold, one needs to find the field-theoretic involutions of other toric phases of $Q^{1,1,1}/\mathbb{Z}_2$ leading to $\U(N)^2\times \SO(N)^4$ gauge theories, whose geometric involution is the same as \eqref{eq:Q111Z2D-Meson-invol}. Scanning the 14 toric phases of $Q^{1,1,1}/\mathbb{Z}_2$, we found that only 5 of them (including phase D) admit $\mathcal{N}=(0,1)$ orientifolds with $\U(N)^2\times \SO(N)^4$ gauge symmetry. Let us first present the $\mathcal{N}=(0,2)$ triality web for these 5 phases in Figure~\ref{fig:02trialityq111z2}, which can be regarded as a portion of the whole triality web for $Q^{1,1,1}/\mathbb{Z}_2$ in \cite{Franco:2018qsc}.

\newpage 

Colored arrows connecting different phases indicate $\mathcal{N}=(0,2)$ triality transformations between them. Furthermore, the quiver node on which triality acts is shown in the same color as the corresponding arrow. Note that from phase D to phase H there are two triality steps, where the intermediate stage is the so-called phase C in \cite{Franco:2018qsc}. However, since phase C does not give rise to a $\U(N)^2\times \SO(N)^4$ orientifold, we do not show its quiver here.

Similarly to phase D, we consider the anti-holomorphic involutions of phases E, H, J and L which act on their quivers shown in Figure~\ref{fig:02trialityq111z2} as reflections with respect to the vertical plane that contains nodes 3, 4, 5 and 6. Then, the nodes on the plane map to themselves, while the pairs $1\leftrightarrow 7$ and $2\leftrightarrow 8$ get identified. In all these cases, we choose the $\gamma_{\Omega_i}$ matrices as for phase D, so they have $\U(N)^2\times \SO(N)^4$ gauge group. The construction of the $\mathcal{N}=(0,1)$ theories associated with the $\Spin(7)$ orientifold for these phases is detailed in Appendix~\ref{app:Q111Z2-details}. The crucial point is that they all correspond to the same $\Spin(7)$ orientifold of $Q^{1,1,1}/\mathbb{Z}_2$, since they are all associated to the same geometric involution as that of phase D, given in \eqref{eq:Q111Z2D-Meson-invol}.

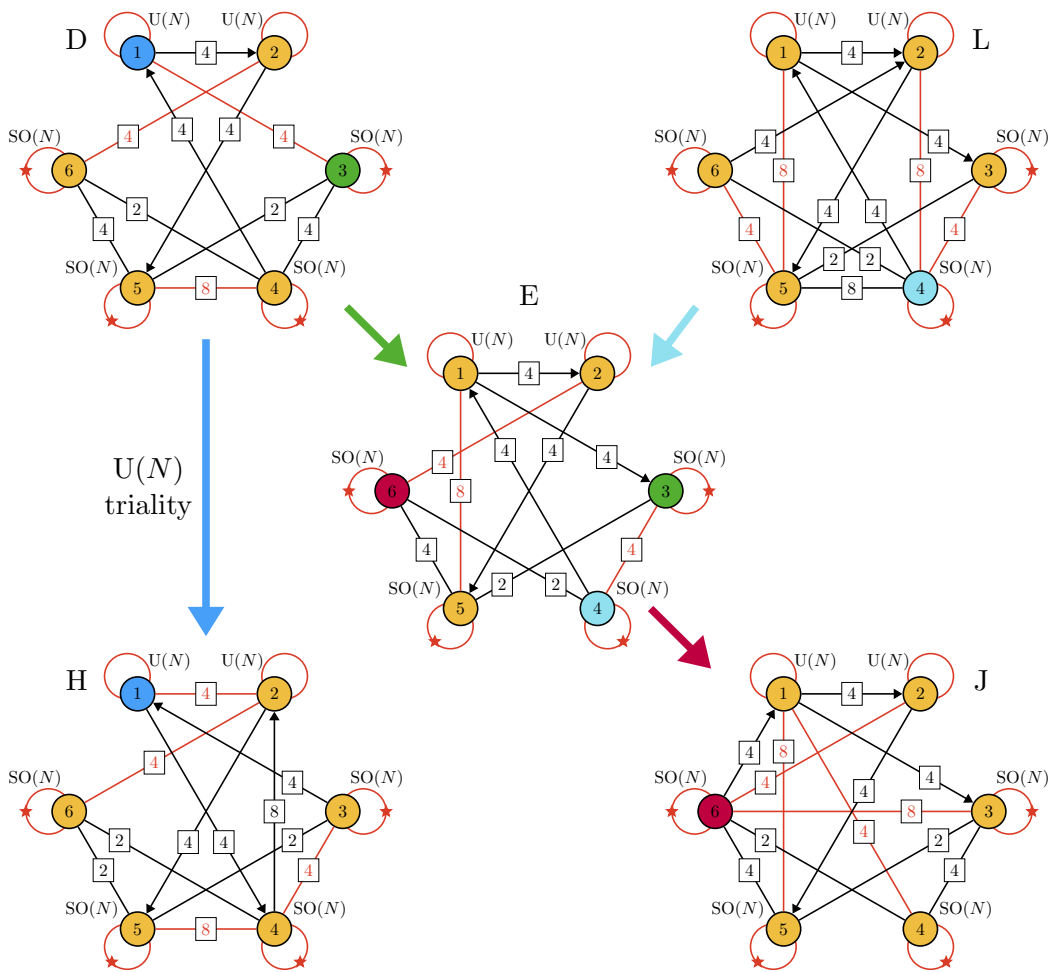
\begin{figure}[ht]
    \centering
	      \begin{tikzpicture}[scale=2]
	   \def\L{2.125};
	   \def\d{0.6};
	   \node (Q111Z2D) at (-\L,\L) {
	   \scalebox{\d}{
	  	\begin{tikzpicture}[scale=2]
	\def\R{1.5};
	     \draw[line width=1pt,redX] (120:\R*1.15) circle (0.25);
	\draw[line width=1pt,redX] (60:\R*1.15) circle (0.25);
	    \draw[line width=1pt,redX] (0:\R*1.15) circle (0.25) node[xshift=0.5cm,star,star points=5, star point ratio=2.25, inner sep=1pt, fill=redX, draw] {};
	    \draw[line width=1pt,redX] (300:\R*1.15) circle (0.25) node[xshift=0.34cm,yshift=-0.34cm,star,star points=5, star point ratio=2.25, inner sep=1pt, fill=redX, draw] {};
	    \draw[line width=1pt,redX] (240:\R*1.15) circle (0.25) node[xshift=-0.34cm,yshift=-0.34cm,star,star points=5, star point ratio=2.25, inner sep=1pt, fill=redX, draw] {};
	    \draw[line width=1pt,redX] (180:\R*1.15) circle (0.25) node[xshift=-0.5cm,star,star points=5, star point ratio=2.25, inner sep=1pt, fill=redX, draw] {};
	       \node[draw=black,line width=1pt,circle,fill=blueX,minimum width=0.75cm,inner sep=1pt,label={[xshift=0.7cm,yshift=-0cm]:$\U(N)$}] (v1) at  (120:\R) {$1$};
	        \node[draw=black,line width=1pt,circle,fill=greenX,minimum width=0.75cm,inner sep=1pt,label={[xshift=0.75cm,yshift=0cm]:$\SO(N)$}] (v3) at (0:\R) {$3$};
	        \node[draw=black,line width=1pt,circle,fill=yellowX,minimum width=0.75cm,inner sep=1pt,label={[xshift=1cm,yshift=-0.3cm]:$\SO(N)$}] (v4) at (300:\R) {$4$};
	        \node[draw=black,line width=1pt,circle,fill=yellowX,minimum width=0.75cm,inner sep=1pt,label={[xshift=-1cm,yshift=-0.3cm]:$\SO(N)$}] (v5) at (240:\R) {$5$};
	        \node[draw=black,line width=1pt,circle,fill=yellowX,minimum width=0.75cm,inner sep=1pt,label={[xshift=-0.75cm,yshift=0cm]:$\SO(N)$}] (v6) at (180:\R) {$6$};
	        \node[draw=black,line width=1pt,circle,fill=yellowX,minimum width=0.75cm,inner sep=1pt,label={[xshift=-0.7cm,yshift=-0cm]:$\U(N)$}] (v7) at  (60:\R) {$2$};
	        \draw[line width=1pt,redX] (v1) to node[fill=white,text opacity=1,fill opacity=1,draw=black,rectangle,thin,pos=0.75] {\color{redX}{$4$}} (v3);
	        \draw[line width=1pt,redX] (v4) to node[fill=white,text opacity=1,fill opacity=1,draw=black,rectangle,thin,pos=0.5] {\color{redX}{$8$}} (v5);
	        \draw[line width=1pt,redX] (v6) to node[fill=white,text opacity=1,fill opacity=1,draw=black,rectangle,thin,pos=0.25] {\color{redX}{$4$}} (v7);
	        \draw[line width=1pt,-Triangle] (v1) to node[fill=white,text opacity=1,fill opacity=1,draw=black,rectangle,thin,pos=0.5] {$4$} (v7);
	        \draw[line width=1pt] (v3) to node[fill=white,text opacity=1,fill opacity=1,draw=black,rectangle,thin,pos=0.5] {$4$} (v4);
	        \draw[line width=1pt,-Triangle] (v4) to node[fill=white,text opacity=1,fill opacity=1,draw=black,rectangle,thin,pos=0.7] {$4$} (v1);
	        \draw[line width=1pt] (v5) to node[fill=white,text opacity=1,fill opacity=1,draw=black,rectangle,thin,pos=0.7] {$2$} (v3);
	        \draw[line width=1pt] (v5) to node[fill=white,text opacity=1,fill opacity=1,draw=black,rectangle,thin,pos=0.5] {$4$} (v6);
	        \draw[line width=1pt] (v6) to node[fill=white,text opacity=1,fill opacity=1,draw=black,rectangle,thin,pos=0.3] {$2$} (v4);
	        \draw[line width=1pt,-Triangle] (v7) to node[fill=white,text opacity=1,fill opacity=1,draw=black,rectangle,thin,pos=0.3] {$4$} (v5);
	\end{tikzpicture}
	        }
	   };
	   \node (Q111Z2E) at (0,0) {
	   \scalebox{\d}{
	  	\begin{tikzpicture}[scale=2]
	\def\R{1.5};
	 \draw[line width=1pt,redX] (120:\R*1.15) circle (0.25);
	\draw[line width=1pt,redX] (60:\R*1.15) circle (0.25);
	    \draw[line width=1pt,redX] (0:\R*1.15) circle (0.25) node[xshift=0.5cm,star,star points=5, star point ratio=2.25, inner sep=1pt, fill=redX, draw] {};
	    \draw[line width=1pt,redX] (300:\R*1.15) circle (0.25) node[xshift=0.34cm,yshift=-0.34cm,star,star points=5, star point ratio=2.25, inner sep=1pt, fill=redX, draw] {};
	    \draw[line width=1pt,redX] (240:\R*1.15) circle (0.25) node[xshift=-0.34cm,yshift=-0.34cm,star,star points=5, star point ratio=2.25, inner sep=1pt, fill=redX, draw] {};
	    \draw[line width=1pt,redX] (180:\R*1.15) circle (0.25) node[xshift=-0.5cm,star,star points=5, star point ratio=2.25, inner sep=1pt, fill=redX, draw] {};
	       \node[draw=black,line width=1pt,circle,fill=yellowX,minimum width=0.75cm,inner sep=1pt,label={[xshift=0.7cm,yshift=-0cm]:$\U(N)$}] (v1) at  (120:\R) {$1$};
	        \node[draw=black,line width=1pt,circle,fill=greenX,minimum width=0.75cm,inner sep=1pt,label={[xshift=0.75cm,yshift=0cm]:$\SO(N)$}] (v3) at (0:\R) {$3$};
	        \node[draw=black,line width=1pt,circle,fill=cyanX,minimum width=0.75cm,inner sep=1pt,label={[xshift=1cm,yshift=-0.3cm]:$\SO(N)$}] (v4) at (300:\R) {$4$};
	        \node[draw=black,line width=1pt,circle,fill=yellowX,minimum width=0.75cm,inner sep=1pt,label={[xshift=-1cm,yshift=-0.3cm]:$\SO(N)$}] (v5) at (240:\R) {$5$};
	        \node[draw=black,line width=1pt,circle,fill=purple,minimum width=0.75cm,inner sep=1pt,label={[xshift=-0.75cm,yshift=0cm]:$\SO(N)$}] (v6) at (180:\R) {$6$};
	        \node[draw=black,line width=1pt,circle,fill=yellowX,minimum width=0.75cm,inner sep=1pt,label={[xshift=-0.7cm,yshift=-0cm]:$\U(N)$}] (v7) at  (60:\R) {$2$};
	        \draw[line width=1pt,redX] (v1) to node[fill=white,text opacity=1,fill opacity=1,draw=black,rectangle,thin,pos=0.5] {\color{redX}{$8$}} (v5);
	        \draw[line width=1pt,redX] (v7) to node[fill=white,text opacity=1,fill opacity=1,draw=black,rectangle,thin,pos=0.8] {\color{redX}{$4$}} (v6);
	        \draw[line width=1pt,redX] (v4) to node[fill=white,text opacity=1,fill opacity=1,draw=black,rectangle,thin,pos=0.5] {\color{redX}{$4$}} (v3);
	        \draw[line width=1pt,-Triangle] (v1) to node[fill=white,text opacity=1,fill opacity=1,draw=black,rectangle,thin,pos=0.5] {$4$} (v7);
	        \draw[line width=1pt,-Triangle] (v1) to node[fill=white,text opacity=1,fill opacity=1,draw=black,rectangle,thin,pos=0.75] {$4$} (v3);
	        \draw[line width=1pt,-Triangle] (v7) to node[fill=white,text opacity=1,fill opacity=1,draw=black,rectangle,thin,pos=0.3] {$4$} (v5);
	        \draw[line width=1pt,-Triangle] (v4) to node[fill=white,text opacity=1,fill opacity=1,draw=black,rectangle,thin,pos=0.7] {$4$} (v1);
	        \draw[line width=1pt] (v5) to node[fill=white,text opacity=1,fill opacity=1,draw=black,rectangle,thin,pos=0.5] {$4$} (v6);
	        \draw[line width=1pt] (v6) to node[fill=white,text opacity=1,fill opacity=1,draw=black,rectangle,thin,pos=0.85] {$2$} (v4);
	        \draw[line width=1pt] (v3) to node[fill=white,text opacity=1,fill opacity=1,draw=black,rectangle,thin,pos=0.85] {$2$} (v5);
	\end{tikzpicture}
	}
	   };
	   \node (Q111Z2H) at (-\L,-\L) {
	   \scalebox{\d}{
		\begin{tikzpicture}[scale=2]
	\def\R{1.5};
	   \draw[line width=1pt,redX] (120:\R*1.15) circle (0.25);
	\draw[line width=1pt,redX] (60:\R*1.15) circle (0.25);
	    \draw[line width=1pt,redX] (0:\R*1.15) circle (0.25) node[xshift=0.5cm,star,star points=5, star point ratio=2.25, inner sep=1pt, fill=redX, draw] {};
	    \draw[line width=1pt,redX] (300:\R*1.15) circle (0.25) node[xshift=0.34cm,yshift=-0.34cm,star,star points=5, star point ratio=2.25, inner sep=1pt, fill=redX, draw] {};
	    \draw[line width=1pt,redX] (240:\R*1.15) circle (0.25) node[xshift=-0.34cm,yshift=-0.34cm,star,star points=5, star point ratio=2.25, inner sep=1pt, fill=redX, draw] {};
	    \draw[line width=1pt,redX] (180:\R*1.15) circle (0.25) node[xshift=-0.5cm,star,star points=5, star point ratio=2.25, inner sep=1pt, fill=redX, draw] {};
	       \node[draw=black,line width=1pt,circle,fill=blueX,minimum width=0.75cm,inner sep=1pt,label={[xshift=0.7cm,yshift=-0cm]:$\U(N)$}] (v1) at  (120:\R) {$1$};
	        \node[draw=black,line width=1pt,circle,fill=yellowX,minimum width=0.75cm,inner sep=1pt,label={[xshift=0.75cm,yshift=0cm]:$\SO(N)$}] (v3) at (0:\R) {$3$};
	        \node[draw=black,line width=1pt,circle,fill=yellowX,minimum width=0.75cm,inner sep=1pt,label={[xshift=1cm,yshift=-0.3cm]:$\SO(N)$}] (v4) at (300:\R) {$4$};
	        \node[draw=black,line width=1pt,circle,fill=yellowX,minimum width=0.75cm,inner sep=1pt,label={[xshift=-1cm,yshift=-0.3cm]:$\SO(N)$}] (v5) at (240:\R) {$5$};
	        \node[draw=black,line width=1pt,circle,fill=yellowX,minimum width=0.75cm,inner sep=1pt,label={[xshift=-0.75cm,yshift=0cm]:$\SO(N)$}] (v6) at (180:\R) {$6$};
	        \node[draw=black,line width=1pt,circle,fill=yellowX,minimum width=0.75cm,inner sep=1pt,label={[xshift=-0.7cm,yshift=-0cm]:$\U(N)$}] (v7) at  (60:\R) {$2$};
	        \draw[line width=1pt,redX] (v6) to node[fill=white,text opacity=1,fill opacity=1,draw=black,rectangle,thin,pos=0.4] {\color{redX}{$4$}} (v7);
	        \draw[line width=1pt,redX] (v4) to node[fill=white,text opacity=1,fill opacity=1,draw=black,rectangle,thin,pos=0.5] {\color{redX}{$4$}} (v3);
	        \draw[line width=1pt,redX] (v5) to node[fill=white,text opacity=1,fill opacity=1,draw=black,rectangle,thin,pos=0.5] {\color{redX}{$8$}} (v4);
	        \draw[line width=1pt,redX] (v1) to node[fill=white,text opacity=1,fill opacity=1,draw=black,rectangle,thin,pos=0.5] {\color{redX}{$4$}} (v7);
	        \draw[line width=1pt,-Triangle] (v3) to node[fill=white,text opacity=1,fill opacity=1,draw=black,rectangle,thin,pos=0.2] {$4$} (v1);
	        \draw[line width=1pt,-Triangle] (v4) to node[fill=white,text opacity=1,fill opacity=1,draw=black,rectangle,thin,pos=0.5] {$8$} (v7);
	        \draw[line width=1pt,-Triangle] (v1) to node[fill=white,text opacity=1,fill opacity=1,draw=black,rectangle,thin,pos=0.65] {$4$} (v4);
	        \draw[line width=1pt,-Triangle] (v7) to node[fill=white,text opacity=1,fill opacity=1,draw=black,rectangle,thin,pos=0.65] {$4$} (v5);
	        \draw[line width=1pt] (v4) to node[fill=white,text opacity=1,fill opacity=1,draw=black,rectangle,thin,pos=0.8] {$2$} (v6);
	        \draw[line width=1pt] (v5) to node[fill=white,text opacity=1,fill opacity=1,draw=black,rectangle,thin,pos=0.5] {$2$} (v6);
	        \draw[line width=1pt] (v5) to node[fill=white,text opacity=1,fill opacity=1,draw=black,rectangle,thin,pos=0.8] {$2$} (v3);
	\end{tikzpicture}
	}
	   };
	   \node (Q111Z2J) at (\L,-\L) {
	   \scalebox{\d}{
	 	\begin{tikzpicture}[scale=2]
	\def\R{1.5};
	\draw[line width=1pt,redX] (120:\R*1.15) circle (0.25);
	\draw[line width=1pt,redX] (60:\R*1.15) circle (0.25);
	    \draw[line width=1pt,redX] (0:\R*1.15) circle (0.25) node[xshift=0.5cm,star,star points=5, star point ratio=2.25, inner sep=1pt, fill=redX, draw] {};
	    \draw[line width=1pt,redX] (300:\R*1.15) circle (0.25) node[xshift=0.34cm,yshift=-0.34cm,star,star points=5, star point ratio=2.25, inner sep=1pt, fill=redX, draw] {};
	    \draw[line width=1pt,redX] (240:\R*1.15) circle (0.25) node[xshift=-0.34cm,yshift=-0.34cm,star,star points=5, star point ratio=2.25, inner sep=1pt, fill=redX, draw] {};
	    \draw[line width=1pt,redX] (180:\R*1.15) circle (0.25) node[xshift=-0.5cm,star,star points=5, star point ratio=2.25, inner sep=1pt, fill=redX, draw] {};
	       \node[draw=black,line width=1pt,circle,fill=yellowX,minimum width=0.75cm,inner sep=1pt,label={[xshift=0.7cm,yshift=-0cm]:$\U(N)$}] (v1) at  (120:\R) {$1$};
	        \node[draw=black,line width=1pt,circle,fill=yellowX,minimum width=0.75cm,inner sep=1pt,label={[xshift=0.75cm,yshift=0cm]:$\SO(N)$}] (v3) at (0:\R) {$3$};
	        \node[draw=black,line width=1pt,circle,fill=yellowX,minimum width=0.75cm,inner sep=1pt,label={[xshift=1cm,yshift=-0.3cm]:$\SO(N)$}] (v4) at (300:\R) {$4$};
	        \node[draw=black,line width=1pt,circle,fill=yellowX,minimum width=0.75cm,inner sep=1pt,label={[xshift=-1cm,yshift=-0.3cm]:$\SO(N)$}] (v5) at (240:\R) {$5$};
	        \node[draw=black,line width=1pt,circle,fill=purple,minimum width=0.75cm,inner sep=1pt,label={[xshift=-0.75cm,yshift=0cm]:$\SO(N)$}] (v6) at (180:\R) {$6$};
	        \node[draw=black,line width=1pt,circle,fill=yellowX,minimum width=0.75cm,inner sep=1pt,label={[xshift=-0.7cm,yshift=-0cm]:$\U(N)$}] (v2) at  (60:\R) {$2$};
	        \draw[line width=1pt,redX] (v1) to node[fill=white,text opacity=1,fill opacity=1,draw=black,rectangle,thin,pos=0.6] {\color{redX}{$4$}} (v4);
	        \draw[line width=1pt,redX] (v1) to node[fill=white,text opacity=1,fill opacity=1,draw=black,rectangle,thin,pos=0.2] {\color{redX}{$8$}} (v5);
	        \draw[line width=1pt,redX] (v2) to node[fill=white,text opacity=1,fill opacity=1,draw=black,rectangle,thin,pos=0.8] {\color{redX}{$4$}} (v6);
	        \draw[line width=1pt,redX] (v3) to node[fill=white,text opacity=1,fill opacity=1,draw=black,rectangle,thin,pos=0.25] {\color{redX}{$8$}} (v6);
	        \draw[line width=1pt,-Triangle] (v1) to node[fill=white,text opacity=1,fill opacity=1,draw=black,rectangle,thin,pos=0.5] {$4$} (v2);
	        \draw[line width=1pt,-Triangle] (v1) to node[fill=white,text opacity=1,fill opacity=1,draw=black,rectangle,thin,pos=0.75] {$4$} (v3);
	        \draw[line width=1pt,-Triangle] (v2) to node[fill=white,text opacity=1,fill opacity=1,draw=black,rectangle,thin,pos=0.4] {$4$} (v5);
	        \draw[line width=1pt,-Triangle] (v6) to node[fill=white,text opacity=1,fill opacity=1,draw=black,rectangle,thin,pos=0.5] {$4$} (v1);
	        \draw[line width=1pt] (v3) to node[fill=white,text opacity=1,fill opacity=1,draw=black,rectangle,thin,pos=0.5] {$4$} (v4);
	        \draw[line width=1pt] (v3) to node[fill=white,text opacity=1,fill opacity=1,draw=black,rectangle,thin,pos=0.2] {$2$} (v5);
	        \draw[line width=1pt] (v4) to node[fill=white,text opacity=1,fill opacity=1,draw=black,rectangle,thin,pos=0.8] {$2$} (v6);
	        \draw[line width=1pt] (v5) to node[fill=white,text opacity=1,fill opacity=1,draw=black,rectangle,thin,pos=0.5] {$4$} (v6);
	\end{tikzpicture}
	}
	   };
	   \node (Q111Z2L) at (\L,\L) {
	   \scalebox{\d}{
	  	\begin{tikzpicture}[scale=2]
	\def\R{1.5};
	  \draw[line width=1pt,redX] (120:\R*1.15) circle (0.25);
	\draw[line width=1pt,redX] (60:\R*1.15) circle (0.25);
	    \draw[line width=1pt,redX] (0:\R*1.15) circle (0.25) node[xshift=0.5cm,star,star points=5, star point ratio=2.25, inner sep=1pt, fill=redX, draw] {};
	    \draw[line width=1pt,redX] (300:\R*1.15) circle (0.25) node[xshift=0.34cm,yshift=-0.34cm,star,star points=5, star point ratio=2.25, inner sep=1pt, fill=redX, draw] {};
	    \draw[line width=1pt,redX] (240:\R*1.15) circle (0.25) node[xshift=-0.34cm,yshift=-0.34cm,star,star points=5, star point ratio=2.25, inner sep=1pt, fill=redX, draw] {};
	    \draw[line width=1pt,redX] (180:\R*1.15) circle (0.25) node[xshift=-0.5cm,star,star points=5, star point ratio=2.25, inner sep=1pt, fill=redX, draw] {};
	       \node[draw=black,line width=1pt,circle,fill=yellowX,minimum width=0.75cm,inner sep=1pt,label={[xshift=0.7cm,yshift=-0cm]:$\U(N)$}] (v1) at  (120:\R) {$1$};
	        \node[draw=black,line width=1pt,circle,fill=yellowX,minimum width=0.75cm,inner sep=1pt,label={[xshift=0.75cm,yshift=0cm]:$\SO(N)$}] (v3) at (0:\R) {$3$};
	        \node[draw=black,line width=1pt,circle,fill=cyanX,minimum width=0.75cm,inner sep=1pt,label={[xshift=1cm,yshift=-0.3cm]:$\SO(N)$}] (v4) at (300:\R) {$4$};
	        \node[draw=black,line width=1pt,circle,fill=yellowX,minimum width=0.75cm,inner sep=1pt,label={[xshift=-1cm,yshift=-0.3cm]:$\SO(N)$}] (v5) at (240:\R) {$5$};
	        \node[draw=black,line width=1pt,circle,fill=yellowX,minimum width=0.75cm,inner sep=1pt,label={[xshift=-0.75cm,yshift=0cm]:$\SO(N)$}] (v6) at (180:\R) {$6$};
	        \node[draw=black,line width=1pt,circle,fill=yellowX,minimum width=0.75cm,inner sep=1pt,label={[xshift=-0.7cm,yshift=-0cm]:$\U(N)$}] (v2) at  (60:\R) {$2$};
	        \draw[line width=1pt,redX] (v1) to node[fill=white,text opacity=1,fill opacity=1,draw=black,rectangle,thin,pos=0.5] {\color{redX}{$8$}} (v5);
	        \draw[line width=1pt,redX] (v2) to node[fill=white,text opacity=1,fill opacity=1,draw=black,rectangle,thin,pos=0.5] {\color{redX}{$8$}} (v4);
	        \draw[line width=1pt,redX] (v5) to node[fill=white,text opacity=1,fill opacity=1,draw=black,rectangle,thin,pos=0.5] {\color{redX}{$4$}} (v6);
	        \draw[line width=1pt,redX] (v4) to node[fill=white,text opacity=1,fill opacity=1,draw=black,rectangle,thin,pos=0.5] {\color{redX}{$4$}} (v3);
	        \draw[line width=1pt,-Triangle] (v1) to node[fill=white,text opacity=1,fill opacity=1,draw=black,rectangle,thin,pos=0.5] {$4$} (v2);
	        \draw[line width=1pt,-Triangle] (v1) to node[fill=white,text opacity=1,fill opacity=1,draw=black,rectangle,thin,pos=0.8] {$4$} (v3);
	        \draw[line width=1pt,-Triangle] (v4) to node[fill=white,text opacity=1,fill opacity=1,draw=black,rectangle,thin,pos=0.3] {$4$} (v1);
	        \draw[line width=1pt,-Triangle] (v2) to node[fill=white,text opacity=1,fill opacity=1,draw=black,rectangle,thin,pos=0.7] {$4$} (v5);
	        \draw[line width=1pt,-Triangle] (v6) to node[fill=white,text opacity=1,fill opacity=1,draw=black,rectangle,thin,pos=0.2] {$4$} (v2);
	        \draw[line width=1pt] (v3) to node[fill=white,text opacity=1,fill opacity=1,draw=black,rectangle,thin,pos=0.8] {$2$} (v5);
	        \draw[line width=1pt] (v4) to node[fill=white,text opacity=1,fill opacity=1,draw=black,rectangle,thin,pos=0.2] {$2$} (v6);
	        \draw[line width=1pt] (v4) to node[fill=white,text opacity=1,fill opacity=1,draw=black,rectangle,thin,pos=0.5] {$8$} (v5);
	\end{tikzpicture}
	}
	   };
	   \node at (-\L*1.4,\L*1.4) {D};
	   \node at (-\L*0,\L*0.6) {E};
	   \node at (-\L*1.4,-\L*0.6) {H};
	   \node at (\L*1.4,-\L*0.6) {J};
	   \node at (\L*1.4,\L*1.4) {L};
	   \draw[line width=1mm,blueX,-Triangle] (Q111Z2D) -- node[midway,left,black] {\shortstack{$\U(N)$\\triality}}
	   (Q111Z2H);
	   (Q111Z2H);
	   \draw[line width=1mm,greenX,-Triangle] (-1.2,1.2) -- 
	   (-0.8,0.8);
	   \draw[line width=1mm,cyanX,-Triangle] (1.1,1.2) -- 
	   (0.8,0.8);
	   \draw[line width=1mm,purple,-Triangle] (0.8,-0.8) -- 
	   (1.2,-1.2);
	    \end{tikzpicture}
	    \caption{Triality web for the $\mathcal{N}=(0,1)$ theories associated with the Spin(7) orientifolds under consideration for phases D, E, H, J and L of $Q^{1,1,1}/\ZZ_2$.}
	    \label{fig:01trialityq111z2}
\end{figure}

From a field theory perspective, we find that the orientifolds of phases D, E, J and L are connected by $\mathcal{N}=(0,1)$ triality transformation on various $\SO(N)$ gauge groups (with the obvious generalization to more general flavor groups). These are the first examples of $\mathcal{N}=(0,1)$ triality in the presence of $\U(N)$ gauge groups. Interestingly, the orientifolds of phases D and H are not connected by the usual $\mathcal{N}=(0,1)$ triality on an $\SO(N)$ node, but by triality on node 1, which is of $\U(N)$ type. This transformation locally follows the rules of $\mathcal{N}=(0,2)$ triality. Such $\U(N)$ triality in $\mathcal{N}=(0,1)$ gauge theories is a new phenomenon which, to the best of our knowledge, has not appeared in the literature before. Following our earlier discussion, it can be nicely interpreted as $\mathcal{N}=(0,2)$ triality in the presence of an $\mathcal{N}=(0,1)$ sector. In our $\Spin(7)$ orientifold construction, the $\U(N)$ triality has a clear origin: the two $\mathcal{N}=(0,2)$ trialities that connect phases D and H passing through phase C are projected onto a single $\U(N)$ triality connecting the orientifolds of phases D and H. In the case of nodes that are not mapped to themselves, an even number of trialities in the parent is necessary in order to get a new phase that is also symmetric under the involution. Figure~\ref{fig:01trialityq111z2} summarizes the web of trialities for the Spin(7) orientifolds under consideration.

\section{Conclusions}
\label{sec:conclusions}

D1-branes probing singularities provide a powerful framework for engineering 2d gauge theories. In our previous work \cite{Franco:2021ixh}, these constructions were extended to $\mathcal{N}=(0,1)$ theories with the introduction of Spin(7) orientifolds.

In this paper we introduced a new, geometric, perspective on the triality of 2d $\mathcal{N}=(0,1)$ gauge theories, by showing that it arises from the non-uniqueness of the correspondence between Spin(7) orientifolds and the gauge theories on D1-brane probes. 

Let us reflect on how 2d trialities with different amounts of SUSY are manifested in D1-branes at singularities. $\mathcal{N}=(0,2)$ triality similarly arises from the fact that multiple gauge theories can be associated to the same underlying CY$_4$ \cite{Franco:2016nwv}. We explained that Spin(7) orientifolds based on the universal involution give rise to exactly the $\mathcal{N}=(0,1)$ triality of \cite{Gukov:2019lzi}. But our work shows that the space of possibilities is far richer. Indeed, general Spin(7) orientifolds extend triality to theories that can be regarded as consisting of coupled $\mathcal{N}=(0,2)$ and $(0,1)$ sectors. The geometric construction of these theories therefore leads to extensions of triality that interpolate between the pure $\mathcal{N}=(0,2)$ and $(0,1)$ cases.

On the practical side, Spin(7) orientifolds also give a precise prescription for how scalar flavors transform under triality in general quivers, which is inherited from the transformation of the corresponding chiral flavors in the parent. 

	
\acknowledgments

We would like to thank Sergei Gukov, Du Pei and Pavel Putrov for enjoyable discussions. The research of S. F. was supported by the U.S. National Science Foundation grants PHY-1820721 and PHY-2112729. The work of A. M. is supported in part by Deutsche Forschungsgemeinschaft under Germany's Excellence Strategy EXC 2121 Quantum Universe 390833306. The work of A. U. is supported by the Spanish Research Agency (Agencia Espanola de Investigaci\'on) through the grants IFT Centro de Excelencia Severo Ochoa SEV-2016-0597, and the grant GC2018-095976-B-C21 from MCIU/AEI/FEDER, UE. S. F. and X. Y. would like to thank the Simons Center for Geometry and Physics for hospitality during part of this work. X.Y. also thanks the UPenn Theory Group and Caltech Particle Theory Group for hospitality during part of this work.

\appendix 

\bigskip

\section{2d $\mathcal{N}=(0,1)$ Formalism}
\label{app:N01multiplets}

In this appendix we review the $2$d $\mathcal{N}=(0,1)$ field theory formalism as we did in~\cite{Franco:2021ixh}. Let us introduce the $2$d $\mathcal{N}=(0,1)$ superspace $\left(x^0,x^1,\theta^+\right)$, on which we can define three types of supermultiplets:
\begin{itemize}
\item Vector multiplet:
\begin{equation}\label{(0,1) vector}
	\begin{split}
		V_+&=\theta^+(A_0(x)+A_1(x))\coma\\
		V_-&=A_0(x)-A_1(x)+\theta^+\lambda_-(x)\fstop
	\end{split}
\end{equation}
It contains a gauge boson $A_{\pm}$ and a left-moving Majorana-Weyl fermion $\lambda_-$ in the adjoint representation. 

\item Scalar multiplet:
\begin{equation}
	\Phi(x,\theta)=\phi(x)+\theta^+\psi_+(x)\fstop
\end{equation}
It has a real scalar field $\phi$ and a right-moving Majorana-Weyl fermion $\psi_+$. 

\item Fermi multiplet:
\begin{equation}\label{(0,1) Fermi}
	\Lambda(x,\theta)=\psi_-(x)+\theta^+F(x)\fstop
\end{equation}
It has a left-moving Majorana-Weyl spinor as its only on-shell degree of freedom. Here $F$ is an auxiliary field. 
\end{itemize}

The kinetic terms for matter fields and their gauge couplings are given by
\begin{equation}
	\begin{split}
	\mathcal{L}_s+\mathcal{L}_F=\int d\theta^+~ \left(\frac{i}{2}\sum_i(\mathcal{D}_+\Phi_i\mathcal{D}_-\Phi_i)-\frac{1}{2}\sum_a(\Lambda_a\mathcal{D}_+\Lambda_a) \right)\coma
	\end{split}
	\label{eq:kingauge01mattfields}
\end{equation}
where $\mathcal{D}_{\pm}$ are super-covariant derivatives~\cite{Gukov:2019lzi}. 

We need also to introduce the $\mathcal{N}=(0,1)$ analog of the $\mathcal{N}=(0,2)$ $J$-term interaction, which is given by
\begin{equation}
	\mathcal{L}_J\equiv \int d \theta^+W^{(0,1)}=\int d\theta^+\sum_a (\Lambda_aJ^a(\Phi_i))\coma
	\label{eq:W01super}
\end{equation}
where $J^a(\Phi_i)$ are real functions of scalar fields. We refer to $W^{(0,1)}$ as the \emph{superpotential}. Both the field content and gauge symmetry (i.e., the quiver for the theories considered in this paper) and $W^{(0,1)}$ are necessary for fully specifying an $\mathcal{N}=(0,1)$ gauge theory. 

After integrating out the auxiliary fields $F_a$, $\mathcal{L}_J$ produces various interactions, including Yukawa-like couplings 
\begin{equation}
	\sum_a \lambda_{-a}\frac{\partial J^a}{\partial \phi_i}\psi_{+i}\coma
\end{equation}
as well as a scalar potential 
\begin{equation}
	\frac{1}{2}\sum_a(J^a(\phi_i))^2\fstop
\end{equation}

\subsection{$\mathcal{N}=(0,2)$ Gauge Theories in $\mathcal{N}=(0,1)$ Superspace}
\label{app:N02inN01Superspace}

For the construction of Spin(7) manifolds, it is useful to express $\mathcal{N}=(0,2)$ gauge theories in $\mathcal{N}=(0,1)$ language. Here, we briefly sketch the decomposition, referring to \cite{Franco:2021ixh} for details:

\bigskip

\begin{enumerate}
    \item An $\mathcal{N}=(0,2)$ vector multiplet $V_i^{(0,2)}$ decomposes into an $\mathcal{N}=(0,1)$ vector multiplet $V_i$ and an $\mathcal{N}=(0,1)$ Fermi multiplet $\Lambda_i^R$.
    \item An $\mathcal{N}=(0,2)$ chiral multiplet $\Phi_m^{(0,2)}$ decomposes into two $\mathcal{N}=(0,1)$ scalar multiplets $\Phi^a_m$ with $a=1,2$. It can then be further re-expressed in an $\mathcal{N}=(0,1)$ complex scalar multiplet $\Phi_m$.
    \item An $\mathcal{N}=(0,2)$ Fermi multiplet $\Lambda_m^{(0,2)}$ decomposes into two $\mathcal{N}=(0,1)$ Fermi multiplets $\Lambda_m^a$, with $a=1,2$, that form an $\mathcal{N}=(0,1)$ complex Fermi multiplet $\Lambda_m$.  
\end{enumerate}

\bigskip

The $J$- and $E$-terms of the $\mathcal{N}=(0,2)$ gauge theory become part of $W^{(0,1)}$ upon the decomposition of Fermi and chiral multiplets in $\mathcal{N}=(0,1)$ language. The interactions between $\mathcal{N}=(0,2)$ vector and chiral multiplets also contribute to $W^{(0,1)}$ couplings between scalar multiplets and $\mathcal{N}=(0,1)$ Fermi multiplets $\Lambda_i^R$ coming from the $\mathcal{N}=(0,2)$ vector multiplets. The full $\mathcal{N}=(0,1)$ superpotential reads
\begin{equation}
\begin{split}
     W^{(0,1)} & = \sum_a\int d\theta^+[\Lambda_a(J^a(\Phi_m)+E^{\dagger a}(\Phi^\dagger_m))+\Lambda^{\dagger a}(E_a(\Phi_m)+J^\dagger_a(\Phi_m^\dagger))]+\sum_i \sum_n \Lambda^R_i \Phi^\dagger_n\Phi_n\coma\\
\end{split}
\end{equation}
where $n$ runs over all complex scalar multiplets transforming under a given gauge group $i$.

\section{Anomalies}
\label{app:AnomalyContr}

\begin{figure}[H]
	\centering
	\begin{tikzpicture}[scale=0.75]
	\draw[line width=1pt] (0,0) circle (1);
	\draw[line width=1pt,decoration=snake,decorate] (-3,0) -- (-1,0);
	\draw[line width=1pt,decoration=snake,decorate] (3,0) -- (1,0);
	\end{tikzpicture}
	\caption{Generic $1$-loop diagram associated with $2$d anomalies.}
	\label{fig:anom1loop}
\end{figure}
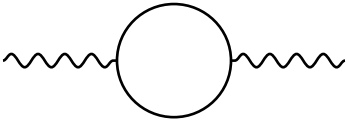

Here, we list the possible contributions to $2$d gauge anomalies coming from fields in the representations considered in this paper. Generically, $2$d anomalies are obtained by a $1$-loop diagram as shown in Figure~\ref{fig:anom1loop}, where left- and right-moving fermions running in the loop contribute oppositely. 

In the case of gauge groups, anomalies must vanish for consistency of the theory at the quantum level. This leads to important constraints in our construction of $2$d $\mathcal{N}=(0,1)$ theories, which may require the introduction of extra flavors to cancel anomalies.

Unlike gauge symmetries, global symmetries may indeed be anomalous. They are also preserved by RG flows, so they are useful for testing dualities between two or more theories. Examples of using global anomalies to check dualities in $2$d $\mathcal{N}=(0,1)$ theories can be found in \cite{Gukov:2019lzi}.

Generically, the gauge theories on D1-branes probing Spin(7) orientifolds that we construct in this paper have non-vanishing Abelian gauge anomalies. However, similarly to the discussion in \cite{Franco:2015tna, Franco:2017cjj}, we expect that such anomalies are canceled by the bulk fields in the closed string sector via a generalized Green-Schwarz (GS) mechanism (see \cite{Ibanez:1998qp,Mohri:1997ef} for derivations in 4d ${\cal N}=1$ and $2$d ${\cal N}=(0,2)$ theories realized at orbifolds/orientifold singularities). For this reason, we mainly focus on non-Abelian anomalies.

Let us consider pure non-Abelian $G^2$ gauge or global anomalies, where $G$ is $\SU(N)$, $\SO(N)$ or $\USp(N)$ group.
The corresponding anomaly is given by
\begin{equation}
	\text{Tr}[\gamma^3J_GJ_G]\coma
\end{equation}
where $\gamma^3$ is the chirality matrix in $2$d and $J_G$ is the current associated to $G$. The resulting anomaly from a field in representation $\rho$ of $G$ can be computed in terms of the Dynkin index $T(\rho)$: 
\begin{equation}
	T(\rho)=C_2(\rho)\frac{d(\rho)}{d(\text{adjoint})}\coma
\end{equation}
where $C_2(\rho)$ is the quadratic Casimir for representation $\rho$. In Table~\ref{tab:SUanomaly} we present anomaly contributions for superfields in the most common representations of $\SU(N)$. In Table~\ref{tab:SOUSpanomaly}, we present anomaly contributions for various representations of $\SO(N)$ and $\USp(N)$ groups, computed using Dynkin indices listed in \cite{Yamatsu:2015npn}. 

\begin{table}[H]
	\centering
\begin{tabular}{|Sc|Sc|Sc|Sc|Sc|}
	\hline
	$\SU(N)$	& fundamental & adjoint & antisymmetric & symmetric\\
	\hhline{|=|=|=|=|=|} 
	vector multiplet & $\times$ & $-N$ & $\times$ &$\times$ \\
	\hline
	Fermi multiplet & $-\dfrac{1}{2}$ & $-N$ & $\dfrac{-N+2}{2}$ & $\dfrac{-N-2}{2}$\\
	\hline
	scalar multiplet & $\dfrac{1}{2}$ & $N$ &$\dfrac{N-2}{2}$ & $\dfrac{N+2}{2}$\\
	\hline
\end{tabular}
\caption{Anomaly contributions of the $2$d $\mathcal{N}=(0,1)$ multiplets in various representations of $\SU(N)$. Since anomalies are quadratic in 2d, the same contributions apply for the conjugate representations.}
\label{tab:SUanomaly}
\end{table}

\begin{table}[H]
	\centering
\begin{tabular}{|Sc|Sc|Sc|Sc|}
	\hline
	$\SO(N)$	& fundamental & antisymmetric (adjoint) & symmetric\\
	\hhline{|=|=|=|=|} 
	vector multiplet & $\times$ & $-N+2$ & $\times$ \\
	\hline
	Fermi multiplet & $-1$ & $-N+2$ & $-N-2$\\
	\hline
	scalar multiplet & $1$ & $N-2$ &$N+2$\\
	\hhline{|=|=|=|=|} 
	$\USp(N)$	& fundamental & antisymmetric  & symmetric (adjoint)\\
	\hhline{|=|=|=|=|} 
	vector multiplet & $\times$ & $\times$ & $-N-2$ \\
	\hline
	Fermi multiplet & $-1$ & $-N+2$ & $-N-2$\\
	\hline
	scalar multiplet & $1$ & $N-2$ &$N+2$\\
	\hline
\end{tabular}
\caption{Anomaly contributions of the $2$d $\mathcal{N}=(0,1)$ multiplets in various representations of $\SO(N)$ and $\USp(N)$.}
\label{tab:SOUSpanomaly}
\end{table}

\section{Details on $Q^{1,1,1}/\ZZ_2$}
\label{app:Q111Z2-details}
	
In Section \ref{sec:Q111Z2}, we introduced a web of trialities that contains a Spin(7) orientifold of phase D of $Q^{1,1,1}/\ZZ_2$ and summarized it in Figure \ref{fig:01trialityq111z2}. In this appendix, we collect all the relevant information for the other theories in this web.

\subsection{Phase E}
\label{app:Q111Z2PhaseE}

The quiver for phase E is shown in Figure \ref{fig:Q111Z2quiverE}. 

	\begin{figure}[H]
		\centering
	\begin{tikzpicture}[scale=2]
	\def\R{1.75};
	\node[draw=black,line width=1pt,circle,fill=yellowX,minimum width=0.75cm,inner sep=1pt] (v1) at (-\R,0,-\R) {$1$};
	\node[draw=black,line width=1pt,circle,fill=yellowX,minimum width=0.75cm,inner sep=1pt] (v2) at (\R,0,-\R) {$7$};
    \node[draw=black,line width=1pt,circle,fill=yellowX,minimum width=0.75cm,inner sep=1pt] (v3) at (0,0,-\R) {$3$};
    \node[draw=black,line width=1pt,circle,fill=yellowX,minimum width=0.75cm,inner sep=1pt] (v4) at (0,\R,0) {$4$};
    \node[draw=black,line width=1pt,circle,fill=yellowX,minimum width=0.75cm,inner sep=1pt] (v5) at (0,-\R,0) {$5$};
    \node[draw=black,line width=1pt,circle,fill=yellowX,minimum width=0.75cm,inner sep=1pt] (v6) at (0,0,\R) {$6$};
    \node[draw=black,line width=1pt,circle,fill=yellowX,minimum width=0.75cm,inner sep=1pt] (v7) at (\R,0,\R) {$2$};
    \node[draw=black,line width=1pt,circle,fill=yellowX,minimum width=0.75cm,inner sep=1pt] (v8) at (-\R,0,\R) {$8$};
    \draw[line width=1pt,redX] (v8) to node[fill=white,text opacity=1,fill opacity=1,draw=black,rectangle,thin,pos=0.4] {\color{redX}{$2$}} (v6);
    \draw[line width=1pt,redX] (v1) to[bend right] node[fill=white,text opacity=1,fill opacity=1,draw=black,rectangle,thin,pos=0.2] {\color{redX}{$4$}} (v5);
    \draw[line width=1pt,redX] (v3) to node[fill=white,text opacity=1,fill opacity=1,draw=black,rectangle,thin,pos=0.5] {\color{redX}{$4$}} (v4);
    \draw[line width=1pt,redX] (v2) to[bend left] node[fill=white,text opacity=1,fill opacity=1,draw=black,rectangle,thin,pos=0.5] {\color{redX}{$4$}} (v5);
    \draw[line width=1pt,redX] (v6) to node[fill=white,text opacity=1,fill opacity=1,draw=black,rectangle,thin,pos=0.4] {\color{redX}{$2$}} (v7);
    \draw[line width=1pt,Triangle-] (v1) to node[fill=white,text opacity=1,fill opacity=1,draw=black,rectangle,thin,pos=0.5] {$2$} (v4);
    \draw[line width=1pt,Triangle-] (v3) to node[fill=white,text opacity=1,fill opacity=1,draw=black,rectangle,thin,pos=0.4] {$2$} (v1);
    \draw[line width=1pt,Triangle-] (v3) to node[fill=white,text opacity=1,fill opacity=1,draw=black,rectangle,thin,pos=0.5] {$2$} (v2);
    \draw[line width=1pt,Triangle-] (v4) to node[fill=white,text opacity=1,fill opacity=1,draw=black,rectangle,thin,pos=0.2] {$2$} (v6);
    \draw[line width=1pt,Triangle-] (v2) to node[fill=white,text opacity=1,fill opacity=1,draw=black,rectangle,thin,pos=0.5] {$2$} (v4);
    \draw[line width=1pt,Triangle-] (v8) to node[fill=white,text opacity=1,fill opacity=1,draw=black,rectangle,thin,pos=0.5] {$2$} (v1);
    \draw[line width=1pt,Triangle-] (v5) to node[fill=white,text opacity=1,fill opacity=1,draw=black,rectangle,thin,pos=0.5] {$2$} (v8);
    \draw[line width=1pt,Triangle-] (v5) to node[fill=white,text opacity=1,fill opacity=1,draw=black,rectangle,thin,pos=0.25] {$2$} (v3);
    \draw[line width=1pt,Triangle-] (v5) to node[fill=white,text opacity=1,fill opacity=1,draw=black,rectangle,thin,pos=0.5] {$2$} (v7);
    \draw[line width=1pt,Triangle-] (v6) to node[fill=white,text opacity=1,fill opacity=1,draw=black,rectangle,thin,pos=0.5] {$4$} (v5);
    \draw[line width=1pt,Triangle-] (v7) to node[fill=white,text opacity=1,fill opacity=1,draw=black,rectangle,thin,pos=0.5] {$2$} (v2);
	\end{tikzpicture}
	\caption{Quiver diagram for phase E of $Q^{1,1,1}/\ZZ_2$.}
	\label{fig:Q111Z2quiverE}
	\end{figure}

The corresponding $J$- and $E$-terms are given by
		\begin{alignat}{4}
	\renewcommand{\arraystretch}{1.1}
    & \centermathcell{J}                           &\text{\hspace{.5cm}}& \centermathcell{E                               }\nonumber \\
\Lambda_{86}^1 \,:\,& \centermathcell{X_{64} Y_{41} Y_{18}-Y_{64} Y_{41} X_{18} }& & \centermathcell{Y_{85} X_{56}-X_{85} Z_{56} }\nonumber \\
\Lambda_{86}^2 \,:\,& \centermathcell{Y_{64} X_{41} X_{18}-X_{64} X_{41} Y_{18} }& & \centermathcell{Y_{85} W_{56}-X_{85} Y_{56} }\nonumber \\
\Lambda_{26}^1 \,:\,& \centermathcell{Y_{64} Y_{47} X_{72}-X_{64} Y_{47} Y_{72} }& & \centermathcell{Y_{25} X_{56}-X_{25} W_{56} }\nonumber \\
\Lambda_{26}^2 \,:\,& \centermathcell{X_{64} X_{47} Y_{72}-Y_{64} X_{47} X_{72} }& & \centermathcell{Y_{25} Z_{56}-X_{25} Y_{56} }\nonumber \\
\Lambda_{75}^1 \,:\,& \centermathcell{Y_{56} Y_{64} X_{47}-W_{56} Y_{64} Y_{47} }& & \centermathcell{X_{73} X_{35}-X_{72} X_{25} }\nonumber \\
\Lambda_{75}^2 \,:\,& \centermathcell{Z_{56} X_{64} X_{47}-X_{56} X_{64} Y_{47} }& & \centermathcell{Y_{73} Y_{35}-Y_{72} Y_{25} }\nonumber \\
\Lambda_{75}^3 \,:\,& \centermathcell{X_{56} Y_{64} Y_{47}-Z_{56} Y_{64} X_{47} }& & \centermathcell{Y_{73} X_{35}-X_{72} Y_{25} }\nonumber \\
\Lambda_{75}^4 \,:\,& \centermathcell{W_{56} X_{64} Y_{47}-X_{64} Y_{56} X_{47} }& & \centermathcell{X_{73} Y_{35}-Y_{72} X_{25} }\label{eq:Q111Z2JEtermsE} \\
\Lambda_{43}^1 \,:\,& \centermathcell{Y_{35} Y_{56} X_{64}-X_{35} Y_{56} Y_{64} }& & \centermathcell{X_{47} X_{73}-X_{41} X_{13} }\nonumber \\
\Lambda_{43}^2 \,:\,& \centermathcell{Y_{35} X_{56} X_{64}-X_{35} X_{56} Y_{64} }& & \centermathcell{Y_{47} Y_{73}-Y_{41} Y_{13} }\nonumber  \\
\Lambda_{43}^3 \,:\,& \centermathcell{X_{35} W_{56} Y_{64}-Y_{35} W_{56} X_{64} }& & \centermathcell{Y_{47} X_{73}-X_{41} Y_{13} }\nonumber \\
\Lambda_{43}^4 \,:\,& \centermathcell{X_{35} Z_{56} Y_{64}-Y_{35} Z_{56} X_{64} }& & \centermathcell{X_{47} Y_{73}-Y_{41} X_{13} }\nonumber \\
\Lambda_{15}^1 \,:\,& \centermathcell{Z_{56} Y_{64} Y_{41}-Y_{56} Y_{64} X_{41} }& & \centermathcell{X_{13} X_{35}-X_{18} X_{85} }\nonumber \\
\Lambda_{15}^2 \,:\,& \centermathcell{X_{56} X_{64} Y_{41}-W_{56} X_{64} X_{41} }& & \centermathcell{Y_{13} Y_{35}-Y_{18} Y_{85} }\nonumber \\
\Lambda_{15}^3 \,:\,& \centermathcell{W_{56} Y_{64} X_{41}-X_{56} Y_{64} Y_{41} }& & \centermathcell{Y_{13} X_{35}-X_{18} Y_{85} }\nonumber \\
\Lambda_{15}^4 \,:\,& \centermathcell{X_{64} Y_{56} X_{41}-X_{64} Z_{56} Y_{41} }& & \centermathcell{X_{13} Y_{35}-Y_{18} X_{85} }\nonumber 	
\end{alignat}
	
	Finally, the generators of the moduli space expressed in terms of the chiral fields are listed in Table~\ref{tab:GenQ111Z2PhaseE}. 

\medskip
	
\paragraph{$\U(N)^2\times \SO(N)^4$ orientifold} \mbox{}

\medskip

Let us consider an anti-holomorphic involution of phase E which acts on the nodes in Figure \ref{fig:Q111Z2quiverE} as $1\leftrightarrow 7$, $2\leftrightarrow 8$ and maps all other nodes mapped to themselves. Chiral fields transform according to
  \begin{equation}
	\begin{array}{cccccccccccc}
		X_{41} & \rightarrow & \gamma_{\Omega_4}\bar{X}_{47}\gamma_{\Omega_7}^{-1}\coma &
		Y_{41} & \rightarrow & \gamma_{\Omega_4}\bar{Y}_{47}\gamma_{\Omega_7}^{-1}\coma &
		X_{72} & \rightarrow & \gamma_{\Omega_1}\bar{Y}_{18}\gamma_{\Omega_8}^{-1}\coma &
		Y_{72} & \rightarrow & \gamma_{\Omega_1}\bar{X}_{18}\gamma_{\Omega_8}^{-1}\coma \\ 
		X_{18} & \rightarrow & \gamma_{\Omega_7}\bar{Y}_{72}\gamma_{\Omega_2}^{-1}\coma &
		Y_{18} & \rightarrow & \gamma_{\Omega_7}\bar{X}_{72}\gamma_{\Omega_2}^{-1}\coma &
		X_{64} & \rightarrow & \gamma_{\Omega_6}\bar{Y}_{64}\gamma_{\Omega_4}^{-1}\coma &
		Y_{64} & \rightarrow & \gamma_{\Omega_6}\bar{X}_{64}\gamma_{\Omega_4}^{-1}\coma \\
		X_{25} & \rightarrow & \gamma_{\Omega_8}\bar{X}_{85}\gamma_{\Omega_5}^{-1}\coma &
		Y_{25} & \rightarrow & \gamma_{\Omega_8}\bar{Y}_{85}\gamma_{\Omega_5}^{-1}\coma &
		X_{85} & \rightarrow & \gamma_{\Omega_2}\bar{X}_{25}\gamma_{\Omega_5}^{-1}\coma &
		Y_{85} & \rightarrow & \gamma_{\Omega_2}\bar{Y}_{25}\gamma_{\Omega_5}^{-1}\coma \\
		X_{35} & \rightarrow & \gamma_{\Omega_3}\bar{Y}_{35}\gamma_{\Omega_5}^{-1}\coma &
		Y_{35} & \rightarrow & \gamma_{\Omega_3}\bar{X}_{35}\gamma_{\Omega_5}^{-1}\coma &
		X_{47} & \rightarrow & \gamma_{\Omega_4}\bar{X}_{41}\gamma_{\Omega_1}^{-1}\coma &
		Y_{47} & \rightarrow & \gamma_{\Omega_4}\bar{Y}_{41}\gamma_{\Omega_1}^{-1}\coma \\
		X_{56} & \rightarrow & \gamma_{\Omega_5}\bar{X}_{56}\gamma_{\Omega_6}^{-1}\coma &
		Y_{56} & \rightarrow & \gamma_{\Omega_5}\bar{Y}_{56}\gamma_{\Omega_6}^{-1}\coma &
		Z_{56} & \rightarrow & \gamma_{\Omega_5}\bar{W}_{56}\gamma_{\Omega_6}^{-1}\coma &
		W_{56} & \rightarrow & \gamma_{\Omega_5}\bar{Z}_{56}\gamma_{\Omega_6}^{-1}\coma \\
		X_{13} & \rightarrow & \gamma_{\Omega_7}\bar{X}_{73}\gamma_{\Omega_3}^{-1}\coma &
		Y_{13} & \rightarrow & \gamma_{\Omega_7}\bar{Y}_{73}\gamma_{\Omega_3}^{-1}\coma &
		X_{73} & \rightarrow & \gamma_{\Omega_1}\bar{X}_{13}\gamma_{\Omega_3}^{-1}\coma &
		Y_{73} & \rightarrow & \gamma_{\Omega_1}\bar{Y}_{13}\gamma_{\Omega_3}^{-1}\fstop
	\end{array}  
	\label{eq:Q111Z2E-chiral-invol}
\end{equation}

Requiring the invariance of $W^{0,1}$, the Fermi fields transform as
\begin{equation}
	\begin{array}{cccccccccccc}
		\Lambda_{86}^1 & \rightarrow &  \gamma_{\Omega_2}\bar{\Lambda}^1_{26}\gamma_{\Omega_6}^{-1}\coma &
		\Lambda_{86}^2 & \rightarrow &  \gamma_{\Omega_2}\bar{\Lambda}^2_{26}\gamma_{\Omega_6}^{-1}\coma &
		\Lambda_{26}^1 & \rightarrow &  \gamma_{\Omega_8}\bar{\Lambda}^1_{86}\gamma_{\Omega_6}^{-1}\coma &
		\Lambda_{26}^2 & \rightarrow &  \gamma_{\Omega_8}\bar{\Lambda}^2_{86}\gamma_{\Omega_6}^{-1}\coma \\ 
		\Lambda_{75}^1 & \rightarrow &  \gamma_{\Omega_1}\bar{\Lambda}^4_{15}\gamma_{\Omega_5}^{-1}\coma &
		\Lambda_{75}^2 & \rightarrow &  \gamma_{\Omega_1}\bar{\Lambda}^3_{15}\gamma_{\Omega_5}^{-1}\coma &
		\Lambda_{75}^3 & \rightarrow &  \gamma_{\Omega_1}\bar{\Lambda}^2_{15}\gamma_{\Omega_5}^{-1}\coma &
		\Lambda_{75}^4 & \rightarrow &  \gamma_{\Omega_1}\bar{\Lambda}^1_{15}\gamma_{\Omega_5}^{-1}\coma \\
		\Lambda_{43}^1 & \rightarrow & -\gamma_{\Omega_4}\bar{\Lambda}^2_{43}\gamma_{\Omega_3}^{-1}\coma &
		\Lambda_{43}^2 & \rightarrow & -\gamma_{\Omega_4}\bar{\Lambda}^1_{43}\gamma_{\Omega_3}^{-1}\coma &
		\Lambda_{43}^3 & \rightarrow & -\gamma_{\Omega_4}\bar{\Lambda}^4_{43}\gamma_{\Omega_3}^{-1}\coma &
		\Lambda_{43}^4 & \rightarrow & -\gamma_{\Omega_4}\bar{\Lambda}^3_{43}\gamma_{\Omega_3}^{-1}\coma \\
		\Lambda_{15}^1 & \rightarrow &  \gamma_{\Omega_7}\bar{\Lambda}^4_{75}\gamma_{\Omega_5}^{-1}\coma &
		\Lambda_{15}^2 & \rightarrow &  \gamma_{\Omega_7}\bar{\Lambda}^3_{75}\gamma_{\Omega_5}^{-1}\coma &
		\Lambda_{15}^3 & \rightarrow &  \gamma_{\Omega_7}\bar{\Lambda}^2_{75}\gamma_{\Omega_5}^{-1}\coma &
		\Lambda_{15}^4 & \rightarrow &  \gamma_{\Omega_7}\bar{\Lambda}^1_{75}\gamma_{\Omega_5}^{-1}\coma
	\end{array}   
	\label{eq:Q111Z2E-Fermi-invol}
\end{equation}
and
\begin{equation}
	\begin{array}{cccccccccccc}
		\Lambda_{11}^R & \rightarrow & \gamma_{\Omega_7}\Lambda_{77}^{R\,\,T}\gamma_{\Omega_7}^{-1} \coma &
		\Lambda_{22}^R & \rightarrow & \gamma_{\Omega_8}\Lambda_{88}^{R\,\,T}\gamma_{\Omega_8}^{-1} \coma &
		\Lambda_{33}^R & \rightarrow & \gamma_{\Omega_3}\Lambda_{33}^{R\,\,T}\gamma_{\Omega_3}^{-1} \coma &
		\Lambda_{44}^R & \rightarrow & \gamma_{\Omega_4}\Lambda_{44}^{R\,\,T}\gamma_{\Omega_4}^{-1} \coma \\
		\Lambda_{55}^R & \rightarrow & \gamma_{\Omega_5}\Lambda_{55}^{R\,\,T}\gamma_{\Omega_5}^{-1} \coma &
		\Lambda_{66}^R & \rightarrow & \gamma_{\Omega_6}\Lambda_{66}^{R\,\,T}\gamma_{\Omega_6}^{-1} \coma &
		\Lambda_{77}^R & \rightarrow & \gamma_{\Omega_1}\Lambda_{11}^{R\,\,T}\gamma_{\Omega_1}^{-1} \coma &
		\Lambda_{88}^R & \rightarrow & \gamma_{\Omega_2}\Lambda_{22}^{R\,\,T}\gamma_{\Omega_2}^{-1} \fstop
	\end{array}   
	\label{eq:Q111Z2E-RFermi-invol}
\end{equation}

Using Table \ref{tab:GenQ111Z2PhaseE}, the corresponding geometric involution acting on the generators reads 
\begin{equation}
     \begin{array}{ccccccccccccccc}
	M_{1}  & \rightarrow & \bar{M}_{27} \coma &
	M_{2}  & \rightarrow & \bar{M}_{24} \coma &
	M_{3}  & \rightarrow & \bar{M}_{21} \coma &
	M_{4}  & \rightarrow & \bar{M}_{18} \coma &
	M_{5}  & \rightarrow & \bar{M}_{12} \coma \\
	M_{6}  & \rightarrow & \bar{M}_{6} \coma &
	M_{7}  & \rightarrow & \bar{M}_{26} \coma &
	M_{8}  & \rightarrow & \bar{M}_{23} \coma &
	M_{9}  & \rightarrow & \bar{M}_{20} \coma &
	M_{10} & \rightarrow & \bar{M}_{17} \coma \\
	M_{11} & \rightarrow & \bar{M}_{11} \coma &
	M_{12} & \rightarrow & \bar{M}_{5} \coma &
	M_{13} & \rightarrow & \bar{M}_{25} \coma &
	M_{14} & \rightarrow & \bar{M}_{22} \coma &
	M_{15} & \rightarrow & \bar{M}_{19} \coma \\
	M_{16} & \rightarrow & \bar{M}_{16} \coma &
	M_{17} & \rightarrow & \bar{M}_{10} \coma &
	M_{18} & \rightarrow & \bar{M}_{4} \coma &
	M_{19} & \rightarrow & \bar{M}_{15} \coma &
	M_{20} & \rightarrow & \bar{M}_{9} \coma \\ 
	M_{21} & \rightarrow & \bar{M}_{3}\coma &
	M_{22} & \rightarrow & \bar{M}_{14}\coma &
	M_{23} & \rightarrow & \bar{M}_{8}\coma &
	M_{24} & \rightarrow & \bar{M}_{2}\coma &
	M_{25} & \rightarrow & \bar{M}_{13}\coma \\
	\multicolumn{15}{c}{M_{26}   \rightarrow  \bar{M}_{7}\coma 
	M_{27}  \rightarrow  \bar{M}_{1}\fstop}
	\end{array}   
	\label{eq:Q111Z2E-Meson-invol}
    \end{equation}
This geometric involution is the same of~\eqref{eq:Q111Z2D-Meson-invol}.
 
The $\gamma_{\Omega_i}$ matrices are constrained as in \eqref{gamma_matrices_Q111/Z2}. As for phase D, we choose
\begin{equation}
\gamma_{\Omega_3}=\gamma_{\Omega_4}=\gamma_{\Omega_5}=\gamma_{\Omega_6}=\ID_N \fstop 
\end{equation}

The resulting orientifold is shown in Figure \ref{fig:orien_Q111Z2quiverE}.

\begin{figure}[H]
		\centering
		\begin{tikzpicture}[scale=2]
	\def\R{1.5};
	     \draw[line width=1pt,redX] (120:\R*1.15) circle (0.25);
	\draw[line width=1pt,redX] (60:\R*1.15) circle (0.25);
	    \draw[line width=1pt,redX] (0:\R*1.15) circle (0.25) node[xshift=0.5cm,star,star points=5, star point ratio=2.25, inner sep=1pt, fill=redX, draw] {};
	    \draw[line width=1pt,redX] (300:\R*1.15) circle (0.25) node[xshift=0.34cm,yshift=-0.34cm,star,star points=5, star point ratio=2.25, inner sep=1pt, fill=redX, draw] {};
	    \draw[line width=1pt,redX] (240:\R*1.15) circle (0.25) node[xshift=-0.34cm,yshift=-0.34cm,star,star points=5, star point ratio=2.25, inner sep=1pt, fill=redX, draw] {};
	    \draw[line width=1pt,redX] (180:\R*1.15) circle (0.25) node[xshift=-0.5cm,star,star points=5, star point ratio=2.25, inner sep=1pt, fill=redX, draw] {};
	       \node[draw=black,line width=1pt,circle,fill=yellowX,minimum width=0.75cm,inner sep=1pt,label={[xshift=0.7cm,yshift=-0cm]:$\U(N)$}] (v1) at  (120:\R) {$1$};
	        \node[draw=black,line width=1pt,circle,fill=yellowX,minimum width=0.75cm,inner sep=1pt,label={[xshift=0.75cm,yshift=0cm]:$\SO(N)$}] (v3) at (0:\R) {$3$};
	        \node[draw=black,line width=1pt,circle,fill=yellowX,minimum width=0.75cm,inner sep=1pt,label={[xshift=1cm,yshift=-0.3cm]:$\SO(N)$}] (v4) at (300:\R) {$4$};
	        \node[draw=black,line width=1pt,circle,fill=yellowX,minimum width=0.75cm,inner sep=1pt,label={[xshift=-1cm,yshift=-0.3cm]:$\SO(N)$}] (v5) at (240:\R) {$5$};
	        \node[draw=black,line width=1pt,circle,fill=yellowX,minimum width=0.75cm,inner sep=1pt,label={[xshift=-0.75cm,yshift=0cm]:$\SO(N)$}] (v6) at (180:\R) {$6$};
	        \node[draw=black,line width=1pt,circle,fill=yellowX,minimum width=0.75cm,inner sep=1pt,label={[xshift=-0.7cm,yshift=-0cm]:$\U(N)$}] (v7) at  (60:\R) {$2$};
	        \draw[line width=1pt,redX] (v1) to node[fill=white,text opacity=1,fill opacity=1,draw=black,rectangle,thin,pos=0.5] {\color{redX}{$8$}} (v5);
	        \draw[line width=1pt,redX] (v7) to node[fill=white,text opacity=1,fill opacity=1,draw=black,rectangle,thin,pos=0.8] {\color{redX}{$4$}} (v6);
	        \draw[line width=1pt,redX] (v4) to node[fill=white,text opacity=1,fill opacity=1,draw=black,rectangle,thin,pos=0.5] {\color{redX}{$4$}} (v3);
	        \draw[line width=1pt,-Triangle] (v1) to node[fill=white,text opacity=1,fill opacity=1,draw=black,rectangle,thin,pos=0.5] {$4$} (v7);
	        \draw[line width=1pt,-Triangle] (v1) to node[fill=white,text opacity=1,fill opacity=1,draw=black,rectangle,thin,pos=0.75] {$4$} (v3);
	        \draw[line width=1pt,-Triangle] (v7) to node[fill=white,text opacity=1,fill opacity=1,draw=black,rectangle,thin,pos=0.3] {$4$} (v5);
	        \draw[line width=1pt,-Triangle] (v4) to node[fill=white,text opacity=1,fill opacity=1,draw=black,rectangle,thin,pos=0.7] {$4$} (v1);
	        \draw[line width=1pt] (v5) to node[fill=white,text opacity=1,fill opacity=1,draw=black,rectangle,thin,pos=0.5] {$4$} (v6);
	        \draw[line width=1pt] (v6) to node[fill=white,text opacity=1,fill opacity=1,draw=black,rectangle,thin,pos=0.85] {$2$} (v4);
	        \draw[line width=1pt] (v3) to node[fill=white,text opacity=1,fill opacity=1,draw=black,rectangle,thin,pos=0.85] {$2$} (v5);
	\end{tikzpicture}
	\caption{Quiver diagram for the Spin(7) orientifold of phase E of $Q^{1,1,1}/\ZZ_2$ using the involution in \cref{eq:Q111Z2E-chiral-invol,eq:Q111Z2E-Fermi-invol,eq:Q111Z2E-RFermi-invol}, together with our choice of $\gamma_{\Omega_i}$ matrices.}
		\label{fig:orien_Q111Z2quiverE}
	\end{figure}

\subsection{Phase H}
\label{app:Q111Z2PhaseH}

The quiver for phase H is shown in Figure \ref{fig:Q111Z2quiverH}. 

\begin{figure}[H]
		\centering
		\begin{tikzpicture}[scale=2]
	\def\R{1.75};
	 \node[draw=black,line width=1pt,circle,fill=yellowX,minimum width=0.75cm,inner sep=1pt] (v1) at (-\R,0,-\R) {$1$};
	        \node[draw=black,line width=1pt,circle,fill=yellowX,minimum width=0.75cm,inner sep=1pt] (v2) at (\R,0,-\R) {$7$};
	        \node[draw=black,line width=1pt,circle,fill=yellowX,minimum width=0.75cm,inner sep=1pt] (v3) at (0,0,-\R) {$3$};
	        \node[draw=black,line width=1pt,circle,fill=yellowX,minimum width=0.75cm,inner sep=1pt] (v4) at (0,\R,0) {$4$};
	        \node[draw=black,line width=1pt,circle,fill=yellowX,minimum width=0.75cm,inner sep=1pt] (v5) at (0,-\R,0) {$5$};
	        \node[draw=black,line width=1pt,circle,fill=yellowX,minimum width=0.75cm,inner sep=1pt] (v6) at (0,0,\R) {$6$};
	        \node[draw=black,line width=1pt,circle,fill=yellowX,minimum width=0.75cm,inner sep=1pt] (v7) at (\R,0,\R) {$2$};
	        \node[draw=black,line width=1pt,circle,fill=yellowX,minimum width=0.75cm,inner sep=1pt] (v8) at (-\R,0,\R) {$8$};
	        \draw[line width=1pt,redX] (v7) to node[fill=white,text opacity=1,fill opacity=1,draw=black,rectangle,thin,pos=0.8] {\color{redX}{$2$}} (v6);
	        \draw[line width=1pt,redX] (v7) to node[fill=white,text opacity=1,fill opacity=1,draw=black,rectangle,thin,pos=0.5] {\color{redX}{$2$}} (v2);
	        \draw[line width=1pt,redX] (v6) to node[fill=white,text opacity=1,fill opacity=1,draw=black,rectangle,thin,pos=0.5] {\color{redX}{$2$}} (v8);
	        \draw[line width=1pt,redX] (v8) to node[fill=white,text opacity=1,fill opacity=1,draw=black,rectangle,thin,pos=0.5] {\color{redX}{$2$}} (v1);
	        \draw[line width=1pt,redX] (v5) to node[fill=white,text opacity=1,fill opacity=1,draw=black,rectangle,thin,pos=0.5] {\color{redX}{$8$}} (v4);
	        \draw[line width=1pt,redX] (v4) to node[fill=white,text opacity=1,fill opacity=1,draw=black,rectangle,thin,pos=0.5] {\color{redX}{$4$}} (v3);
	        \draw[line width=1pt,-Triangle] (v7) to node[fill=white,text opacity=1,fill opacity=1,draw=black,rectangle,thin,pos=0.5] {$2$} (v5);
	        \draw[line width=1pt,-Triangle] (v8) to node[fill=white,text opacity=1,fill opacity=1,draw=black,rectangle,thin,pos=0.5] {$2$} (v5);
	        \draw[line width=1pt,-Triangle] (v5) to node[fill=white,text opacity=1,fill opacity=1,draw=black,rectangle,thin,pos=0.5] {$4$} (v6);
	        \draw[line width=1pt,-Triangle] (v5) to node[fill=white,text opacity=1,fill opacity=1,draw=black,rectangle,thin,pos=0.75] {$2$} (v3);
	        \draw[line width=1pt,-Triangle] (v6) to node[fill=white,text opacity=1,fill opacity=1,draw=black,rectangle,thin,pos=0.25] {$2$} (v4);
	        \draw[line width=1pt,-Triangle] (v3) to node[fill=white,text opacity=1,fill opacity=1,draw=black,rectangle,thin,pos=0.75] {$2$} (v1);
	        \draw[line width=1pt,-Triangle] (v3) to node[fill=white,text opacity=1,fill opacity=1,draw=black,rectangle,thin,pos=0.5] {$2$} (v2);
	        \draw[line width=1pt,-Triangle] (v4) to[bend left] node[fill=white,text opacity=1,fill opacity=1,draw=black,rectangle,thin,pos=0.75] {$4$} (v7);
	        \draw[line width=1pt,-Triangle] (v4) to[bend right] node[fill=white,text opacity=1,fill opacity=1,draw=black,rectangle,thin,pos=0.5] {$4$} (v8);
	        \draw[line width=1pt,-Triangle] (v1) to node[fill=white,text opacity=1,fill opacity=1,draw=black,rectangle,thin,pos=0.5] {$2$} (v4);
	        \draw[line width=1pt,-Triangle] (v2) to node[fill=white,text opacity=1,fill opacity=1,draw=black,rectangle,thin,pos=0.5] {$2$} (v4);
	\end{tikzpicture}
	\caption{Quiver diagram for phase H of $Q^{1,1,1}/\ZZ_2$.}
	\label{fig:Q111Z2quiverH}
	\end{figure}

The $J$- and $E$-terms are
		\begin{alignat}{4}
	\renewcommand{\arraystretch}{1.1}
    & \centermathcell{J}                           &\text{\hspace{.5cm}}& \centermathcell{E                               }\nonumber \\
\Lambda_{26}^{1} \,:\,& \centermathcell{Y_{64} Z_{42}-X_{64} W_{42} }& & \centermathcell{Y_{25} X_{56}-X_{25} Z_{56} }\nonumber\\
\Lambda_{26}^{7} \,:\,& \centermathcell{Y_{64} X_{42}-Y_{42} X_{64} }& & \centermathcell{X_{25} W_{56}-Y_{25} Y_{56} }\nonumber\\
\Lambda_{27}^{1} \,:\,& \centermathcell{X_{74} W_{42}-Y_{74} Y_{42} }& & \centermathcell{Y_{25} X_{53} X_{37}-X_{25} X_{53} Y_{37} }\nonumber\\
\Lambda_{27}^{7} \,:\,& \centermathcell{X_{74} Z_{42}-Y_{74} X_{42} }& & \centermathcell{X_{25} Y_{53} Y_{37}-Y_{25} Y_{53} X_{37} }\nonumber\\
\Lambda_{68}^{1} \,:\,& \centermathcell{X_{85} W_{56}-Y_{85} Z_{56} }& & \centermathcell{X_{64} Y_{48}-Y_{64} X_{48} }\nonumber\\
\Lambda_{68}^{7} \,:\,& \centermathcell{X_{85} Y_{56}-Y_{85} X_{56} }& & \centermathcell{Y_{64} Z_{48}-X_{64} W_{48} }\nonumber\\
\Lambda_{81}^{1} \,:\,& \centermathcell{X_{14} W_{48}-Y_{14} Y_{48} }& & \centermathcell{X_{85} X_{53} Y_{31}-Y_{85} X_{53} X_{31} }\nonumber\\
\Lambda_{81}^{7} \,:\,& \centermathcell{X_{14} Z_{48}-Y_{14} X_{48} }& & \centermathcell{Y_{85} Y_{53} X_{31}-X_{85} Y_{53} Y_{31} }\nonumber\\
\Lambda_{54}^{1} \,:\,& \centermathcell{W_{48} Y_{85}-W_{42} Y_{25} }& & \centermathcell{X_{53} X_{31} X_{14}-X_{56} X_{64} }\nonumber\\
\Lambda_{54}^{7} \,:\,& \centermathcell{Z_{48} Y_{85}-Z_{42} Y_{25} }& & \centermathcell{X_{56} Y_{64}-Y_{53} X_{31} X_{14} }\label{eq:Q111Z2JEtermsH}\\
\Lambda_{54}^{3} \,:\,& \centermathcell{W_{48} X_{85}-Y_{42} Y_{25} }& & \centermathcell{Y_{56} X_{64}-X_{53} Y_{31} X_{14} }\nonumber\\
\Lambda_{54}^{4} \,:\,& \centermathcell{Z_{48} X_{85}-X_{42} Y_{25} }& & \centermathcell{Y_{53} X_{37} Y_{74}-Y_{56} Y_{64} }\nonumber\\
\Lambda_{54}^{5} \,:\,& \centermathcell{Y_{48} Y_{85}-W_{42} X_{25} }& & \centermathcell{Z_{56}X_{64}-X_{53} X_{31} Y_{14} }\nonumber\\
\Lambda_{54}^{6} \,:\,& \centermathcell{X_{48} Y_{85}-Z_{42} X_{25} }& & \centermathcell{Y_{53} Y_{37} X_{74}-Z_{56} Y_{64} }\nonumber\\
\Lambda_{54}^{2} \,:\,& \centermathcell{Y_{48} X_{85}-Y_{42} X_{25} }& & \centermathcell{X_{53} Y_{37} Y_{74}-W_{56} X_{64} }\nonumber\\
\Lambda_{54}^{8} \,:\,& \centermathcell{X_{48} X_{85}-X_{42} X_{25} }& & \centermathcell{W_{56} Y_{64}-Y_{53} Y_{37} Y_{74}  }\nonumber\\
\Lambda_{43}^{1} \,:\,& \centermathcell{Y_{31} Y_{14}-Y_{37} Y_{74} }& & \centermathcell{Y_{48} X_{85} X_{53}-X_{48} X_{85} Y_{53} }\nonumber\\
\Lambda_{43}^{7} \,:\,& \centermathcell{X_{31} Y_{14}-Y_{37} X_{74} }& & \centermathcell{X_{48} Y_{85} Y_{53}-W_{42} X_{25} X_{53} }\nonumber\\
\Lambda_{43}^{3} \,:\,& \centermathcell{Y_{31} X_{14}-X_{37} Y_{74} }& & \centermathcell{Z_{48} X_{85} Y_{53}-Y_{42} Y_{25} X_{53} }\nonumber\\
\Lambda_{43}^{4} \,:\,& \centermathcell{X_{31} X_{14}-X_{37} X_{74} }& & \centermathcell{W_{42} Y_{25} X_{53}-Z_{42} Y_{25} Y_{53} }\nonumber	
\end{alignat}
The generators of the moduli space expressed in terms of the chiral fields are listed in Table \ref{tab:GenQ111Z2PhaseH}.

\medskip

\paragraph{$\U(N)^2\times \SO(N)^4$ orientifold}\mbox{}

\medskip 

Let us consider an anti-holomorphic involution of phase H which acts on the nodes in Figure \ref{fig:Q111Z2quiverH} as $1\leftrightarrow 7$ and $2\leftrightarrow 8$ and maps all other nodes mapped to themselves. Chiral fields transform according to
\begin{equation}
	\begin{array}{cccccccccccc}
		Y_{64}& \rightarrow & \gamma_{\Omega_6}\bar{X}_{64}\gamma_{\Omega_4}^{-2} \coma &
		X_{64}& \rightarrow & \gamma_{\Omega_6}\bar{Y}_{64}\gamma_{\Omega_4}^{-2} \coma &
		Z_{42}& \rightarrow & \gamma_{\Omega_4}\bar{Y}_{48}\gamma_{\Omega_8}^{-2} \coma &
		Y_{48}& \rightarrow & \gamma_{\Omega_4}\bar{Z}_{42}\gamma_{\Omega_2}^{-2} \coma \\ 
		W_{42}& \rightarrow & \gamma_{\Omega_4}\bar{X}_{48}\gamma_{\Omega_8}^{-2} \coma &
		X_{48}& \rightarrow & \gamma_{\Omega_4}\bar{W}_{42}\gamma_{\Omega_2}^{-2} \coma &
		Y_{25}& \rightarrow & \gamma_{\Omega_8}\bar{X}_{85}\gamma_{\Omega_5}^{-2} \coma &
		X_{85}& \rightarrow & \gamma_{\Omega_2}\bar{Y}_{25}\gamma_{\Omega_5}^{-2} \coma \\
		X_{56}& \rightarrow & \gamma_{\Omega_5}\bar{W}_{56}\gamma_{\Omega_6}^{-2} \coma &
		W_{56}& \rightarrow & \gamma_{\Omega_5}\bar{X}_{56}\gamma_{\Omega_6}^{-2} \coma &
		X_{25}& \rightarrow & \gamma_{\Omega_8}\bar{Y}_{85}\gamma_{\Omega_5}^{-2} \coma &
		Y_{85}& \rightarrow & \gamma_{\Omega_2}\bar{X}_{25}\gamma_{\Omega_5}^{-2} \coma \\
		Z_{56}& \rightarrow & \gamma_{\Omega_5}\bar{Z}_{56}\gamma_{\Omega_6}^{-2} \coma &
		X_{42}& \rightarrow & \gamma_{\Omega_4}\bar{W}_{48}\gamma_{\Omega_8}^{-2} \coma &
		W_{48}& \rightarrow & \gamma_{\Omega_4}\bar{X}_{42}\gamma_{\Omega_2}^{-2} \coma &
		Y_{42}& \rightarrow & \gamma_{\Omega_4}\bar{Z}_{48}\gamma_{\Omega_8}^{-2} \coma \\
		Z_{48}& \rightarrow & \gamma_{\Omega_4}\bar{Y}_{42}\gamma_{\Omega_2}^{-2} \coma &
		Y_{56}& \rightarrow & \gamma_{\Omega_5}\bar{Y}_{56}\gamma_{\Omega_6}^{-2} \coma &
		X_{74}& \rightarrow & \gamma_{\Omega_1}\bar{Y}_{14}\gamma_{\Omega_4}^{-2} \coma &
		Y_{14}& \rightarrow & \gamma_{\Omega_7}\bar{X}_{74}\gamma_{\Omega_4}^{-2} \coma \\
		Y_{74}& \rightarrow & \gamma_{\Omega_1}\bar{X}_{14}\gamma_{\Omega_4}^{-2} \coma &
		X_{14}& \rightarrow & \gamma_{\Omega_7}\bar{Y}_{74}\gamma_{\Omega_4}^{-2} \coma &
		X_{53}& \rightarrow & \gamma_{\Omega_5}\bar{Y}_{53}\gamma_{\Omega_3}^{-2} \coma &
		Y_{53}& \rightarrow & \gamma_{\Omega_5}\bar{X}_{53}\gamma_{\Omega_3}^{-2} \coma \\
		X_{37}& \rightarrow & \gamma_{\Omega_3}\bar{Y}_{31}\gamma_{\Omega_1}^{-2} \coma &
		Y_{31}& \rightarrow & \gamma_{\Omega_3}\bar{X}_{37}\gamma_{\Omega_7}^{-2} \coma &
		Y_{37}& \rightarrow & \gamma_{\Omega_3}\bar{X}_{31}\gamma_{\Omega_1}^{-2} \coma &
		X_{31}& \rightarrow & \gamma_{\Omega_3}\bar{Y}_{37}\gamma_{\Omega_7}^{-2} \fstop
	\end{array}  
	\label{eq:Q111Z2H-chiral-invol}
\end{equation}

Requiring the invariance of $W^{(0,1)}$, the Fermi fields transform as
\begin{equation}
	\begin{array}{cccccccccccc}
		\Lambda_{26}^{2} & \rightarrow & \gamma_{\Omega_6}\Lambda^1_{68}\gamma_{\Omega_8}^{-2} \coma &
		\Lambda_{26}^{8} & \rightarrow & -\gamma_{\Omega_6}\Lambda^2_{68}\gamma_{\Omega_8}^{-2} \coma &
		\Lambda_{27}^{2} & \rightarrow & -\gamma_{\Omega_8}\bar{\Lambda}^2_{81}\gamma_{\Omega_1}^{-2} \coma &
		\Lambda_{27}^{8} & \rightarrow & -\gamma_{\Omega_8}\bar{\Lambda}^1_{81}\gamma_{\Omega_1}^{-2} \coma \\
		\Lambda_{68}^{2} & \rightarrow & \gamma_{\Omega_2}\Lambda^1_{26}\gamma_{\Omega_6}^{-2} \coma &
		\Lambda_{68}^{8} & \rightarrow & -\gamma_{\Omega_2}\Lambda^2_{26}\gamma_{\Omega_6}^{-2} \coma &
		\Lambda_{81}^{2} & \rightarrow & -\gamma_{\Omega_2}\bar{\Lambda}^2_{27}\gamma_{\Omega_7}^{-2} \coma &
		\Lambda_{81}^{8} & \rightarrow & -\gamma_{\Omega_2}\bar{\Lambda}^1_{27}\gamma_{\Omega_7}^{-2} \coma \\
		\Lambda_{54}^{2} & \rightarrow & -\gamma_{\Omega_5}\bar{\Lambda}^8_{54}\gamma_{\Omega_4}^{-2} \coma &
		\Lambda_{54}^{8} & \rightarrow & -\gamma_{\Omega_5}\bar{\Lambda}^7_{54}\gamma_{\Omega_4}^{-2} \coma &
		\Lambda_{54}^{5} & \rightarrow & -\gamma_{\Omega_5}\bar{\Lambda}^4_{54}\gamma_{\Omega_4}^{-2} \coma &
		\Lambda_{54}^{6} & \rightarrow & -\gamma_{\Omega_5}\bar{\Lambda}^3_{54}\gamma_{\Omega_4}^{-2} \coma \\
		\Lambda_{54}^{3} & \rightarrow & -\gamma_{\Omega_5}\bar{\Lambda}^6_{54}\gamma_{\Omega_4}^{-2} \coma &
		\Lambda_{54}^{4} & \rightarrow & -\gamma_{\Omega_5}\bar{\Lambda}^5_{54}\gamma_{\Omega_4}^{-2} \coma &
		\Lambda_{54}^{1} & \rightarrow & -\gamma_{\Omega_5}\bar{\Lambda}^2_{54}\gamma_{\Omega_4}^{-2} \coma &
		\Lambda_{54}^{7} & \rightarrow & -\gamma_{\Omega_5}\bar{\Lambda}^1_{54}\gamma_{\Omega_4}^{-2} \coma \\
		\Lambda_{43}^{2} & \rightarrow & -\gamma_{\Omega_4}\bar{\Lambda}^4_{43}\gamma_{\Omega_3}^{-2} \coma &
		\Lambda_{43}^{8} & \rightarrow & -\gamma_{\Omega_4}\bar{\Lambda}^2_{43}\gamma_{\Omega_3}^{-2} \coma &
		\Lambda_{43}^{5} & \rightarrow & -\gamma_{\Omega_4}\bar{\Lambda}^3_{43}\gamma_{\Omega_3}^{-2} \coma &
		\Lambda_{43}^{6} & \rightarrow & -\gamma_{\Omega_4}\bar{\Lambda}^1_{43}\gamma_{\Omega_3}^{-2} \coma 
	\end{array}   
	\label{eq:Q111Z2H-Fermi-invol}
\end{equation}
and
	\begin{equation}
	 \begin{array}{cccccccccccc}
	\Lambda_{11}^R & \rightarrow & \gamma_{\Omega_7}\Lambda_{77}^{R\,\,T}\gamma_{\Omega_7}^{-1} \coma &
    \Lambda_{22}^R & \rightarrow & \gamma_{\Omega_8}\Lambda_{88}^{R\,\,T}\gamma_{\Omega_8}^{-1} \coma &
	\Lambda_{33}^R & \rightarrow & \gamma_{\Omega_3}\Lambda_{33}^{R\,\,T}\gamma_{\Omega_3}^{-1} \coma &
	\Lambda_{44}^R & \rightarrow & \gamma_{\Omega_4}\Lambda_{44}^{R\,\,T}\gamma_{\Omega_4}^{-1} \coma \\
	\Lambda_{55}^R & \rightarrow & \gamma_{\Omega_5}\Lambda_{55}^{R\,\,T}\gamma_{\Omega_5}^{-1} \coma &
	\Lambda_{66}^R & \rightarrow & \gamma_{\Omega_6}\Lambda_{66}^{R\,\,T}\gamma_{\Omega_6}^{-1} \coma &
	\Lambda_{77}^R & \rightarrow & \gamma_{\Omega_1}\Lambda_{11}^{R\,\,T}\gamma_{\Omega_1}^{-1} \coma &
	\Lambda_{88}^R & \rightarrow & \gamma_{\Omega_2}\Lambda_{22}^{R\,\,T}\gamma_{\Omega_2}^{-1} \fstop
	\end{array}   
	\label{eq:Q111Z2H-RFermi-invol}
	\end{equation}
	
Using Table \ref{tab:GenQ111Z2PhaseH}, the corresponding geometric involution acting on the generators reads 
\begin{equation}
     \begin{array}{ccccccccccccccc}
	M_{1}  & \rightarrow & \bar{M}_{27} \coma &
	M_{2}  & \rightarrow & \bar{M}_{24} \coma &
	M_{3}  & \rightarrow & \bar{M}_{21} \coma &
	M_{4}  & \rightarrow & \bar{M}_{18} \coma &
	M_{5}  & \rightarrow & \bar{M}_{12} \coma \\
	M_{6}  & \rightarrow & \bar{M}_{6} \coma &
	M_{7}  & \rightarrow & \bar{M}_{26} \coma &
	M_{8}  & \rightarrow & \bar{M}_{23} \coma &
	M_{9}  & \rightarrow & \bar{M}_{20} \coma &
	M_{10} & \rightarrow & \bar{M}_{17} \coma \\
	M_{11} & \rightarrow & \bar{M}_{11} \coma &
	M_{12} & \rightarrow & \bar{M}_{5} \coma &
	M_{13} & \rightarrow & \bar{M}_{25} \coma &
	M_{14} & \rightarrow & \bar{M}_{22} \coma &
	M_{15} & \rightarrow & \bar{M}_{19} \coma \\
	M_{16} & \rightarrow & \bar{M}_{16} \coma &
	M_{17} & \rightarrow & \bar{M}_{10} \coma &
	M_{18} & \rightarrow & \bar{M}_{4} \coma &
	M_{19} & \rightarrow & \bar{M}_{15} \coma &
	M_{20} & \rightarrow & \bar{M}_{9} \coma \\ 
	M_{21} & \rightarrow & \bar{M}_{3}\coma &
	M_{22} & \rightarrow & \bar{M}_{14}\coma &
	M_{23} & \rightarrow & \bar{M}_{8}\coma &
	M_{24} & \rightarrow & \bar{M}_{2}\coma &
	M_{25} & \rightarrow & \bar{M}_{13}\coma \\
	\multicolumn{15}{c}{M_{26}   \rightarrow  \bar{M}_{7}\coma 
	M_{27}  \rightarrow  \bar{M}_{1}\fstop}
	\end{array}   
	\label{eq:Q111Z2H-Meson-invol}
    \end{equation}
Once again, this is the same involution of phase D in \eqref{eq:Q111Z2D-Meson-invol}. 
    
    The $\gamma_{\Omega_i}$ matrices are constrained as in \eqref{gamma_matrices_Q111/Z2}. As for phase D, we choose
\begin{equation}
\gamma_{\Omega_3}=\gamma_{\Omega_4}=\gamma_{\Omega_5}=\gamma_{\Omega_6}=\ID_N \fstop 
\end{equation}
    
Figure \ref{fig:orien_Q111Z2quiverH} shows the quiver for the resulting orientifold of phase H.
  
    \begin{figure}[H]
		\centering
		\begin{tikzpicture}[scale=2]
	\def\R{1.5};
	     \draw[line width=1pt,redX] (120:\R*1.15) circle (0.25);
	\draw[line width=1pt,redX] (60:\R*1.15) circle (0.25);
	    \draw[line width=1pt,redX] (0:\R*1.15) circle (0.25) node[xshift=0.5cm,star,star points=5, star point ratio=2.25, inner sep=1pt, fill=redX, draw] {};
	    \draw[line width=1pt,redX] (300:\R*1.15) circle (0.25) node[xshift=0.34cm,yshift=-0.34cm,star,star points=5, star point ratio=2.25, inner sep=1pt, fill=redX, draw] {};
	    \draw[line width=1pt,redX] (240:\R*1.15) circle (0.25) node[xshift=-0.34cm,yshift=-0.34cm,star,star points=5, star point ratio=2.25, inner sep=1pt, fill=redX, draw] {};
	    \draw[line width=1pt,redX] (180:\R*1.15) circle (0.25) node[xshift=-0.5cm,star,star points=5, star point ratio=2.25, inner sep=1pt, fill=redX, draw] {};
	       \node[draw=black,line width=1pt,circle,fill=yellowX,minimum width=0.75cm,inner sep=1pt,label={[xshift=0.7cm,yshift=-0cm]:$\U(N)$}] (v1) at  (120:\R) {$1$};
	        \node[draw=black,line width=1pt,circle,fill=yellowX,minimum width=0.75cm,inner sep=1pt,label={[xshift=0.75cm,yshift=0cm]:$\SO(N)$}] (v3) at (0:\R) {$3$};
	        \node[draw=black,line width=1pt,circle,fill=yellowX,minimum width=0.75cm,inner sep=1pt,label={[xshift=1cm,yshift=-0.3cm]:$\SO(N)$}] (v4) at (300:\R) {$4$};
	        \node[draw=black,line width=1pt,circle,fill=yellowX,minimum width=0.75cm,inner sep=1pt,label={[xshift=-1cm,yshift=-0.3cm]:$\SO(N)$}] (v5) at (240:\R) {$5$};
	        \node[draw=black,line width=1pt,circle,fill=yellowX,minimum width=0.75cm,inner sep=1pt,label={[xshift=-0.75cm,yshift=0cm]:$\SO(N)$}] (v6) at (180:\R) {$6$};
	        \node[draw=black,line width=1pt,circle,fill=yellowX,minimum width=0.75cm,inner sep=1pt,label={[xshift=-0.7cm,yshift=-0cm]:$\U(N)$}] (v7) at  (60:\R) {$2$};
	        \draw[line width=1pt,redX] (v6) to node[fill=white,text opacity=1,fill opacity=1,draw=black,rectangle,thin,pos=0.4] {\color{redX}{$4$}} (v7);
	        \draw[line width=1pt,redX] (v4) to node[fill=white,text opacity=1,fill opacity=1,draw=black,rectangle,thin,pos=0.5] {\color{redX}{$4$}} (v3);
	        \draw[line width=1pt,redX] (v5) to node[fill=white,text opacity=1,fill opacity=1,draw=black,rectangle,thin,pos=0.5] {\color{redX}{$8$}} (v4);
	        \draw[line width=1pt,redX] (v1) to node[fill=white,text opacity=1,fill opacity=1,draw=black,rectangle,thin,pos=0.5] {\color{redX}{$4$}} (v7);
	        \draw[line width=1pt,-Triangle] (v3) to node[fill=white,text opacity=1,fill opacity=1,draw=black,rectangle,thin,pos=0.2] {$4$} (v1);
	        \draw[line width=1pt,-Triangle] (v4) to node[fill=white,text opacity=1,fill opacity=1,draw=black,rectangle,thin,pos=0.5] {$8$} (v7);
	        \draw[line width=1pt,-Triangle] (v1) to node[fill=white,text opacity=1,fill opacity=1,draw=black,rectangle,thin,pos=0.65] {$4$} (v4);
	        \draw[line width=1pt,-Triangle] (v7) to node[fill=white,text opacity=1,fill opacity=1,draw=black,rectangle,thin,pos=0.65] {$4$} (v5);
	        \draw[line width=1pt] (v4) to node[fill=white,text opacity=1,fill opacity=1,draw=black,rectangle,thin,pos=0.8] {$2$} (v6);
	        \draw[line width=1pt] (v5) to node[fill=white,text opacity=1,fill opacity=1,draw=black,rectangle,thin,pos=0.5] {$2$} (v6);
	        \draw[line width=1pt] (v5) to node[fill=white,text opacity=1,fill opacity=1,draw=black,rectangle,thin,pos=0.8] {$2$} (v3);
	\end{tikzpicture}
	\caption{Quiver diagram for the Spin(7) orientifold of phase H of $Q^{1,1,1}/\ZZ_2$ using the involution in \cref{eq:Q111Z2H-chiral-invol,eq:Q111Z2H-Fermi-invol,eq:Q111Z2H-RFermi-invol}, together with our choice of $\gamma_{\Omega_i}$ matrices.}
		\label{fig:orien_Q111Z2quiverH}
	\end{figure}

\subsection{Phase J}
\label{app:Q111Z2PhaseJ}

The quiver for phase J is shown in Figure \ref{fig:Q111Z2quiverJ}. 

\begin{figure}[H]
		\centering
		\begin{tikzpicture}[scale=2]
	\def\R{1.75};
	        \node[draw=black,line width=1pt,circle,fill=yellowX,minimum width=0.75cm,inner sep=1pt] (v1) at (-\R,0,-\R) {$1$};
	\node[draw=black,line width=1pt,circle,fill=yellowX,minimum width=0.75cm,inner sep=1pt] (v2) at (\R,0,-\R) {$7$};
    \node[draw=black,line width=1pt,circle,fill=yellowX,minimum width=0.75cm,inner sep=1pt] (v3) at (0,0,-\R) {$3$};
    \node[draw=black,line width=1pt,circle,fill=yellowX,minimum width=0.75cm,inner sep=1pt] (v4) at (0,\R,0) {$4$};
    \node[draw=black,line width=1pt,circle,fill=yellowX,minimum width=0.75cm,inner sep=1pt] (v5) at (0,-\R,0) {$5$};
    \node[draw=black,line width=1pt,circle,fill=yellowX,minimum width=0.75cm,inner sep=1pt] (v6) at (0,0,\R) {$6$};
    \node[draw=black,line width=1pt,circle,fill=yellowX,minimum width=0.75cm,inner sep=1pt] (v7) at (\R,0,\R) {$2$};
    \node[draw=black,line width=1pt,circle,fill=yellowX,minimum width=0.75cm,inner sep=1pt] (v8) at (-\R,0,\R) {$8$};
	        \draw[line width=1pt,redX] (v1) to node[fill=white,text opacity=1,fill opacity=1,draw=black,rectangle,thin,pos=0.5] {\color{redX}{$2$}} (v4);
	        \draw[line width=1pt,redX] (v1) to[bend right] node[fill=white,text opacity=1,fill opacity=1,draw=black,rectangle,thin,pos=0.3] {\color{redX}{$4$}} (v5);
	        \draw[line width=1pt,redX] (v6) to node[fill=white,text opacity=1,fill opacity=1,draw=black,rectangle,thin,pos=0.5] {\color{redX}{$8$}} (v3);
	        \draw[line width=1pt,redX] (v6) to node[fill=white,text opacity=1,fill opacity=1,draw=black,rectangle,thin,pos=0.4] {\color{redX}{$2$}} (v7);
	        \draw[line width=1pt,redX] (v8) to node[fill=white,text opacity=1,fill opacity=1,draw=black,rectangle,thin,pos=0.5] {\color{redX}{$2$}} (v6);
	        \draw[line width=1pt,redX] (v4) to node[fill=white,text opacity=1,fill opacity=1,draw=black,rectangle,thin,pos=0.5] {\color{redX}{$2$}} (v2);
	        \draw[line width=1pt,redX] (v5) to[bend right] node[fill=white,text opacity=1,fill opacity=1,draw=black,rectangle,thin,pos=0.5] {\color{redX}{$4$}} (v2);
	        \draw[line width=1pt,-Triangle] (v1) to node[fill=white,text opacity=1,fill opacity=1,draw=black,rectangle,thin,pos=0.5] {$2$} (v8);
	        \draw[line width=1pt,-Triangle] (v1) to node[fill=white,text opacity=1,fill opacity=1,draw=black,rectangle,thin,pos=0.6] {$2$} (v3);
	        \draw[line width=1pt,-Triangle] (v6) to node[fill=white,text opacity=1,fill opacity=1,draw=black,rectangle,thin,pos=0.5] {$4$} (v1);
	        \draw[line width=1pt,-Triangle] (v6) to node[fill=white,text opacity=1,fill opacity=1,draw=black,rectangle,thin,pos=0.5] {$4$} (v2);
	        \draw[line width=1pt,-Triangle] (v8) to node[fill=white,text opacity=1,fill opacity=1,draw=black,rectangle,thin,pos=0.5] {$2$} (v5);
	        \draw[line width=1pt,-Triangle] (v4) to node[fill=white,text opacity=1,fill opacity=1,draw=black,rectangle,thin,pos=0.6] {$2$} (v6);
	        \draw[line width=1pt,-Triangle] (v5) to node[fill=white,text opacity=1,fill opacity=1,draw=black,rectangle,thin,pos=0.5] {$4$} (v6);
	        \draw[line width=1pt,-Triangle] (v2) to node[fill=white,text opacity=1,fill opacity=1,draw=black,rectangle,thin,pos=0.5] {$2$} (v7);
	        \draw[line width=1pt,-Triangle] (v2) to node[fill=white,text opacity=1,fill opacity=1,draw=black,rectangle,thin,pos=0.5] {$2$} (v3);
	        \draw[line width=1pt,-Triangle] (v7) to node[fill=white,text opacity=1,fill opacity=1,draw=black,rectangle,thin,pos=0.5] {$2$} (v5);
	        \draw[line width=1pt,-Triangle] (v3) to node[fill=white,text opacity=1,fill opacity=1,draw=black,rectangle,thin,pos=0.5] {$4$} (v4);
	        \draw[line width=1pt,-Triangle] (v3) to node[fill=white,text opacity=1,fill opacity=1,draw=black,rectangle,thin,pos=0.75] {$2$} (v5);
	\end{tikzpicture}
	\caption{Quiver diagram for phase J of $Q^{1,1,1}/\ZZ_2$.}
	\label{fig:Q111Z2quiverJ}
	\end{figure}

The $J$- and $E$-terms are
		\begin{alignat}{4}
	\renewcommand{\arraystretch}{1.1}
    & \centermathcell{J}                           &\text{\hspace{.5cm}}& \centermathcell{E                               }\nonumber \\
\Lambda^1_{14} \,:\, & \centermathcell{Y_{46} Y_{61}-X_{46} W_{61} }& & \centermathcell{Y_{13} X_{34}-X_{13} Y_{34}}\nonumber \\
\Lambda^2_{14} \,:\, & \centermathcell{Y_{46} X_{61}-X_{46} Z_{61} }& & \centermathcell{X_{13} W_{34}-Y_{13} Z_{34}}\nonumber \\
\Lambda^1_{15} \,:\, & \centermathcell{Y_{56} W_{61}-W_{56} Z_{61} }& & \centermathcell{X_{18} X_{85}-X_{13} X_{35}}\nonumber \\
\Lambda^2_{15} \,:\, & \centermathcell{Z_{56} Z_{61}-X_{56} W_{61} }& & \centermathcell{X_{18} Y_{85}-Y_{13} X_{35}}\nonumber \\
\Lambda^3_{15} \,:\, & \centermathcell{Y_{56} Y_{61}-W_{56} X_{61} }& & \centermathcell{X_{13} Y_{35}-Y_{18} X_{85}}\nonumber \\
\Lambda^4_{15} \,:\, & \centermathcell{X_{56} Y_{61}-Z_{56} X_{61} }& & \centermathcell{Y_{18} Y_{85}-Y_{13} Y_{35}}\nonumber \\
\Lambda^1_{86} \,:\, & \centermathcell{Y_{61} Y_{18}-W_{61} X_{18} }& & \centermathcell{X_{85} Y_{56}-Y_{85} X_{56}}\nonumber \\
\Lambda^2_{86} \,:\, & \centermathcell{X_{61} Y_{18}-Z_{61} X_{18} }& & \centermathcell{Y_{85} Z_{56}-X_{85} W_{56}}\nonumber \\
\Lambda^1_{47} \,:\, & \centermathcell{X_{73} W_{34}-Y_{73} Y_{34} }& & \centermathcell{X_{46} Z_{67}-Y_{46} X_{67}}\nonumber \\
\Lambda^2_{47} \,:\, & \centermathcell{X_{73} Z_{34}-Y_{73} X_{34} }& & \centermathcell{Y_{46} Y_{67}-X_{46} W_{67}}\nonumber \\
\Lambda^1_{63} \,:\, & \centermathcell{Y_{35} W_{56}-W_{34} Y_{46} }& & \centermathcell{X_{61} X_{13}-X_{67} X_{73}}\nonumber \\
\Lambda^2_{63} \,:\, & \centermathcell{Y_{35} Z_{56}-Z_{34} Y_{46} }& & \centermathcell{Y_{67} X_{73}-X_{61} Y_{13}}\label{eq:Q111Z2JEtermsJ} \\
\Lambda^3_{63} \,:\, & \centermathcell{Y_{35} Y_{56}-Y_{34} Y_{46} }& & \centermathcell{X_{67} Y_{73}-Y_{61} X_{13}}\nonumber \\
\Lambda^4_{63} \,:\, & \centermathcell{Y_{35} X_{56}-X_{34} Y_{46} }& & \centermathcell{Y_{61} Y_{13}-Y_{67} Y_{73}}\nonumber \\
\Lambda^5_{63} \,:\, & \centermathcell{X_{35} W_{56}-W_{34} X_{46} }& & \centermathcell{Z_{67} X_{73}-Z_{61} X_{13}}\nonumber \\
\Lambda^6_{63} \,:\, & \centermathcell{X_{35} Z_{56}-Z_{34} X_{46} }& & \centermathcell{Z_{61} Y_{13}-W_{67} X_{73}}\nonumber \\
\Lambda^7_{63} \,:\, & \centermathcell{X_{35} Y_{56}-Y_{34} X_{46} }& & \centermathcell{W_{61} X_{13}-Z_{67} Y_{73}}\nonumber \\
\Lambda^8_{63} \,:\, & \centermathcell{X_{35} X_{56}-X_{34} X_{46} }& & \centermathcell{W_{67} Y_{73}-W_{61} Y_{13}}\nonumber \\
\Lambda^1_{62} \,:\, & \centermathcell{X_{25} W_{56}-Y_{25} Y_{56} }& & \centermathcell{X_{67} Y_{72}-Z_{67} X_{72}}\nonumber \\
\Lambda^2_{62} \,:\, & \centermathcell{X_{25} Z_{56}-Y_{25} X_{56} }& & \centermathcell{W_{67} X_{72}-Y_{67} Y_{72}}\nonumber \\
\Lambda^1_{57} \,:\, & \centermathcell{Y_{72} Y_{25}-Y_{73} Y_{35} }& & \centermathcell{Y_{56} X_{67}-X_{56} Y_{67}}\nonumber \\
\Lambda^2_{57} \,:\, & \centermathcell{Y_{72} X_{25}-X_{73} Y_{35} }& & \centermathcell{Z_{56} Y_{67}-W_{56} X_{67}}\nonumber \\
\Lambda^3_{57} \,:\, & \centermathcell{X_{72} Y_{25}-Y_{73} X_{35} }& & \centermathcell{X_{56} W_{67}-Y_{56} Z_{67}}\nonumber \\
\Lambda^4_{57} \,:\, & \centermathcell{X_{72} X_{25}-X_{73} X_{35} }& & \centermathcell{W_{56} Z_{67}-Z_{56} W_{67}}\nonumber 	
\end{alignat}
Once again, the generators can be found in Table \ref{tab:GenQ111Z2PhaseJ}. 
\medskip

\paragraph{$\U(N)^2\times \SO(N)^4$ orientifold}\mbox{}

\medskip

Let us consider an anti-holomorphic involution of phase J which acts on the nodes in Figure \ref{fig:Q111Z2quiverJ} as $1\leftrightarrow 7$ and $2\leftrightarrow 8$ and maps all other nodes mapped to themselves. Chiral fields transform according to
\begin{equation}
	\begin{array}{cccccccccccc}
		Y_{46} & \rightarrow & \gamma_{\Omega_4} \bar{X}_{46}\gamma_{\Omega_6}^{-1} \coma &
		X_{46} & \rightarrow & \gamma_{\Omega_4} \bar{Y}_{46}\gamma_{\Omega_6}^{-1} \coma &
		Y_{61} & \rightarrow & \gamma_{\Omega_6} \bar{Z}_{67}\gamma_{\Omega_7}^{-1} \coma &
		Z_{67} & \rightarrow & \gamma_{\Omega_6} \bar{Y}_{61}\gamma_{\Omega_1}^{-1} \coma \\
		W_{61} & \rightarrow & \gamma_{\Omega_6} \bar{X}_{67}\gamma_{\Omega_7}^{-1} \coma &
		X_{67} & \rightarrow & \gamma_{\Omega_6} \bar{W}_{61}\gamma_{\Omega_1}^{-1} \coma &
		Y_{13} & \rightarrow & \gamma_{\Omega_7} \bar{X}_{73}\gamma_{\Omega_3}^{-1} \coma &
		X_{73} & \rightarrow & \gamma_{\Omega_1} \bar{Y}_{13}\gamma_{\Omega_3}^{-1} \coma \\
		X_{13} & \rightarrow & \gamma_{\Omega_7} \bar{Y}_{73}\gamma_{\Omega_3}^{-1} \coma &
		Y_{73} & \rightarrow & \gamma_{\Omega_1} \bar{X}_{13}\gamma_{\Omega_3}^{-1} \coma &
		X_{34} & \rightarrow & \gamma_{\Omega_3} \bar{W}_{34}\gamma_{\Omega_4}^{-1} \coma &
		W_{34} & \rightarrow & \gamma_{\Omega_3} \bar{X}_{34}\gamma_{\Omega_4}^{-1} \coma \\
		Y_{34} & \rightarrow & \gamma_{\Omega_3} \bar{Y}_{34}\gamma_{\Omega_4}^{-1} \coma &
		X_{61} & \rightarrow & \gamma_{\Omega_6} \bar{W}_{67}\gamma_{\Omega_7}^{-1} \coma &
		W_{67} & \rightarrow & \gamma_{\Omega_6} \bar{X}_{61}\gamma_{\Omega_1}^{-1} \coma &
		Z_{61} & \rightarrow & \gamma_{\Omega_6} \bar{Y}_{67}\gamma_{\Omega_7}^{-1} \coma \\ 
		Y_{67} & \rightarrow & \gamma_{\Omega_6} \bar{Z}_{61}\gamma_{\Omega_1}^{-1} \coma &
		Z_{34} & \rightarrow & \gamma_{\Omega_3} \bar{Z}_{34}\gamma_{\Omega_4}^{-1} \coma &
		Y_{56} & \rightarrow & \gamma_{\Omega_5} \bar{Y}_{56}\gamma_{\Omega_6}^{-1} \coma &
		W_{56} & \rightarrow & \gamma_{\Omega_5} \bar{X}_{56}\gamma_{\Omega_6}^{-1} \coma \\ 
		X_{56} & \rightarrow & \gamma_{\Omega_5} \bar{W}_{56}\gamma_{\Omega_6}^{-1} \coma &
		X_{35} & \rightarrow & \gamma_{\Omega_3} \bar{Y}_{35}\gamma_{\Omega_5}^{-1} \coma &
		Y_{35} & \rightarrow & \gamma_{\Omega_3} \bar{X}_{35}\gamma_{\Omega_5}^{-1} \coma &
		X_{18} & \rightarrow & \gamma_{\Omega_7} \bar{Y}_{72}\gamma_{\Omega_2}^{-1} \coma \\
		Y_{72} & \rightarrow & \gamma_{\Omega_1} \bar{X}_{18}\gamma_{\Omega_8}^{-1} \coma &
		X_{85} & \rightarrow & \gamma_{\Omega_2} \bar{Y}_{25}\gamma_{\Omega_5}^{-1} \coma &
		Y_{25} & \rightarrow & \gamma_{\Omega_8} \bar{X}_{85}\gamma_{\Omega_5}^{-1} \coma &
		Z_{56} & \rightarrow & \gamma_{\Omega_5} \bar{Z}_{56}\gamma_{\Omega_6}^{-1} \coma \\ 
		Y_{18} & \rightarrow & \gamma_{\Omega_7} \bar{X}_{72}\gamma_{\Omega_2}^{-1} \coma &
		X_{72} & \rightarrow & \gamma_{\Omega_1} \bar{Y}_{18}\gamma_{\Omega_8}^{-1} \coma &
		Y_{85} & \rightarrow & \gamma_{\Omega_2} \bar{X}_{25}\gamma_{\Omega_5}^{-1} \coma &
		X_{25} & \rightarrow & \gamma_{\Omega_8} \bar{Y}_{85}\gamma_{\Omega_5}^{-1} \coma \\
		& & & Y_{46}  & \rightarrow & \gamma_{\Omega_4}  \bar{X}_{46}\gamma_{\Omega_4}^{-1} \coma &
		X_{46} & \rightarrow & \gamma_{\Omega_4} \bar{Y}_{46}\gamma_{\Omega_6}^{-1} \fstop
	\end{array}  
	\label{eq:Q111Z2J-chiral-invol}
\end{equation}

Requiring the invariance of $W^{(0,1)}$, the Fermi fields transform as
\begin{equation}
	\begin{array}{cccccccccccc}
		\Lambda^1_{14} & \rightarrow & \gamma_{\Omega_4} \Lambda^1_{47} \gamma_{\Omega_7}^{-1} \coma &
		\Lambda^2_{14} & \rightarrow & -\gamma_{\Omega_4} \Lambda^2_{47} \gamma_{\Omega_7}^{-1} \coma &
		\Lambda^1_{15} & \rightarrow & \gamma_{\Omega_5} \Lambda^1_{57} \gamma_{\Omega_7}^{-1} \coma &
		\Lambda^2_{15} & \rightarrow & \gamma_{\Omega_5} \Lambda^2_{57} \gamma_{\Omega_7}^{-1} \coma \\
		\Lambda^3_{15} & \rightarrow & -\gamma_{\Omega_5} \Lambda^3_{57} \gamma_{\Omega_7}^{-1} \coma &
		\Lambda^4_{15} & \rightarrow & \gamma_{\Omega_5} \Lambda^4_{57} \gamma_{\Omega_7}^{-1} \coma &
		\Lambda^1_{86} & \rightarrow & -\gamma_{\Omega_6} \Lambda^1_{62} \gamma_{\Omega_2}^{-1} \coma &
		\Lambda^2_{86} & \rightarrow & \gamma_{\Omega_6} \Lambda^2_{62} \gamma_{\Omega_2}^{-1} \coma \\
		\Lambda^1_{47} & \rightarrow & \gamma_{\Omega_1} \Lambda^1_{14} \gamma_{\Omega_4}^{-1} \coma &
		\Lambda^2_{47} & \rightarrow & -\gamma_{\Omega_4} \Lambda^2_{47} \gamma_{\Omega_7}^{-1} \coma &
		\Lambda^1_{63} & \rightarrow & \gamma_{\Omega_6} \bar{\Lambda}^8_{63} \gamma_{\Omega_3}^{-1} \coma &
		\Lambda^2_{63} & \rightarrow & \gamma_{\Omega_6} \bar{\Lambda}^6_{63} \gamma_{\Omega_3}^{-1} \coma \\
		\Lambda^3_{63} & \rightarrow & \gamma_{\Omega_6} \bar{\Lambda}^7_{63} \gamma_{\Omega_3}^{-1} \coma &
		\Lambda^4_{63} & \rightarrow & \gamma_{\Omega_6} \bar{\Lambda}^5_{63} \gamma_{\Omega_3}^{-1} \coma &
		\Lambda^5_{63} & \rightarrow & \gamma_{\Omega_6} \bar{\Lambda}^4_{63} \gamma_{\Omega_3}^{-1} \coma &
		\Lambda^6_{63} & \rightarrow & \gamma_{\Omega_6} \bar{\Lambda}^2_{63} \gamma_{\Omega_3}^{-1} \coma \\
		\Lambda^7_{63} & \rightarrow & \gamma_{\Omega_6} \bar{\Lambda}^3_{63} \gamma_{\Omega_3}^{-1} \coma &
		\Lambda^8_{63} & \rightarrow & \gamma_{\Omega_6} \bar{\Lambda}^1_{63} \gamma_{\Omega_3}^{-1} \coma &
		\Lambda^1_{62} & \rightarrow & -\gamma_{\Omega_8} \Lambda^1_{86} \gamma_{\Omega_6}^{-1} \coma &
		\Lambda^2_{62} & \rightarrow & \gamma_{\Omega_8} \Lambda^2_{86} \gamma_{\Omega_6}^{-1} \coma \\
		\Lambda^1_{57} & \rightarrow & \gamma_{\Omega_1} \Lambda^1_{15} \gamma_{\Omega_5}^{-1} \coma &
		\Lambda^2_{57} & \rightarrow & \gamma_{\Omega_1} \Lambda^2_{15} \gamma_{\Omega_5}^{-1} \coma &
		\Lambda^3_{57} & \rightarrow & -\gamma_{\Omega_1} \Lambda^3_{15} \gamma_{\Omega_5}^{-1} \coma &
		\Lambda^4_{57} & \rightarrow & \gamma_{\Omega_1} \Lambda^4_{15} \gamma_{\Omega_5}^{-1} \fstop
	\end{array}   
	\label{eq:Q111Z2J-Fermi-invol}
\end{equation}
and
\begin{equation}
	\begin{array}{cccccccccccc}
		\Lambda_{11}^R & \rightarrow & \gamma_{\Omega_7}\Lambda_{77}^{R\,\,T}\gamma_{\Omega_7}^{-1} \coma &
		\Lambda_{22}^R & \rightarrow & \gamma_{\Omega_8}\Lambda_{88}^{R\,\,T}\gamma_{\Omega_8}^{-1} \coma &
		\Lambda_{33}^R & \rightarrow & \gamma_{\Omega_3}\Lambda_{33}^{R\,\,T}\gamma_{\Omega_3}^{-1} \coma &
		\Lambda_{44}^R & \rightarrow & \gamma_{\Omega_4}\Lambda_{44}^{R\,\,T}\gamma_{\Omega_4}^{-1} \coma \\
		\Lambda_{55}^R & \rightarrow & \gamma_{\Omega_5}\Lambda_{55}^{R\,\,T}\gamma_{\Omega_5}^{-1} \coma &
		\Lambda_{66}^R & \rightarrow & \gamma_{\Omega_6}\Lambda_{66}^{R\,\,T}\gamma_{\Omega_6}^{-1} \coma &
		\Lambda_{77}^R & \rightarrow & \gamma_{\Omega_1}\Lambda_{11}^{R\,\,T}\gamma_{\Omega_1}^{-1} \coma &	
		\Lambda_{88}^R & \rightarrow & \gamma_{\Omega_2}\Lambda_{22}^{R\,\,T}\gamma_{\Omega_2}^{-1} \fstop
	\end{array}  
	\label{eq:Q111Z2J-RFermi-invol}
\end{equation}

Using Table \ref{tab:GenQ111Z2PhaseJ}, the corresponding geometric involution acting on the generators reads
\begin{equation}
     \begin{array}{ccccccccccccccc}
	M_{1}  & \rightarrow & \bar{M}_{27} \coma &
	M_{2}  & \rightarrow & \bar{M}_{24} \coma &
	M_{3}  & \rightarrow & \bar{M}_{21} \coma &
	M_{4}  & \rightarrow & \bar{M}_{18} \coma &
	M_{5}  & \rightarrow & \bar{M}_{12} \coma \\
	M_{6}  & \rightarrow & \bar{M}_{6} \coma &
	M_{7}  & \rightarrow & \bar{M}_{26} \coma &
	M_{8}  & \rightarrow & \bar{M}_{23} \coma &
	M_{9}  & \rightarrow & \bar{M}_{20} \coma &
	M_{10} & \rightarrow & \bar{M}_{17} \coma \\
	M_{11} & \rightarrow & \bar{M}_{11} \coma &
	M_{12} & \rightarrow & \bar{M}_{5} \coma &
	M_{13} & \rightarrow & \bar{M}_{25} \coma &
	M_{14} & \rightarrow & \bar{M}_{22} \coma &
	M_{15} & \rightarrow & \bar{M}_{19} \coma \\
	M_{16} & \rightarrow & \bar{M}_{16} \coma &
	M_{17} & \rightarrow & \bar{M}_{10} \coma &
	M_{18} & \rightarrow & \bar{M}_{4} \coma &
	M_{19} & \rightarrow & \bar{M}_{15} \coma &
	M_{20} & \rightarrow & \bar{M}_{9} \coma \\ 
	M_{21} & \rightarrow & \bar{M}_{3}\coma &
	M_{22} & \rightarrow & \bar{M}_{14}\coma &
	M_{23} & \rightarrow & \bar{M}_{8}\coma &
	M_{24} & \rightarrow & \bar{M}_{2}\coma &
	M_{25} & \rightarrow & \bar{M}_{13}\coma \\
	\multicolumn{15}{c}{M_{26}   \rightarrow  \bar{M}_{7}\coma 
	M_{27}  \rightarrow  \bar{M}_{1}\fstop}
	\end{array}   
	\label{eq:Q111Z2J-Meson-invol}
    \end{equation}
Notice, again, that this is the same geometric action that we have found for phase D in \eqref{eq:Q111Z2D-Meson-invol}. 

The $\gamma_{\Omega_i}$ matrices are constrained as in \eqref{gamma_matrices_Q111/Z2}. As for phase D, we choose
\begin{equation}
\gamma_{\Omega_3}=\gamma_{\Omega_4}=\gamma_{\Omega_5}=\gamma_{\Omega_6}=\ID_N \fstop 
\end{equation}

The resulting orientifold of phase J is shown in Figure \ref{fig:orien_Q111Z2quiverJ}.

 \begin{figure}[H]
		\centering
		\begin{tikzpicture}[scale=2]
	\def\R{1.5};
	    \draw[line width=1pt,redX] (120:\R*1.15) circle (0.25);
	\draw[line width=1pt,redX] (60:\R*1.15) circle (0.25);
	    \draw[line width=1pt,redX] (0:\R*1.15) circle (0.25) node[xshift=0.5cm,star,star points=5, star point ratio=2.25, inner sep=1pt, fill=redX, draw] {};
	    \draw[line width=1pt,redX] (300:\R*1.15) circle (0.25) node[xshift=0.34cm,yshift=-0.34cm,star,star points=5, star point ratio=2.25, inner sep=1pt, fill=redX, draw] {};
	    \draw[line width=1pt,redX] (240:\R*1.15) circle (0.25) node[xshift=-0.34cm,yshift=-0.34cm,star,star points=5, star point ratio=2.25, inner sep=1pt, fill=redX, draw] {};
	    \draw[line width=1pt,redX] (180:\R*1.15) circle (0.25) node[xshift=-0.5cm,star,star points=5, star point ratio=2.25, inner sep=1pt, fill=redX, draw] {};
	       \node[draw=black,line width=1pt,circle,fill=yellowX,minimum width=0.75cm,inner sep=1pt,label={[xshift=0.7cm,yshift=-0cm]:$\U(N)$}] (v1) at  (120:\R) {$1$};
	        \node[draw=black,line width=1pt,circle,fill=yellowX,minimum width=0.75cm,inner sep=1pt,label={[xshift=0.75cm,yshift=0cm]:$\SO(N)$}] (v3) at (0:\R) {$3$};
	        \node[draw=black,line width=1pt,circle,fill=yellowX,minimum width=0.75cm,inner sep=1pt,label={[xshift=1cm,yshift=-0.3cm]:$\SO(N)$}] (v4) at (300:\R) {$4$};
	        \node[draw=black,line width=1pt,circle,fill=yellowX,minimum width=0.75cm,inner sep=1pt,label={[xshift=-1cm,yshift=-0.3cm]:$\SO(N)$}] (v5) at (240:\R) {$5$};
	        \node[draw=black,line width=1pt,circle,fill=yellowX,minimum width=0.75cm,inner sep=1pt,label={[xshift=-0.75cm,yshift=0cm]:$\SO(N)$}] (v6) at (180:\R) {$6$};
	        \node[draw=black,line width=1pt,circle,fill=yellowX,minimum width=0.75cm,inner sep=1pt,label={[xshift=-0.7cm,yshift=-0cm]:$\U(N)$}] (v2) at  (60:\R) {$2$};
	        \draw[line width=1pt,redX] (v1) to node[fill=white,text opacity=1,fill opacity=1,draw=black,rectangle,thin,pos=0.6] {\color{redX}{$4$}} (v4);
	        \draw[line width=1pt,redX] (v1) to node[fill=white,text opacity=1,fill opacity=1,draw=black,rectangle,thin,pos=0.2] {\color{redX}{$8$}} (v5);
	        \draw[line width=1pt,redX] (v2) to node[fill=white,text opacity=1,fill opacity=1,draw=black,rectangle,thin,pos=0.8] {\color{redX}{$4$}} (v6);
	        \draw[line width=1pt,redX] (v3) to node[fill=white,text opacity=1,fill opacity=1,draw=black,rectangle,thin,pos=0.25] {\color{redX}{$8$}} (v6);
	        \draw[line width=1pt,-Triangle] (v1) to node[fill=white,text opacity=1,fill opacity=1,draw=black,rectangle,thin,pos=0.5] {$4$} (v2);
	        \draw[line width=1pt,-Triangle] (v1) to node[fill=white,text opacity=1,fill opacity=1,draw=black,rectangle,thin,pos=0.75] {$4$} (v3);
	        \draw[line width=1pt,-Triangle] (v2) to node[fill=white,text opacity=1,fill opacity=1,draw=black,rectangle,thin,pos=0.4] {$4$} (v5);
	        \draw[line width=1pt,-Triangle] (v6) to node[fill=white,text opacity=1,fill opacity=1,draw=black,rectangle,thin,pos=0.5] {$4$} (v1);
	        \draw[line width=1pt] (v3) to node[fill=white,text opacity=1,fill opacity=1,draw=black,rectangle,thin,pos=0.5] {$4$} (v4);
	        \draw[line width=1pt] (v3) to node[fill=white,text opacity=1,fill opacity=1,draw=black,rectangle,thin,pos=0.2] {$2$} (v5);
	        \draw[line width=1pt] (v4) to node[fill=white,text opacity=1,fill opacity=1,draw=black,rectangle,thin,pos=0.8] {$2$} (v6);
	        \draw[line width=1pt] (v5) to node[fill=white,text opacity=1,fill opacity=1,draw=black,rectangle,thin,pos=0.5] {$4$} (v6);
	\end{tikzpicture}
	\caption{Quiver diagram for the Spin(7) orientifold of phase J of $Q^{1,1,1}/\ZZ_2$ using the involution in \cref{eq:Q111Z2J-chiral-invol,eq:Q111Z2J-Fermi-invol,eq:Q111Z2J-RFermi-invol}, together with our choice of $\gamma_{\Omega_i}$ matrices.}
		\label{fig:orien_Q111Z2quiverJ}
	\end{figure}

\subsection{Phase L}
\label{sec:Q111Z2PhaseL}

The last $\mathcal{N}=(0,2)$ quiver we consider is phase L, shown in Figure \ref{fig:Q111Z2quiverL}. 

\begin{figure}[H]
		\centering
		\begin{tikzpicture}[scale=2]
	\def\R{1.75};
	 \node[draw=black,line width=1pt,circle,fill=yellowX,minimum width=0.75cm,inner sep=1pt] (v1) at (-\R,0,-\R) {$1$};
	        \node[draw=black,line width=1pt,circle,fill=yellowX,minimum width=0.75cm,inner sep=1pt] (v2) at (\R,0,-\R) {$7$};
	        \node[draw=black,line width=1pt,circle,fill=yellowX,minimum width=0.75cm,inner sep=1pt] (v3) at (0,0,-\R) {$3$};
	        \node[draw=black,line width=1pt,circle,fill=yellowX,minimum width=0.75cm,inner sep=1pt] (v4) at (0,\R,0) {$4$};
	        \node[draw=black,line width=1pt,circle,fill=yellowX,minimum width=0.75cm,inner sep=1pt] (v5) at (0,-\R,0) {$5$};
	        \node[draw=black,line width=1pt,circle,fill=yellowX,minimum width=0.75cm,inner sep=1pt] (v6) at (0,0,\R) {$6$};
	        \node[draw=black,line width=1pt,circle,fill=yellowX,minimum width=0.75cm,inner sep=1pt] (v7) at (\R,0,\R) {$2$};
	        \node[draw=black,line width=1pt,circle,fill=yellowX,minimum width=0.75cm,inner sep=1pt] (v8) at (-\R,0,\R) {$8$};
	        \draw[line width=1pt,redX] (v1) to[bend right] node[fill=white,text opacity=1,fill opacity=1,draw=black,rectangle,thin,pos=0.25] {\color{redX}{$4$}} (v5);
	        \draw[line width=1pt,redX] (v2) to[bend left] node[fill=white,text opacity=1,fill opacity=1,draw=black,rectangle,thin,pos=0.5] {\color{redX}{$4$}} (v5);
	        \draw[line width=1pt,redX] (v3) to node[fill=white,text opacity=1,fill opacity=1,draw=black,rectangle,thin,pos=0.5] {\color{redX}{$4$}} (v4);
	        \draw[line width=1pt,redX] (v4) to[bend right] node[fill=white,text opacity=1,fill opacity=1,draw=black,rectangle,thin,pos=0.5] {\color{redX}{$4$}} (v8);
	        \draw[line width=1pt,redX] (v4) to[bend left] node[fill=white,text opacity=1,fill opacity=1,draw=black,rectangle,thin,pos=0.75] {\color{redX}{$4$}} (v7);
	        \draw[line width=1pt,redX] (v5) to node[fill=white,text opacity=1,fill opacity=1,draw=black,rectangle,thin,pos=0.5] {\color{redX}{$4$}} (v6);
	        \draw[line width=1pt,-Triangle] (v1) to node[fill=white,text opacity=1,fill opacity=1,draw=black,rectangle,thin,pos=0.15] {$2$} (v3);
	        \draw[line width=1pt,-Triangle] (v1) to node[fill=white,text opacity=1,fill opacity=1,draw=black,rectangle,thin,pos=0.5] {$2$} (v8);
	        \draw[line width=1pt,-Triangle] (v2) to node[fill=white,text opacity=1,fill opacity=1,draw=black,rectangle,thin,pos=0.5] {$2$} (v3);
	        \draw[line width=1pt,-Triangle] (v2) to node[fill=white,text opacity=1,fill opacity=1,draw=black,rectangle,thin,pos=0.5] {$2$} (v7);
	        \draw[line width=1pt,-Triangle] (v3) to[bend left=10] node[fill=white,text opacity=1,fill opacity=1,draw=black,rectangle,thin,pos=0.25] {$2$} (v5);
	        \draw[line width=1pt,-Triangle] (v4) to node[fill=white,text opacity=1,fill opacity=1,draw=black,rectangle,thin,pos=0.5] {$2$} (v1);
	        \draw[line width=1pt,-Triangle] (v4) to node[fill=white,text opacity=1,fill opacity=1,draw=black,rectangle,thin,pos=0.5] {$2$} (v2);
	        \draw[line width=1pt,-Triangle] (v4) to[bend right=10] node[fill=white,text opacity=1,fill opacity=1,draw=black,rectangle,thin,pos=0.7] {$2$} (v6);
	        \draw[line width=1pt,-Triangle] (v5) to node[fill=white,text opacity=1,fill opacity=1,draw=black,rectangle,thin,pos=0.5] {$8$} (v4);
	        \draw[line width=1pt,-Triangle] (v6) to node[fill=white,text opacity=1,fill opacity=1,draw=black,rectangle,thin,pos=0.5] {$2$} (v7);
	        \draw[line width=1pt,-Triangle] (v6) to node[fill=white,text opacity=1,fill opacity=1,draw=black,rectangle,thin,pos=0.5] {$2$} (v8);
	        \draw[line width=1pt,-Triangle] (v7) to node[fill=white,text opacity=1,fill opacity=1,draw=black,rectangle,thin,pos=0.5] {$2$} (v5);
	        \draw[line width=1pt,-Triangle] (v8) to node[fill=white,text opacity=1,fill opacity=1,draw=black,rectangle,thin,pos=0.5] {$2$} (v5);
	\end{tikzpicture}
	\caption{Quiver diagram for phase L of $Q^{1,1,1}/\ZZ_2$.}
	\label{fig:Q111Z2quiverL}
	\end{figure}

The $J$- and $E$-terms are
		\begin{alignat}{4}
	\renewcommand{\arraystretch}{1.1}
    & \centermathcell{J}                           &\text{\hspace{.5cm}}& \centermathcell{E                               }\nonumber \\
\Lambda^1_{15} \,:\, & \centermathcell{B_{54} Y_{41}-D_{54} X_{41} }& & \centermathcell{X_{13} X_{35}-X_{18} X_{85}}\nonumber \\
\Lambda^2_{15} \,:\, & \centermathcell{C_{54} X_{41}-A_{54} Y_{41} }& & \centermathcell{X_{13} Y_{35}-Y_{18} X_{85}}\nonumber \\
\Lambda^3_{15} \,:\, & \centermathcell{W_{54} X_{41}-Y_{54} Y_{41} }& & \centermathcell{Y_{13} X_{35}-X_{18} Y_{85}}\nonumber \\
\Lambda^4_{15} \,:\, & \centermathcell{Z_{54} X_{41}-X_{54} Y_{41} }& & \centermathcell{Y_{18} Y_{85}-Y_{13} Y_{35}}\nonumber \\
\Lambda^1_{34} \,:\, & \centermathcell{Y_{41} Y_{13}-Y_{47} Y_{73} }& & \centermathcell{X_{35} Y_{54}-Y_{35} X_{54}}\nonumber \\
\Lambda^2_{34} \,:\, & \centermathcell{X_{41} Y_{13}-Y_{47} X_{73} }& & \centermathcell{Y_{35} Z_{54}-X_{35} W_{54}}\nonumber \\
\Lambda^3_{34} \,:\, & \centermathcell{Y_{41} X_{13}-X_{47} Y_{73} }& & \centermathcell{Y_{35} A_{54}-X_{35} B_{54}}\nonumber \\
\Lambda^4_{34} \,:\, & \centermathcell{X_{41} X_{13}-X_{47} X_{73} }& & \centermathcell{X_{35} D_{54}-Y_{35} C_{54}}\nonumber \\
\Lambda^1_{48} \,:\, & \centermathcell{X_{85} D_{54}-Y_{85} W_{54} }& & \centermathcell{X_{46} X_{68}-X_{41} X_{18}}\nonumber \\
\Lambda^2_{48} \,:\, & \centermathcell{X_{85} C_{54}-Y_{85} Z_{54} }& & \centermathcell{X_{41} Y_{18}-Y_{46} X_{68}}\nonumber \\
\Lambda^3_{48} \,:\, & \centermathcell{X_{85} B_{54}-Y_{85} Y_{54} }& & \centermathcell{Y_{41} X_{18}-X_{46} Y_{68}}\nonumber \\
\Lambda^4_{48} \,:\, & \centermathcell{X_{85} A_{54}-Y_{85} X_{54} }& & \centermathcell{Y_{46} Y_{68}-Y_{41} Y_{18}}\label{eq:Q111Z2JEtermsL} \\
\Lambda^1_{42} \,:\, & \centermathcell{X_{25} D_{54}-Y_{25} B_{54} }& & \centermathcell{X_{47} X_{72}-X_{46} X_{62}}\nonumber \\
\Lambda^2_{42} \,:\, & \centermathcell{X_{25} C_{54}-Y_{25} A_{54} }& & \centermathcell{Y_{46} X_{62}-X_{47} Y_{72}}\nonumber \\
\Lambda^3_{42} \,:\, & \centermathcell{X_{25} W_{54}-Y_{25} Y_{54} }& & \centermathcell{X_{46} Y_{62}-Y_{47} X_{72}}\nonumber \\
\Lambda^4_{42} \,:\, & \centermathcell{X_{25} Z_{54}-Y_{25} X_{54} }& & \centermathcell{Y_{47} Y_{72}-Y_{46} Y_{62}}\nonumber \\
\Lambda^1_{75} \,:\, & \centermathcell{W_{54} Y_{47}-D_{54} X_{47} }& & \centermathcell{X_{72} X_{25}-X_{73} X_{35}}\nonumber \\
\Lambda^2_{75} \,:\, & \centermathcell{Z_{54} Y_{47}-C_{54} X_{47} }& & \centermathcell{X_{73} Y_{35}-Y_{72} X_{25}}\nonumber \\
\Lambda^3_{75} \,:\, & \centermathcell{Y_{54} Y_{47}-B_{54} X_{47} }& & \centermathcell{Y_{73} X_{35}-X_{72} Y_{25}}\nonumber \\
\Lambda^4_{75} \,:\, & \centermathcell{X_{54} Y_{47}-A_{54} X_{47} }& & \centermathcell{Y_{72} Y_{25}-Y_{73} Y_{35}}\nonumber \\
\Lambda^1_{56} \,:\, & \centermathcell{Y_{68} Y_{85}-Y_{62} Y_{25} }& & \centermathcell{X_{54} Y_{46}-Y_{54} X_{46}}\nonumber \\
\Lambda^2_{56} \,:\, & \centermathcell{X_{68} Y_{85}-Y_{62} X_{25} }& & \centermathcell{W_{54} X_{46}-Z_{54} Y_{46}}\nonumber \\
\Lambda^3_{56} \,:\, & \centermathcell{Y_{68} X_{85}-X_{62} Y_{25} }& & \centermathcell{B_{54} X_{46}-A_{54} Y_{46}}\nonumber \\
\Lambda^4_{56} \,:\, & \centermathcell{X_{68} X_{85}-X_{62} X_{25} }& & \centermathcell{C_{54} Y_{46}-D_{54} X_{46}}\nonumber 	
\end{alignat}

\paragraph{$\U(N)^2\times \SO(N)^4$ orientifold}\mbox{}

We can, again, look for an involution that maps the nodes in Figure \ref{fig:Q111Z2quiverL} $1\leftrightarrow 7$ and $2\leftrightarrow 8$ and all the other nodes to themselves. The map on the fields is
\begin{equation}
	\begin{array}{cccccccccccc}
		B_{54} & \rightarrow & \gamma_{\Omega_5}\bar{Z}_{54}\gamma_{\Omega_4}^{-1} \coma &
		Z_{54} & \rightarrow & \gamma_{\Omega_5}\bar{B}_{54}\gamma_{\Omega_4}^{-1} \coma &
		D_{54} & \rightarrow & \gamma_{\Omega_5}\bar{C}_{54}\gamma_{\Omega_4}^{-1} \coma & 
		C_{54} & \rightarrow & \gamma_{\Omega_5}\bar{D}_{54}\gamma_{\Omega_4}^{-1} \coma \\
		Y_{41} & \rightarrow & \gamma_{\Omega_4}\bar{Y}_{47}\gamma_{\Omega_7}^{-1} \coma &
		Y_{47} & \rightarrow & \gamma_{\Omega_4}\bar{Y}_{41}\gamma_{\Omega_1}^{-1} \coma &
		X_{41} & \rightarrow & \gamma_{\Omega_4}\bar{X}_{47}\gamma_{\Omega_7}^{-1} \coma & 
		X_{47} & \rightarrow & \gamma_{\Omega_4}\bar{X}_{41}\gamma_{\Omega_1}^{-1} \coma \\
		X_{13} & \rightarrow & \gamma_{\Omega_7}\bar{X}_{73}\gamma_{\Omega_3}^{-1} \coma & 
		X_{73} & \rightarrow & \gamma_{\Omega_1}\bar{X}_{13}\gamma_{\Omega_3}^{-1} \coma & 
		X_{35} & \rightarrow & \gamma_{\Omega_3}\bar{Y}_{35}\gamma_{\Omega_5}^{-1} \coma & 
		Y_{35} & \rightarrow & \gamma_{\Omega_3}\bar{X}_{35}\gamma_{\Omega_5}^{-1} \coma \\ 
		X_{18} & \rightarrow & \gamma_{\Omega_7}\bar{Y}_{72}\gamma_{\Omega_2}^{-1} \coma & 
		Y_{72} & \rightarrow & \gamma_{\Omega_1}\bar{X}_{18}\gamma_{\Omega_8}^{-1} \coma & 
		X_{85} & \rightarrow & \gamma_{\Omega_2}\bar{X}_{25}\gamma_{\Omega_5}^{-1} \coma & 
		X_{25} & \rightarrow & \gamma_{\Omega_8}\bar{X}_{85}\gamma_{\Omega_5}^{-1} \coma \\ 
		A_{54} & \rightarrow & \gamma_{\Omega_5}\bar{W}_{54}\gamma_{\Omega_4}^{-1} \coma & 
		W_{54} & \rightarrow & \gamma_{\Omega_5}\bar{A}_{54}\gamma_{\Omega_4}^{-1} \coma & 
		Y_{18} & \rightarrow & \gamma_{\Omega_7}\bar{X}_{72}\gamma_{\Omega_2}^{-1} \coma & 
		X_{72} & \rightarrow & \gamma_{\Omega_1}\bar{Y}_{18}\gamma_{\Omega_8}^{-1} \coma \\ 
		Y_{13} & \rightarrow & \gamma_{\Omega_7}\bar{Y}_{73}\gamma_{\Omega_3}^{-1} \coma & 
		Y_{73} & \rightarrow & \gamma_{\Omega_1}\bar{Y}_{13}\gamma_{\Omega_3}^{-1} \coma & 
		Y_{85} & \rightarrow & \gamma_{\Omega_2}\bar{Y}_{25}\gamma_{\Omega_5}^{-1} \coma & 
		Y_{25} & \rightarrow & \gamma_{\Omega_8}\bar{Y}_{85}\gamma_{\Omega_5}^{-1} \coma \\ 
		Y_{54} & \rightarrow & \gamma_{\Omega_5}\bar{X}_{54}\gamma_{\Omega_4}^{-1} \coma & 
		X_{54} & \rightarrow & \gamma_{\Omega_5}\bar{Y}_{54}\gamma_{\Omega_4}^{-1} \coma & 
		X_{46} & \rightarrow & \gamma_{\Omega_4}\bar{Y}_{46}\gamma_{\Omega_6}^{-1} \coma & 
		Y_{46} & \rightarrow & \gamma_{\Omega_4}\bar{X}_{46}\gamma_{\Omega_6}^{-1} \coma \\ 
		Y_{68} & \rightarrow & \gamma_{\Omega_6}\bar{Y}_{62}\gamma_{\Omega_2}^{-1} \coma & 
		Y_{62} & \rightarrow & \gamma_{\Omega_6}\bar{Y}_{68}\gamma_{\Omega_8}^{-1} \coma & 
		X_{68} & \rightarrow & \gamma_{\Omega_6}\bar{X}_{62}\gamma_{\Omega_2}^{-1} \coma &
		X_{62} & \rightarrow & \gamma_{\Omega_6}\bar{X}_{68}\gamma_{\Omega_8}^{-1} \fstop
	\end{array}  
	\label{eq:Q111Z2L-chiral-invol}
\end{equation}

Requiring the invariance of the $W^{(0,1)}$ superpotential, we obtain also the following transformations for the Fermi fields:
\begin{equation}
	\begin{array}{cccccccccccc}
		\Lambda^1_{15} & \rightarrow & \gamma_{\Omega_7}\bar{\Lambda}^2_{75}\gamma_{\Omega_5}^{-1} \coma &
		\Lambda^2_{15} & \rightarrow & -\gamma_{\Omega_7}\bar{\Lambda}^1_{75}\gamma_{\Omega_5}^{-1} \coma &
		\Lambda^3_{15} & \rightarrow & -\gamma_{\Omega_7}\bar{\Lambda}^4_{75}\gamma_{\Omega_5}^{-1} \coma &
		\Lambda^4_{15} & \rightarrow & -\gamma_{\Omega_7}\bar{\Lambda}^3_{75}\gamma_{\Omega_5}^{-1} \coma \\
		\Lambda^1_{34} & \rightarrow & -\gamma_{\Omega_3}\bar{\Lambda}^1_{34}\gamma_{\Omega_4}^{-1} \coma &
		\Lambda^2_{34} & \rightarrow & -\gamma_{\Omega_3}\bar{\Lambda}^3_{34}\gamma_{\Omega_4}^{-1} \coma &
		\Lambda^3_{34} & \rightarrow & -\gamma_{\Omega_3}\bar{\Lambda}^2_{34}\gamma_{\Omega_4}^{-1} \coma &
		\Lambda^4_{34} & \rightarrow & -\gamma_{\Omega_3}\bar{\Lambda}^4_{34}\gamma_{\Omega_4}^{-1} \coma \\ 
		\Lambda^1_{48} & \rightarrow & \gamma_{\Omega_4}\bar{\Lambda}^2_{42}\gamma_{\Omega_2}^{-1} \coma &
		\Lambda^2_{48} & \rightarrow & \gamma_{\Omega_4}\bar{\Lambda}^1_{42}\gamma_{\Omega_2}^{-1} \coma &
		\Lambda^3_{48} & \rightarrow & \gamma_{\Omega_4}\bar{\Lambda}^4_{42}\gamma_{\Omega_2}^{-1} \coma &
		\Lambda^4_{48} & \rightarrow & \gamma_{\Omega_4}\bar{\Lambda}^3_{42}\gamma_{\Omega_2}^{-1} \coma \\
		\Lambda^1_{42} & \rightarrow & \gamma_{\Omega_4}\bar{\Lambda}^2_{48}\gamma_{\Omega_8}^{-1} \coma &
		\Lambda^2_{42} & \rightarrow & \gamma_{\Omega_4}\bar{\Lambda}^1_{48}\gamma_{\Omega_8}^{-1} \coma &
		\Lambda^3_{42} & \rightarrow & \gamma_{\Omega_4}\bar{\Lambda}^4_{48}\gamma_{\Omega_8}^{-1} \coma &
		\Lambda^4_{42} & \rightarrow & \gamma_{\Omega_4}\bar{\Lambda}^3_{48}\gamma_{\Omega_8}^{-1} \coma \\
		\Lambda^1_{75} & \rightarrow & -\gamma_{\Omega_1}\bar{\Lambda}^2_{15}\gamma_{\Omega_5}^{-1} \coma &
		\Lambda^2_{75} & \rightarrow & \gamma_{\Omega_1}\bar{\Lambda}^1_{15}\gamma_{\Omega_5}^{-1} \coma &
		\Lambda^3_{75} & \rightarrow & -\gamma_{\Omega_1}\bar{\Lambda}^4_{15}\gamma_{\Omega_5}^{-1} \coma &
		\Lambda^4_{75} & \rightarrow & -\gamma_{\Omega_1}\bar{\Lambda}^3_{15}\gamma_{\Omega_5}^{-1} \coma \\
		\Lambda^1_{56} & \rightarrow & -\gamma_{\Omega_5}\bar{\Lambda}^1_{56}\gamma_{\Omega_6}^{-1} \coma &
		\Lambda^2_{56} & \rightarrow & -\gamma_{\Omega_5}\bar{\Lambda}^3_{56}\gamma_{\Omega_6}^{-1} \coma &
		\Lambda^3_{56} & \rightarrow & -\gamma_{\Omega_5}\bar{\Lambda}^2_{56}\gamma_{\Omega_6}^{-1} \coma &
		\Lambda^4_{56} & \rightarrow & -\gamma_{\Omega_5}\bar{\Lambda}^4_{56}\gamma_{\Omega_6}^{-1} \coma
	\end{array}   
	\label{eq:Q111Z2L-Fermi-invol}
\end{equation}
and
\begin{equation}
	\begin{array}{cccccccccccc}
		\Lambda_{11}^R & \rightarrow & \gamma_{\Omega_7}\Lambda_{77}^{R\,\,T}\gamma_{\Omega_7}^{-1} \coma &
		\Lambda_{22}^R & \rightarrow & \gamma_{\Omega_8}\Lambda_{88}^{R\,\,T}\gamma_{\Omega_8}^{-1} \coma &
		\Lambda_{33}^R & \rightarrow & \gamma_{\Omega_3}\Lambda_{33}^{R\,\,T}\gamma_{\Omega_3}^{-1} \coma &
		\Lambda_{44}^R & \rightarrow & \gamma_{\Omega_4}\Lambda_{44}^{R\,\,T}\gamma_{\Omega_4}^{-1} \coma \\
		\Lambda_{55}^R & \rightarrow & \gamma_{\Omega_5}\Lambda_{55}^{R\,\,T}\gamma_{\Omega_5}^{-1} \coma &
		\Lambda_{66}^R & \rightarrow & \gamma_{\Omega_1}\Lambda_{11}^{R\,\,T}\gamma_{\Omega_1}^{-1} \coma &
		\Lambda_{77}^R & \rightarrow & \gamma_{\Omega_1}\Lambda_{11}^{R\,\,T}\gamma_{\Omega_1}^{-1} \coma &
		\Lambda_{88}^R & \rightarrow & \gamma_{\Omega_2}\Lambda_{22}^{R\,\,T}\gamma_{\Omega_2}^{-1} \fstop
	\end{array}   
	\label{eq:Q111Z2L-RFermi-invol}
\end{equation} 

Using Table \ref{tab:GenQ111Z2PhaseL}, the corresponding geometric involution acting on the generators reads
\begin{equation}
     \begin{array}{ccccccccccccccc}
	M_{1}  & \rightarrow & \bar{M}_{27} \coma &
	M_{2}  & \rightarrow & \bar{M}_{24} \coma &
	M_{3}  & \rightarrow & \bar{M}_{21} \coma &
	M_{4}  & \rightarrow & \bar{M}_{18} \coma &
	M_{5}  & \rightarrow & \bar{M}_{12} \coma \\
	M_{6}  & \rightarrow & \bar{M}_{6} \coma &
	M_{7}  & \rightarrow & \bar{M}_{26} \coma &
	M_{8}  & \rightarrow & \bar{M}_{23} \coma &
	M_{9}  & \rightarrow & \bar{M}_{20} \coma &
	M_{10} & \rightarrow & \bar{M}_{17} \coma \\
	M_{11} & \rightarrow & \bar{M}_{11} \coma &
	M_{12} & \rightarrow & \bar{M}_{5} \coma &
	M_{13} & \rightarrow & \bar{M}_{25} \coma &
	M_{14} & \rightarrow & \bar{M}_{22} \coma &
	M_{15} & \rightarrow & \bar{M}_{19} \coma \\
	M_{16} & \rightarrow & \bar{M}_{16} \coma &
	M_{17} & \rightarrow & \bar{M}_{10} \coma &
	M_{18} & \rightarrow & \bar{M}_{4} \coma &
	M_{19} & \rightarrow & \bar{M}_{15} \coma &
	M_{20} & \rightarrow & \bar{M}_{9} \coma \\ 
	M_{21} & \rightarrow & \bar{M}_{3}\coma &
	M_{22} & \rightarrow & \bar{M}_{14}\coma &
	M_{23} & \rightarrow & \bar{M}_{8}\coma &
	M_{24} & \rightarrow & \bar{M}_{2}\coma &
	M_{25} & \rightarrow & \bar{M}_{13}\coma \\
	\multicolumn{15}{c}{M_{26}   \rightarrow  \bar{M}_{7}\coma 
	M_{27}  \rightarrow  \bar{M}_{1}\fstop}
	\end{array}   
	\label{eq:Q111Z2L-Meson-invol}
    \end{equation}
This is, once again, the same geometric involution. 

The $\gamma_{\Omega_i}$ matrices are constrained as in \eqref{gamma_matrices_Q111/Z2}. As for phase D, we choose
\begin{equation}
\gamma_{\Omega_3}=\gamma_{\Omega_4}=\gamma_{\Omega_5}=\gamma_{\Omega_6}=\ID_N \fstop 
\end{equation}

The resulting orientifold is shown in Figure \ref{fig:orien_Q111Z2quiverL}. 

\begin{figure}[H]
		\centering
		\begin{tikzpicture}[scale=2]
	\def\R{1.5};
	   \draw[line width=1pt,redX] (120:\R*1.15) circle (0.25);
	\draw[line width=1pt,redX] (60:\R*1.15) circle (0.25);
	    \draw[line width=1pt,redX] (0:\R*1.15) circle (0.25) node[xshift=0.5cm,star,star points=5, star point ratio=2.25, inner sep=1pt, fill=redX, draw] {};
	    \draw[line width=1pt,redX] (300:\R*1.15) circle (0.25) node[xshift=0.34cm,yshift=-0.34cm,star,star points=5, star point ratio=2.25, inner sep=1pt, fill=redX, draw] {};
	    \draw[line width=1pt,redX] (240:\R*1.15) circle (0.25) node[xshift=-0.34cm,yshift=-0.34cm,star,star points=5, star point ratio=2.25, inner sep=1pt, fill=redX, draw] {};
	    \draw[line width=1pt,redX] (180:\R*1.15) circle (0.25) node[xshift=-0.5cm,star,star points=5, star point ratio=2.25, inner sep=1pt, fill=redX, draw] {};
	       \node[draw=black,line width=1pt,circle,fill=yellowX,minimum width=0.75cm,inner sep=1pt,label={[xshift=0.7cm,yshift=-0cm]:$\U(N)$}] (v1) at  (120:\R) {$1$};
	        \node[draw=black,line width=1pt,circle,fill=yellowX,minimum width=0.75cm,inner sep=1pt,label={[xshift=0.75cm,yshift=0cm]:$\SO(N)$}] (v3) at (0:\R) {$3$};
	        \node[draw=black,line width=1pt,circle,fill=yellowX,minimum width=0.75cm,inner sep=1pt,label={[xshift=1cm,yshift=-0.3cm]:$\SO(N)$}] (v4) at (300:\R) {$4$};
	        \node[draw=black,line width=1pt,circle,fill=yellowX,minimum width=0.75cm,inner sep=1pt,label={[xshift=-1cm,yshift=-0.3cm]:$\SO(N)$}] (v5) at (240:\R) {$5$};
	        \node[draw=black,line width=1pt,circle,fill=yellowX,minimum width=0.75cm,inner sep=1pt,label={[xshift=-0.75cm,yshift=0cm]:$\SO(N)$}] (v6) at (180:\R) {$6$};
	        \node[draw=black,line width=1pt,circle,fill=yellowX,minimum width=0.75cm,inner sep=1pt,label={[xshift=-0.7cm,yshift=-0cm]:$\U(N)$}] (v2) at  (60:\R) {$2$};
	        \draw[line width=1pt,redX] (v1) to node[fill=white,text opacity=1,fill opacity=1,draw=black,rectangle,thin,pos=0.5] {\color{redX}{$8$}} (v5);
	        \draw[line width=1pt,redX] (v2) to node[fill=white,text opacity=1,fill opacity=1,draw=black,rectangle,thin,pos=0.5] {\color{redX}{$8$}} (v4);
	        \draw[line width=1pt,redX] (v5) to node[fill=white,text opacity=1,fill opacity=1,draw=black,rectangle,thin,pos=0.5] {\color{redX}{$4$}} (v6);
	        \draw[line width=1pt,redX] (v4) to node[fill=white,text opacity=1,fill opacity=1,draw=black,rectangle,thin,pos=0.5] {\color{redX}{$4$}} (v3);
	        \draw[line width=1pt,-Triangle] (v1) to node[fill=white,text opacity=1,fill opacity=1,draw=black,rectangle,thin,pos=0.5] {$4$} (v2);
	        \draw[line width=1pt,-Triangle] (v1) to node[fill=white,text opacity=1,fill opacity=1,draw=black,rectangle,thin,pos=0.8] {$4$} (v3);
	        \draw[line width=1pt,-Triangle] (v4) to node[fill=white,text opacity=1,fill opacity=1,draw=black,rectangle,thin,pos=0.3] {$4$} (v1);
	        \draw[line width=1pt,-Triangle] (v2) to node[fill=white,text opacity=1,fill opacity=1,draw=black,rectangle,thin,pos=0.7] {$4$} (v5);
	        \draw[line width=1pt,-Triangle] (v6) to node[fill=white,text opacity=1,fill opacity=1,draw=black,rectangle,thin,pos=0.2] {$4$} (v2);
	        \draw[line width=1pt] (v3) to node[fill=white,text opacity=1,fill opacity=1,draw=black,rectangle,thin,pos=0.8] {$2$} (v5);
	        \draw[line width=1pt] (v4) to node[fill=white,text opacity=1,fill opacity=1,draw=black,rectangle,thin,pos=0.2] {$2$} (v6);
	        \draw[line width=1pt] (v4) to node[fill=white,text opacity=1,fill opacity=1,draw=black,rectangle,thin,pos=0.5] {$8$} (v5);
	\end{tikzpicture}
	\caption{Quiver diagram for the Spin(7) orientifold of phase L of $Q^{1,1,1}/\ZZ_2$ using the involution in \cref{eq:Q111Z2L-chiral-invol,eq:Q111Z2L-Fermi-invol,eq:Q111Z2L-RFermi-invol}, together with our choice of $\gamma_{\Omega_i}$ matrices.}
		\label{fig:orien_Q111Z2quiverL}
	\end{figure}

\subsection{Generators of $Q^{1,1,1}/\ZZ_2$}
\label{app:Q111Z2-gener}
	
In \cref{tab:GenQ111Z2PhaseD,tab:GenQ111Z2PhaseE,tab:GenQ111Z2PhaseH,tab:GenQ111Z2PhaseJ,tab:GenQ111Z2PhaseL} we list the generators of  $Q^{1,1,1}/\ZZ_2$ in terms of the chiral fields of phases D, E, H, J and L.

The relations among the generators are the same for all phases, and they are:
	\begin{align}
\scriptstyle \mathcal{I} = &\,\scriptstyle \left\langle M_{1} M_{3}=M_{2}^2 \coma M_{1} M_{5}=M_{2} M_{4} \coma M_{2} M_{6}=M_{3} M_{5} \coma M_{1} M_{8}=M_{2} M_{7} \coma M_{2} M_{9}=M_{3} M_{8} \coma M_{1} M_{10}=M_{4} M_{7} \coma \right.\nonumber\\ 
& \scriptstyle \left. M_{3} M_{12}=M_{6} M_{9} \coma M_{1} M_{13}=M_{7}^2 \coma M_{7} M_{14}=M_{8} M_{13} \coma M_{3} M_{15}=M_{9}^2 \coma M_{8} M_{15}=M_{9} M_{14} \coma M_{13} M_{15}=M_{14}^2 \coma \right.\nonumber\\ 
& \scriptstyle \left. M_{7} M_{16}=M_{10} M_{13} \coma M_{13} M_{17}=M_{14} M_{16} \coma M_{9} M_{18}=M_{12} M_{15} \coma M_{14} M_{18}=M_{15} M_{17} \coma M_{1} M_{19}=M_{4}^2 \coma \right.\nonumber\\ 
& \scriptstyle \left. M_{4} M_{20}=M_{5} M_{19} \coma M_{3} M_{21}=M_{6}^2 \coma M_{5} M_{21}=M_{6} M_{20} \coma M_{19} M_{21}=M_{20}^2 \coma M_{4} M_{22}=M_{10} M_{19} \coma \right.\nonumber\\ 
& \scriptstyle \left. M_{19} M_{23}=M_{20} M_{22} \coma M_{6} M_{24}=M_{12} M_{21} \coma M_{20} M_{24}=M_{21} M_{23} \coma M_{10} M_{25}=M_{16} M_{22} \coma M_{13} M_{25}=M_{16}^2 \coma \right.\nonumber\\ 
& \scriptstyle \left. M_{19} M_{25}=M_{22}^2 \coma M_{16} M_{26}=M_{17} M_{25} \coma M_{22} M_{26}=M_{23} M_{25} \coma M_{12} M_{27}=M_{18} M_{24} \coma M_{15} M_{27}=M_{18}^2 \coma \right.\nonumber\\ 
& \scriptstyle \left. M_{17} M_{27}=M_{18} M_{26} \coma M_{21} M_{27}=M_{24}^2 \coma M_{23} M_{27}=M_{24} M_{26} \coma M_{25} M_{27}=M_{26}^2 \coma M_{1} M_{6}=M_{2} M_{5}=M_{3} M_{4} \coma \right.\nonumber\\ 
& \scriptstyle \left. M_{1} M_{9}=M_{2} M_{8}=M_{3} M_{7} \coma M_{1} M_{14}=M_{2} M_{13}=M_{7} M_{8} \coma M_{2} M_{15}=M_{3} M_{14}=M_{8} M_{9} \coma \right.\nonumber\\ 
& \scriptstyle \left. M_{7} M_{15}=M_{8} M_{14}=M_{9} M_{13} \coma M_{1} M_{16}=M_{4} M_{13}=M_{7} M_{10} \coma M_{3} M_{18}=M_{6} M_{15}=M_{9} M_{12} \coma \right.\nonumber\\ 
& \scriptstyle \left. M_{13} M_{18}=M_{14} M_{17}=M_{15} M_{16} \coma M_{1} M_{20}=M_{2} M_{19}=M_{4} M_{5} \coma M_{2} M_{21}=M_{3} M_{20}=M_{5} M_{6} \coma \right.\nonumber\\ 
& \scriptstyle \left. M_{4} M_{21}=M_{5} M_{20}=M_{6} M_{19} \coma M_{1} M_{22}=M_{4} M_{10}=M_{7} M_{19} \coma M_{3} M_{24}=M_{6} M_{12}=M_{9} M_{21} \coma \right.\nonumber\\ 
& \scriptstyle \left. M_{19} M_{24}=M_{20} M_{23}=M_{21} M_{22} \coma M_{4} M_{25}=M_{10} M_{22}=M_{16} M_{19} \coma M_{7} M_{25}=M_{10} M_{16}=M_{13} M_{22} \coma \right.\nonumber\\ 
& \scriptstyle \left. M_{13} M_{26}=M_{14} M_{25}=M_{16} M_{17} \coma M_{19} M_{26}=M_{20} M_{25}=M_{22} M_{23} \coma M_{6} M_{27}=M_{12} M_{24}=M_{18} M_{21} \coma \right.\nonumber\\ 
& \scriptstyle \left. M_{9} M_{27}=M_{12} M_{18}=M_{15} M_{24} \coma M_{14} M_{27}=M_{15} M_{26}=M_{17} M_{18} \coma M_{16} M_{27}=M_{17} M_{26}=M_{18} M_{25} \coma \right.\nonumber\\ 
& \scriptstyle \left. M_{20} M_{27}=M_{21} M_{26}=M_{23} M_{24} \coma M_{22} M_{27}=M_{23} M_{26}=M_{24} M_{25} \coma M_{1} M_{11}=M_{2} M_{10}=M_{4} M_{8}=M_{5} M_{7} \coma \right.\nonumber\\ 
& \scriptstyle \left. M_{2} M_{12}=M_{3} M_{11}=M_{5} M_{9}=M_{6}M_{8} \coma M_{7} M_{17}=M_{8} M_{16}=M_{10} M_{14}=M_{11} M_{13} \coma \right.\nonumber\\ 
& \scriptstyle \left. M_{8} M_{18}=M_{9} M_{17}=M_{11} M_{15}=M_{12} M_{14} \coma M_{4} M_{23}=M_{5} M_{22}=M_{10} M_{20}=M_{11} M_{19} \coma \right.\\ 
& \scriptstyle \left. M_{5} M_{24}=M_{6} M_{23}=M_{11} M_{21}=M_{12} M_{20} \coma M_{10} M_{26}=M_{11} M_{25}=M_{16} M_{23}=M_{17} M_{22} \coma \right.\nonumber\\ 
& \scriptstyle \left. M_{11} M_{27}=M_{12} M_{26}=M_{17} M_{24}=M_{18} M_{23} \coma M_{1} M_{15}=M_{2} M_{14}=M_{3} M_{13}=M_{7} M_{9}=M_{8}^2 \coma \right.\nonumber\\ 
& \scriptstyle \left. M_{1} M_{21}=M_{2} M_{20}=M_{3} M_{19}=M_{4} M_{6}=M_{5}^2 \coma M_{1} M_{25}=M_{4} M_{16}=M_{7} M_{22}=M_{10}^2=M_{13} M_{19} \coma \right.\nonumber\\ 
& \scriptstyle \left. M_{3} M_{27}=M_{6} M_{18}=M_{9} M_{24}=M_{12}^2=M_{15} M_{21} \coma M_{13} M_{27}=M_{14} M_{26}=M_{15} M_{25}=M_{16} M_{18}=M_{17}^2 \coma \right.\nonumber\\ 
& \scriptstyle \left. M_{19} M_{27}=M_{20} M_{26}=M_{21} M_{25}=M_{22} M_{24}=M_{23}^2 \coma M_{1} M_{12}=M_{2} M_{11}=M_{3} M_{10}=M_{4} M_{9}=M_{5} M_{8}=M_{6} M_{7} \coma \right.\nonumber\\ 
& \scriptstyle \left. M_{1} M_{17}=M_{2} M_{16}=M_{4} M_{14}=M_{5} M_{13}=M_{7} M_{11}=M_{8} M_{10} \coma M_{2} M_{18}=M_{3} M_{17}=M_{5} M_{15}=M_{6} M_{14}=M_{8} M_{12}= \right.\nonumber\\ 
& \scriptstyle \left. =M_{9} M_{11} \coma M_{7} M_{18}=M_{8} M_{17}=M_{9} M_{16}=M_{10} M_{15}=M_{11} M_{14}=M_{12} M_{13} \coma M_{1} M_{23}=M_{2} M_{22}=M_{4} M_{11}=\right.\nonumber\\ 
& \scriptstyle \left. =M_{5} M_{10}=M_{7} M_{20}=M_{8} M_{19} \coma M_{2} M_{24}=M_{3} M_{23}=M_{5} M_{12}=M_{6} M_{11}=M_{8} M_{21}=M_{9} M_{20} \coma M_{4} M_{24}=M_{5} M_{23}=\right.\nonumber\\ 
& \scriptstyle \left. =M_{6} M_{22}=M_{10} M_{21}=M_{11} M_{20}=M_{12} M_{19} \coma M_{4} M_{26}=M_{5} M_{25}=M_{10} M_{23}=M_{11} M_{22}=M_{16} M_{20}=M_{17} M_{19} \coma \right.\nonumber\\ 
& \scriptstyle \left. M_{7} M_{26}=M_{8} M_{25}=M_{10} M_{17}=M_{11} M_{16}=M_{13} M_{23}=M_{14} M_{22} \coma M_{5} M_{27}=M_{6} M_{26}=M_{11} M_{24}=M_{12} M_{23}=M_{17} M_{21}=\right.\nonumber\\ 
& \scriptstyle \left. =M_{18} M_{20} \coma M_{8} M_{27}=M_{9} M_{26}=M_{11} M_{18}=M_{12} M_{17}=M_{14} M_{24}=M_{15} M_{23} \coma M_{10} M_{27}=M_{11} M_{26}=M_{12} M_{25}=\right.\nonumber\\ 
& \scriptstyle \left. =M_{16} M_{24}=M_{17} M_{23}=M_{18} M_{22} \coma M_{1} M_{18}=M_{2} M_{17}=M_{3} M_{16}=M_{4} M_{15}=M_{5} M_{14}=M_{6} M_{13}=M_{7} M_{12}=M_{8} M_{11}=\right.\nonumber\\ 
& \scriptstyle \left. =M_{9} M_{10} \coma M_{1} M_{24}=M_{2} M_{23}=M_{3} M_{22}=M_{4} M_{12}=M_{5} M_{11}=M_{6} M_{10}=M_{7} M_{21}=M_{8} M_{20}=M_{9} M_{19} \coma M_{1} M_{26}=\right.\nonumber\\ 
& \scriptstyle \left. =M_{2} M_{25}=M_{4} M_{17}=M_{5} M_{16}=M_{7} M_{23}=M_{8} M_{22}=M_{10} M_{11}=M_{13} M_{20}=M_{14} M_{19} \coma M_{2} M_{27}=M_{3} M_{26}=\right.\nonumber\\ 
& \scriptstyle \left. =M_{5} M_{18}=M_{6} M_{17}=M_{8} M_{24}=M_{9} M_{23}=M_{11} M_{12}=M_{14} M_{21}=M_{15} M_{20} \coma M_{4} M_{27}=M_{5} M_{26}=M_{6} M_{25}=M_{10}M_{24}=\right.\nonumber\\ 
& \scriptstyle \left. =M_{11} M_{23}=M_{12} M_{22}=M_{16} M_{21}=M_{17} M_{20}=M_{18} M_{19} \coma M_{7} M_{27}=M_{8} M_{26}=M_{9} M_{25}=M_{10} M_{18}=M_{11} M_{17}=\right.\nonumber\\ 
& \scriptstyle \left. =M_{12} M_{16}=M_{13} M_{24}=M_{14} M_{23}=M_{15} M_{22} \coma M_{1} M_{27}=M_{2} M_{26}=M_{3} M_{25}=M_{4} M_{18}=M_{5} M_{17}=M_{6} M_{16}=M_{7} M_{24}=\right.\nonumber\\ 
& \scriptstyle \left. =M_{8} M_{23}=M_{9} M_{22}=M_{10} M_{12}=M_{11}^2=M_{13} M_{21}=M_{14} M_{20}=M_{15} M_{19}\right\rangle\nonumber
	\end{align}

\begin{center}
		\renewcommand{\arraystretch}{1.1}
	\begin{longtable}{c|c}
		\caption{Generators of $Q^{1,1,1}/\ZZ_2$ in Phase D.}\\
		Field & Chiral superfields\\
		\hline
		\endhead
		\label{tab:GenQ111Z2PhaseD}$M_{1}$ & $X_{72} X_{47} X_{64} Y_{56} Y_{25}$ $=$ $X_{72} X_{47} X_{53} Y_{34} Y_{25}$ $=$ $X_{18} X_{64} X_{85} Y_{41} Y_{56}$ $=$  \\ 
 & $=$ $X_{18} X_{53} X_{85} Y_{34} Y_{41}$ \\
\hline
 $M_{2}$ & $X_{72} X_{47} X_{64} X_{25} Y_{56}$ $=$ $X_{72} X_{47} X_{56} X_{64} Y_{25}$ $=$ $X_{72} X_{47} X_{53} X_{25} Y_{34}$ $=$  \\ 
 & $=$ $X_{72} X_{34} X_{47} X_{53} Y_{25}$ $=$ $X_{18} X_{56} X_{64} X_{85} Y_{41}$ $=$ $X_{18} X_{41} X_{64} X_{85} Y_{56}$ $=$  \\ 
 & $=$ $X_{18} X_{41} X_{53} X_{85} Y_{34}$ $=$ $X_{18} X_{34} X_{53} X_{85} Y_{41}$ \\
\hline
 $M_{3}$ & $X_{72} X_{47} X_{56} X_{64} X_{25}$ $=$ $X_{72} X_{34} X_{47} X_{53} X_{25}$ $=$ $X_{18} X_{41} X_{56} X_{64} X_{85}$ $=$  \\ 
 & $=$ $X_{18} X_{34} X_{41} X_{53} X_{85}$ \\
\hline
 $M_{4}$ & $X_{64} X_{85} Y_{18} Y_{41} Y_{56}$ $=$ $X_{53} X_{85} Y_{18} Y_{34} Y_{41}$ $=$ $X_{47} X_{64} Y_{72} Y_{56} Y_{25}$ $=$  \\ 
 & $=$ $X_{47} X_{53} Y_{72} Y_{34} Y_{25}$ $=$ $X_{72} X_{47} Y_{56} Y_{64} Y_{25}$ $=$ $X_{72} X_{47} Y_{34} Y_{53} Y_{25}$ $=$  \\ 
 & $=$ $X_{18} X_{85} Y_{41} Y_{56} Y_{64}$ $=$ $X_{18} X_{85} Y_{34} Y_{41} Y_{53}$ \\
\hline
 $M_{5}$ & $X_{56} X_{64} X_{85} Y_{18} Y_{41}$ $=$ $X_{47} X_{64} X_{25} Y_{72} Y_{56}$ $=$ $X_{47} X_{56} X_{64} Y_{72} Y_{25}$ $=$  \\ 
 & $=$ $X_{47} X_{53} X_{25} Y_{72} Y_{34}$ $=$ $X_{41} X_{64} X_{85} Y_{18} Y_{56}$ $=$ $X_{41} X_{53} X_{85} Y_{18} Y_{34}$ $=$  \\ 
 & $=$ $X_{34} X_{53} X_{85} Y_{18} Y_{41}$ $=$ $X_{34} X_{47} X_{53} Y_{72} Y_{25}$ $=$ $X_{72} X_{47} X_{25} Y_{56} Y_{64}$ $=$  \\ 
 & $=$ $X_{72} X_{47} X_{25} Y_{34} Y_{53}$ $=$ $X_{72} X_{47} X_{56} Y_{64} Y_{25}$ $=$ $X_{72} X_{34} X_{47} Y_{53} Y_{25}$ $=$  \\ 
 & $=$ $X_{18} X_{56} X_{85} Y_{41} Y_{64}$ $=$ $X_{18} X_{41} X_{85} Y_{56} Y_{64}$ $=$ $X_{18} X_{41} X_{85} Y_{34} Y_{53}$ $=$  \\ 
 & $=$ $X_{18} X_{34} X_{85} Y_{41} Y_{53}$ \\
\hline
 $M_{6}$ & $X_{47} X_{56} X_{64} X_{25} Y_{72}$ $=$ $X_{41} X_{56} X_{64} X_{85} Y_{18}$ $=$ $X_{34} X_{47} X_{53} X_{25} Y_{72}$ $=$  \\ 
 & $=$ $X_{34} X_{41} X_{53} X_{85} Y_{18}$ $=$ $X_{72} X_{47} X_{56} X_{25} Y_{64}$ $=$ $X_{72} X_{34} X_{47} X_{25} Y_{53}$ $=$  \\ 
 & $=$ $X_{18} X_{41} X_{56} X_{85} Y_{64}$ $=$ $X_{18} X_{34} X_{41} X_{85} Y_{53}$ \\
\hline
 $M_{2}$ & $X_{72} X_{64} Y_{47} Y_{56} Y_{25}$ $=$ $X_{72} X_{53} Y_{34} Y_{47} Y_{25}$ $=$ $X_{18} X_{64} Y_{41} Y_{56} Y_{85}$ $=$  \\ 
 & $=$ $X_{18} X_{53} Y_{34} Y_{41} Y_{85}$ $=$ $W_{56} X_{72} X_{47} X_{64} Y_{25}$ $=$ $W_{56} X_{18} X_{64} X_{85} Y_{41}$ $=$  \\ 
 & $=$ $W_{34} X_{72} X_{47} X_{53} Y_{25}$ $=$ $W_{34} X_{18} X_{53} X_{85} Y_{41}$ \\
\hline
 $M_{8}$ & $X_{72} X_{64} X_{25} Y_{47} Y_{56}$ $=$ $X_{72} X_{56} X_{64} Y_{47} Y_{25}$ $=$ $X_{72} X_{53} X_{25} Y_{34} Y_{47}$ $=$  \\ 
 & $=$ $X_{72} X_{47} X_{64} Y_{25} Z_{56}$ $=$ $X_{72} X_{47} X_{53} Y_{25} Z_{34}$ $=$ $X_{72} X_{34} X_{53} Y_{47} Y_{25}$ $=$  \\ 
 & $=$ $X_{18} X_{64} X_{85} Y_{41} Z_{56}$ $=$ $X_{18} X_{56} X_{64} Y_{41} Y_{85}$ $=$ $X_{18} X_{53} X_{85} Y_{41} Z_{34}$ $=$  \\ 
 & $=$ $X_{18} X_{41} X_{64} Y_{56} Y_{85}$ $=$ $X_{18} X_{41} X_{53} Y_{34} Y_{85}$ $=$ $X_{18} X_{34} X_{53} Y_{41} Y_{85}$ $=$  \\ 
 & $=$ $W_{56} X_{72} X_{47} X_{64} X_{25}$ $=$ $W_{56} X_{18} X_{41} X_{64} X_{85}$ $=$ $W_{34} X_{72} X_{47} X_{53} X_{25}$ $=$  \\ 
 & $=$ $W_{34} X_{18} X_{41} X_{53} X_{85}$ \\
\hline
 $M_{9}$ & $X_{72} X_{56} X_{64} X_{25} Y_{47}$ $=$ $X_{72} X_{47} X_{64} X_{25} Z_{56}$ $=$ $X_{72} X_{47} X_{53} X_{25} Z_{34}$ $=$  \\ 
 & $=$ $X_{72} X_{34} X_{53} X_{25} Y_{47}$ $=$ $X_{18} X_{41} X_{64} X_{85} Z_{56}$ $=$ $X_{18} X_{41} X_{56} X_{64} Y_{85}$ $=$  \\ 
 & $=$ $X_{18} X_{41} X_{53} X_{85} Z_{34}$ $=$ $X_{18} X_{34} X_{41} X_{53} Y_{85}$ \\
\hline
 $M_{10}$ & $X_{64} Y_{72} Y_{47} Y_{56} Y_{25}$ $=$ $X_{64} Y_{18} Y_{41} Y_{56} Y_{85}$ $=$ $X_{53} Y_{72} Y_{34} Y_{47} Y_{25}$ $=$  \\ 
 & $=$ $X_{53} Y_{18} Y_{34} Y_{41} Y_{85}$ $=$ $X_{72} Y_{47} Y_{56} Y_{64} Y_{25}$ $=$ $X_{72} Y_{34} Y_{47} Y_{53} Y_{25}$ $=$  \\ 
 & $=$ $X_{18} Y_{41} Y_{56} Y_{64} Y_{85}$ $=$ $X_{18} Y_{34} Y_{41} Y_{53} Y_{85}$ $=$ $W_{56} X_{64} X_{85} Y_{18} Y_{41}$ $=$  \\ 
 & $=$ $W_{56} X_{47} X_{64} Y_{72} Y_{25}$ $=$ $W_{56} X_{72} X_{47} Y_{64} Y_{25}$ $=$ $W_{56} X_{18} X_{85} Y_{41} Y_{64}$ $=$  \\ 
 & $=$ $W_{34} X_{53} X_{85} Y_{18} Y_{41}$ $=$ $W_{34} X_{47} X_{53} Y_{72} Y_{25}$ $=$ $W_{34} X_{72} X_{47} Y_{53} Y_{25}$ $=$  \\ 
 & $=$ $W_{34} X_{18} X_{85} Y_{41} Y_{53}$ \\
\hline
 $M_{11}$ & $X_{64} X_{85} Y_{18} Y_{41} Z_{56}$ $=$ $X_{64} X_{25} Y_{72} Y_{47} Y_{56}$ $=$ $X_{56} X_{64} Y_{72} Y_{47} Y_{25}$ $=$  \\ 
 & $=$ $X_{56} X_{64} Y_{18} Y_{41} Y_{85}$ $=$ $X_{53} X_{85} Y_{18} Y_{41} Z_{34}$ $=$ $X_{53} X_{25} Y_{72} Y_{34} Y_{47}$ $=$  \\ 
 & $=$ $X_{47} X_{64} Y_{72} Y_{25} Z_{56}$ $=$ $X_{47} X_{53} Y_{72} Y_{25} Z_{34}$ $=$ $X_{41} X_{64} Y_{18} Y_{56} Y_{85}$ $=$  \\ 
 & $=$ $X_{41} X_{53} Y_{18} Y_{34} Y_{85}$ $=$ $X_{34} X_{53} Y_{72} Y_{47} Y_{25}$ $=$ $X_{34} X_{53} Y_{18} Y_{41} Y_{85}$ $=$  \\ 
 & $=$ $X_{72} X_{25} Y_{47} Y_{56} Y_{64}$ $=$ $X_{72} X_{25} Y_{34} Y_{47} Y_{53}$ $=$ $X_{72} X_{56} Y_{47} Y_{64} Y_{25}$ $=$  \\ 
 & $=$ $X_{72} X_{47} Y_{64} Y_{25} Z_{56}$ $=$ $X_{72} X_{47} Y_{53} Y_{25} Z_{34}$ $=$ $X_{72} X_{34} Y_{47} Y_{53} Y_{25}$ $=$  \\ 
 & $=$ $X_{18} X_{85} Y_{41} Y_{64} Z_{56}$ $=$ $X_{18} X_{85} Y_{41} Y_{53} Z_{34}$ $=$ $X_{18} X_{56} Y_{41} Y_{64} Y_{85}$ $=$  \\ 
 & $=$ $X_{18} X_{41} Y_{56} Y_{64} Y_{85}$ $=$ $X_{18} X_{41} Y_{34} Y_{53} Y_{85}$ $=$ $X_{18} X_{34} Y_{41} Y_{53} Y_{85}$ $=$  \\ 
 & $=$ $W_{56} X_{47} X_{64} X_{25} Y_{72}$ $=$ $W_{56} X_{41} X_{64} X_{85} Y_{18}$ $=$ $W_{56} X_{72} X_{47} X_{25} Y_{64}$ $=$  \\ 
 & $=$ $W_{56} X_{18} X_{41} X_{85} Y_{64}$ $=$ $W_{34} X_{47} X_{53} X_{25} Y_{72}$ $=$ $W_{34} X_{41} X_{53} X_{85} Y_{18}$ $=$  \\ 
 & $=$ $W_{34} X_{72} X_{47} X_{25} Y_{53}$ $=$ $W_{34} X_{18} X_{41} X_{85} Y_{53}$ \\
\hline
 $M_{12}$ & $X_{56} X_{64} X_{25} Y_{72} Y_{47}$ $=$ $X_{47} X_{64} X_{25} Y_{72} Z_{56}$ $=$ $X_{47} X_{53} X_{25} Y_{72} Z_{34}$ $=$  \\ 
 & $=$ $X_{41} X_{64} X_{85} Y_{18} Z_{56}$ $=$ $X_{41} X_{56} X_{64} Y_{18} Y_{85}$ $=$ $X_{41} X_{53} X_{85} Y_{18} Z_{34}$ $=$  \\ 
 & $=$ $X_{34} X_{53} X_{25} Y_{72} Y_{47}$ $=$ $X_{34} X_{41} X_{53} Y_{18} Y_{85}$ $=$ $X_{72} X_{56} X_{25} Y_{47} Y_{64}$ $=$  \\ 
 & $=$ $X_{72} X_{47} X_{25} Y_{64} Z_{56}$ $=$ $X_{72} X_{47} X_{25} Y_{53} Z_{34}$ $=$ $X_{72} X_{34} X_{25} Y_{47} Y_{53}$ $=$  \\ 
 & $=$ $X_{18} X_{41} X_{85} Y_{64} Z_{56}$ $=$ $X_{18} X_{41} X_{85} Y_{53} Z_{34}$ $=$ $X_{18} X_{41} X_{56} Y_{64} Y_{85}$ $=$  \\ 
 & $=$ $X_{18} X_{34} X_{41} Y_{53} Y_{85}$ \\
\hline
 $M_{13}$ & $W_{56} X_{72} X_{64} Y_{47} Y_{25}$ $=$ $W_{56} X_{18} X_{64} Y_{41} Y_{85}$ $=$ $W_{34} X_{72} X_{53} Y_{47} Y_{25}$ $=$  \\ 
 & $=$ $W_{34} X_{18} X_{53} Y_{41} Y_{85}$ \\
\hline
 $M_{14}$ & $X_{72} X_{64} Y_{47} Y_{25} Z_{56}$ $=$ $X_{72} X_{53} Y_{47} Y_{25} Z_{34}$ $=$ $X_{18} X_{64} Y_{41} Y_{85} Z_{56}$ $=$  \\ 
 & $=$ $X_{18} X_{53} Y_{41} Y_{85} Z_{34}$ $=$ $W_{56} X_{72} X_{64} X_{25} Y_{47}$ $=$ $W_{56} X_{18} X_{41} X_{64} Y_{85}$ $=$  \\ 
 & $=$ $W_{34} X_{72} X_{53} X_{25} Y_{47}$ $=$ $W_{34} X_{18} X_{41} X_{53} Y_{85}$ \\
\hline
 $M_{15}$ & $X_{72} X_{64} X_{25} Y_{47} Z_{56}$ $=$ $X_{72} X_{53} X_{25} Y_{47} Z_{34}$ $=$ $X_{18} X_{41} X_{64} Y_{85} Z_{56}$ $=$  \\ 
 & $=$ $X_{18} X_{41} X_{53} Y_{85} Z_{34}$ \\
\hline
 $M_{16}$ & $W_{56} X_{64} Y_{72} Y_{47} Y_{25}$ $=$ $W_{56} X_{64} Y_{18} Y_{41} Y_{85}$ $=$ $W_{56} X_{72} Y_{47} Y_{64} Y_{25}$ $=$  \\ 
 & $=$ $W_{56} X_{18} Y_{41} Y_{64} Y_{85}$ $=$ $W_{34} X_{53} Y_{72} Y_{47} Y_{25}$ $=$ $W_{34} X_{53} Y_{18} Y_{41} Y_{85}$ $=$  \\ 
 & $=$ $W_{34} X_{72} Y_{47} Y_{53} Y_{25}$ $=$ $W_{34} X_{18} Y_{41} Y_{53} Y_{85}$ \\
\hline
 $M_{17}$ & $X_{64} Y_{72} Y_{47} Y_{25} Z_{56}$ $=$ $X_{64} Y_{18} Y_{41} Y_{85} Z_{56}$ $=$ $X_{53} Y_{72} Y_{47} Y_{25} Z_{34}$ $=$  \\ 
 & $=$ $X_{53} Y_{18} Y_{41} Y_{85} Z_{34}$ $=$ $X_{72} Y_{47} Y_{64} Y_{25} Z_{56}$ $=$ $X_{72} Y_{47} Y_{53} Y_{25} Z_{34}$ $=$  \\ 
 & $=$ $X_{18} Y_{41} Y_{64} Y_{85} Z_{56}$ $=$ $X_{18} Y_{41} Y_{53} Y_{85} Z_{34}$ $=$ $W_{56} X_{64} X_{25} Y_{72} Y_{47}$ $=$  \\ 
 & $=$ $W_{56} X_{41} X_{64} Y_{18} Y_{85}$ $=$ $W_{56} X_{72} X_{25} Y_{47} Y_{64}$ $=$ $W_{56} X_{18} X_{41} Y_{64} Y_{85}$ $=$  \\ 
 & $=$ $W_{34} X_{53} X_{25} Y_{72} Y_{47}$ $=$ $W_{34} X_{41} X_{53} Y_{18} Y_{85}$ $=$ $W_{34} X_{72} X_{25} Y_{47} Y_{53}$ $=$  \\ 
 & $=$ $W_{34} X_{18} X_{41} Y_{53} Y_{85}$ \\
\hline
 $M_{18}$ & $X_{64} X_{25} Y_{72} Y_{47} Z_{56}$ $=$ $X_{53} X_{25} Y_{72} Y_{47} Z_{34}$ $=$ $X_{41} X_{64} Y_{18} Y_{85} Z_{56}$ $=$  \\ 
 & $=$ $X_{41} X_{53} Y_{18} Y_{85} Z_{34}$ $=$ $X_{72} X_{25} Y_{47} Y_{64} Z_{56}$ $=$ $X_{72} X_{25} Y_{47} Y_{53} Z_{34}$ $=$  \\ 
 & $=$ $X_{18} X_{41} Y_{64} Y_{85} Z_{56}$ $=$ $X_{18} X_{41} Y_{53} Y_{85} Z_{34}$ \\
\hline
 $M_{19}$ & $X_{85} Y_{18} Y_{41} Y_{56} Y_{64}$ $=$ $X_{85} Y_{18} Y_{34} Y_{41} Y_{53}$ $=$ $X_{47} Y_{72} Y_{56} Y_{64} Y_{25}$ $=$  \\ 
 & $=$ $X_{47} Y_{72} Y_{34} Y_{53} Y_{25}$ \\
\hline
 $M_{20}$ & $X_{56} X_{85} Y_{18} Y_{41} Y_{64}$ $=$ $X_{47} X_{25} Y_{72} Y_{56} Y_{64}$ $=$ $X_{47} X_{25} Y_{72} Y_{34} Y_{53}$ $=$  \\ 
 & $=$ $X_{47} X_{56} Y_{72} Y_{64} Y_{25}$ $=$ $X_{41} X_{85} Y_{18} Y_{56} Y_{64}$ $=$ $X_{41} X_{85} Y_{18} Y_{34} Y_{53}$ $=$  \\ 
 & $=$ $X_{34} X_{85} Y_{18} Y_{41} Y_{53}$ $=$ $X_{34} X_{47} Y_{72} Y_{53} Y_{25}$ \\
\hline
 $M_{21}$ & $X_{47} X_{56} X_{25} Y_{72} Y_{64}$ $=$ $X_{41} X_{56} X_{85} Y_{18} Y_{64}$ $=$ $X_{34} X_{47} X_{25} Y_{72} Y_{53}$ $=$  \\ 
 & $=$ $X_{34} X_{41} X_{85} Y_{18} Y_{53}$ \\
\hline
 $M_{27}$ & $Y_{72} Y_{47} Y_{56} Y_{64} Y_{25}$ $=$ $Y_{72} Y_{34} Y_{47} Y_{53} Y_{25}$ $=$ $Y_{18} Y_{41} Y_{56} Y_{64} Y_{85}$ $=$  \\ 
 & $=$ $Y_{18} Y_{34} Y_{41} Y_{53} Y_{85}$ $=$ $W_{56} X_{85} Y_{18} Y_{41} Y_{64}$ $=$ $W_{56} X_{47} Y_{72} Y_{64} Y_{25}$ $=$  \\ 
 & $=$ $W_{34} X_{85} Y_{18} Y_{41} Y_{53}$ $=$ $W_{34} X_{47} Y_{72} Y_{53} Y_{25}$ \\
\hline
 $M_{23}$ & $X_{85} Y_{18} Y_{41} Y_{64} Z_{56}$ $=$ $X_{85} Y_{18} Y_{41} Y_{53} Z_{34}$ $=$ $X_{25} Y_{72} Y_{47} Y_{56} Y_{64}$ $=$  \\ 
 & $=$ $X_{25} Y_{72} Y_{34} Y_{47} Y_{53}$ $=$ $X_{56} Y_{72} Y_{47} Y_{64} Y_{25}$ $=$ $X_{56} Y_{18} Y_{41} Y_{64} Y_{85}$ $=$  \\ 
 & $=$ $X_{47} Y_{72} Y_{64} Y_{25} Z_{56}$ $=$ $X_{47} Y_{72} Y_{53} Y_{25} Z_{34}$ $=$ $X_{41} Y_{18} Y_{56} Y_{64} Y_{85}$ $=$  \\ 
 & $=$ $X_{41} Y_{18} Y_{34} Y_{53} Y_{85}$ $=$ $X_{34} Y_{72} Y_{47} Y_{53} Y_{25}$ $=$ $X_{34} Y_{18} Y_{41} Y_{53} Y_{85}$ $=$  \\ 
 & $=$ $W_{56} X_{47} X_{25} Y_{72} Y_{64}$ $=$ $W_{56} X_{41} X_{85} Y_{18} Y_{64}$ $=$ $W_{34} X_{47} X_{25} Y_{72} Y_{53}$ $=$  \\ 
 & $=$ $W_{34} X_{41} X_{85} Y_{18} Y_{53}$ \\
\hline
 $M_{24}$ & $X_{56} X_{25} Y_{72} Y_{47} Y_{64}$ $=$ $X_{47} X_{25} Y_{72} Y_{64} Z_{56}$ $=$ $X_{47} X_{25} Y_{72} Y_{53} Z_{34}$ $=$  \\ 
 & $=$ $X_{41} X_{85} Y_{18} Y_{64} Z_{56}$ $=$ $X_{41} X_{85} Y_{18} Y_{53} Z_{34}$ $=$ $X_{41} X_{56} Y_{18} Y_{64} Y_{85}$ $=$  \\ 
 & $=$ $X_{34} X_{25} Y_{72} Y_{47} Y_{53}$ $=$ $X_{34} X_{41} Y_{18} Y_{53} Y_{85}$ \\
\hline
 $M_{25}$ & $W_{56} Y_{72} Y_{47} Y_{64} Y_{25}$ $=$ $W_{56} Y_{18} Y_{41} Y_{64} Y_{85}$ $=$ $W_{34} Y_{72} Y_{47} Y_{53} Y_{25}$ $=$  \\ 
 & $=$ $W_{34} Y_{18} Y_{41} Y_{53} Y_{85}$ \\
\hline
 $M_{26}$ & $Y_{72} Y_{47} Y_{64} Y_{25} Z_{56}$ $=$ $Y_{72} Y_{47} Y_{53} Y_{25} Z_{34}$ $=$ $Y_{18} Y_{41} Y_{64} Y_{85} Z_{56}$ $=$  \\ 
 & $=$ $Y_{18} Y_{41} Y_{53} Y_{85} Z_{34}$ $=$ $W_{56} X_{25} Y_{72} Y_{47} Y_{64}$ $=$ $W_{56} X_{41} Y_{18} Y_{64} Y_{85}$ $=$  \\ 
 & $=$ $W_{34} X_{25} Y_{72} Y_{47} Y_{53}$ $=$ $W_{34} X_{41} Y_{18} Y_{53} Y_{85}$ \\
\hline
 $M_{27}$ & $X_{25} Y_{72} Y_{47} Y_{64} Z_{56}$ $=$ $X_{25} Y_{72} Y_{47} Y_{53} Z_{34}$ $=$ $X_{41} Y_{18} Y_{64} Y_{85} Z_{56}$ $=$  \\ 
 & $=$ $X_{41} Y_{18} Y_{53} Y_{85} Z_{34}$ 
	\end{longtable}
\end{center}
	

\begin{center}
		\renewcommand{\arraystretch}{1.1}
	\begin{longtable}{c|c}
		\caption{Generators of $Q^{1,1,1}/\ZZ_2$ in Phase E.}\\
		Field & Chiral superfields\\
		\hline
		\endhead
		\label{tab:GenQ111Z2PhaseE}$M_{1}$ & $W_{56}X_{35}X_{47}X_{64}Y_{73}$ $=$ $W_{56}X_{18}X_{64}X_{85}Y_{41}$ $=$ $W_{56}X_{13}X_{35}X_{64}Y_{41}$ $=$ \\
& $=$ $W_{56}X_{72}X_{47}X_{64}Y_{25}$ \\ 
\hline
$M_{2}$ & $X_{35}X_{47}X_{64}Y_{73}Y_{56}$ $=$ $W_{56}X_{35}X_{64}Y_{73}Y_{47}$ $=$ $X_{18}X_{64}X_{85}Y_{41}Y_{56}$ $=$ \\
& $=$ $X_{13}X_{35}X_{64}Y_{41}Y_{56}$ $=$ $W_{56}X_{18}X_{64}Y_{41}Y_{85}$ $=$ $X_{72}X_{47}X_{64}Y_{56}Y_{25}$ $=$ \\
& $=$ $W_{56}X_{72}X_{64}Y_{47}Y_{25}$ $=$ $W_{56}X_{35}X_{64}Y_{13}Y_{41}$ \\ 
\hline
$M_{3}$ & $X_{35}X_{64}Y_{73}Y_{47}Y_{56}$ $=$ $X_{18}X_{64}Y_{41}Y_{56}Y_{85}$ $=$ $X_{72}X_{64}Y_{47}Y_{56}Y_{25}$ $=$ \\
& $=$ $X_{35}X_{64}Y_{13}Y_{41}Y_{56}$ \\ 
\hline
$M_{4}$ & $W_{56}X_{47}X_{64}Y_{73}Y_{35}$ $=$ $W_{56}X_{35}X_{47}Y_{73}Y_{64}$ $=$ $W_{56}X_{13}X_{64}Y_{35}Y_{41}$ $=$ \\
& $=$ $W_{56}X_{18}X_{85}Y_{41}Y_{64}$ $=$ $W_{56}X_{13}X_{35}Y_{41}Y_{64}$ $=$ $W_{56}X_{72}X_{47}Y_{64}Y_{25}$ $=$ \\
& $=$ $W_{56}X_{47}X_{64}Y_{72}Y_{25}$ $=$ $W_{56}X_{64}X_{85}Y_{18}Y_{41}$ \\ 
\hline
$M_{5}$ & $X_{47}X_{64}Y_{73}Y_{35}Y_{56}$ $=$ $X_{35}X_{47}Y_{73}Y_{56}Y_{64}$ $=$ $W_{56}X_{64}Y_{73}Y_{35}Y_{47}$ $=$ \\
& $=$ $W_{56}X_{35}Y_{73}Y_{47}Y_{64}$ $=$ $X_{13}X_{64}Y_{35}Y_{41}Y_{56}$ $=$ $X_{18}X_{85}Y_{41}Y_{56}Y_{64}$ $=$ \\
& $=$ $X_{13}X_{35}Y_{41}Y_{56}Y_{64}$ $=$ $W_{56}X_{18}Y_{41}Y_{64}Y_{85}$ $=$ $X_{72}X_{47}Y_{56}Y_{64}Y_{25}$ $=$ \\
& $=$ $X_{47}X_{64}Y_{72}Y_{56}Y_{25}$ $=$ $W_{56}X_{72}Y_{47}Y_{64}Y_{25}$ $=$ $W_{56}X_{64}Y_{72}Y_{47}Y_{25}$ $=$ \\
& $=$ $W_{56}X_{64}Y_{13}Y_{35}Y_{41}$ $=$ $W_{56}X_{35}Y_{13}Y_{41}Y_{64}$ $=$ $X_{64}X_{85}Y_{18}Y_{41}Y_{56}$ $=$ \\
& $=$ $W_{56}X_{64}Y_{18}Y_{41}Y_{85}$ \\ 
\hline
$M_{6}$ & $X_{64}Y_{73}Y_{35}Y_{47}Y_{56}$ $=$ $X_{35}Y_{73}Y_{47}Y_{56}Y_{64}$ $=$ $X_{18}Y_{41}Y_{56}Y_{64}Y_{85}$ $=$ \\
& $=$ $X_{72}Y_{47}Y_{56}Y_{64}Y_{25}$ $=$ $X_{64}Y_{72}Y_{47}Y_{56}Y_{25}$ $=$ $X_{64}Y_{13}Y_{35}Y_{41}Y_{56}$ $=$ \\
& $=$ $X_{35}Y_{13}Y_{41}Y_{56}Y_{64}$ $=$ $X_{64}Y_{18}Y_{41}Y_{56}Y_{85}$ \\ 
\hline
$M_{2}$ & $X_{35}X_{47}X_{56}X_{64}Y_{73}$ $=$ $W_{56}X_{18}X_{41}X_{64}X_{85}$ $=$ $W_{56}X_{13}X_{35}X_{41}X_{64}$ $=$ \\
& $=$ $W_{56}X_{72}X_{47}X_{64}X_{25}$ $=$ $W_{56}X_{73}X_{35}X_{47}X_{64}$ $=$ $X_{18}X_{56}X_{64}X_{85}Y_{41}$ $=$ \\
& $=$ $X_{13}X_{35}X_{56}X_{64}Y_{41}$ $=$ $X_{72}X_{47}X_{56}X_{64}Y_{25}$ \\ 
\hline
$M_{8}$ & $X_{18}X_{41}X_{64}X_{85}Y_{56}$ $=$ $X_{13}X_{35}X_{41}X_{64}Y_{56}$ $=$ $X_{72}X_{47}X_{64}X_{25}Y_{56}$ $=$ \\
& $=$ $X_{73}X_{35}X_{47}X_{64}Y_{56}$ $=$ $X_{35}X_{47}X_{64}Y_{73}Z_{56}$ $=$ $X_{35}X_{56}X_{64}Y_{73}Y_{47}$ $=$ \\
& $=$ $W_{56}X_{72}X_{64}X_{25}Y_{47}$ $=$ $W_{56}X_{73}X_{35}X_{64}Y_{47}$ $=$ $X_{18}X_{64}X_{85}Y_{41}Z_{56}$ $=$ \\
& $=$ $X_{13}X_{35}X_{64}Y_{41}Z_{56}$ $=$ $W_{56}X_{18}X_{41}X_{64}Y_{85}$ $=$ $X_{18}X_{56}X_{64}Y_{41}Y_{85}$ $=$ \\
& $=$ $X_{72}X_{47}X_{64}Y_{25}Z_{56}$ $=$ $X_{72}X_{56}X_{64}Y_{47}Y_{25}$ $=$ $W_{56}X_{35}X_{41}X_{64}Y_{13}$ $=$ \\
& $=$ $X_{35}X_{56}X_{64}Y_{13}Y_{41}$ \\ 
\hline
$M_{9}$ & $X_{72}X_{64}X_{25}Y_{47}Y_{56}$ $=$ $X_{73}X_{35}X_{64}Y_{47}Y_{56}$ $=$ $X_{35}X_{64}Y_{73}Y_{47}Z_{56}$ $=$ \\
& $=$ $X_{18}X_{41}X_{64}Y_{56}Y_{85}$ $=$ $X_{18}X_{64}Y_{41}Y_{85}Z_{56}$ $=$ $X_{72}X_{64}Y_{47}Y_{25}Z_{56}$ $=$ \\
& $=$ $X_{35}X_{41}X_{64}Y_{13}Y_{56}$ $=$ $X_{35}X_{64}Y_{13}Y_{41}Z_{56}$ \\ 
\hline
$M_{10}$ & $X_{47}X_{56}X_{64}Y_{73}Y_{35}$ $=$ $X_{35}X_{47}X_{56}Y_{73}Y_{64}$ $=$ $W_{56}X_{13}X_{41}X_{64}Y_{35}$ $=$ \\
& $=$ $W_{56}X_{73}X_{47}X_{64}Y_{35}$ $=$ $W_{56}X_{18}X_{41}X_{85}Y_{64}$ $=$ $W_{56}X_{13}X_{35}X_{41}Y_{64}$ $=$ \\
& $=$ $W_{56}X_{72}X_{47}X_{25}Y_{64}$ $=$ $W_{56}X_{73}X_{35}X_{47}Y_{64}$ $=$ $W_{56}X_{47}X_{64}X_{25}Y_{72}$ $=$ \\
& $=$ $X_{13}X_{56}X_{64}Y_{35}Y_{41}$ $=$ $X_{18}X_{56}X_{85}Y_{41}Y_{64}$ $=$ $X_{13}X_{35}X_{56}Y_{41}Y_{64}$ $=$ \\
& $=$ $X_{72}X_{47}X_{56}Y_{64}Y_{25}$ $=$ $X_{47}X_{56}X_{64}Y_{72}Y_{25}$ $=$ $W_{56}X_{41}X_{64}X_{85}Y_{18}$ $=$ \\
& $=$ $X_{56}X_{64}X_{85}Y_{18}Y_{41}$ \\ 
\hline
$M_{11}$ & $X_{13}X_{41}X_{64}Y_{35}Y_{56}$ $=$ $X_{73}X_{47}X_{64}Y_{35}Y_{56}$ $=$ $X_{18}X_{41}X_{85}Y_{56}Y_{64}$ $=$ \\
& $=$ $X_{13}X_{35}X_{41}Y_{56}Y_{64}$ $=$ $X_{72}X_{47}X_{25}Y_{56}Y_{64}$ $=$ $X_{73}X_{35}X_{47}Y_{56}Y_{64}$ $=$ \\
& $=$ $X_{47}X_{64}X_{25}Y_{72}Y_{56}$ $=$ $X_{47}X_{64}Y_{73}Y_{35}Z_{56}$ $=$ $X_{35}X_{47}Y_{73}Y_{64}Z_{56}$ $=$ \\
& $=$ $X_{56}X_{64}Y_{73}Y_{35}Y_{47}$ $=$ $X_{35}X_{56}Y_{73}Y_{47}Y_{64}$ $=$ $W_{56}X_{73}X_{64}Y_{35}Y_{47}$ $=$ \\
& $=$ $W_{56}X_{72}X_{25}Y_{47}Y_{64}$ $=$ $W_{56}X_{73}X_{35}Y_{47}Y_{64}$ $=$ $W_{56}X_{64}X_{25}Y_{72}Y_{47}$ $=$ \\
& $=$ $X_{13}X_{64}Y_{35}Y_{41}Z_{56}$ $=$ $X_{18}X_{85}Y_{41}Y_{64}Z_{56}$ $=$ $X_{13}X_{35}Y_{41}Y_{64}Z_{56}$ $=$ \\
& $=$ $W_{56}X_{18}X_{41}Y_{64}Y_{85}$ $=$ $X_{18}X_{56}Y_{41}Y_{64}Y_{85}$ $=$ $X_{72}X_{47}Y_{64}Y_{25}Z_{56}$ $=$ \\
& $=$ $X_{47}X_{64}Y_{72}Y_{25}Z_{56}$ $=$ $X_{72}X_{56}Y_{47}Y_{64}Y_{25}$ $=$ $X_{56}X_{64}Y_{72}Y_{47}Y_{25}$ $=$ \\
& $=$ $W_{56}X_{41}X_{64}Y_{13}Y_{35}$ $=$ $W_{56}X_{35}X_{41}Y_{13}Y_{64}$ $=$ $X_{56}X_{64}Y_{13}Y_{35}Y_{41}$ $=$ \\
& $=$ $X_{35}X_{56}Y_{13}Y_{41}Y_{64}$ $=$ $X_{41}X_{64}X_{85}Y_{18}Y_{56}$ $=$ $X_{64}X_{85}Y_{18}Y_{41}Z_{56}$ $=$ \\
& $=$ $W_{56}X_{41}X_{64}Y_{18}Y_{85}$ $=$ $X_{56}X_{64}Y_{18}Y_{41}Y_{85}$ \\ 
\hline
$M_{12}$ & $X_{73}X_{64}Y_{35}Y_{47}Y_{56}$ $=$ $X_{72}X_{25}Y_{47}Y_{56}Y_{64}$ $=$ $X_{73}X_{35}Y_{47}Y_{56}Y_{64}$ $=$ \\
& $=$ $X_{64}X_{25}Y_{72}Y_{47}Y_{56}$ $=$ $X_{64}Y_{73}Y_{35}Y_{47}Z_{56}$ $=$ $X_{35}Y_{73}Y_{47}Y_{64}Z_{56}$ $=$ \\
& $=$ $X_{18}X_{41}Y_{56}Y_{64}Y_{85}$ $=$ $X_{18}Y_{41}Y_{64}Y_{85}Z_{56}$ $=$ $X_{72}Y_{47}Y_{64}Y_{25}Z_{56}$ $=$ \\
& $=$ $X_{64}Y_{72}Y_{47}Y_{25}Z_{56}$ $=$ $X_{41}X_{64}Y_{13}Y_{35}Y_{56}$ $=$ $X_{35}X_{41}Y_{13}Y_{56}Y_{64}$ $=$ \\
& $=$ $X_{64}Y_{13}Y_{35}Y_{41}Z_{56}$ $=$ $X_{35}Y_{13}Y_{41}Y_{64}Z_{56}$ $=$ $X_{41}X_{64}Y_{18}Y_{56}Y_{85}$ $=$ \\
& $=$ $X_{64}Y_{18}Y_{41}Y_{85}Z_{56}$ \\ 
\hline
$M_{13}$ & $X_{18}X_{41}X_{56}X_{64}X_{85}$ $=$ $X_{13}X_{35}X_{41}X_{56}X_{64}$ $=$ $X_{72}X_{47}X_{56}X_{64}X_{25}$ $=$ \\
& $=$ $X_{73}X_{35}X_{47}X_{56}X_{64}$ \\ 
\hline
$M_{14}$ & $X_{18}X_{41}X_{64}X_{85}Z_{56}$ $=$ $X_{13}X_{35}X_{41}X_{64}Z_{56}$ $=$ $X_{72}X_{47}X_{64}X_{25}Z_{56}$ $=$ \\
& $=$ $X_{73}X_{35}X_{47}X_{64}Z_{56}$ $=$ $X_{72}X_{56}X_{64}X_{25}Y_{47}$ $=$ $X_{73}X_{35}X_{56}X_{64}Y_{47}$ $=$ \\
& $=$ $X_{18}X_{41}X_{56}X_{64}Y_{85}$ $=$ $X_{35}X_{41}X_{56}X_{64}Y_{13}$ \\ 
\hline
$M_{15}$ & $X_{72}X_{64}X_{25}Y_{47}Z_{56}$ $=$ $X_{73}X_{35}X_{64}Y_{47}Z_{56}$ $=$ $X_{18}X_{41}X_{64}Y_{85}Z_{56}$ $=$ \\
& $=$ $X_{35}X_{41}X_{64}Y_{13}Z_{56}$ \\ 
\hline
$M_{16}$ & $X_{13}X_{41}X_{56}X_{64}Y_{35}$ $=$ $X_{73}X_{47}X_{56}X_{64}Y_{35}$ $=$ $X_{18}X_{41}X_{56}X_{85}Y_{64}$ $=$ \\
& $=$ $X_{13}X_{35}X_{41}X_{56}Y_{64}$ $=$ $X_{72}X_{47}X_{56}X_{25}Y_{64}$ $=$ $X_{73}X_{35}X_{47}X_{56}Y_{64}$ $=$ \\
& $=$ $X_{47}X_{56}X_{64}X_{25}Y_{72}$ $=$ $X_{41}X_{56}X_{64}X_{85}Y_{18}$ \\ 
\hline
$M_{17}$ & $X_{13}X_{41}X_{64}Y_{35}Z_{56}$ $=$ $X_{73}X_{47}X_{64}Y_{35}Z_{56}$ $=$ $X_{18}X_{41}X_{85}Y_{64}Z_{56}$ $=$ \\
& $=$ $X_{13}X_{35}X_{41}Y_{64}Z_{56}$ $=$ $X_{72}X_{47}X_{25}Y_{64}Z_{56}$ $=$ $X_{73}X_{35}X_{47}Y_{64}Z_{56}$ $=$ \\
& $=$ $X_{47}X_{64}X_{25}Y_{72}Z_{56}$ $=$ $X_{73}X_{56}X_{64}Y_{35}Y_{47}$ $=$ $X_{72}X_{56}X_{25}Y_{47}Y_{64}$ $=$ \\
& $=$ $X_{73}X_{35}X_{56}Y_{47}Y_{64}$ $=$ $X_{56}X_{64}X_{25}Y_{72}Y_{47}$ $=$ $X_{18}X_{41}X_{56}Y_{64}Y_{85}$ $=$ \\
& $=$ $X_{41}X_{56}X_{64}Y_{13}Y_{35}$ $=$ $X_{35}X_{41}X_{56}Y_{13}Y_{64}$ $=$ $X_{41}X_{64}X_{85}Y_{18}Z_{56}$ $=$ \\
& $=$ $X_{41}X_{56}X_{64}Y_{18}Y_{85}$ \\ 
\hline
$M_{18}$ & $X_{73}X_{64}Y_{35}Y_{47}Z_{56}$ $=$ $X_{72}X_{25}Y_{47}Y_{64}Z_{56}$ $=$ $X_{73}X_{35}Y_{47}Y_{64}Z_{56}$ $=$ \\
& $=$ $X_{64}X_{25}Y_{72}Y_{47}Z_{56}$ $=$ $X_{18}X_{41}Y_{64}Y_{85}Z_{56}$ $=$ $X_{41}X_{64}Y_{13}Y_{35}Z_{56}$ $=$ \\
& $=$ $X_{35}X_{41}Y_{13}Y_{64}Z_{56}$ $=$ $X_{41}X_{64}Y_{18}Y_{85}Z_{56}$ \\ 
\hline
$M_{19}$ & $W_{56}X_{47}Y_{73}Y_{35}Y_{64}$ $=$ $W_{56}X_{13}Y_{35}Y_{41}Y_{64}$ $=$ $W_{56}X_{47}Y_{72}Y_{64}Y_{25}$ $=$ \\
& $=$ $W_{56}X_{85}Y_{18}Y_{41}Y_{64}$ \\ 
\hline
$M_{20}$ & $X_{47}Y_{73}Y_{35}Y_{56}Y_{64}$ $=$ $W_{56}Y_{73}Y_{35}Y_{47}Y_{64}$ $=$ $X_{13}Y_{35}Y_{41}Y_{56}Y_{64}$ $=$ \\
& $=$ $X_{47}Y_{72}Y_{56}Y_{64}Y_{25}$ $=$ $W_{56}Y_{72}Y_{47}Y_{64}Y_{25}$ $=$ $W_{56}Y_{13}Y_{35}Y_{41}Y_{64}$ $=$ \\
& $=$ $X_{85}Y_{18}Y_{41}Y_{56}Y_{64}$ $=$ $W_{56}Y_{18}Y_{41}Y_{64}Y_{85}$ \\
\hline
$M_{21}$ & $Y_{73}Y_{35}Y_{47}Y_{56}Y_{64}$ $=$ $Y_{72}Y_{47}Y_{56}Y_{64}Y_{25}$ $=$ $Y_{13}Y_{35}Y_{41}Y_{56}Y_{64}$ $=$ \\
& $=$ $Y_{18}Y_{41}Y_{56}Y_{64}Y_{85}$ \\ 
\hline
$M_{27}$ & $X_{47}X_{56}Y_{73}Y_{35}Y_{64}$ $=$ $W_{56}X_{13}X_{41}Y_{35}Y_{64}$ $=$ $W_{56}X_{73}X_{47}Y_{35}Y_{64}$ $=$ \\
& $=$ $W_{56}X_{47}X_{25}Y_{72}Y_{64}$ $=$ $X_{13}X_{56}Y_{35}Y_{41}Y_{64}$ $=$ $X_{47}X_{56}Y_{72}Y_{64}Y_{25}$ $=$ \\
& $=$ $W_{56}X_{41}X_{85}Y_{18}Y_{64}$ $=$ $X_{56}X_{85}Y_{18}Y_{41}Y_{64}$ \\ 
\hline
$M_{23}$ & $X_{13}X_{41}Y_{35}Y_{56}Y_{64}$ $=$ $X_{73}X_{47}Y_{35}Y_{56}Y_{64}$ $=$ $X_{47}X_{25}Y_{72}Y_{56}Y_{64}$ $=$ \\
& $=$ $X_{47}Y_{73}Y_{35}Y_{64}Z_{56}$ $=$ $X_{56}Y_{73}Y_{35}Y_{47}Y_{64}$ $=$ $W_{56}X_{73}Y_{35}Y_{47}Y_{64}$ $=$ \\
& $=$ $W_{56}X_{25}Y_{72}Y_{47}Y_{64}$ $=$ $X_{13}Y_{35}Y_{41}Y_{64}Z_{56}$ $=$ $X_{47}Y_{72}Y_{64}Y_{25}Z_{56}$ $=$ \\
& $=$ $X_{56}Y_{72}Y_{47}Y_{64}Y_{25}$ $=$ $W_{56}X_{41}Y_{13}Y_{35}Y_{64}$ $=$ $X_{56}Y_{13}Y_{35}Y_{41}Y_{64}$ $=$ \\
& $=$ $X_{41}X_{85}Y_{18}Y_{56}Y_{64}$ $=$ $X_{85}Y_{18}Y_{41}Y_{64}Z_{56}$ $=$ $W_{56}X_{41}Y_{18}Y_{64}Y_{85}$ $=$ \\
& $=$ $X_{56}Y_{18}Y_{41}Y_{64}Y_{85}$ \\ 
\hline
$M_{24}$ & $X_{73}Y_{35}Y_{47}Y_{56}Y_{64}$ $=$ $X_{25}Y_{72}Y_{47}Y_{56}Y_{64}$ $=$ $Y_{73}Y_{35}Y_{47}Y_{64}Z_{56}$ $=$ \\
& $=$ $Y_{72}Y_{47}Y_{64}Y_{25}Z_{56}$ $=$ $X_{41}Y_{13}Y_{35}Y_{56}Y_{64}$ $=$ $Y_{13}Y_{35}Y_{41}Y_{64}Z_{56}$ $=$ \\
& $=$ $X_{41}Y_{18}Y_{56}Y_{64}Y_{85}$ $=$ $Y_{18}Y_{41}Y_{64}Y_{85}Z_{56}$ \\ 
\hline
$M_{25}$ & $X_{13}X_{41}X_{56}Y_{35}Y_{64}$ $=$ $X_{73}X_{47}X_{56}Y_{35}Y_{64}$ $=$ $X_{47}X_{56}X_{25}Y_{72}Y_{64}$ $=$ \\
& $=$ $X_{41}X_{56}X_{85}Y_{18}Y_{64}$ \\ 
\hline
$M_{26}$ & $X_{13}X_{41}Y_{35}Y_{64}Z_{56}$ $=$ $X_{73}X_{47}Y_{35}Y_{64}Z_{56}$ $=$ $X_{47}X_{25}Y_{72}Y_{64}Z_{56}$ $=$ \\
& $=$ $X_{73}X_{56}Y_{35}Y_{47}Y_{64}$ $=$ $X_{56}X_{25}Y_{72}Y_{47}Y_{64}$ $=$ $X_{41}X_{56}Y_{13}Y_{35}Y_{64}$ $=$ \\
& $=$ $X_{41}X_{85}Y_{18}Y_{64}Z_{56}$ $=$ $X_{41}X_{56}Y_{18}Y_{64}Y_{85}$ \\ 
\hline
$M_{27}$ & $X_{73}Y_{35}Y_{47}Y_{64}Z_{56}$ $=$ $X_{25}Y_{72}Y_{47}Y_{64}Z_{56}$ $=$ $X_{41}Y_{13}Y_{35}Y_{64}Z_{56}$ $=$ \\
& $=$ $X_{41}Y_{18}Y_{64}Y_{85}Z_{56}$
	\end{longtable}
\end{center}

\begin{center}
		\renewcommand{\arraystretch}{1.1}
	\begin{longtable}{c|c}
		\caption{Generators of $Q^{1,1,1}/\ZZ_2$ in Phase H.}\\
		Field & Chiral superfields\\
		\hline
		\endhead
		\label{tab:GenQ111Z2PhaseH}$M_{1}$ & $X_{42}X_{56}X_{64}X_{25}$ $=$ $X_{48}X_{56}X_{64}X_{85}$ $=$ $X_{14}X_{31}X_{42}X_{53}X_{25}$ $=$ \\
& $=$ $X_{74}X_{37}X_{42}X_{53}X_{25}$ $=$ $X_{14}X_{31}X_{48}X_{53}X_{85}$ $=$ $X_{74}X_{37}X_{48}X_{53}X_{85}$ \\ 
\hline
	$M_{2}$ & $X_{37}X_{42}X_{53}X_{25}Y_{74}$ $=$ $X_{37}X_{48}X_{53}X_{85}Y_{74}$ $=$ $X_{56}X_{64}X_{25}Z_{42}$ $=$ \\
& $=$ $X_{14}X_{31}X_{53}X_{25}Z_{42}$ $=$ $X_{74}X_{37}X_{53}X_{25}Z_{42}$ $=$ $X_{14}X_{42}X_{53}X_{25}Y_{31}$ $=$ \\
& $=$ $X_{14}X_{48}X_{53}X_{85}Y_{31}$ $=$ $X_{42}X_{64}X_{25}Y_{56}$ $=$ $X_{48}X_{64}X_{85}Y_{56}$ $=$ \\
& $=$ $X_{48}X_{56}X_{64}Y_{85}$ $=$ $X_{14}X_{31}X_{48}X_{53}Y_{85}$ $=$ $X_{74}X_{37}X_{48}X_{53}Y_{85}$ \\ 
	\hline
	$M_{3}$ & $X_{37}X_{53}X_{25}Y_{74}Z_{42}$ $=$ $X_{14}X_{53}X_{25}Y_{31}Z_{42}$ $=$ $X_{64}X_{25}Y_{56}Z_{42}$ $=$ \\
& $=$ $X_{37}X_{48}X_{53}Y_{74}Y_{85}$ $=$ $X_{14}X_{48}X_{53}Y_{31}Y_{85}$ $=$ $X_{48}X_{64}Y_{56}Y_{85}$ \\ 
	\hline
	$M_{4}$ & $X_{56}X_{64}X_{85}Y_{48}$ $=$ $X_{14}X_{31}X_{53}X_{85}Y_{48}$ $=$ $X_{74}X_{37}X_{53}X_{85}Y_{48}$ $=$ \\
& $=$ $X_{56}X_{64}X_{25}Y_{42}$ $=$ $X_{14}X_{31}X_{53}X_{25}Y_{42}$ $=$ $X_{74}X_{37}X_{53}X_{25}Y_{42}$ $=$ \\
& $=$ $X_{42}X_{56}X_{25}Y_{64}$ $=$ $X_{48}X_{56}X_{85}Y_{64}$ $=$ $X_{14}X_{31}X_{42}X_{25}Y_{53}$ $=$ \\
& $=$ $X_{74}X_{37}X_{42}X_{25}Y_{53}$ $=$ $X_{14}X_{31}X_{48}X_{85}Y_{53}$ $=$ $X_{74}X_{37}X_{48}X_{85}Y_{53}$ \\ 
	\hline
	$M_{5}$ & $X_{37}X_{53}X_{85}Y_{74}Y_{48}$ $=$ $X_{37}X_{53}X_{25}Y_{74}Y_{42}$ $=$ $W_{42}X_{56}X_{64}X_{25}$ $=$ \\
& $=$ $W_{42}X_{14}X_{31}X_{53}X_{25}$ $=$ $W_{42}X_{74}X_{37}X_{53}X_{25}$ $=$ $X_{14}X_{53}X_{85}Y_{31}Y_{48}$ $=$ \\
& $=$ $X_{14}X_{53}X_{25}Y_{31}Y_{42}$ $=$ $X_{56}X_{25}Y_{64}Z_{42}$ $=$ $X_{37}X_{42}X_{25}Y_{74}Y_{53}$ $=$ \\
& $=$ $X_{37}X_{48}X_{85}Y_{74}Y_{53}$ $=$ $X_{14}X_{31}X_{25}Y_{53}Z_{42}$ $=$ $X_{74}X_{37}X_{25}Y_{53}Z_{42}$ $=$ \\
& $=$ $X_{14}X_{42}X_{25}Y_{31}Y_{53}$ $=$ $X_{14}X_{48}X_{85}Y_{31}Y_{53}$ $=$ $X_{64}X_{85}Y_{48}Y_{56}$ $=$ \\
& $=$ $X_{64}X_{25}Y_{42}Y_{56}$ $=$ $X_{42}X_{25}Y_{56}Y_{64}$ $=$ $X_{48}X_{85}Y_{56}Y_{64}$ $=$ \\
& $=$ $X_{56}X_{64}Y_{48}Y_{85}$ $=$ $X_{14}X_{31}X_{53}Y_{48}Y_{85}$ $=$ $X_{74}X_{37}X_{53}Y_{48}Y_{85}$ $=$ \\
& $=$ $X_{48}X_{56}Y_{64}Y_{85}$ $=$ $X_{14}X_{31}X_{48}Y_{53}Y_{85}$ $=$ $X_{74}X_{37}X_{48}Y_{53}Y_{85}$ \\ 
	\hline
	$M_{6}$ & $W_{42}X_{37}X_{53}X_{25}Y_{74}$ $=$ $W_{42}X_{14}X_{53}X_{25}Y_{31}$ $=$ $X_{37}X_{25}Y_{74}Y_{53}Z_{42}$ $=$ \\
& $=$ $X_{14}X_{25}Y_{31}Y_{53}Z_{42}$ $=$ $W_{42}X_{64}X_{25}Y_{56}$ $=$ $X_{25}Y_{56}Y_{64}Z_{42}$ $=$ \\
& $=$ $X_{37}X_{53}Y_{74}Y_{48}Y_{85}$ $=$ $X_{14}X_{53}Y_{31}Y_{48}Y_{85}$ $=$ $X_{37}X_{48}Y_{74}Y_{53}Y_{85}$ $=$ \\
& $=$ $X_{14}X_{48}Y_{31}Y_{53}Y_{85}$ $=$ $X_{64}Y_{48}Y_{56}Y_{85}$ $=$ $X_{48}Y_{56}Y_{64}Y_{85}$ \\ 
	\hline
	$M_{2}$ & $X_{31}X_{42}X_{53}X_{25}Y_{14}$ $=$ $X_{31}X_{48}X_{53}X_{85}Y_{14}$ $=$ $X_{56}X_{64}X_{85}Z_{48}$ $=$ \\
& $=$ $X_{14}X_{31}X_{53}X_{85}Z_{48}$ $=$ $X_{74}X_{37}X_{53}X_{85}Z_{48}$ $=$ $X_{74}X_{42}X_{53}X_{25}Y_{37}$ $=$ \\
& $=$ $X_{74}X_{48}X_{53}X_{85}Y_{37}$ $=$ $X_{42}X_{64}X_{25}Z_{56}$ $=$ $X_{48}X_{64}X_{85}Z_{56}$ $=$ \\
& $=$ $X_{42}X_{56}X_{64}Y_{25}$ $=$ $X_{14}X_{31}X_{42}X_{53}Y_{25}$ $=$ $X_{74}X_{37}X_{42}X_{53}Y_{25}$ \\ 
	\hline
	$M_{8}$ & $X_{37}X_{53}X_{85}Y_{74}Z_{48}$ $=$ $X_{31}X_{53}X_{25}Y_{14}Z_{42}$ $=$ $X_{42}X_{53}X_{25}Y_{74}Y_{37}$ $=$ \\
& $=$ $X_{48}X_{53}X_{85}Y_{74}Y_{37}$ $=$ $X_{74}X_{53}X_{25}Y_{37}Z_{42}$ $=$ $X_{42}X_{53}X_{25}Y_{14}Y_{31}$ $=$ \\
& $=$ $X_{48}X_{53}X_{85}Y_{14}Y_{31}$ $=$ $X_{14}X_{53}X_{85}Y_{31}Z_{48}$ $=$ $X_{64}X_{85}Y_{56}Z_{48}$ $=$ \\
& $=$ $X_{64}X_{25}Z_{42}Z_{56}$ $=$ $W_{56}X_{42}X_{64}X_{25}$ $=$ $W_{56}X_{48}X_{64}X_{85}$ $=$ \\
& $=$ $X_{31}X_{48}X_{53}Y_{14}Y_{85}$ $=$ $X_{56}X_{64}Y_{85}Z_{48}$ $=$ $X_{14}X_{31}X_{53}Y_{85}Z_{48}$ $=$ \\
& $=$ $X_{74}X_{37}X_{53}Y_{85}Z_{48}$ $=$ $X_{74}X_{48}X_{53}Y_{37}Y_{85}$ $=$ $X_{48}X_{64}Y_{85}Z_{56}$ $=$ \\
& $=$ $X_{37}X_{42}X_{53}Y_{74}Y_{25}$ $=$ $X_{56}X_{64}Y_{25}Z_{42}$ $=$ $X_{14}X_{31}X_{53}Y_{25}Z_{42}$ $=$ \\
& $=$ $X_{74}X_{37}X_{53}Y_{25}Z_{42}$ $=$ $X_{14}X_{42}X_{53}Y_{31}Y_{25}$ $=$ $X_{42}X_{64}Y_{56}Y_{25}$ \\ 
	\hline
	$M_{9}$ & $X_{53}X_{25}Y_{74}Y_{37}Z_{42}$ $=$ $X_{53}X_{25}Y_{14}Y_{31}Z_{42}$ $=$ $W_{56}X_{64}X_{25}Z_{42}$ $=$ \\
& $=$ $X_{37}X_{53}Y_{74}Y_{85}Z_{48}$ $=$ $X_{48}X_{53}Y_{74}Y_{37}Y_{85}$ $=$ $X_{48}X_{53}Y_{14}Y_{31}Y_{85}$ $=$ \\
& $=$ $X_{14}X_{53}Y_{31}Y_{85}Z_{48}$ $=$ $X_{64}Y_{56}Y_{85}Z_{48}$ $=$ $W_{56}X_{48}X_{64}Y_{85}$ $=$ \\
& $=$ $X_{37}X_{53}Y_{74}Y_{25}Z_{42}$ $=$ $X_{14}X_{53}Y_{31}Y_{25}Z_{42}$ $=$ $X_{64}Y_{56}Y_{25}Z_{42}$ \\ 
	\hline
	$M_{10}$ & $X_{31}X_{53}X_{85}Y_{14}Y_{48}$ $=$ $W_{48}X_{56}X_{64}X_{85}$ $=$ $W_{48}X_{14}X_{31}X_{53}X_{85}$ $=$ \\
& $=$ $W_{48}X_{74}X_{37}X_{53}X_{85}$ $=$ $X_{31}X_{53}X_{25}Y_{14}Y_{42}$ $=$ $X_{74}X_{53}X_{85}Y_{37}Y_{48}$ $=$ \\
& $=$ $X_{74}X_{53}X_{25}Y_{37}Y_{42}$ $=$ $X_{56}X_{85}Y_{64}Z_{48}$ $=$ $X_{31}X_{42}X_{25}Y_{14}Y_{53}$ $=$ \\
& $=$ $X_{31}X_{48}X_{85}Y_{14}Y_{53}$ $=$ $X_{14}X_{31}X_{85}Y_{53}Z_{48}$ $=$ $X_{74}X_{37}X_{85}Y_{53}Z_{48}$ $=$ \\
& $=$ $X_{74}X_{42}X_{25}Y_{37}Y_{53}$ $=$ $X_{74}X_{48}X_{85}Y_{37}Y_{53}$ $=$ $X_{64}X_{85}Y_{48}Z_{56}$ $=$ \\
& $=$ $X_{64}X_{25}Y_{42}Z_{56}$ $=$ $X_{42}X_{25}Y_{64}Z_{56}$ $=$ $X_{48}X_{85}Y_{64}Z_{56}$ $=$ \\
& $=$ $X_{56}X_{64}Y_{42}Y_{25}$ $=$ $X_{14}X_{31}X_{53}Y_{42}Y_{25}$ $=$ $X_{74}X_{37}X_{53}Y_{42}Y_{25}$ $=$ \\
& $=$ $X_{42}X_{56}Y_{64}Y_{25}$ $=$ $X_{14}X_{31}X_{42}Y_{53}Y_{25}$ $=$ $X_{74}X_{37}X_{42}Y_{53}Y_{25}$ \\ 
	\hline
	$M_{11}$ & $W_{48}X_{37}X_{53}X_{85}Y_{74}$ $=$ $W_{42}X_{31}X_{53}X_{25}Y_{14}$ $=$ $X_{53}X_{85}Y_{74}Y_{37}Y_{48}$ $=$ \\
& $=$ $X_{53}X_{25}Y_{74}Y_{37}Y_{42}$ $=$ $W_{42}X_{74}X_{53}X_{25}Y_{37}$ $=$ $X_{53}X_{85}Y_{14}Y_{31}Y_{48}$ $=$ \\
& $=$ $W_{48}X_{14}X_{53}X_{85}Y_{31}$ $=$ $X_{53}X_{25}Y_{14}Y_{31}Y_{42}$ $=$ $X_{37}X_{85}Y_{74}Y_{53}Z_{48}$ $=$ \\
& $=$ $X_{31}X_{25}Y_{14}Y_{53}Z_{42}$ $=$ $X_{42}X_{25}Y_{74}Y_{37}Y_{53}$ $=$ $X_{48}X_{85}Y_{74}Y_{37}Y_{53}$ $=$ \\
& $=$ $X_{74}X_{25}Y_{37}Y_{53}Z_{42}$ $=$ $X_{42}X_{25}Y_{14}Y_{31}Y_{53}$ $=$ $X_{48}X_{85}Y_{14}Y_{31}Y_{53}$ $=$ \\
& $=$ $X_{14}X_{85}Y_{31}Y_{53}Z_{48}$ $=$ $W_{48}X_{64}X_{85}Y_{56}$ $=$ $X_{85}Y_{56}Y_{64}Z_{48}$ $=$ \\
& $=$ $W_{42}X_{64}X_{25}Z_{56}$ $=$ $X_{25}Y_{64}Z_{42}Z_{56}$ $=$ $W_{56}X_{64}X_{85}Y_{48}$ $=$ \\
& $=$ $W_{56}X_{64}X_{25}Y_{42}$ $=$ $W_{56}X_{42}X_{25}Y_{64}$ $=$ $W_{56}X_{48}X_{85}Y_{64}$ $=$ \\
& $=$ $X_{31}X_{53}Y_{14}Y_{48}Y_{85}$ $=$ $W_{48}X_{56}X_{64}Y_{85}$ $=$ $W_{48}X_{14}X_{31}X_{53}Y_{85}$ $=$ \\
& $=$ $W_{48}X_{74}X_{37}X_{53}Y_{85}$ $=$ $X_{74}X_{53}Y_{37}Y_{48}Y_{85}$ $=$ $X_{56}Y_{64}Y_{85}Z_{48}$ $=$ \\
& $=$ $X_{31}X_{48}Y_{14}Y_{53}Y_{85}$ $=$ $X_{14}X_{31}Y_{53}Y_{85}Z_{48}$ $=$ $X_{74}X_{37}Y_{53}Y_{85}Z_{48}$ $=$ \\
& $=$ $X_{74}X_{48}Y_{37}Y_{53}Y_{85}$ $=$ $X_{64}Y_{48}Y_{85}Z_{56}$ $=$ $X_{48}Y_{64}Y_{85}Z_{56}$ $=$ \\
& $=$ $X_{37}X_{53}Y_{74}Y_{42}Y_{25}$ $=$ $W_{42}X_{56}X_{64}Y_{25}$ $=$ $W_{42}X_{14}X_{31}X_{53}Y_{25}$ $=$ \\
& $=$ $W_{42}X_{74}X_{37}X_{53}Y_{25}$ $=$ $X_{14}X_{53}Y_{31}Y_{42}Y_{25}$ $=$ $X_{56}Y_{64}Y_{25}Z_{42}$ $=$ \\
& $=$ $X_{37}X_{42}Y_{74}Y_{53}Y_{25}$ $=$ $X_{14}X_{31}Y_{53}Y_{25}Z_{42}$ $=$ $X_{74}X_{37}Y_{53}Y_{25}Z_{42}$ $=$ \\
& $=$ $X_{14}X_{42}Y_{31}Y_{53}Y_{25}$ $=$ $X_{64}Y_{42}Y_{56}Y_{25}$ $=$ $X_{42}Y_{56}Y_{64}Y_{25}$ \\ 
	\hline
	$M_{12}$ & $W_{42}X_{53}X_{25}Y_{74}Y_{37}$ $=$ $W_{42}X_{53}X_{25}Y_{14}Y_{31}$ $=$ $X_{25}Y_{74}Y_{37}Y_{53}Z_{42}$ $=$ \\
& $=$ $X_{25}Y_{14}Y_{31}Y_{53}Z_{42}$ $=$ $W_{42}W_{56}X_{64}X_{25}$ $=$ $W_{56}X_{25}Y_{64}Z_{42}$ $=$ \\
& $=$ $W_{48}X_{37}X_{53}Y_{74}Y_{85}$ $=$ $X_{53}Y_{74}Y_{37}Y_{48}Y_{85}$ $=$ $X_{53}Y_{14}Y_{31}Y_{48}Y_{85}$ $=$ \\
& $=$ $W_{48}X_{14}X_{53}Y_{31}Y_{85}$ $=$ $X_{37}Y_{74}Y_{53}Y_{85}Z_{48}$ $=$ $X_{48}Y_{74}Y_{37}Y_{53}Y_{85}$ $=$ \\
& $=$ $X_{48}Y_{14}Y_{31}Y_{53}Y_{85}$ $=$ $X_{14}Y_{31}Y_{53}Y_{85}Z_{48}$ $=$ $W_{48}X_{64}Y_{56}Y_{85}$ $=$ \\
& $=$ $Y_{56}Y_{64}Y_{85}Z_{48}$ $=$ $W_{56}X_{64}Y_{48}Y_{85}$ $=$ $W_{56}X_{48}Y_{64}Y_{85}$ $=$ \\
& $=$ $W_{42}X_{37}X_{53}Y_{74}Y_{25}$ $=$ $W_{42}X_{14}X_{53}Y_{31}Y_{25}$ $=$ $X_{37}Y_{74}Y_{53}Y_{25}Z_{42}$ $=$ \\
& $=$ $X_{14}Y_{31}Y_{53}Y_{25}Z_{42}$ $=$ $W_{42}X_{64}Y_{56}Y_{25}$ $=$ $Y_{56}Y_{64}Y_{25}Z_{42}$ \\ 
	\hline
	$M_{13}$ & $X_{31}X_{53}X_{85}Y_{14}Z_{48}$ $=$ $X_{74}X_{53}X_{85}Y_{37}Z_{48}$ $=$ $X_{64}X_{85}Z_{48}Z_{56}$ $=$ \\
& $=$ $X_{31}X_{42}X_{53}Y_{14}Y_{25}$ $=$ $X_{74}X_{42}X_{53}Y_{37}Y_{25}$ $=$ $X_{42}X_{64}Y_{25}Z_{56}$ \\ 
	\hline
	$M_{14}$ & $X_{53}X_{85}Y_{74}Y_{37}Z_{48}$ $=$ $X_{53}X_{85}Y_{14}Y_{31}Z_{48}$ $=$ $W_{56}X_{64}X_{85}Z_{48}$ $=$ \\
& $=$ $X_{31}X_{53}Y_{14}Y_{85}Z_{48}$ $=$ $X_{74}X_{53}Y_{37}Y_{85}Z_{48}$ $=$ $X_{64}Y_{85}Z_{48}Z_{56}$ $=$ \\
& $=$ $X_{31}X_{53}Y_{14}Y_{25}Z_{42}$ $=$ $X_{42}X_{53}Y_{74}Y_{37}Y_{25}$ $=$ $X_{74}X_{53}Y_{37}Y_{25}Z_{42}$ $=$ \\
& $=$ $X_{42}X_{53}Y_{14}Y_{31}Y_{25}$ $=$ $X_{64}Y_{25}Z_{42}Z_{56}$ $=$ $W_{56}X_{42}X_{64}Y_{25}$ \\ 
	\hline
	$M_{15}$ & $X_{53}Y_{74}Y_{37}Y_{85}Z_{48}$ $=$ $X_{53}Y_{14}Y_{31}Y_{85}Z_{48}$ $=$ $W_{56}X_{64}Y_{85}Z_{48}$ $=$ \\
& $=$ $X_{53}Y_{74}Y_{37}Y_{25}Z_{42}$ $=$ $X_{53}Y_{14}Y_{31}Y_{25}Z_{42}$ $=$ $W_{56}X_{64}Y_{25}Z_{42}$ \\ 
	\hline
	$M_{16}$ & $W_{48}X_{31}X_{53}X_{85}Y_{14}$ $=$ $W_{48}X_{74}X_{53}X_{85}Y_{37}$ $=$ $X_{31}X_{85}Y_{14}Y_{53}Z_{48}$ $=$ \\
& $=$ $X_{74}X_{85}Y_{37}Y_{53}Z_{48}$ $=$ $W_{48}X_{64}X_{85}Z_{56}$ $=$ $X_{85}Y_{64}Z_{48}Z_{56}$ $=$ \\
& $=$ $X_{31}X_{53}Y_{14}Y_{42}Y_{25}$ $=$ $X_{74}X_{53}Y_{37}Y_{42}Y_{25}$ $=$ $X_{31}X_{42}Y_{14}Y_{53}Y_{25}$ $=$ \\
& $=$ $X_{74}X_{42}Y_{37}Y_{53}Y_{25}$ $=$ $X_{64}Y_{42}Y_{25}Z_{56}$ $=$ $X_{42}Y_{64}Y_{25}Z_{56}$ \\ 
	\hline
	$M_{17}$ & $W_{48}X_{53}X_{85}Y_{74}Y_{37}$ $=$ $W_{48}X_{53}X_{85}Y_{14}Y_{31}$ $=$ $X_{85}Y_{74}Y_{37}Y_{53}Z_{48}$ $=$ \\
& $=$ $X_{85}Y_{14}Y_{31}Y_{53}Z_{48}$ $=$ $W_{48}W_{56}X_{64}X_{85}$ $=$ $W_{56}X_{85}Y_{64}Z_{48}$ $=$ \\
& $=$ $W_{48}X_{31}X_{53}Y_{14}Y_{85}$ $=$ $W_{48}X_{74}X_{53}Y_{37}Y_{85}$ $=$ $X_{31}Y_{14}Y_{53}Y_{85}Z_{48}$ $=$ \\
& $=$ $X_{74}Y_{37}Y_{53}Y_{85}Z_{48}$ $=$ $W_{48}X_{64}Y_{85}Z_{56}$ $=$ $Y_{64}Y_{85}Z_{48}Z_{56}$ $=$ \\
& $=$ $W_{42}X_{31}X_{53}Y_{14}Y_{25}$ $=$ $X_{53}Y_{74}Y_{37}Y_{42}Y_{25}$ $=$ $W_{42}X_{74}X_{53}Y_{37}Y_{25}$ $=$ \\
& $=$ $X_{53}Y_{14}Y_{31}Y_{42}Y_{25}$ $=$ $X_{31}Y_{14}Y_{53}Y_{25}Z_{42}$ $=$ $X_{42}Y_{74}Y_{37}Y_{53}Y_{25}$ $=$ \\
& $=$ $X_{74}Y_{37}Y_{53}Y_{25}Z_{42}$ $=$ $X_{42}Y_{14}Y_{31}Y_{53}Y_{25}$ $=$ $W_{42}X_{64}Y_{25}Z_{56}$ $=$ \\
& $=$ $Y_{64}Y_{25}Z_{42}Z_{56}$ $=$ $W_{56}X_{64}Y_{42}Y_{25}$ $=$ $W_{56}X_{42}Y_{64}Y_{25}$ \\ 
	\hline
	$M_{18}$ & $W_{48}X_{53}Y_{74}Y_{37}Y_{85}$ $=$ $W_{48}X_{53}Y_{14}Y_{31}Y_{85}$ $=$ $Y_{74}Y_{37}Y_{53}Y_{85}Z_{48}$ $=$ \\
& $=$ $Y_{14}Y_{31}Y_{53}Y_{85}Z_{48}$ $=$ $W_{48}W_{56}X_{64}Y_{85}$ $=$ $W_{56}Y_{64}Y_{85}Z_{48}$ $=$ \\
& $=$ $W_{42}X_{53}Y_{74}Y_{37}Y_{25}$ $=$ $W_{42}X_{53}Y_{14}Y_{31}Y_{25}$ $=$ $Y_{74}Y_{37}Y_{53}Y_{25}Z_{42}$ $=$ \\
& $=$ $Y_{14}Y_{31}Y_{53}Y_{25}Z_{42}$ $=$ $W_{42}W_{56}X_{64}Y_{25}$ $=$ $W_{56}Y_{64}Y_{25}Z_{42}$ \\ 
	\hline
	$M_{19}$ & $X_{56}X_{85}Y_{48}Y_{64}$ $=$ $X_{56}X_{25}Y_{42}Y_{64}$ $=$ $X_{14}X_{31}X_{85}Y_{48}Y_{53}$ $=$ \\
& $=$ $X_{74}X_{37}X_{85}Y_{48}Y_{53}$ $=$ $X_{14}X_{31}X_{25}Y_{42}Y_{53}$ $=$ $X_{74}X_{37}X_{25}Y_{42}Y_{53}$ \\ 
	\hline
	$M_{20}$ & $W_{42}X_{56}X_{25}Y_{64}$ $=$ $X_{37}X_{85}Y_{74}Y_{48}Y_{53}$ $=$ $X_{37}X_{25}Y_{74}Y_{42}Y_{53}$ $=$ \\
& $=$ $W_{42}X_{14}X_{31}X_{25}Y_{53}$ $=$ $W_{42}X_{74}X_{37}X_{25}Y_{53}$ $=$ $X_{14}X_{85}Y_{31}Y_{48}Y_{53}$ $=$ \\
& $=$ $X_{14}X_{25}Y_{31}Y_{42}Y_{53}$ $=$ $X_{85}Y_{48}Y_{56}Y_{64}$ $=$ $X_{25}Y_{42}Y_{56}Y_{64}$ $=$ \\
& $=$ $X_{56}Y_{48}Y_{64}Y_{85}$ $=$ $X_{14}X_{31}Y_{48}Y_{53}Y_{85}$ $=$ $X_{74}X_{37}Y_{48}Y_{53}Y_{85}$ \\ 
	\hline
	$M_{21}$ & $W_{42}X_{37}X_{25}Y_{74}Y_{53}$ $=$ $W_{42}X_{14}X_{25}Y_{31}Y_{53}$ $=$ $W_{42}X_{25}Y_{56}Y_{64}$ $=$ \\
& $=$ $X_{37}Y_{74}Y_{48}Y_{53}Y_{85}$ $=$ $X_{14}Y_{31}Y_{48}Y_{53}Y_{85}$ $=$ $Y_{48}Y_{56}Y_{64}Y_{85}$ \\ 
	\hline
	$M_{27}$ & $W_{48}X_{56}X_{85}Y_{64}$ $=$ $X_{31}X_{85}Y_{14}Y_{48}Y_{53}$ $=$ $W_{48}X_{14}X_{31}X_{85}Y_{53}$ $=$ \\
& $=$ $W_{48}X_{74}X_{37}X_{85}Y_{53}$ $=$ $X_{31}X_{25}Y_{14}Y_{42}Y_{53}$ $=$ $X_{74}X_{85}Y_{37}Y_{48}Y_{53}$ $=$ \\
& $=$ $X_{74}X_{25}Y_{37}Y_{42}Y_{53}$ $=$ $X_{85}Y_{48}Y_{64}Z_{56}$ $=$ $X_{25}Y_{42}Y_{64}Z_{56}$ $=$ \\
& $=$ $X_{56}Y_{42}Y_{64}Y_{25}$ $=$ $X_{14}X_{31}Y_{42}Y_{53}Y_{25}$ $=$ $X_{74}X_{37}Y_{42}Y_{53}Y_{25}$ \\ 
	\hline
	$M_{23}$ & $W_{48}X_{37}X_{85}Y_{74}Y_{53}$ $=$ $W_{42}X_{31}X_{25}Y_{14}Y_{53}$ $=$ $X_{85}Y_{74}Y_{37}Y_{48}Y_{53}$ $=$ \\
& $=$ $X_{25}Y_{74}Y_{37}Y_{42}Y_{53}$ $=$ $W_{42}X_{74}X_{25}Y_{37}Y_{53}$ $=$ $X_{85}Y_{14}Y_{31}Y_{48}Y_{53}$ $=$ \\
& $=$ $W_{48}X_{14}X_{85}Y_{31}Y_{53}$ $=$ $X_{25}Y_{14}Y_{31}Y_{42}Y_{53}$ $=$ $W_{48}X_{85}Y_{56}Y_{64}$ $=$ \\
& $=$ $W_{42}X_{25}Y_{64}Z_{56}$ $=$ $W_{56}X_{85}Y_{48}Y_{64}$ $=$ $W_{56}X_{25}Y_{42}Y_{64}$ $=$ \\
& $=$ $W_{48}X_{56}Y_{64}Y_{85}$ $=$ $X_{31}Y_{14}Y_{48}Y_{53}Y_{85}$ $=$ $W_{48}X_{14}X_{31}Y_{53}Y_{85}$ $=$ \\
& $=$ $W_{48}X_{74}X_{37}Y_{53}Y_{85}$ $=$ $X_{74}Y_{37}Y_{48}Y_{53}Y_{85}$ $=$ $Y_{48}Y_{64}Y_{85}Z_{56}$ $=$ \\
& $=$ $W_{42}X_{56}Y_{64}Y_{25}$ $=$ $X_{37}Y_{74}Y_{42}Y_{53}Y_{25}$ $=$ $W_{42}X_{14}X_{31}Y_{53}Y_{25}$ $=$ \\
& $=$ $W_{42}X_{74}X_{37}Y_{53}Y_{25}$ $=$ $X_{14}Y_{31}Y_{42}Y_{53}Y_{25}$ $=$ $Y_{42}Y_{56}Y_{64}Y_{25}$ \\ 
	\hline
	$M_{24}$ & $W_{42}X_{25}Y_{74}Y_{37}Y_{53}$ $=$ $W_{42}X_{25}Y_{14}Y_{31}Y_{53}$ $=$ $W_{42}W_{56}X_{25}Y_{64}$ $=$ \\
& $=$ $W_{48}X_{37}Y_{74}Y_{53}Y_{85}$ $=$ $Y_{74}Y_{37}Y_{48}Y_{53}Y_{85}$ $=$ $Y_{14}Y_{31}Y_{48}Y_{53}Y_{85}$ $=$ \\
& $=$ $W_{48}X_{14}Y_{31}Y_{53}Y_{85}$ $=$ $W_{48}Y_{56}Y_{64}Y_{85}$ $=$ $W_{56}Y_{48}Y_{64}Y_{85}$ $=$ \\
& $=$ $W_{42}X_{37}Y_{74}Y_{53}Y_{25}$ $=$ $W_{42}X_{14}Y_{31}Y_{53}Y_{25}$ $=$ $W_{42}Y_{56}Y_{64}Y_{25}$ \\ 
	\hline
	$M_{25}$ & $W_{48}X_{31}X_{85}Y_{14}Y_{53}$ $=$ $W_{48}X_{74}X_{85}Y_{37}Y_{53}$ $=$ $W_{48}X_{85}Y_{64}Z_{56}$ $=$ \\
& $=$ $X_{31}Y_{14}Y_{42}Y_{53}Y_{25}$ $=$ $X_{74}Y_{37}Y_{42}Y_{53}Y_{25}$ $=$ $Y_{42}Y_{64}Y_{25}Z_{56}$ \\ 
	\hline
	$M_{26}$ & $W_{48}X_{85}Y_{74}Y_{37}Y_{53}$ $=$ $W_{48}X_{85}Y_{14}Y_{31}Y_{53}$ $=$ $W_{48}W_{56}X_{85}Y_{64}$ $=$ \\
& $=$ $W_{48}X_{31}Y_{14}Y_{53}Y_{85}$ $=$ $W_{48}X_{74}Y_{37}Y_{53}Y_{85}$ $=$ $W_{48}Y_{64}Y_{85}Z_{56}$ $=$ \\
& $=$ $W_{42}X_{31}Y_{14}Y_{53}Y_{25}$ $=$ $Y_{74}Y_{37}Y_{42}Y_{53}Y_{25}$ $=$ $W_{42}X_{74}Y_{37}Y_{53}Y_{25}$ $=$ \\
& $=$ $Y_{14}Y_{31}Y_{42}Y_{53}Y_{25}$ $=$ $W_{42}Y_{64}Y_{25}Z_{56}$ $=$ $W_{56}Y_{42}Y_{64}Y_{25}$ \\ 
	\hline
	$M_{27}$ & $W_{48}Y_{74}Y_{37}Y_{53}Y_{85}$ $=$ $W_{48}Y_{14}Y_{31}Y_{53}Y_{85}$ $=$ $W_{48}W_{56}Y_{64}Y_{85}$ $=$ \\
& $=$ $W_{42}Y_{74}Y_{37}Y_{53}Y_{25}$ $=$ $W_{42}Y_{14}Y_{31}Y_{53}Y_{25}$ $=$ $W_{42}W_{56}Y_{64}Y_{25}$
	\end{longtable}
\end{center}

\begin{center}
	\renewcommand{\arraystretch}{1.1}
	\begin{longtable}{c|c}
		\caption{Generators of $Q^{1,1,1}/\ZZ_2$ in Phase J.}\\
		Field & Chiral superfields\\
		\hline
		\endhead
		\label{tab:GenQ111Z2PhaseJ}$M_{1}$ & $X_{18}X_{56}X_{61}X_{85}$ $=$ $X_{13}X_{34}X_{46}X_{61}$ $=$ $X_{13}X_{35}X_{56}X_{61}$ $=$ $X_{73}X_{34}X_{46}X_{67}$ $=$ \\
& $=$ $X_{72}X_{56}X_{67}X_{25}$ $=$ $X_{73}X_{35}X_{56}X_{67}$ \\ 
\hline
$M_{2}$ & $X_{13}X_{46}X_{61}Z_{34}$ $=$ $X_{73}X_{46}X_{67}Z_{34}$ $=$ $X_{72}X_{56}X_{67}Y_{25}$ $=$ $X_{34}X_{46}X_{67}Y_{73}$ $=$ \\
& $=$ $X_{35}X_{56}X_{67}Y_{73}$ $=$ $X_{18}X_{61}X_{85}Z_{56}$ $=$ $X_{13}X_{35}X_{61}Z_{56}$ $=$ $X_{72}X_{67}X_{25}Z_{56}$ $=$ \\
& $=$ $X_{73}X_{35}X_{67}Z_{56}$ $=$ $X_{18}X_{56}X_{85}Y_{61}$ $=$ $X_{13}X_{34}X_{46}Y_{61}$ $=$ $X_{13}X_{35}X_{56}Y_{61}$ \\ 
\hline
$M_{3}$ & $X_{46}X_{67}Y_{73}Z_{34}$ $=$ $X_{72}X_{67}Y_{25}Z_{56}$ $=$ $X_{35}X_{67}Y_{73}Z_{56}$ $=$ $X_{13}X_{46}Y_{61}Z_{34}$ $=$ \\
& $=$ $X_{18}X_{85}Y_{61}Z_{56}$ $=$ $X_{13}X_{35}Y_{61}Z_{56}$ \\ 
\hline
$M_{4}$ & $X_{13}X_{56}X_{61}Y_{35}$ $=$ $X_{73}X_{56}X_{67}Y_{35}$ $=$ $X_{56}X_{67}X_{25}Y_{72}$ $=$ $X_{13}X_{34}X_{61}Y_{46}$ $=$ \\
& $=$ $X_{73}X_{34}X_{67}Y_{46}$ $=$ $X_{73}X_{34}X_{46}Z_{67}$ $=$ $X_{72}X_{56}X_{25}Z_{67}$ $=$ $X_{73}X_{35}X_{56}Z_{67}$ $=$ \\
& $=$ $X_{18}X_{56}X_{85}Z_{61}$ $=$ $X_{13}X_{34}X_{46}Z_{61}$ $=$ $X_{13}X_{35}X_{56}Z_{61}$ $=$ $X_{56}X_{61}X_{85}Y_{18}$ \\ 
\hline
$M_{5}$ & $X_{56}X_{67}Y_{73}Y_{35}$ $=$ $X_{56}X_{67}Y_{72}Y_{25}$ $=$ $X_{13}X_{61}Y_{35}Z_{56}$ $=$ $X_{73}X_{67}Y_{35}Z_{56}$ $=$ \\
& $=$ $X_{67}X_{25}Y_{72}Z_{56}$ $=$ $X_{13}X_{61}Y_{46}Z_{34}$ $=$ $X_{73}X_{67}Y_{46}Z_{34}$ $=$ $X_{34}X_{67}Y_{73}Y_{46}$ $=$ \\
& $=$ $X_{73}X_{46}Z_{34}Z_{67}$ $=$ $X_{72}X_{56}Y_{25}Z_{67}$ $=$ $X_{34}X_{46}Y_{73}Z_{67}$ $=$ $X_{35}X_{56}Y_{73}Z_{67}$ $=$ \\
& $=$ $X_{72}X_{25}Z_{56}Z_{67}$ $=$ $X_{73}X_{35}Z_{56}Z_{67}$ $=$ $X_{13}X_{56}Y_{35}Y_{61}$ $=$ $X_{13}X_{34}Y_{46}Y_{61}$ $=$ \\
& $=$ $X_{13}X_{46}Z_{34}Z_{61}$ $=$ $X_{18}X_{85}Z_{56}Z_{61}$ $=$ $X_{13}X_{35}Z_{56}Z_{61}$ $=$ $W_{61}X_{18}X_{56}X_{85}$ $=$ \\
& $=$ $W_{61}X_{13}X_{34}X_{46}$ $=$ $W_{61}X_{13}X_{35}X_{56}$ $=$ $X_{61}X_{85}Y_{18}Z_{56}$ $=$ $X_{56}X_{85}Y_{18}Y_{61}$ \\ 
\hline
$M_{6}$ & $X_{67}Y_{73}Y_{35}Z_{56}$ $=$ $X_{67}Y_{72}Y_{25}Z_{56}$ $=$ $X_{67}Y_{73}Y_{46}Z_{34}$ $=$ $X_{46}Y_{73}Z_{34}Z_{67}$ $=$ \\
& $=$ $X_{72}Y_{25}Z_{56}Z_{67}$ $=$ $X_{35}Y_{73}Z_{56}Z_{67}$ $=$ $X_{13}Y_{35}Y_{61}Z_{56}$ $=$ $X_{13}Y_{46}Y_{61}Z_{34}$ $=$ \\
& $=$ $W_{61}X_{13}X_{46}Z_{34}$ $=$ $W_{61}X_{18}X_{85}Z_{56}$ $=$ $W_{61}X_{13}X_{35}Z_{56}$ $=$ $X_{85}Y_{18}Y_{61}Z_{56}$ \\ 
\hline
$M_{2}$ & $X_{13}X_{46}X_{61}Y_{34}$ $=$ $X_{73}X_{46}X_{67}Y_{34}$ $=$ $X_{18}X_{61}X_{85}Y_{56}$ $=$ $X_{13}X_{35}X_{61}Y_{56}$ $=$ \\
& $=$ $X_{72}X_{67}X_{25}Y_{56}$ $=$ $X_{73}X_{35}X_{67}Y_{56}$ $=$ $X_{18}X_{56}X_{61}Y_{85}$ $=$ $X_{73}X_{34}X_{46}Y_{67}$ $=$ \\
& $=$ $X_{72}X_{56}X_{25}Y_{67}$ $=$ $X_{73}X_{35}X_{56}Y_{67}$ $=$ $X_{34}X_{46}X_{61}Y_{13}$ $=$ $X_{35}X_{56}X_{61}Y_{13}$ \\ 
\hline
$M_{8}$ & $W_{34}X_{13}X_{46}X_{61}$ $=$ $W_{34}X_{73}X_{46}X_{67}$ $=$ $X_{46}X_{67}Y_{73}Y_{34}$ $=$ $X_{72}X_{67}Y_{56}Y_{25}$ $=$ \\
& $=$ $X_{35}X_{67}Y_{73}Y_{56}$ $=$ $W_{56}X_{18}X_{61}X_{85}$ $=$ $W_{56}X_{13}X_{35}X_{61}$ $=$ $W_{56}X_{72}X_{67}X_{25}$ $=$ \\
& $=$ $W_{56}X_{73}X_{35}X_{67}$ $=$ $X_{18}X_{61}Y_{85}Z_{56}$ $=$ $X_{73}X_{46}Y_{67}Z_{34}$ $=$ $X_{72}X_{56}Y_{67}Y_{25}$ $=$ \\
& $=$ $X_{34}X_{46}Y_{73}Y_{67}$ $=$ $X_{35}X_{56}Y_{73}Y_{67}$ $=$ $X_{72}X_{25}Y_{67}Z_{56}$ $=$ $X_{73}X_{35}Y_{67}Z_{56}$ $=$ \\
& $=$ $X_{13}X_{46}Y_{34}Y_{61}$ $=$ $X_{18}X_{85}Y_{56}Y_{61}$ $=$ $X_{13}X_{35}Y_{56}Y_{61}$ $=$ $X_{18}X_{56}Y_{61}Y_{85}$ $=$ \\
& $=$ $X_{46}X_{61}Y_{13}Z_{34}$ $=$ $X_{35}X_{61}Y_{13}Z_{56}$ $=$ $X_{34}X_{46}Y_{13}Y_{61}$ $=$ $X_{35}X_{56}Y_{13}Y_{61}$ \\ 
\hline
$M_{9}$ & $W_{34}X_{46}X_{67}Y_{73}$ $=$ $W_{56}X_{72}X_{67}Y_{25}$ $=$ $W_{56}X_{35}X_{67}Y_{73}$ $=$ $X_{46}Y_{73}Y_{67}Z_{34}$ $=$ \\
& $=$ $X_{72}Y_{67}Y_{25}Z_{56}$ $=$ $X_{35}Y_{73}Y_{67}Z_{56}$ $=$ $W_{34}X_{13}X_{46}Y_{61}$ $=$ $W_{56}X_{18}X_{85}Y_{61}$ $=$ \\
& $=$ $W_{56}X_{13}X_{35}Y_{61}$ $=$ $X_{18}Y_{61}Y_{85}Z_{56}$ $=$ $X_{46}Y_{13}Y_{61}Z_{34}$ $=$ $X_{35}Y_{13}Y_{61}Z_{56}$ \\ 
\hline
$M_{10}$ & $X_{13}X_{61}Y_{35}Y_{56}$ $=$ $X_{73}X_{67}Y_{35}Y_{56}$ $=$ $X_{67}X_{25}Y_{72}Y_{56}$ $=$ $X_{13}X_{61}Y_{34}Y_{46}$ $=$ \\
& $=$ $X_{73}X_{67}Y_{34}Y_{46}$ $=$ $X_{73}X_{56}Y_{35}Y_{67}$ $=$ $X_{56}X_{25}Y_{72}Y_{67}$ $=$ $X_{73}X_{34}Y_{46}Y_{67}$ $=$ \\
& $=$ $X_{73}X_{46}Y_{34}Z_{67}$ $=$ $X_{72}X_{25}Y_{56}Z_{67}$ $=$ $X_{73}X_{35}Y_{56}Z_{67}$ $=$ $W_{67}X_{73}X_{34}X_{46}$ $=$ \\
& $=$ $W_{67}X_{72}X_{56}X_{25}$ $=$ $W_{67}X_{73}X_{35}X_{56}$ $=$ $X_{13}X_{46}Y_{34}Z_{61}$ $=$ $X_{18}X_{85}Y_{56}Z_{61}$ $=$ \\
& $=$ $X_{13}X_{35}Y_{56}Z_{61}$ $=$ $X_{18}X_{56}Y_{85}Z_{61}$ $=$ $X_{56}X_{61}Y_{13}Y_{35}$ $=$ $X_{34}X_{61}Y_{13}Y_{46}$ $=$ \\
& $=$ $X_{34}X_{46}Y_{13}Z_{61}$ $=$ $X_{35}X_{56}Y_{13}Z_{61}$ $=$ $X_{61}X_{85}Y_{18}Y_{56}$ $=$ $X_{56}X_{61}Y_{18}Y_{85}$ \\ 
\hline
$M_{11}$ & $X_{67}Y_{73}Y_{35}Y_{56}$ $=$ $X_{67}Y_{72}Y_{56}Y_{25}$ $=$ $W_{56}X_{13}X_{61}Y_{35}$ $=$ $W_{56}X_{73}X_{67}Y_{35}$ $=$ \\
& $=$ $W_{56}X_{67}X_{25}Y_{72}$ $=$ $W_{34}X_{13}X_{61}Y_{46}$ $=$ $W_{34}X_{73}X_{67}Y_{46}$ $=$ $X_{67}Y_{73}Y_{34}Y_{46}$ $=$ \\
& $=$ $X_{56}Y_{73}Y_{35}Y_{67}$ $=$ $X_{56}Y_{72}Y_{67}Y_{25}$ $=$ $X_{73}Y_{35}Y_{67}Z_{56}$ $=$ $X_{25}Y_{72}Y_{67}Z_{56}$ $=$ \\
& $=$ $X_{73}Y_{46}Y_{67}Z_{34}$ $=$ $X_{34}Y_{73}Y_{46}Y_{67}$ $=$ $W_{34}X_{73}X_{46}Z_{67}$ $=$ $X_{46}Y_{73}Y_{34}Z_{67}$ $=$ \\
& $=$ $X_{72}Y_{56}Y_{25}Z_{67}$ $=$ $X_{35}Y_{73}Y_{56}Z_{67}$ $=$ $W_{56}X_{72}X_{25}Z_{67}$ $=$ $W_{56}X_{73}X_{35}Z_{67}$ $=$ \\
& $=$ $W_{67}X_{73}X_{46}Z_{34}$ $=$ $W_{67}X_{72}X_{56}Y_{25}$ $=$ $W_{67}X_{34}X_{46}Y_{73}$ $=$ $W_{67}X_{35}X_{56}Y_{73}$ $=$ \\
& $=$ $W_{67}X_{72}X_{25}Z_{56}$ $=$ $W_{67}X_{73}X_{35}Z_{56}$ $=$ $X_{13}Y_{35}Y_{56}Y_{61}$ $=$ $X_{13}Y_{34}Y_{46}Y_{61}$ $=$ \\
& $=$ $W_{34}X_{13}X_{46}Z_{61}$ $=$ $W_{56}X_{18}X_{85}Z_{61}$ $=$ $W_{56}X_{13}X_{35}Z_{61}$ $=$ $X_{18}Y_{85}Z_{56}Z_{61}$ $=$ \\
& $=$ $W_{61}X_{13}X_{46}Y_{34}$ $=$ $W_{61}X_{18}X_{85}Y_{56}$ $=$ $W_{61}X_{13}X_{35}Y_{56}$ $=$ $W_{61}X_{18}X_{56}Y_{85}$ $=$ \\
& $=$ $X_{61}Y_{13}Y_{35}Z_{56}$ $=$ $X_{61}Y_{13}Y_{46}Z_{34}$ $=$ $X_{56}Y_{13}Y_{35}Y_{61}$ $=$ $X_{34}Y_{13}Y_{46}Y_{61}$ $=$ \\
& $=$ $X_{46}Y_{13}Z_{34}Z_{61}$ $=$ $X_{35}Y_{13}Z_{56}Z_{61}$ $=$ $W_{61}X_{34}X_{46}Y_{13}$ $=$ $W_{61}X_{35}X_{56}Y_{13}$ $=$ \\
& $=$ $W_{56}X_{61}X_{85}Y_{18}$ $=$ $X_{61}Y_{18}Y_{85}Z_{56}$ $=$ $X_{85}Y_{18}Y_{56}Y_{61}$ $=$ $X_{56}Y_{18}Y_{61}Y_{85}$ \\ 
\hline
$M_{12}$ & $W_{56}X_{67}Y_{73}Y_{35}$ $=$ $W_{56}X_{67}Y_{72}Y_{25}$ $=$ $W_{34}X_{67}Y_{73}Y_{46}$ $=$ $Y_{73}Y_{35}Y_{67}Z_{56}$ $=$ \\
& $=$ $Y_{72}Y_{67}Y_{25}Z_{56}$ $=$ $Y_{73}Y_{46}Y_{67}Z_{34}$ $=$ $W_{34}X_{46}Y_{73}Z_{67}$ $=$ $W_{56}X_{72}Y_{25}Z_{67}$ $=$ \\
& $=$ $W_{56}X_{35}Y_{73}Z_{67}$ $=$ $W_{67}X_{46}Y_{73}Z_{34}$ $=$ $W_{67}X_{72}Y_{25}Z_{56}$ $=$ $W_{67}X_{35}Y_{73}Z_{56}$ $=$ \\
& $=$ $W_{56}X_{13}Y_{35}Y_{61}$ $=$ $W_{34}X_{13}Y_{46}Y_{61}$ $=$ $W_{34}W_{61}X_{13}X_{46}$ $=$ $W_{56}W_{61}X_{18}X_{85}$ $=$ \\
& $=$ $W_{56}W_{61}X_{13}X_{35}$ $=$ $W_{61}X_{18}Y_{85}Z_{56}$ $=$ $Y_{13}Y_{35}Y_{61}Z_{56}$ $=$ $Y_{13}Y_{46}Y_{61}Z_{34}$ $=$ \\
& $=$ $W_{61}X_{46}Y_{13}Z_{34}$ $=$ $W_{61}X_{35}Y_{13}Z_{56}$ $=$ $W_{56}X_{85}Y_{18}Y_{61}$ $=$ $Y_{18}Y_{61}Y_{85}Z_{56}$ \\ 
\hline
$M_{13}$ & $X_{18}X_{61}Y_{56}Y_{85}$ $=$ $X_{73}X_{46}Y_{34}Y_{67}$ $=$ $X_{72}X_{25}Y_{56}Y_{67}$ $=$ $X_{73}X_{35}Y_{56}Y_{67}$ $=$ \\
& $=$ $X_{46}X_{61}Y_{13}Y_{34}$ $=$ $X_{35}X_{61}Y_{13}Y_{56}$ \\ 
\hline
$M_{14}$ & $W_{56}X_{18}X_{61}Y_{85}$ $=$ $W_{34}X_{73}X_{46}Y_{67}$ $=$ $X_{46}Y_{73}Y_{34}Y_{67}$ $=$ $X_{72}Y_{56}Y_{67}Y_{25}$ $=$ \\
& $=$ $X_{35}Y_{73}Y_{56}Y_{67}$ $=$ $W_{56}X_{72}X_{25}Y_{67}$ $=$ $W_{56}X_{73}X_{35}Y_{67}$ $=$ $X_{18}Y_{56}Y_{61}Y_{85}$ $=$ \\
& $=$ $W_{34}X_{46}X_{61}Y_{13}$ $=$ $W_{56}X_{35}X_{61}Y_{13}$ $=$ $X_{46}Y_{13}Y_{34}Y_{61}$ $=$ $X_{35}Y_{13}Y_{56}Y_{61}$ \\ 
\hline
$M_{15}$ & $W_{34}X_{46}Y_{73}Y_{67}$ $=$ $W_{56}X_{72}Y_{67}Y_{25}$ $=$ $W_{56}X_{35}Y_{73}Y_{67}$ $=$ $W_{56}X_{18}Y_{61}Y_{85}$ $=$ \\
& $=$ $W_{34}X_{46}Y_{13}Y_{61}$ $=$ $W_{56}X_{35}Y_{13}Y_{61}$ \\ 
\hline
$M_{16}$ & $X_{73}Y_{35}Y_{56}Y_{67}$ $=$ $X_{25}Y_{72}Y_{56}Y_{67}$ $=$ $X_{73}Y_{34}Y_{46}Y_{67}$ $=$ $W_{67}X_{73}X_{46}Y_{34}$ $=$ \\
& $=$ $W_{67}X_{72}X_{25}Y_{56}$ $=$ $W_{67}X_{73}X_{35}Y_{56}$ $=$ $X_{18}Y_{56}Y_{85}Z_{61}$ $=$ $X_{61}Y_{13}Y_{35}Y_{56}$ $=$ \\
& $=$ $X_{61}Y_{13}Y_{34}Y_{46}$ $=$ $X_{46}Y_{13}Y_{34}Z_{61}$ $=$ $X_{35}Y_{13}Y_{56}Z_{61}$ $=$ $X_{61}Y_{18}Y_{56}Y_{85}$ \\ 
\hline
$M_{17}$ & $Y_{73}Y_{35}Y_{56}Y_{67}$ $=$ $Y_{72}Y_{56}Y_{67}Y_{25}$ $=$ $W_{56}X_{73}Y_{35}Y_{67}$ $=$ $W_{56}X_{25}Y_{72}Y_{67}$ $=$ \\
& $=$ $W_{34}X_{73}Y_{46}Y_{67}$ $=$ $Y_{73}Y_{34}Y_{46}Y_{67}$ $=$ $W_{34}W_{67}X_{73}X_{46}$ $=$ $W_{67}X_{46}Y_{73}Y_{34}$ $=$ \\
& $=$ $W_{67}X_{72}Y_{56}Y_{25}$ $=$ $W_{67}X_{35}Y_{73}Y_{56}$ $=$ $W_{56}W_{67}X_{72}X_{25}$ $=$ $W_{56}W_{67}X_{73}X_{35}$ $=$ \\
& $=$ $W_{56}X_{18}Y_{85}Z_{61}$ $=$ $W_{61}X_{18}Y_{56}Y_{85}$ $=$ $W_{56}X_{61}Y_{13}Y_{35}$ $=$ $W_{34}X_{61}Y_{13}Y_{46}$ $=$ \\
& $=$ $Y_{13}Y_{35}Y_{56}Y_{61}$ $=$ $Y_{13}Y_{34}Y_{46}Y_{61}$ $=$ $W_{34}X_{46}Y_{13}Z_{61}$ $=$ $W_{56}X_{35}Y_{13}Z_{61}$ $=$ \\
& $=$ $W_{61}X_{46}Y_{13}Y_{34}$ $=$ $W_{61}X_{35}Y_{13}Y_{56}$ $=$ $W_{56}X_{61}Y_{18}Y_{85}$ $=$ $Y_{18}Y_{56}Y_{61}Y_{85}$ \\ 
\hline
$M_{18}$ & $W_{56}Y_{73}Y_{35}Y_{67}$ $=$ $W_{56}Y_{72}Y_{67}Y_{25}$ $=$ $W_{34}Y_{73}Y_{46}Y_{67}$ $=$ $W_{34}W_{67}X_{46}Y_{73}$ $=$ \\
& $=$ $W_{56}W_{67}X_{72}Y_{25}$ $=$ $W_{56}W_{67}X_{35}Y_{73}$ $=$ $W_{56}W_{61}X_{18}Y_{85}$ $=$ $W_{56}Y_{13}Y_{35}Y_{61}$ $=$ \\
& $=$ $W_{34}Y_{13}Y_{46}Y_{61}$ $=$ $W_{34}W_{61}X_{46}Y_{13}$ $=$ $W_{56}W_{61}X_{35}Y_{13}$ $=$ $W_{56}Y_{18}Y_{61}Y_{85}$ \\ 
\hline
$M_{19}$ & $X_{73}X_{56}Y_{35}Z_{67}$ $=$ $X_{56}X_{25}Y_{72}Z_{67}$ $=$ $X_{73}X_{34}Y_{46}Z_{67}$ $=$ $X_{13}X_{56}Y_{35}Z_{61}$ $=$ \\
& $=$ $X_{13}X_{34}Y_{46}Z_{61}$ $=$ $X_{56}X_{85}Y_{18}Z_{61}$ \\ 
\hline
$M_{20}$ & $X_{56}Y_{73}Y_{35}Z_{67}$ $=$ $X_{56}Y_{72}Y_{25}Z_{67}$ $=$ $X_{73}Y_{35}Z_{56}Z_{67}$ $=$ $X_{25}Y_{72}Z_{56}Z_{67}$ $=$ \\
& $=$ $X_{73}Y_{46}Z_{34}Z_{67}$ $=$ $X_{34}Y_{73}Y_{46}Z_{67}$ $=$ $X_{13}Y_{35}Z_{56}Z_{61}$ $=$ $X_{13}Y_{46}Z_{34}Z_{61}$ $=$ \\
& $=$ $W_{61}X_{13}X_{56}Y_{35}$ $=$ $W_{61}X_{13}X_{34}Y_{46}$ $=$ $X_{85}Y_{18}Z_{56}Z_{61}$ $=$ $W_{61}X_{56}X_{85}Y_{18}$ \\ 
\hline
$M_{21}$ & $Y_{73}Y_{35}Z_{56}Z_{67}$ $=$ $Y_{72}Y_{25}Z_{56}Z_{67}$ $=$ $Y_{73}Y_{46}Z_{34}Z_{67}$ $=$ $W_{61}X_{13}Y_{35}Z_{56}$ $=$ \\
& $=$ $W_{61}X_{13}Y_{46}Z_{34}$ $=$ $W_{61}X_{85}Y_{18}Z_{56}$ \\ 
\hline
$M_{27}$ & $X_{73}Y_{35}Y_{56}Z_{67}$ $=$ $X_{25}Y_{72}Y_{56}Z_{67}$ $=$ $X_{73}Y_{34}Y_{46}Z_{67}$ $=$ $W_{67}X_{73}X_{56}Y_{35}$ $=$ \\
& $=$ $W_{67}X_{56}X_{25}Y_{72}$ $=$ $W_{67}X_{73}X_{34}Y_{46}$ $=$ $X_{13}Y_{35}Y_{56}Z_{61}$ $=$ $X_{13}Y_{34}Y_{46}Z_{61}$ $=$ \\
& $=$ $X_{56}Y_{13}Y_{35}Z_{61}$ $=$ $X_{34}Y_{13}Y_{46}Z_{61}$ $=$ $X_{85}Y_{18}Y_{56}Z_{61}$ $=$ $X_{56}Y_{18}Y_{85}Z_{61}$ \\ 
\hline
$M_{23}$ & $Y_{73}Y_{35}Y_{56}Z_{67}$ $=$ $Y_{72}Y_{56}Y_{25}Z_{67}$ $=$ $W_{56}X_{73}Y_{35}Z_{67}$ $=$ $W_{56}X_{25}Y_{72}Z_{67}$ $=$ \\
& $=$ $W_{34}X_{73}Y_{46}Z_{67}$ $=$ $Y_{73}Y_{34}Y_{46}Z_{67}$ $=$ $W_{67}X_{56}Y_{73}Y_{35}$ $=$ $W_{67}X_{56}Y_{72}Y_{25}$ $=$ \\
& $=$ $W_{67}X_{73}Y_{35}Z_{56}$ $=$ $W_{67}X_{25}Y_{72}Z_{56}$ $=$ $W_{67}X_{73}Y_{46}Z_{34}$ $=$ $W_{67}X_{34}Y_{73}Y_{46}$ $=$ \\
& $=$ $W_{56}X_{13}Y_{35}Z_{61}$ $=$ $W_{34}X_{13}Y_{46}Z_{61}$ $=$ $W_{61}X_{13}Y_{35}Y_{56}$ $=$ $W_{61}X_{13}Y_{34}Y_{46}$ $=$ \\
& $=$ $Y_{13}Y_{35}Z_{56}Z_{61}$ $=$ $Y_{13}Y_{46}Z_{34}Z_{61}$ $=$ $W_{61}X_{56}Y_{13}Y_{35}$ $=$ $W_{61}X_{34}Y_{13}Y_{46}$ $=$ \\
& $=$ $W_{56}X_{85}Y_{18}Z_{61}$ $=$ $Y_{18}Y_{85}Z_{56}Z_{61}$ $=$ $W_{61}X_{85}Y_{18}Y_{56}$ $=$ $W_{61}X_{56}Y_{18}Y_{85}$ \\ 
\hline
$M_{24}$ & $W_{56}Y_{73}Y_{35}Z_{67}$ $=$ $W_{56}Y_{72}Y_{25}Z_{67}$ $=$ $W_{34}Y_{73}Y_{46}Z_{67}$ $=$ $W_{67}Y_{73}Y_{35}Z_{56}$ $=$ \\
& $=$ $W_{67}Y_{72}Y_{25}Z_{56}$ $=$ $W_{67}Y_{73}Y_{46}Z_{34}$ $=$ $W_{56}W_{61}X_{13}Y_{35}$ $=$ $W_{34}W_{61}X_{13}Y_{46}$ $=$ \\
& $=$ $W_{61}Y_{13}Y_{35}Z_{56}$ $=$ $W_{61}Y_{13}Y_{46}Z_{34}$ $=$ $W_{56}W_{61}X_{85}Y_{18}$ $=$ $W_{61}Y_{18}Y_{85}Z_{56}$ \\ 
\hline
$M_{25}$ & $W_{67}X_{73}Y_{35}Y_{56}$ $=$ $W_{67}X_{25}Y_{72}Y_{56}$ $=$ $W_{67}X_{73}Y_{34}Y_{46}$ $=$ $Y_{13}Y_{35}Y_{56}Z_{61}$ $=$ \\
& $=$ $Y_{13}Y_{34}Y_{46}Z_{61}$ $=$ $Y_{18}Y_{56}Y_{85}Z_{61}$ \\ 
\hline
$M_{26}$ & $W_{67}Y_{73}Y_{35}Y_{56}$ $=$ $W_{67}Y_{72}Y_{56}Y_{25}$ $=$ $W_{56}W_{67}X_{73}Y_{35}$ $=$ $W_{56}W_{67}X_{25}Y_{72}$ $=$ \\
& $=$ $W_{34}W_{67}X_{73}Y_{46}$ $=$ $W_{67}Y_{73}Y_{34}Y_{46}$ $=$ $W_{56}Y_{13}Y_{35}Z_{61}$ $=$ $W_{34}Y_{13}Y_{46}Z_{61}$ $=$ \\
& $=$ $W_{61}Y_{13}Y_{35}Y_{56}$ $=$ $W_{61}Y_{13}Y_{34}Y_{46}$ $=$ $W_{56}Y_{18}Y_{85}Z_{61}$ $=$ $W_{61}Y_{18}Y_{56}Y_{85}$ \\ 
\hline
$M_{27}$ & $W_{56}W_{67}Y_{73}Y_{35}$ $=$ $W_{56}W_{67}Y_{72}Y_{25}$ $=$ $W_{34}W_{67}Y_{73}Y_{46}$ $=$ $W_{56}W_{61}Y_{13}Y_{35}$ $=$ \\
& $=$ $W_{34}W_{61}Y_{13}Y_{46}$ $=$ $W_{56}W_{61}Y_{18}Y_{85}$ 
	\end{longtable}
\end{center}

\begin{center}
	\renewcommand{\arraystretch}{1.1}
	\begin{longtable}{c|c}
		\caption{Generators of $Q^{1,1,1}/\ZZ_2$ in Phase L.}\\
		Field & Chiral superfields\\
		\hline
		\endhead
		\label{tab:GenQ111Z2PhaseL}$M_{1}$ & $A_{54} X_{18} X_{41} Y_{85}$ $=$ $A_{54} X_{46} X_{68} Y_{85}$ $=$ $A_{54} X_{46} X_{25} Y_{62}$ $=$ $A_{54} X_{73} X_{35} Y_{47}$ $=$ \\
& $=$ $A_{54} X_{72} X_{25} Y_{47}$ $=$ $A_{54} X_{35} X_{41} Y_{13}$ \\ 
\hline
$M_{2}$ & $X_{18} X_{41} X_{54} Y_{85}$ $=$ $X_{46} X_{54} X_{68} Y_{85}$ $=$ $X_{46} X_{54} X_{25} Y_{62}$ $=$ $A_{54} X_{13} X_{35} X_{41}$ $=$ \\
& $=$ $A_{54} X_{18} X_{41} X_{85}$ $=$ $A_{54} X_{73} X_{35} X_{47}$ $=$ $A_{54} X_{72} X_{47} X_{25}$ $=$ $A_{54} X_{46} X_{62} X_{25}$ $=$ \\
& $=$ $A_{54} X_{46} X_{68} X_{85}$ $=$ $X_{73} X_{35} X_{54} Y_{47}$ $=$ $X_{72} X_{54} X_{25} Y_{47}$ $=$ $X_{35} X_{41} X_{54} Y_{13}$ \\ 
\hline
$M_{3}$ & $X_{13} X_{35} X_{41} X_{54}$ $=$ $X_{18} X_{41} X_{54} X_{85}$ $=$ $X_{73} X_{35} X_{47} X_{54}$ $=$ $X_{72} X_{47} X_{54} X_{25}$ $=$ \\
& $=$ $X_{46} X_{54} X_{62} X_{25}$ $=$ $X_{46} X_{54} X_{68} X_{85}$ \\ 
\hline
$M_{4}$ & $B_{54} X_{18} X_{41} Y_{85}$ $=$ $B_{54} X_{46} X_{68} Y_{85}$ $=$ $B_{54} X_{46} X_{25} Y_{62}$ $=$ $A_{54} X_{68} Y_{46} Y_{85}$ $=$ \\
& $=$ $A_{54} X_{25} Y_{46} Y_{62}$ $=$ $B_{54} X_{73} X_{35} Y_{47}$ $=$ $B_{54} X_{72} X_{25} Y_{47}$ $=$ $A_{54} X_{73} Y_{35} Y_{47}$ $=$ \\
& $=$ $A_{54} X_{25} Y_{72} Y_{47}$ $=$ $A_{54} X_{41} Y_{18} Y_{85}$ $=$ $B_{54} X_{35} X_{41} Y_{13}$ $=$ $A_{54} X_{41} Y_{13} Y_{35}$ \\
\hline
$M_{5}$ & $X_{18} X_{41} Y_{54} Y_{85}$ $=$ $X_{46} X_{68} Y_{54} Y_{85}$ $=$ $X_{46} X_{25} Y_{54} Y_{62}$ $=$ $B_{54} X_{13} X_{35} X_{41}$ $=$ \\
& $=$ $B_{54} X_{18} X_{41} X_{85}$ $=$ $B_{54} X_{73} X_{35} X_{47}$ $=$ $B_{54} X_{72} X_{47} X_{25}$ $=$ $B_{54} X_{46} X_{62} X_{25}$ $=$ \\
& $=$ $B_{54} X_{46} X_{68} X_{85}$ $=$ $X_{54} X_{68} Y_{46} Y_{85}$ $=$ $X_{54} X_{25} Y_{46} Y_{62}$ $=$ $A_{54} X_{62} X_{25} Y_{46}$ $=$ \\
& $=$ $A_{54} X_{68} X_{85} Y_{46}$ $=$ $X_{73} X_{35} Y_{47} Y_{54}$ $=$ $X_{72} X_{25} Y_{47} Y_{54}$ $=$ $A_{54} X_{13} X_{41} Y_{35}$ $=$ \\
& $=$ $A_{54} X_{73} X_{47} Y_{35}$ $=$ $X_{73} X_{54} Y_{35} Y_{47}$ $=$ $A_{54} X_{47} X_{25} Y_{72}$ $=$ $X_{54} X_{25} Y_{72} Y_{47}$ $=$ \\
& $=$ $X_{41} X_{54} Y_{18} Y_{85}$ $=$ $A_{54} X_{41} X_{85} Y_{18}$ $=$ $X_{35} X_{41} Y_{13} Y_{54}$ $=$ $X_{41} X_{54} Y_{13} Y_{35}$ \\ 
\hline
$M_{6}$ & $X_{13} X_{35} X_{41} Y_{54}$ $=$ $X_{18} X_{41} X_{85} Y_{54}$ $=$ $X_{73} X_{35} X_{47} Y_{54}$ $=$ $X_{72} X_{47} X_{25} Y_{54}$ $=$ \\
& $=$ $X_{46} X_{62} X_{25} Y_{54}$ $=$ $X_{46} X_{68} X_{85} Y_{54}$ $=$ $X_{54} X_{62} X_{25} Y_{46}$ $=$ $X_{54} X_{68} X_{85} Y_{46}$ $=$ \\
& $=$ $X_{13} X_{41} X_{54} Y_{35}$ $=$ $X_{73} X_{47} X_{54} Y_{35}$ $=$ $X_{47} X_{54} X_{25} Y_{72}$ $=$ $X_{41} X_{54} X_{85} Y_{18}$ \\ 
\hline
 $M_{2}$ & $A_{54} X_{46} Y_{68} Y_{85}$ $=$ $A_{54} X_{46} Y_{62} Y_{25}$ $=$ $X_{18} X_{41} Y_{85} C_{54}$ $=$ $X_{46} X_{68} Y_{85} C_{54}$ $=$ \\
& $=$ $X_{46} X_{25} Y_{62} C_{54}$ $=$ $A_{54} X_{72} Y_{47} Y_{25}$ $=$ $X_{73} X_{35} Y_{47} C_{54}$ $=$ $X_{72} X_{25} Y_{47} C_{54}$ $=$ \\
& $=$ $A_{54} X_{18} Y_{41} Y_{85}$ $=$ $A_{54} X_{35} Y_{73} Y_{47}$ $=$ $X_{35} X_{41} Y_{13} C_{54}$ $=$ $A_{54} X_{35} Y_{13} Y_{41}$ \\ 
\hline
$M_{8}$ & $X_{46} X_{54} Y_{68} Y_{85}$ $=$ $X_{46} X_{54} Y_{62} Y_{25}$ $=$ $X_{18} X_{41} Y_{85} Z_{54}$ $=$ $X_{46} X_{68} Y_{85} Z_{54}$ $=$ \\
& $=$ $X_{46} X_{25} Y_{62} Z_{54}$ $=$ $A_{54} X_{72} X_{47} Y_{25}$ $=$ $A_{54} X_{46} X_{62} Y_{25}$ $=$ $A_{54} X_{46} X_{85} Y_{68}$ $=$ \\
& $=$ $X_{13} X_{35} X_{41} C_{54}$ $=$ $X_{18} X_{41} X_{85} C_{54}$ $=$ $X_{73} X_{35} X_{47} C_{54}$ $=$ $X_{72} X_{47} X_{25} C_{54}$ $=$ \\
& $=$ $X_{46} X_{62} X_{25} C_{54}$ $=$ $X_{46} X_{68} X_{85} C_{54}$ $=$ $X_{72} X_{54} Y_{47} Y_{25}$ $=$ $X_{73} X_{35} Y_{47} Z_{54}$ $=$ \\
& $=$ $X_{72} X_{25} Y_{47} Z_{54}$ $=$ $X_{18} X_{54} Y_{41} Y_{85}$ $=$ $A_{54} X_{13} X_{35} Y_{41}$ $=$ $A_{54} X_{18} X_{85} Y_{41}$ $=$ \\
& $=$ $A_{54} X_{35} X_{47} Y_{73}$ $=$ $X_{35} X_{54} Y_{73} Y_{47}$ $=$ $X_{35} X_{41} Y_{13} Z_{54}$ $=$ $X_{35} X_{54} Y_{13} Y_{41}$ \\ 
\hline
$M_{9}$ & $X_{72} X_{47} X_{54} Y_{25}$ $=$ $X_{46} X_{54} X_{62} Y_{25}$ $=$ $X_{46} X_{54} X_{85} Y_{68}$ $=$ $X_{13} X_{35} X_{41} Z_{54}$ $=$ \\
& $=$ $X_{18} X_{41} X_{85} Z_{54}$ $=$ $X_{73} X_{35} X_{47} Z_{54}$ $=$ $X_{72} X_{47} X_{25} Z_{54}$ $=$ $X_{46} X_{62} X_{25} Z_{54}$ $=$ \\
& $=$ $X_{46} X_{68} X_{85} Z_{54}$ $=$ $X_{13} X_{35} X_{54} Y_{41}$ $=$ $X_{18} X_{54} X_{85} Y_{41}$ $=$ $X_{35} X_{47} X_{54} Y_{73}$ \\ 
\hline
$M_{10}$ & $B_{54} X_{46} Y_{68} Y_{85}$ $=$ $B_{54} X_{46} Y_{62} Y_{25}$ $=$ $D_{54} X_{18} X_{41} Y_{85}$ $=$ $D_{54} X_{46} X_{68} Y_{85}$ $=$ \\
& $=$ $D_{54} X_{46} X_{25} Y_{62}$ $=$ $A_{54} Y_{46} Y_{68} Y_{85}$ $=$ $A_{54} Y_{46} Y_{62} Y_{25}$ $=$ $X_{68} Y_{46} Y_{85} C_{54}$ $=$ \\
& $=$ $X_{25} Y_{46} Y_{62} C_{54}$ $=$ $B_{54} X_{72} Y_{47} Y_{25}$ $=$ $D_{54} X_{73} X_{35} Y_{47}$ $=$ $D_{54} X_{72} X_{25} Y_{47}$ $=$ \\
& $=$ $B_{54} X_{18} Y_{41} Y_{85}$ $=$ $X_{73} Y_{35} Y_{47} C_{54}$ $=$ $A_{54} Y_{72} Y_{47} Y_{25}$ $=$ $X_{25} Y_{72} Y_{47} C_{54}$ $=$ \\
& $=$ $B_{54} X_{35} Y_{73} Y_{47}$ $=$ $A_{54} Y_{73} Y_{35} Y_{47}$ $=$ $X_{41} Y_{18} Y_{85} C_{54}$ $=$ $A_{54} Y_{18} Y_{41} Y_{85}$ $=$ \\
& $=$ $D_{54} X_{35} X_{41} Y_{13}$ $=$ $B_{54} X_{35} Y_{13} Y_{41}$ $=$ $X_{41} Y_{13} Y_{35} C_{54}$ $=$ $A_{54} Y_{13} Y_{35} Y_{41}$ \\ 
\hline
$M_{11}$ & $X_{46} Y_{54} Y_{68} Y_{85}$ $=$ $X_{46} Y_{54} Y_{62} Y_{25}$ $=$ $W_{54} X_{18} X_{41} Y_{85}$ $=$ $W_{54} X_{46} X_{68} Y_{85}$ $=$ \\
& $=$ $W_{54} X_{46} X_{25} Y_{62}$ $=$ $B_{54} X_{72} X_{47} Y_{25}$ $=$ $B_{54} X_{46} X_{62} Y_{25}$ $=$ $B_{54} X_{46} X_{85} Y_{68}$ $=$ \\
& $=$ $D_{54} X_{13} X_{35} X_{41}$ $=$ $D_{54} X_{18} X_{41} X_{85}$ $=$ $D_{54} X_{73} X_{35} X_{47}$ $=$ $D_{54} X_{72} X_{47} X_{25}$ $=$ \\
& $=$ $D_{54} X_{46} X_{62} X_{25}$ $=$ $D_{54} X_{46} X_{68} X_{85}$ $=$ $X_{54} Y_{46} Y_{68} Y_{85}$ $=$ $X_{54} Y_{46} Y_{62} Y_{25}$ $=$ \\
& $=$ $X_{68} Y_{46} Y_{85} Z_{54}$ $=$ $X_{25} Y_{46} Y_{62} Z_{54}$ $=$ $A_{54} X_{62} Y_{46} Y_{25}$ $=$ $A_{54} X_{85} Y_{46} Y_{68}$ $=$ \\
& $=$ $X_{62} X_{25} Y_{46} C_{54}$ $=$ $X_{68} X_{85} Y_{46} C_{54}$ $=$ $X_{72} Y_{47} Y_{54} Y_{25}$ $=$ $W_{54} X_{73} X_{35} Y_{47}$ $=$ \\
& $=$ $W_{54} X_{72} X_{25} Y_{47}$ $=$ $X_{18} Y_{41} Y_{54} Y_{85}$ $=$ $B_{54} X_{13} X_{35} Y_{41}$ $=$ $B_{54} X_{18} X_{85} Y_{41}$ $=$ \\
& $=$ $X_{13} X_{41} Y_{35} C_{54}$ $=$ $X_{73} X_{47} Y_{35} C_{54}$ $=$ $X_{73} Y_{35} Y_{47} Z_{54}$ $=$ $A_{54} X_{13} Y_{35} Y_{41}$ $=$ \\
& $=$ $A_{54} X_{47} Y_{72} Y_{25}$ $=$ $X_{47} X_{25} Y_{72} C_{54}$ $=$ $X_{54} Y_{72} Y_{47} Y_{25}$ $=$ $X_{25} Y_{72} Y_{47} Z_{54}$ $=$ \\
& $=$ $B_{54} X_{35} X_{47} Y_{73}$ $=$ $X_{35} Y_{73} Y_{47} Y_{54}$ $=$ $A_{54} X_{47} Y_{73} Y_{35}$ $=$ $X_{54} Y_{73} Y_{35} Y_{47}$ $=$ \\
& $=$ $X_{41} Y_{18} Y_{85} Z_{54}$ $=$ $X_{41} X_{85} Y_{18} C_{54}$ $=$ $X_{54} Y_{18} Y_{41} Y_{85}$ $=$ $A_{54}X_{85} Y_{18} Y_{41}$ $=$ \\
& $=$ $W_{54} X_{35} X_{41} Y_{13}$ $=$ $X_{35} Y_{13} Y_{41} Y_{54}$ $=$ $X_{41} Y_{13} Y_{35} Z_{54}$ $=$ $X_{54} Y_{13} Y_{35} Y_{41}$ \\ 
\hline
$M_{12}$ & $X_{72} X_{47} Y_{54} Y_{25}$ $=$ $X_{46} X_{62} Y_{54} Y_{25}$ $=$ $X_{46} X_{85} Y_{54} Y_{68}$ $=$ $W_{54} X_{13} X_{35} X_{41}$ $=$ \\
& $=$ $W_{54} X_{18} X_{41} X_{85}$ $=$ $W_{54} X_{73} X_{35} X_{47}$ $=$ $W_{54} X_{72} X_{47} X_{25}$ $=$ $W_{54} X_{46} X_{62} X_{25}$ $=$ \\
& $=$ $W_{54} X_{46} X_{68} X_{85}$ $=$ $X_{54} X_{62} Y_{46} Y_{25}$ $=$ $X_{54} X_{85} Y_{46} Y_{68}$ $=$ $X_{62} X_{25} Y_{46} Z_{54}$ $=$ \\
& $=$ $X_{68} X_{85} Y_{46} Z_{54}$ $=$ $X_{13} X_{35} Y_{41} Y_{54}$ $=$ $X_{18} X_{85} Y_{41} Y_{54}$ $=$ $X_{13} X_{41} Y_{35} Z_{54}$ $=$ \\
& $=$ $X_{73} X_{47} Y_{35} Z_{54}$ $=$ $X_{13} X_{54} Y_{35} Y_{41}$ $=$ $X_{47} X_{54} Y_{72} Y_{25}$ $=$ $X_{47} X_{25} Y_{72} Z_{54}$ $=$ \\
& $=$ $X_{35} X_{47} Y_{73} Y_{54}$ $=$ $X_{47} X_{54} Y_{73} Y_{35}$ $=$ $X_{41} X_{85} Y_{18} Z_{54}$ $=$ $X_{54} X_{85} Y_{18} Y_{41}$ \\ 
\hline
$M_{13}$ & $X_{46} Y_{68} Y_{85} C_{54}$ $=$ $X_{46} Y_{62} Y_{25} C_{54}$ $=$ $X_{72} Y_{47} Y_{25} C_{54}$ $=$ $X_{18} Y_{41} Y_{85} C_{54}$ $=$ \\
& $=$ $X_{35} Y_{73} Y_{47} C_{54}$ $=$ $X_{35} Y_{13} Y_{41} C_{54}$ \\ 
\hline
$M_{14}$ & $X_{46} Y_{68} Y_{85} Z_{54}$ $=$ $X_{46} Y_{62} Y_{25} Z_{54}$ $=$ $X_{72} X_{47} Y_{25} C_{54}$ $=$ $X_{46} X_{62} Y_{25} C_{54}$ $=$ \\
& $=$ $X_{46} X_{85} Y_{68} C_{54}$ $=$ $X_{72} Y_{47} Y_{25} Z_{54}$ $=$ $X_{18} Y_{41} Y_{85} Z_{54}$ $=$ $X_{13} X_{35} Y_{41} C_{54}$ $=$ \\
& $=$ $X_{18} X_{85} Y_{41} C_{54}$ $=$ $X_{35} X_{47} Y_{73} C_{54}$ $=$ $X_{35} Y_{73} Y_{47} Z_{54}$ $=$ $X_{35} Y_{13} Y_{41} Z_{54}$ \\ 
\hline
$M_{15}$ & $X_{72} X_{47} Y_{25} Z_{54}$ $=$ $X_{46} X_{62} Y_{25} Z_{54}$ $=$ $X_{46} X_{85} Y_{68} Z_{54}$ $=$ $X_{13} X_{35} Y_{41} Z_{54}$ $=$ \\
& $=$ $X_{18} X_{85} Y_{41} Z_{54}$ $=$ $X_{35} X_{47} Y_{73} Z_{54}$ \\ 
\hline
$M_{16}$ & $D_{54} X_{46} Y_{68} Y_{85}$ $=$ $D_{54} X_{46} Y_{62} Y_{25}$ $=$ $Y_{46} Y_{68} Y_{85} C_{54}$ $=$ $Y_{46} Y_{62} Y_{25} C_{54}$ $=$ \\
& $=$ $D_{54} X_{72} Y_{47} Y_{25}$ $=$ $D_{54} X_{18} Y_{41} Y_{85}$ $=$ $Y_{72} Y_{47} Y_{25} C_{54}$ $=$ $D_{54} X_{35} Y_{73} Y_{47}$ $=$ \\
& $=$ $Y_{73} Y_{35} Y_{47} C_{54}$ $=$ $Y_{18} Y_{41} Y_{85} C_{54}$ $=$ $D_{54} X_{35} Y_{13} Y_{41}$ $=$ $Y_{13} Y_{35} Y_{41} C_{54}$ \\ 
\hline
$M_{17}$ & $W_{54} X_{46} Y_{68} Y_{85}$ $=$ $W_{54} X_{46} Y_{62} Y_{25}$ $=$ $D_{54} X_{72} X_{47} Y_{25}$ $=$ $D_{54} X_{46} X_{62} Y_{25}$ $=$ \\
& $=$ $D_{54} X_{46} X_{85} Y_{68}$ $=$ $Y_{46} Y_{68} Y_{85} Z_{54}$ $=$ $Y_{46} Y_{62} Y_{25} Z_{54}$ $=$ $X_{62} Y_{46} Y_{25} C_{54}$ $=$ \\
& $=$ $X_{85} Y_{46} Y_{68} C_{54}$ $=$ $W_{54} X_{72} Y_{47} Y_{25}$ $=$ $W_{54} X_{18} Y_{41} Y_{85}$ $=$ $D_{54} X_{13} X_{35} Y_{41}$ $=$ \\
& $=$ $D_{54} X_{18} X_{85} Y_{41}$ $=$ $X_{13} Y_{35} Y_{41} C_{54}$ $=$ $X_{47} Y_{72} Y_{25} C_{54}$ $=$ $Y_{72} Y_{47} Y_{25} Z_{54}$ $=$ \\
& $=$ $D_{54} X_{35} X_{47} Y_{73}$ $=$ $W_{54} X_{35} Y_{73} Y_{47}$ $=$ $X_{47} Y_{73} Y_{35} C_{54}$ $=$ $Y_{73} Y_{35} Y_{47} Z_{54}$ $=$ \\
& $=$ $Y_{18} Y_{41} Y_{85} Z_{54}$ $=$ $X_{85} Y_{18} Y_{41} C_{54}$ $=$ $W_{54} X_{35} Y_{13} Y_{41}$ $=$ $Y_{13} Y_{35} Y_{41} Z_{54}$ \\ 
\hline
$M_{18}$ & $W_{54} X_{72} X_{47} Y_{25}$ $=$ $W_{54} X_{46} X_{62} Y_{25}$ $=$ $W_{54} X_{46} X_{85} Y_{68}$ $=$ $X_{62} Y_{46} Y_{25} Z_{54}$ $=$ \\
& $=$ $X_{85} Y_{46} Y_{68} Z_{54}$ $=$ $W_{54} X_{13} X_{35} Y_{41}$ $=$ $W_{54} X_{18} X_{85} Y_{41}$ $=$ $X_{13} Y_{35} Y_{41} Z_{54}$ $=$ \\
& $=$ $X_{47} Y_{72} Y_{25} Z_{54}$ $=$ $W_{54} X_{35} X_{47} Y_{73}$ $=$ $X_{47} Y_{73} Y_{35} Z_{54}$ $=$ $X_{85} Y_{18} Y_{41} Z_{54}$ \\ 
\hline
$M_{19}$ & $B_{54} X_{68} Y_{46} Y_{85}$ $=$ $B_{54} X_{25} Y_{46} Y_{62}$ $=$ $B_{54} X_{73} Y_{35} Y_{47}$ $=$ $B_{54} X_{25} Y_{72} Y_{47}$ $=$ \\
& $=$ $B_{54} X_{41} Y_{18} Y_{85}$ $=$ $B_{54} X_{41} Y_{13} Y_{35}$ \\ 
\hline
$M_{20}$ & $X_{68} Y_{46} Y_{54} Y_{85}$ $=$ $X_{25} Y_{46} Y_{54} Y_{62}$ $=$ $B_{54} X_{62} X_{25} Y_{46}$ $=$ $B_{54} X_{68} X_{85} Y_{46}$ $=$ \\
& $=$ $B_{54} X_{13} X_{41} Y_{35}$ $=$ $B_{54} X_{73} X_{47} Y_{35}$ $=$ $X_{73} Y_{35} Y_{47} Y_{54}$ $=$ $B_{54} X_{47} X_{25} Y_{72}$ $=$ \\
& $=$ $X_{25} Y_{72} Y_{47} Y_{54}$ $=$ $X_{41} Y_{18} Y_{54} Y_{85}$ $=$ $B_{54} X_{41} X_{85} Y_{18}$ $=$ $X_{41} Y_{13} Y_{35} Y_{54}$ \\ 
\hline
$M_{21}$ & $X_{62} X_{25} Y_{46} Y_{54}$ $=$ $X_{68} X_{85} Y_{46} Y_{54}$ $=$ $X_{13} X_{41} Y_{35} Y_{54}$ $=$ $X_{73} X_{47} Y_{35} Y_{54}$ $=$ \\
& $=$ $X_{47} X_{25} Y_{72} Y_{54}$ $=$ $X_{41} X_{85} Y_{18} Y_{54}$ \\ 
\hline
$M_{27}$ & $B_{54} Y_{46} Y_{68} Y_{85}$ $=$ $B_{54} Y_{46} Y_{62} Y_{25}$ $=$ $D_{54} X_{68} Y_{46} Y_{85}$ $=$ $D_{54} X_{25} Y_{46} Y_{62}$ $=$ \\
& $=$ $D_{54} X_{73} Y_{35} Y_{47}$ $=$ $B_{54} Y_{72} Y_{47} Y_{25}$ $=$ $D_{54} X_{25} Y_{72} Y_{47}$ $=$ $B_{54} Y_{73} Y_{35} Y_{47}$ $=$ \\
& $=$ $D_{54} X_{41} Y_{18} Y_{85}$ $=$ $B_{54} Y_{18} Y_{41} Y_{85}$ $=$ $D_{54} X_{41} Y_{13} Y_{35}$ $=$ $B_{54} Y_{13} Y_{35} Y_{41}$ \\ 
\hline
$M_{23}$ & $Y_{46} Y_{54} Y_{68} Y_{85}$ $=$ $Y_{46} Y_{54} Y_{62} Y_{25}$ $=$ $W_{54} X_{68} Y_{46} Y_{85}$ $=$ $W_{54} X_{25} Y_{46} Y_{62}$ $=$ \\
& $=$ $B_{54} X_{62} Y_{46} Y_{25}$ $=$ $B_{54} X_{85} Y_{46} Y_{68}$ $=$ $D_{54} X_{62} X_{25} Y_{46}$ $=$ $D_{54} X_{68} X_{85} Y_{46}$ $=$ \\
& $=$ $D_{54} X_{13} X_{41} Y_{35}$ $=$ $D_{54} X_{73} X_{47} Y_{35}$ $=$ $W_{54} X_{73} Y_{35} Y_{47}$ $=$ $B_{54} X_{13} Y_{35} Y_{41}$ $=$ \\
& $=$ $B_{54} X_{47} Y_{72} Y_{25}$ $=$ $D_{54} X_{47} X_{25} Y_{72}$ $=$ $Y_{72} Y_{47} Y_{54} Y_{25}$ $=$ $W_{54} X_{25} Y_{72} Y_{47}$ $=$ \\
& $=$ $B_{54} X_{47} Y_{73} Y_{35}$ $=$ $Y_{73} Y_{35} Y_{47} Y_{54}$ $=$ $W_{54} X_{41} Y_{18} Y_{85}$ $=$ $D_{54} X_{41} X_{85} Y_{18}$ $=$ \\
& $=$ $Y_{18} Y_{41} Y_{54} Y_{85}$ $=$ $B_{54} X_{85} Y_{18} Y_{41}$ $=$ $W_{54} X_{41} Y_{13} Y_{35}$ $=$ $Y_{13} Y_{35} Y_{41} Y_{54}$ \\ 
\hline
$M_{24}$ & $X_{62} Y_{46} Y_{54} Y_{25}$ $=$ $X_{85} Y_{46} Y_{54} Y_{68}$ $=$ $W_{54} X_{62} X_{25} Y_{46}$ $=$ $W_{54} X_{68} X_{85} Y_{46}$ $=$ \\
& $=$ $W_{54} X_{13} X_{41} Y_{35}$ $=$ $W_{54} X_{73} X_{47} Y_{35}$ $=$ $X_{13} Y_{35} Y_{41} Y_{54}$ $=$ $X_{47} Y_{72} Y_{54} Y_{25}$ $=$ \\
& $=$ $W_{54} X_{47} X_{25} Y_{72}$ $=$ $X_{47} Y_{73} Y_{35} Y_{54}$ $=$ $W_{54} X_{41} X_{85} Y_{18}$ $=$ $X_{85} Y_{18} Y_{41} Y_{54}$ \\ 
\hline
$M_{25}$ & $D_{54} Y_{46} Y_{68} Y_{85}$ $=$ $D_{54} Y_{46} Y_{62} Y_{25}$ $=$ $D_{54} Y_{72} Y_{47} Y_{25}$ $=$ $D_{54} Y_{73} Y_{35} Y_{47}$ $=$ \\
& $=$ $D_{54} Y_{18} Y_{41} Y_{85}$ $=$ $D_{54} Y_{13} Y_{35} Y_{41}$\\
\hline
$M_{26}$ & $W_{54} Y_{46} Y_{68} Y_{85}$ $=$ $W_{54} Y_{46} Y_{62} Y_{25}$ $=$ $D_{54} X_{62} Y_{46} Y_{25}$ $=$ $D_{54} X_{85} Y_{46} Y_{68}$ $=$ \\
& $=$ $D_{54} X_{13} Y_{35} Y_{41}$ $=$ $D_{54} X_{47} Y_{72} Y_{25}$ $=$ $W_{54} Y_{72} Y_{47} Y_{25}$ $=$ $D_{54} X_{47} Y_{73} Y_{35}$ $=$ \\
& $=$ $W_{54} Y_{73} Y_{35} Y_{47}$ $=$ $W_{54} Y_{18} Y_{41} Y_{85}$ $=$ $D_{54} X_{85} Y_{18} Y_{41}$ $=$ $W_{54} Y_{13} Y_{35} Y_{41}$ \\ 
\hline
$M_{27}$ & $W_{54} X_{62} Y_{46} Y_{25}$ $=$ $W_{54} X_{85} Y_{46} Y_{68}$ $=$ $W_{54} X_{13} Y_{35} Y_{41}$ $=$ $W_{54} X_{47} Y_{72} Y_{25}$ $=$ \\
& $=$ $W_{54} X_{47} Y_{73} Y_{35}$ $=$ $W_{54} X_{85} Y_{18} Y_{41}$ 
	\end{longtable}
\end{center}

\bibliographystyle{JHEP}

\bibliography{ref}
	
\end{document}